\documentclass[
preprint,
superscriptaddress,
 amsmath,amssymb,
 aps,
prb,
]{revtex4-2}
\usepackage{graphics}
\usepackage{subfig}
\usepackage{dcolumn}
\usepackage{bm}
\DeclareMathAlphabet{\mathpzc}{OT1}{pzc}{m}{it}
\usepackage{fullpage}
\usepackage{array}
\usepackage{booktabs}
\usepackage{multirow}
\usepackage{esint}
\usepackage{color}
\usepackage{graphicx}
\usepackage{epstopdf}
\usepackage{verbatim}
\usepackage{float}
\usepackage{subfig}
\usepackage{fancyvrb}
\usepackage[export]{adjustbox}
\usepackage{svg} 
\usepackage[version=4]{mhchem}
\usepackage{textcomp, gensymb}
\usepackage{tikz}
\usepackage{pgfplots}%
\usepackage{pgfplotstable}
\pgfplotsset{compat = newest}
\usetikzlibrary{shapes.geometric, arrows}
\usepackage{placeins}

\usepackage[colorlinks   = true, urlcolor     = cyan, linkcolor    = blue, citecolor   = red]{hyperref}
\usepackage[capitalize]{cleveref}

\usepackage{xcolor}
\definecolor{mypink}{rgb}{0.96 0.68 .80}
\definecolor{myyellow}{rgb}{1 0.98 0.70}
\definecolor{mycyan}{rgb}{0.80 0.91 0.84}

\def\R{{\mathbb{R}}}

\newcommand{\bipink}[1]{\raisebox{1.5pt}{\fcolorbox{black}{mypink}{\rule[2pt]{6pt}{0pt} }}}
\newcommand{\semicyan}[1]{\raisebox{1.5pt}{\fcolorbox{black}{mycyan}{\rule[2pt]{6pt}{0pt} }}}
\newcommand{\tiyellow}[1]{\raisebox{1.5pt}{\fcolorbox{black}{myyellow}{\rule[2pt]{6pt}{0pt} }}}

\usepackage{gensymb}

\usepackage{scalerel}
\usepackage{tikz}
\usetikzlibrary{svg.path}
\definecolor{orcidlogocol}{HTML}{A6CE39}
\tikzset{
  orcidlogo/.pic={
    \fill[orcidlogocol] svg{M256,128c0,70.7-57.3,128-128,128C57.3,256,0,198.7,0,128C0,57.3,57.3,0,128,0C198.7,0,256,57.3,256,128z};
    \fill[white] svg{M86.3,186.2H70.9V79.1h15.4v48.4V186.2z}
                 svg{M108.9,79.1h41.6c39.6,0,57,28.3,57,53.6c0,27.5-21.5,53.6-56.8,53.6h-41.8V79.1z M124.3,172.4h24.5c34.9,0,42.9-26.5,42.9-39.7c0-21.5-13.7-39.7-43.7-39.7h-23.7V172.4z}
                 svg{M88.7,56.8c0,5.5-4.5,10.1-10.1,10.1c-5.6,0-10.1-4.6-10.1-10.1c0-5.6,4.5-10.1,10.1-10.1C84.2,46.7,88.7,51.3,88.7,56.8z};
  }
}
\newcommand\orcidicon[1]{\href{https://orcid.org/#1}{\mbox{\scalerel*{
\begin{tikzpicture}[yscale=-1,transform shape]
\pic{orcidlogo};
\end{tikzpicture}
}{|}}}}

\newcommand{\bfa}{{\bf a}}

\newcommand{\bfd}{{\bf d}}
\newcommand{\bfe}{{\bf e}}

\newcommand{\bfk}{{\bf k}}

\newcommand{\bfr}{{\bf r}}

\newcommand{\bfu}{{\bf u}}

\newcommand{\beq}{\begin{equation}}
\newcommand{\eeq}{\end{equation}}
\newcommand{\beqs}{\begin{eqnarray}}
\newcommand{\eeqs}{\end{eqnarray}}
\newcommand{\beql}{\begin{equation} \label}

\newcommand{\calO}{{\cal O}}

\newcommand{\Z}{\mathbb{Z}}

\let\oldFootnote\footnote
\newcommand\nextToken\relax

\renewcommand\footnote[1]{%
    \oldFootnote{#1}\futurelet\nextToken\isFootnote}

\newcommand\isFootnote{%
    \ifx\footnote\nextToken\textsuperscript{,}\fi}

\begin{document}

\title{Strain-Tunable Topological Phase Transitions in Line- and Split-Graph Flat-Band Lattices\texorpdfstring{\footnote{Dedicated to Prof. Richard D. James on his $70^{\text{th}}$ birthday.}}{}}
\author{Shivam Sharma \orcidicon{0000-0001-5607-4752}}
\affiliation{Department of Aerospace Engineering and Mechanics, University of Minnesota, Minneapolis, MN 55455}
\author{Amartya S. Banerjee \orcidicon{0000-0001-5916-9167}}
\email{asbanerjee@ucla.edu}
\affiliation{Department of Materials Science and Engineering, University of California, Los Angeles, CA 90095, USA}
\begin{abstract}
In recent years, materials with topological flat bands have attracted significant attention due to their association with extraordinary transport properties and strongly correlated electrons. Yet, generic principles linking lattice architecture, strain, and band topology remain scarce. Here, using a unified graph-theoretic framework we generate entire families of two-dimensional lattices and, using analytical tight-binding calculations, demonstrate that a single mechanical knob --- uniform in-plane strain --- drives universal transitions between trivial insulating, Dirac semimetal, and quantum spin-Hall phases across all lattices. The framework yields several flat band lattices that were hitherto absent or largely unexplored in the literature --- for example, the checkerboard split-graph and triangular-Kagome lattices --- whose strain-driven topological phase diagrams we establish here for the first time. The design rules implied by our studies provide a blueprint for engineering topological states in a wide variety of 2D materials, photonic crystals, and circuit lattices, and are anticipated to accelerate the discovery of strain-programmable quantum matter.
\end{abstract}
\maketitle
\newpage 
\section{Introduction} 
\label{sec:introduction}
The design, synthesis and characterization of low dimensional materials featuring exotic electronic band structure forms a significant theme of contemporary  materials research. Such materials often exhibit remarkable  physical properties, making them attractive for adoption into emergent quantum technologies, spintronic devices, and next-generation microelectronics \cite{tokura2017emergent,de2021materials,giustino20212021,basov2017towards,mas20112d, aiello2022chirality}. A famous example in two-dimensional (2D) materials is graphene, where linearly dispersive Dirac bands lead to unusual electronic \cite{castro2009electronic}, optical \cite{falkovsky2008optical}, transport\cite{peres2010colloquium} and topological properties \cite{gomes2012designer}. Recent developments in materials physics --- e.g.~observation of dispersion-less states in magic angle twisted bilayer graphene (TBLG) \cite{cao2018unconventional,cao2018correlated} --- have underlined the crucial role of flat bands in hosting emergent strongly correlated electronic phenomena such as superconductivity, ferromagnetism, Wigner crystallization and zero-magnetic-field fractional quantum Hall effect \cite{mielke1993ferromagnetism,hase2018possibility,sharpe2019emergent,saito2021hofstadter,aoki2020theoretical,balents2020superconductivity,liu2021spectroscopy,peri2021fragile,wu2007flat,chen2018ferromagnetism,wang2011nearly,regnault2011fractional,liu2012fractional,bergholtz2013topological,sun2011nearly,tang2011high,neupert2011fractional,neupert2012topological}. This has sparked a surge in interest not only in the theoretical explanation of superconductivity in such materials \cite{tarnopolsky2019origin,po2018origin,isobe2018unconventional}, but also in exploring novel flat bands systems in various other platforms. Recently investigated materials include bulk systems  \cite{kang2020topological,liu2020orbital,yin2019negative,kang2020dirac,lin2018flatbands,di2023flat}, twisted bilayer transition metal dichalcogenides \cite{zhang2020flat}, photonic systems \cite{leykam2018artificial,mukherjee2018experimental,PhysRevLett.124.183901,guzman2014experimental,vicencio2015observation,mukherjee2015observation,ma2020direct,milicevic2019type}, quantum circuits \cite{kollar2019hyperbolic} and ultracold atoms \cite{PhysRevLett.108.045305,PhysRevLett.126.103601}. A particularly interesting thread of research has been the exploration of such states in quasi-one-dimensional (1D) materials --- such as collapsed nanotubes \citep{zhou2024pressure, arroyo2023universality, arroyo2020one} or systems with specialized unit cells \citep{yu2024carbon, ShivamP2C3} --- enabled by recently developed symmetry-adapted electronic structure calculation techniques \citep{banerjee2016cyclic, banerjee2021ab, ghosh2019symmetry, yu2022density, pathrudkar2022machine, agarwal2024solution}.

Despite significant interest and recent progress, systematic rules that relate lattice geometry and external perturbations to the emergence of topological phases in flat-band systems remain underdeveloped. In particular, the explicit relationship between strain and topological phase transitions across families of structurally related lattices is largely unexplored. Our study directly addresses this gap by demonstrating that universal topological phase transitions can be induced across entire classes of flat-band lattices (derived using simple graph-theoretic tools), thus highlighting the interplay between lattice geometry, mechanical deformation, and quantum phases. We anticipate that given the substantial interest in strain-programmable quantum matter \citep{kim2023strain,dai2019strain}, the design rules implied by our work will be valuable for engineering topological states in a wide variety of 2D materials, photonic crystals, and circuit lattices. 

Interest in topological tools to classify emergent quantum phases of matter began nearly four decades ago, driven by studies on the quantum Hall effect \cite{klitzing1980new} and polyacetylene \cite{heeger1988solitons, kivelson1982hubbard}. The research in this area increased significantly, when it was realized that spin-orbit coupling (SOC) can lead to  topological insulating electronic phases \cite{kane2005quantum,kane2005z,hasan2010colloquium, bernevig2006quantum} which were subsequently observed in real materials (see \footnote{Topological insulators (TIs) have SOC induced bulk energy gap, which is topologically distinct from ordinary (trivial) insulators. These two phases are distinguished by the $\Z_2$ invariant which is associated with the bulk band structure. An ordinary insulator is a $\Z_2$-even phase and can be adiabatically (continuously) deformed into an atomic insulator. It is characterized by strictly localized orbitals or exponentially localized Wannier functions. In contrast, the ``topological'' state is $\Z_2$-odd and shows obstruction from any Wannier representation \cite{po2018fragile}. The TI phase in 2D materials is also known as a quantum spin hall (QSH) insulator. It features an odd number of Kramers pairs and gapless edge modes which conduct spin-up and spin-down electrons in opposite directions --- a hallmark of topological materials. These spin edge states are protected by time-reversal symmetry and are robust against small perturbations and weak electron-electron interaction \cite{xu2006stability,wu2006helical}. The two phases are separated by a topological phase transition, which is triggered by the closing of the bulk bandgap at a critical point of the system's parameters.} for further details). Crucially, a topological insulator and an ordinary band insulator can be connected only through a topological phase transition (TPT) where the bulk energy gap closes (or the protecting symmetry is broken), allowing the topological invariant to change. TPTs are often studied within the tight-binding (TB) approximation, which replaces the continuum Hamiltonian with a simplified discrete model \footnote{The tight-binding approach describes hopping between localized electronic orbitals bounded to the atomic sites and generally provides accurate solutions for the low-energy part of the spectrum. This makes it convenient to analytically investigate topological phase transitions in crystals including flat bands lattices such as Kagome, Lieb and decorated honeycombs  \cite{sun2009topological,mojarro2023strain,ruegg2010topological,guo2009topological,weeks2010topological}.}. Concurrently, TB models have also been proposed to generate flat band lattices \cite{flach2014detangling,xu2020building}, with graph theory often playing a prominent role in such studies \citep{kollar2020line, ma2020spin, miyahara2005flat, mielke1991ferromagnetic,mielke1991ferromagnetism}. These models typically exhibit topologically trivial flat bands, spanned by the combination of localized states associated with flat bands and delocalized states from dispersive bands  that touch at a high-symmetry point of the Brillouin zone \cite{bergman2008band}. Generally, in flat band systems with trivial topology, the electrons have quenched kinetic energy and become localized, resembling atomic-like orbitals. Based on the Ginzburg-Landau theory \cite{basov2011manifesto}, this localization results in a vanishing superfluid weight, implying the absence of superconductivity. However, for flat bands with  non-trivial topology, the superfluid weight can be preserved  due to imposition of a lower bound on the superfluid density by a non-zero Chern number \cite{peotta2015superfluidity}, thereby enabling dissipation-less transport. Such topological non-triviality can be introduced by means of SOC in the TB model \cite{kane2005quantum}, whereby the degenerate flat bands can be transformed into isolated quasi-flat bands \footnote{More exotic types of quantum geometrical contributions (i.e.~fragile topology) to the superfluid weight has been shown for some systems \cite{xie2020topology,song2019all,po2019faithful,ahn2019failure,julku2020superfluid} and also explored experimentally \cite{tian2021evidence}. This underscores the critical role of both band topology and strong electronic correlation for realizing novel quantum phenomena \cite{kopnin2011high,lau2021designing}.}.  At the same time, since uniform in‑plane strain directly modulates the nearest‑neighbour hopping amplitudes --- and hence the Berry curvature and quantum‑geometric tensor that bound the superfluid weight --- it offers a natural, experimentally accessible handle to endow intrinsically flat bands with topological character, capable of supporting coherent transport \citep{Chenhaoyue_Transport_1}. This provides the motivation to investigate the competing roles of SOC and strain in a wide variety of flat band lattices, as done here.

In this paper, starting from bipartite parent (root) graphs, we utilize unifying graph theoretic tools to systematically generate a variety of complex 2D Euclidean lattices with flat bands. The number of times line and split graph operations are applied to the parent graph determines the generation number. For instance, first-generation lattices are obtained directly as line and split graphs of the parent, while second-generation lattices include structures like the line graph of a split graph. These generations inherit all the lattice symmetries and band structure features of their predecessors, such as flat bands with quadratic and Dirac band crossings, and isolated Dirac cones, while also adding new bands with interesting characteristics. Consequently, to keep the number of lattices manageable, our study focuses till second-generation graphs and includes lattices with interesting combinations of electronic structure features. While some the structures studied here are very well known in the literature (e.g. Kagome lattice), others presented are understudied (e.g. triangular Kagome), and some are completely new (e.g. checkerboard splitgraph), and our work establishes strain-driven topological phase diagrams for such cases for the first time. 

We investigate, at the TB level, the role of the system's parameters such as inter-atomic hopping amplitude, SOC, on-site potential and strain on TPTs. Strain alters bond lengths and angles, and can effectively change the band structure, thus providing a fine control over quantum phases. Strain engineering has been of particular interest in material science to tune the electronic, magnetic and topological properties of materials \citep{kim2023strain, dai2019strain}.  Additionally, in some cases, external strain (e.g. applied through substrates) has been shown to be  crucial in imparting thermodynamic stability to low dimensional materials \citep{borlido2019structural, abidi2024gentle, chen2020strain, ruf2021strain}.  Here, we systematically analyze the effect of strain on the topological features of energy bands of various lattices generated through our graph-theoretic framework. Although our attention is largely on states close to the Fermi level,  we also discuss electronic bands away from the Fermi level, since these can be relevant in many experimental scenarios. Such bands can be accessed through methods such as  electrostatic gating, non-equilibrium photo-excitation, electro-chemical doping, Floquet engineering, and nonlinear optical experiments.

In addition to its comprehensive scope, the novelty of the present contribution lies in the systematic construction and unified analysis of previously understudied and entirely new flat-band lattices which have not been investigated in the context of topological phase transitions. By elucidating how uniform in-plane strain universally controls the topological character of electronic states across all the lattices, we establish new, broadly applicable design principles that go beyond the isolated examples commonly discussed in earlier literature. These insights have implications not only for electronic condensed matter systems but also extend to photonic, cold-atomic, and circuit-based analogs, thereby considerably expanding the potential impact of our findings.

The rest of the paper is organized as follows. In section \ref{sec:overview}, we will give an overview of graph theoretic  tools used to generate various 2D lattices, the tight binding model as well as the topological and electronic phases observed in these structures. Section \ref{sec:results} presents the results, touching on topological and quantum phase transitions observed in first and second generation of root (parent) graph due external perturbations like strain. We conclude in section \ref{sec:conclusion}.
\section{Preliminaries}
\label{sec:overview}
In this section, we provide a general introduction to the graph theoretic results and tools used to generate different structures from primitive bipartite lattices. More details can be found in standard textbooks on graph theory and in papers \cite{doob1980spectra,kollar2020line,desai1994characterization,ma2020spin}. We also briefly describe the tight-binding model which serves as the framework for investigating the electronic and topological properties of these lattices. First, we layout the notation used throughout  the paper.

In what follows, the lattice $\mathcal{L}\subset \R^2$ is the discrete group of translations which contains the set of periodically arranged lattice points $\mathcal{P}$. We consider $\mathcal{L}$ as a $d$-regular Euclidean lattice (i.e., lattices defined on a Euclidean plane) with $d\geq 3$, where $d$ is the coordination number. The lattice vectors are denoted as $\bfa_1 = a[1,0]$ and $\bfa_2 = a[\cos{\theta},\sin{\theta}]$, where $\theta = \frac{\pi}{2}$ and $\theta = \frac{\pi}{3}$ for the structures with square and hexagonal geometries, respectively. The lattice constant $a$ is set to 1 for all structures considered here. The Pauli matrices denoted as $\boldsymbol{\sigma} = \{\sigma_x,\sigma_y,\sigma_z\}$ span the spin space. For the orbital space, we use $\boldsymbol{\tau} \text{ and }\boldsymbol{\gamma}$, with identity matrices $\sigma_0,\tau_0 \text{ and } \gamma_0$.  We denote the standard orthonormal basis of $\R^3$ as $\bfe_1,\bfe_2,\bfe_3$. Additionally, the matrices $S_{ij}= \bfe_i\otimes\bfe_j$ are  used for constructing tight binding Hamiltonian in orbital space.

\subsection{Graphs theoretic tools and tight binding model}
\begin{figure}[!htbp]
    \centering
    \subfloat[]{\includegraphics[scale =0.25,trim={0cm 0cm 0cm 0.cm},clip]{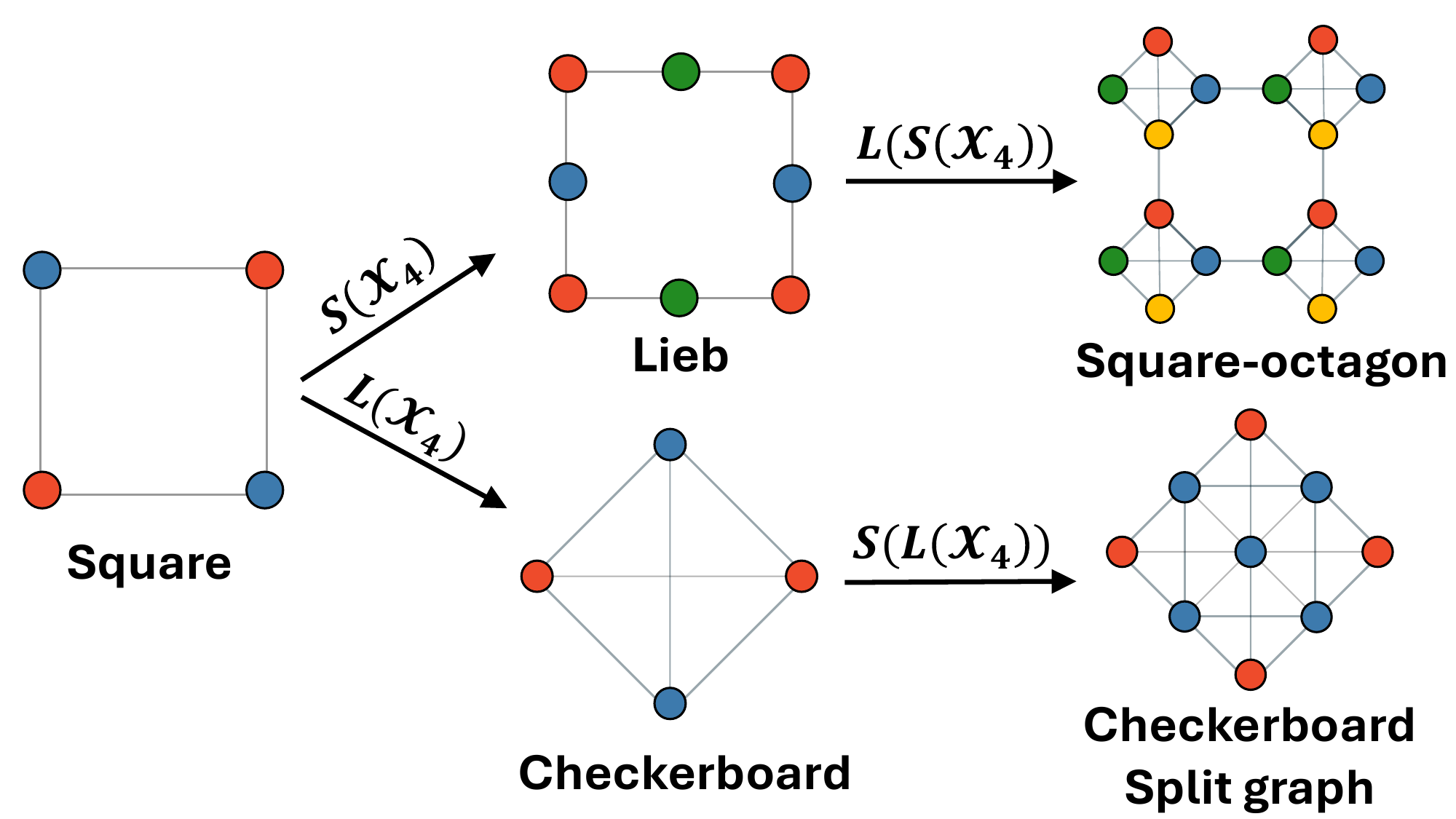}\label{fig:square_lattice_rep}}
     \subfloat[]{\includegraphics[scale =0.15,trim={0cm 0cm 0cm 0.cm},clip]{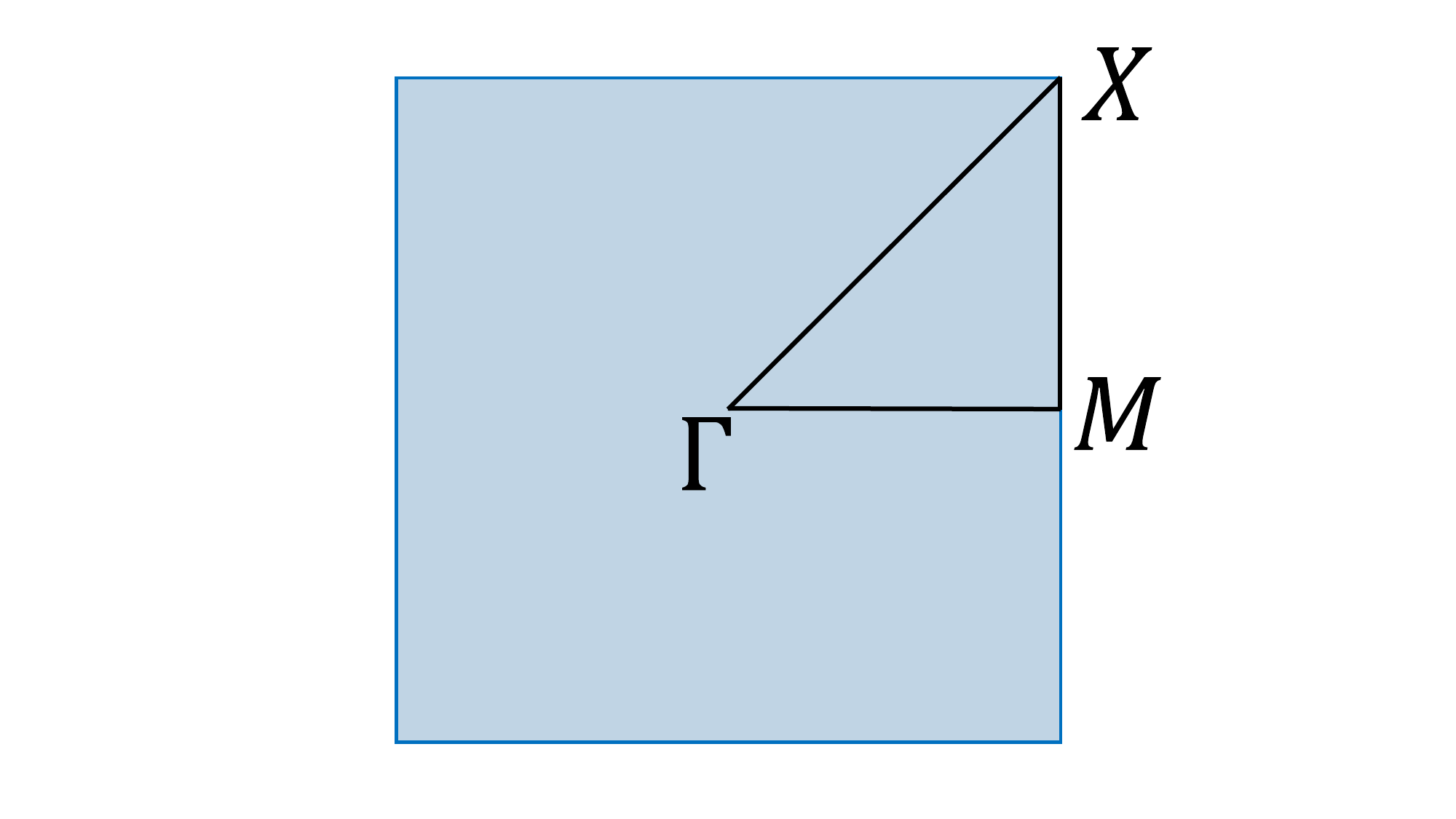}\label{fig:BZ_square}} \\
     \subfloat[]{\includegraphics[scale =0.25,trim={0cm 0cm 0cm 0.cm},clip]{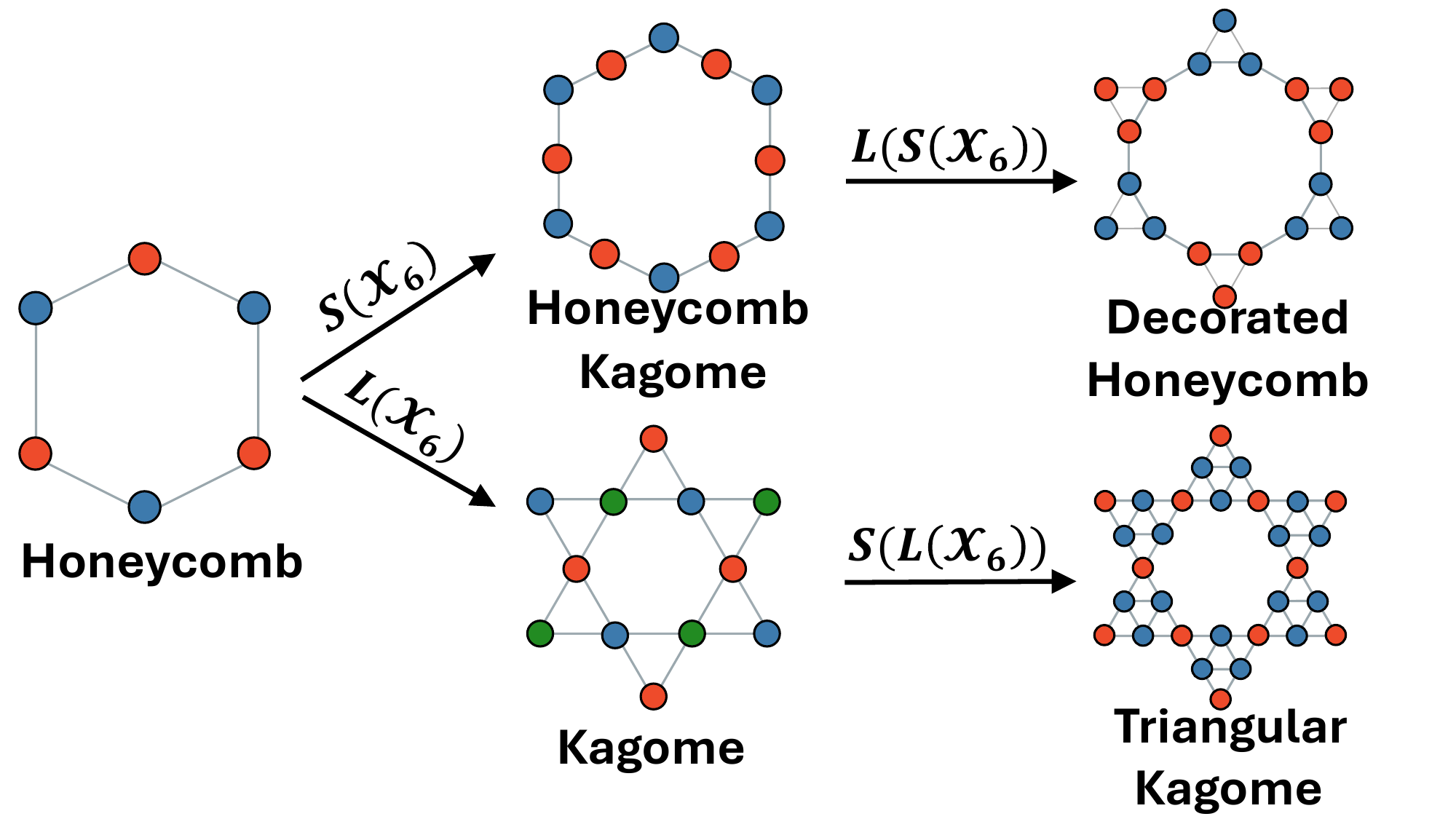}\label{fig:hexagonal_lattice_rep}} 
    \subfloat[]{\includegraphics[scale =0.15,trim={0cm 0cm 0cm 0.cm},clip]{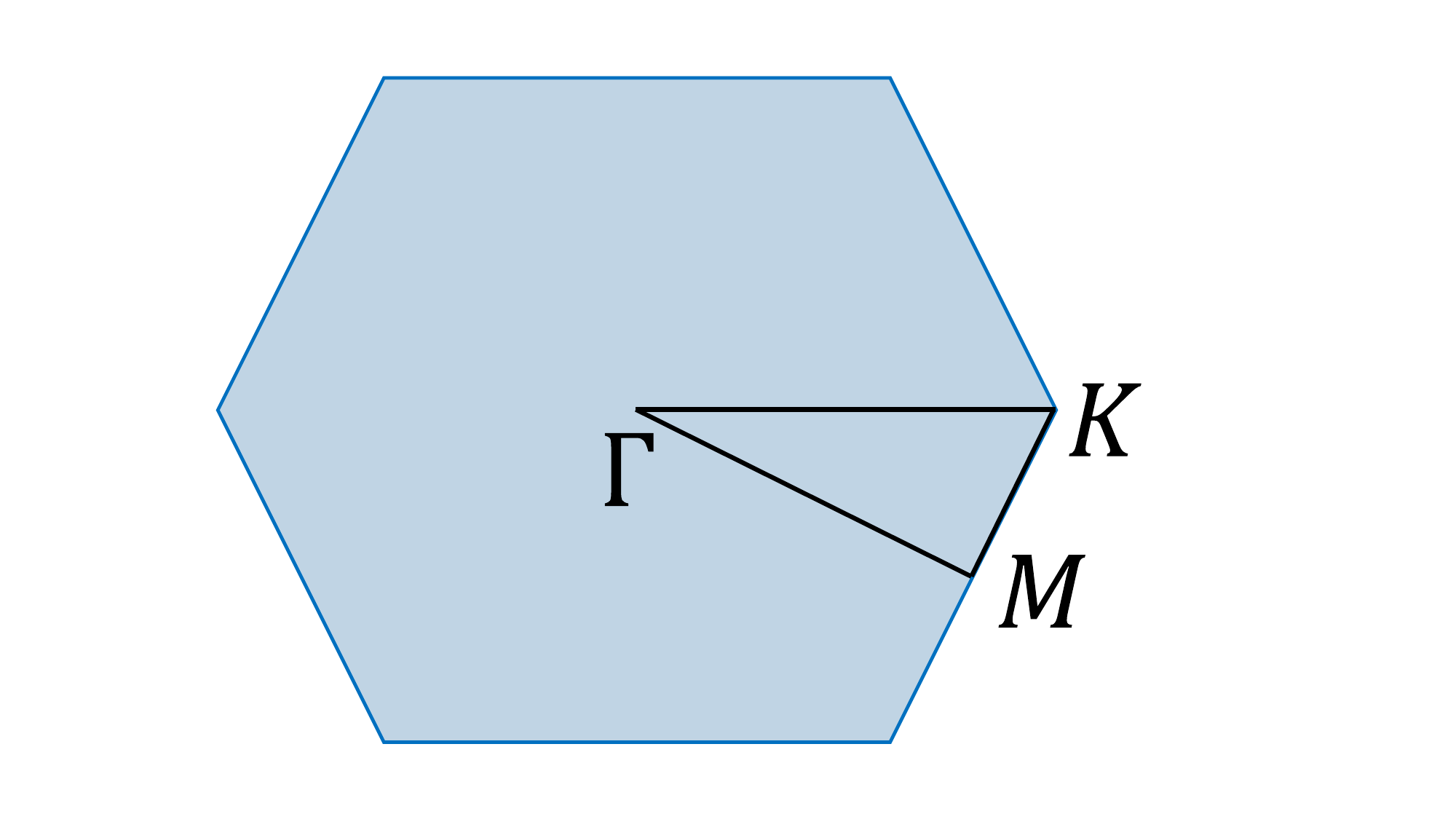}\label{fig:BZ_hexagonal}}
\caption{Schematic showing the application of split graph and line graph operations to generate different type of lattices from bipartite lattice, square $\mathcal{X}_4$ and honeycomb $\mathcal{X}_6$. From left to right, the columns of the image show the root, first and second generation lattices respectively. (a) Starting from the parent (root) bipartite square lattice $\mathcal{X}_4$, the splitgraph $S(\mathcal{X}_4)$ operation gives a  commonly known Lieb lattice and linegraph $L(\mathcal{X}_4)$ gives checkerboard lattice (second column). Further applying linegraph function on Lieb lattice gives square-octagon lattice $L(S(\mathcal{X}_4))$ and splitgraph of checkerboard lattice gives $S(L(\mathcal{X}_4))$. (b) First Brillouin zone of the square lattices, where $\Gamma$, $M$ and $X$ denotes the symmetry points (c) Applying similar procedure, honeycomb-kagome $S(\mathcal{X}_6)$, kagome $L(\mathcal{X}_6)$, decorated honeycomb $L(S(\mathcal{X}_6))$ and triangular kagome $S(L(\mathcal{X}_6))$ lattices can be obtained from the parent honeycomb lattice. (d) First Brillouin zone of the hexagonal lattices, where $\Gamma$, $M$ and $K$ denotes the symmetry points. }
\label{fig:rep_sq_hx}
\end{figure}

For each lattice $\mathcal{L}$ we will associate a graph $\mathcal{X} = (\mathcal{V},\mathcal{E})$, where the set $\mathcal{V}(\mathcal{X})$ contains exactly one vertex related to each of the lattice points in $\mathcal{P}$ and $\mathcal{E}(\mathcal{X})$ is the set containing the edges connecting the nearest neighbor. If $\mathcal{X}$ is bipartite, then $\mathcal{V}(\mathcal{X})$  can be decomposed into two disjoints and independent sets $X$ and $Y$ such that every edge in $\mathcal{E}(\mathcal{X})$ always connects a vertex in $X$ to a vertex in $Y$. The prototypical examples of this in Euclidean lattices are square and hexagonal honeycomb lattices; we will refer to them as parent (root) graph lattices $\mathcal{X}$, where blue and red color atoms can be considered in sets $X$ and $Y$, respectively (illustrated in the first column of Fig.~\ref{fig:rep_sq_hx}). 

A line graph $L(\mathcal{X})$ can be formed by placing a vertex $v_i\in \mathcal{V}(L(\mathcal{X}))$ at the midpoint of an edge $e_i \in \mathcal{E}(\mathcal{X})$ and connecting vertices $v_i$ and $v_j$ for adjacent edges $e_i$ and $e_j$ in $\mathcal{X}$. The new vertices are the new atomic sites and equal to the set of edges of $\mathcal{X}$, $\mathcal{V}(L(\mathcal{X})) = \mathcal{E}(\mathcal{X})$. For example, the line graph of square and honeycomb lattice are checkerboard and kagome lattice, respectively, shown in the second column of the schematic in Fig.~\ref{fig:square_lattice_rep} \& \ref{fig:hexagonal_lattice_rep} . The split graph $S(\mathcal{X})$ is generated by adding an extra vertex on each edge $e_i \in \mathcal{E}(\mathcal{X})$, as depicted as Lieb and honeycomb-kagome latices in Fig.~\ref{fig:square_lattice_rep} \& \ref{fig:hexagonal_lattice_rep}. We call these graphs the first generation of $\mathcal{X}$. Further generations can simply be build by the application of line- and split-graph operations on the previous generation. These subsequent lattices inherit all the periodicity and symmetry properties of their respective parent graphs; e.g. all square and hexagonal lattices possess four fold ($C_4$) and six fold ($C_6$) rotational symmetry. In this paper, we will restrict ourselves till the second generation graphs, as shown in the third column of Fig.~\ref{fig:rep_sq_hx}.

Fermionic physics on the 2D graph lattices, generated from the line and split graph operations  discussed above, can be simply described using a single-orbital per site tight binding (TB) model. The general TB Hamiltonian $H_{\mathcal{X}}$ is given by:
\beq
H_{\mathcal{X}} = \sum_{i,\sigma}\varepsilon_{i}c^\dagger_{i,\sigma} c_{i,\sigma} + \sum_{\langle i,j\rangle,\sigma}t_{ij}c^\dagger_{i,\sigma}c_{j,\sigma}+i \lambda_I \sum_{\langle\langle i,j\rangle\rangle,\alpha\beta} c^\dagger_{i,\alpha} (\bfe_{ij}\cdot \boldsymbol{\sigma})_{\alpha\beta}c_{j,\beta} + \text{H.c.},
\label{eq:hamiltonian}
\eeq
where, $\varepsilon_i$ is the onsite energy of vertex $v_i$,  $c^\dagger_{i,\sigma} \text{ and }c_{j,\sigma}$ are the fermionic creation and annihilation operators with spin $\sigma \in \{\uparrow, \downarrow\}$, $t_{ij}$ is the hopping amplitude between adjacent vertices $v_i$ and $v_j$, $\langle i,j\rangle$ and $\langle\langle i,j\rangle\rangle$ are the nearest-neighbor (NN) and next nearest neighbor (NNN) pairs, respectively.  The third term describes the intrinsic spin-orbit coupling (SOC) between the NNN sites whose relative position is described by the unit vector $\bfe_{ij} = \frac{\bfd^1_{kj}\times \bfd^2_{ik}}{|\bfd^1_{kj}\times \bfd^2_{ik}|}$. Here the bond vector $\bfd_{kj}^1$ points from the vertex $v_j$ to the nearest vertex $v_k$ and the second bond vector $\bfd_{ik}^2$ directs from the vertex $v_k$ to the closest vertex $v_i$. Additionally, $\alpha$ and $\beta$ denote the fermionic spin as $\sigma$ and $\lambda_I$ is the intrinsic SOC parameter. The intrinsic SOC preserves the $s_z$ spin-symmetry but it can uplift the degeneracy between the bands, driving the system into a quantum spin Hall state \cite{kane2005quantum}. The SOC interaction will be considered between the second nearest neighbors throughout the paper, except in the case of the triangular-kagome lattice (Section \ref{subsec:SLX}), where the lack of local inversion symmetry dictates that the nearest neighbor SOC terms be considered as well.

Exploiting the translation symmetry of the 2D lattices, the total Hamiltonian in the momentum space is diagonalized as $H = \sum_{\bfk\sigma}\Psi^\dagger_{\bfk\sigma} \tilde{H}_{\bfk\sigma}\Psi_{\bfk\sigma}$, where $\Psi^\dagger_{\bfk\sigma} = (c^\dagger_{1\bfk\sigma},c^\dagger_{2\bfk\sigma},\dots,c^\dagger_{n\bfk\sigma})$ is the basis representing the number of sites in the unit cell,  $\tilde{H}_{\bfk\sigma}$ is the Hamiltonian in the Fourier or reciprocal space. Here, $\bfk$ is restricted to the fundamental domain in reciprocal space, i.e., the first 2D Brillouin zone (BZ) shown in Fig.~\ref{fig:BZ_hexagonal} \& \ref{fig:BZ_square}.

For every $d-$regular Euclidean lattice with equal hopping amplitude $t_{ij}$ between adjacent vertices, there exists a one-to-one correspondence with any mathematical graph $\mathcal{X}$. This implies that the hopping Hamiltonian (second term in eq.~\ref{eq:hamiltonian}) can also be expressed in terms of the adjacency (transition) operator of the $\mathcal{X}$:
\beq
H_{TB} = tA_\mathcal{X}.
\eeq
Here, the adjacency operator $A_\mathcal{X}=MM^\dagger-D_\mathcal{X}$, with $D_\mathcal{X}$ being the coordination matrix, that has the form $D_{\mathcal{X}}=d\,\mathbb{I}_m$. Furthermore, $M$ is the $m \times n$ dimensional incidence operator of the parent graph $\mathcal{X}$ with $m$ vertices and and $n$ edges, respectively. The entries of $M$ are $1$ when edges and vertices are incident, otherwise $0$. Since the graphs considered here are periodic, we can exploit the Bloch theorem to write $M$ in momentum space \citep{kollar2020line}. For example, for the bipartite honeycomb graph ($\mathcal{X}_6$) $D_{\mathcal{X}_6} = 3\,\mathbb{I}_2$ and the matrix $M$ is $2 \times 3$, since the unit cell has $2$ independent vertices and $3$ edges. The line graph, $L(\mathcal{X})$, is a $2d-2>3$ regular Euclidean lattice and satisfies the relation $M^\dagger M = A_{L(\mathcal{X})}+2\mathbb{I}_m$. The split graph $S(\mathcal{X})$ is a $(d,2)-$ biregular graph (i.e., some vertices has coordination number $d$ while others have coordination number $2$). Consequently, the line graph $L(S(\mathcal{X}))$ is also a $d-$regular Euclidean graph. The adjacency operators for both these generations of graphs can be written in terms of incidence matrix of the parent graph:
\beqs
A_{S(\mathcal{X})} &=& \begin{pmatrix}
    0 & M  \\
    M^\dagger & 0
\end{pmatrix}, \nonumber\\
A_{L(S(\mathcal{X}))} &=& \begin{pmatrix}
    M(\frac{M^\dagger+N^\dagger}{2}) -\mathbb{I}_m & \mathcal{D}\\
    \mathcal{D}^\dagger & M(\frac{M^\dagger+N^\dagger}{2}) -\mathbb{I}_m
\end{pmatrix}.
\eeqs
Above, $\mathbb{I}_l$ is the $l\times l$  identity operator, $\mathcal{D}$ is the diagonal matrix of $\mathcal{X}$ in momentum space and $N$ is $m \times n$ directed incidence matrix of parent graph $\mathcal{X}$:
\beq
N_{ij} = \begin{cases}
    1, \quad \text{if} \; e_j \; \text{enters}\; v_i,\\
    -1, \;  \text{if} \;  e_j\; \text{leaves}\; v_i,\\
    0 \quad \; \text{otherwise}.
\end{cases}
\eeq
 From the above relations and given the spectrum $E_\mathcal{X}(\bfk)$ of the parent graph, we may determine the spectrum of the Hamiltonians of $L(\mathcal{X})$, $S(\mathcal{X})$ and $L(S(\mathcal{X}))$.  In reciprocal space, these are denoted as $E_{L(\mathcal{X})}, E_{S(\mathcal{X})}$ and $E_{L(S(\mathcal{X}))}$, respectively, and using the results stated in ref. \cite{doob1980spectra},  they follow the following relations in reciprocal space:
\beqs
E_{L(\mathcal{X})}(\bfk) &=& \{2\}^{\infty}\cup\{d-2+E_{\mathcal{X}}(\bfk)\} \nonumber\\
E_{S(\mathcal{X})}(\bfk) &=& \begin{cases}
    \pm \sqrt{E_{\mathcal{X}}(\bfk)+d}\\
    0,
\end{cases} \nonumber\\
E_{L(S(\mathcal{X}))}(\bfk) &=& \begin{cases}
    2,\\
    \frac{1\pm\sqrt{1+4(E_{\mathcal{X}}(\bfk)+d)}}{2}\;, \\
    0.
\end{cases}
\label{eq:spectrum}
\eeqs
 The eigenvalues $2$ and $0$ have infinite multiplicity and give rise to flat bands consisting of localized eigenstate with compact support, arising from the destructive interference of hopping amplitudes \cite{bergman2008band}. Since $\mathcal{X}$ is a bipartite lattice, these flat bands are gapless with the dispersive bands touching at high symmetry quasi-momentum points. By introducing SOC the gap can be opened, which may induce $\Z_2$ topology. The dispersion-less bands can be naturally gapped \cite{kollar2020line,PhysRevResearch.2.043414} if and only if $\mathcal{X}$ is a non-bipartite graph. Additionally, the gapped flat bands in non-bipartite lattices can exhibit fragile topology without SOC \cite{PhysRevResearch.2.043414,peri2021fragile,sethi2024graph}. 

\subsection{Modeling strain and capturing topological phase diagrams}
While the graph-theoretic models of lattices featuring equal hopping amplitudes and zero on-site energies are useful, realistic materials (even with with lattice geometries shown in Fig.~\ref{fig:rep_sq_hx}) can deviate substantially from them. Thus, the ideal graph theoretic spectrum discussed above may not be realized in real materials due to specific chemical characteristics such as orbital hybridization (related to the hopping parameters) and different on-site energies. In this paper, we use different values of hopping parameters, different on-site energies, include SOC and investigate the effects of applied strain. Together, these have the effect of reducing the unnatural symmetries associated with the graph theoretic models, originating from their assumption of uniform parameter values. Indeed, different on-site energies and applied deformations can break various lattice symmetries and  distort symmetry protected eigenstates which in turn, can lead to fascinating quantum phase transitions. A well known example related to this is in Kagome lattices, where the flat band with quadratic band touching can evolve into tilted Dirac cones when the six fold symmetry of the lattice is broken \cite{lim2020dirac,jiang2019topological,yu2024carbon,mojarro2023strain,montambaux2018winding,sun2009topological,rhim2019classification}. 

In-plane strain, which can have particularly pronounced effects on the stability and electronic properties of 2D materials \citep{dai2019strain, kim2023strain, peng2020strain, roldan2015strain}, can be incorporated by applying the displacement field $\bfu(\bfr) = (u_{x}(\bfr),u_{y}(\bfr))$ to the atomic position vectors $\bfr$. Due to the applied strains, the bond lengths change and the new lattice sites after deformation are at $\bfr^\prime = \bfr + \bfu(\bfr)$. For uniform strain in the linear regime, the displacement field can be written as $\bfu(\bfr) = \hat{\epsilon} \cdot \bfr$. Here $\hat{\epsilon}$ is the strain tensor:
\begin{align}
\hat{\epsilon} = \epsilon\begin{pmatrix}
    \cos^2\varphi -\nu\sin^2\varphi && (1+\nu)\cos\varphi \sin\varphi \\
    (1+\nu)\cos\varphi \sin\varphi && \sin^2\varphi - \nu \cos^2\varphi
\end{pmatrix}\;.
\end{align}
Here, $\nu$ is the Poisson's ratio, $\epsilon$ is the strain magnitude and $\varphi$ is the direction of applied strain with respect to the $x$-axis. The vectors joining adjacent vertices labeled $i,j$ transform as $\tilde{\bfd}_{ij} = (\mathbb{I}_2+\hat{\epsilon})\cdot\bfd_{ij}$. The hopping amplitude, which depends on the edge length, changes according to:
\beq
\tilde{t}_{ij} = t_{ij} \exp\left[-\beta \left(\frac{|\tilde{\bfd}_{ij}|}{|\bfd_{ij}|}-1\right)\right],
\eeq
where, $i$ and $j$ denoted the vertices and $\beta$ is the Gr\"{u}neisen parameter. For the purpose of the paper, $\beta$ is set to $3$ and $v$ is equal to $0.165$ \cite{mohiuddin2009uniaxial}. These parameters are well known for graphene, but, the corresponding values have not yet been reported for most of the lattices considered here. Keeping with standard practice, we choose the same parameter values as graphene \cite{mojarro2023strain}, for these lattices. 
\begin{figure}[!htbp]
    \centering
    \subfloat[Topological phase diagram]{\includegraphics[scale =0.4,trim={9cm 5.5cm 9cm 4cm},clip]{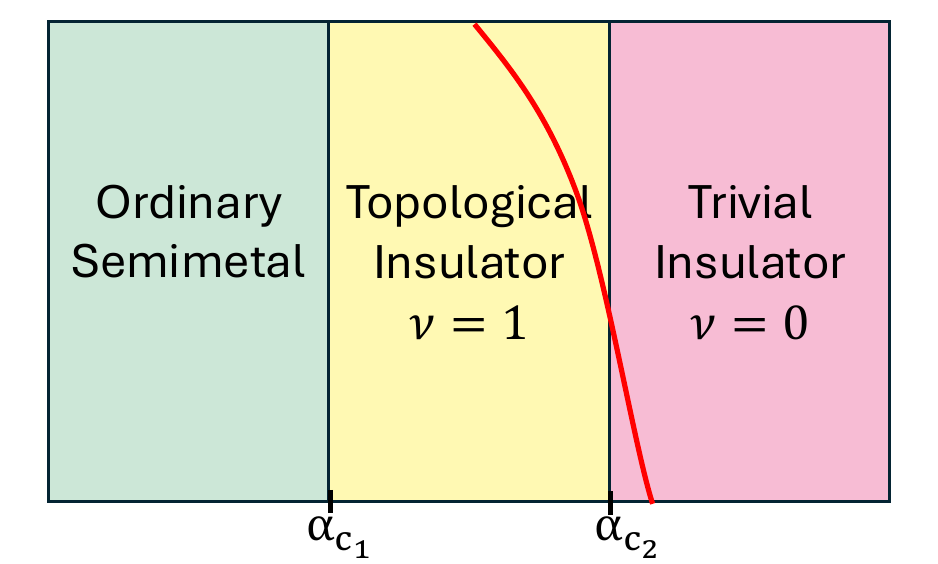}}
     \subfloat[Ordinary semimetal]{\includegraphics[scale =0.4,trim={9.5cm 5.5cm 9.5cm 4.5cm},clip]{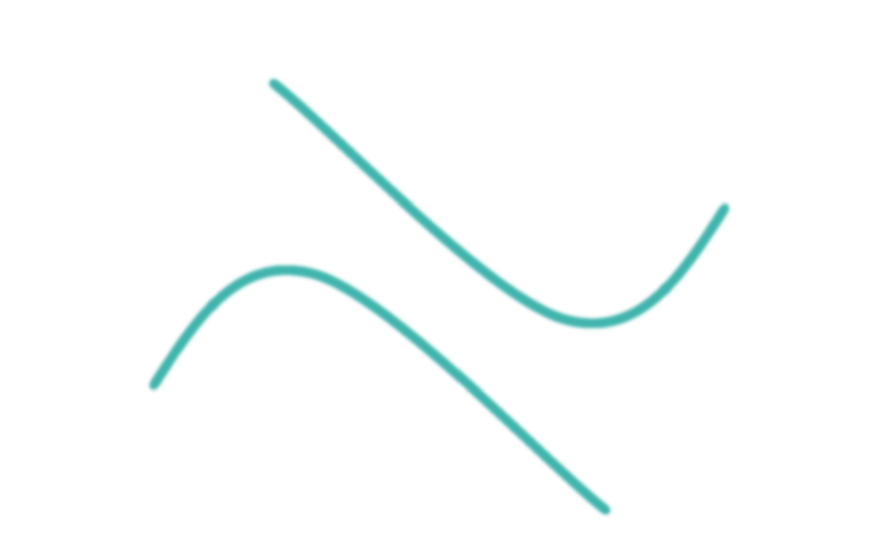}\label{fig:OSM}} \\
    \subfloat[Dirac semimetal]{\includegraphics[scale =0.22,trim={4cm 5.5cm 4cm 4.5cm},clip]{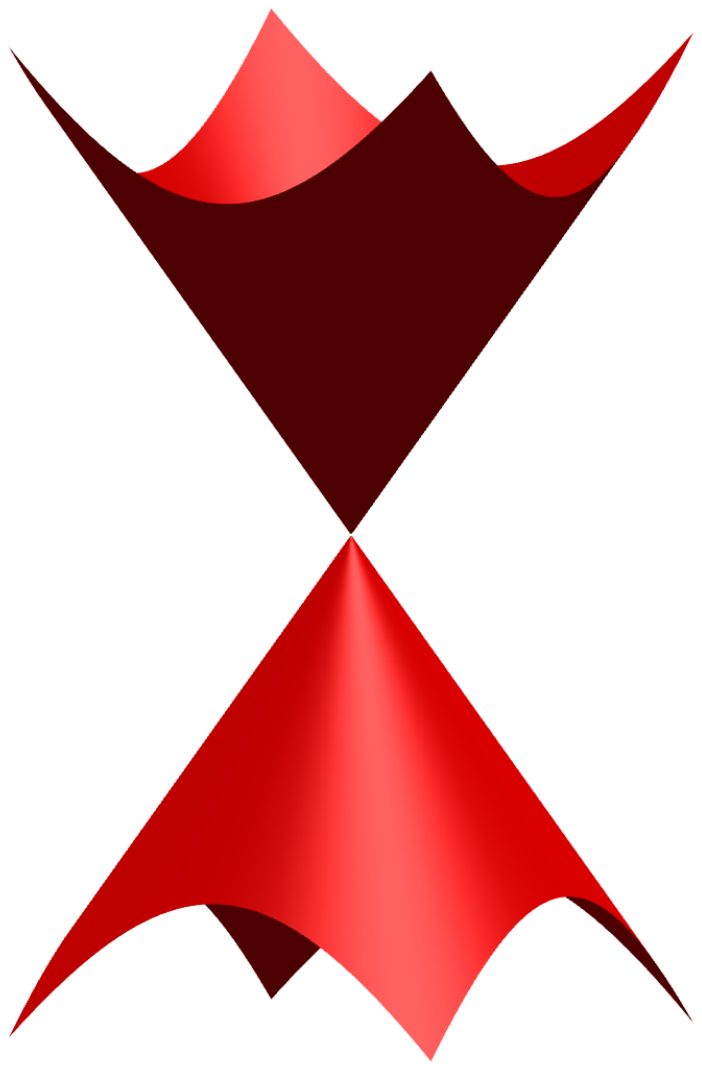}\label{fig:DSM}}
    \subfloat[Semi-Dirac bands]{\includegraphics[scale =0.25,trim={1cm 7cm 1cm 6cm},clip]{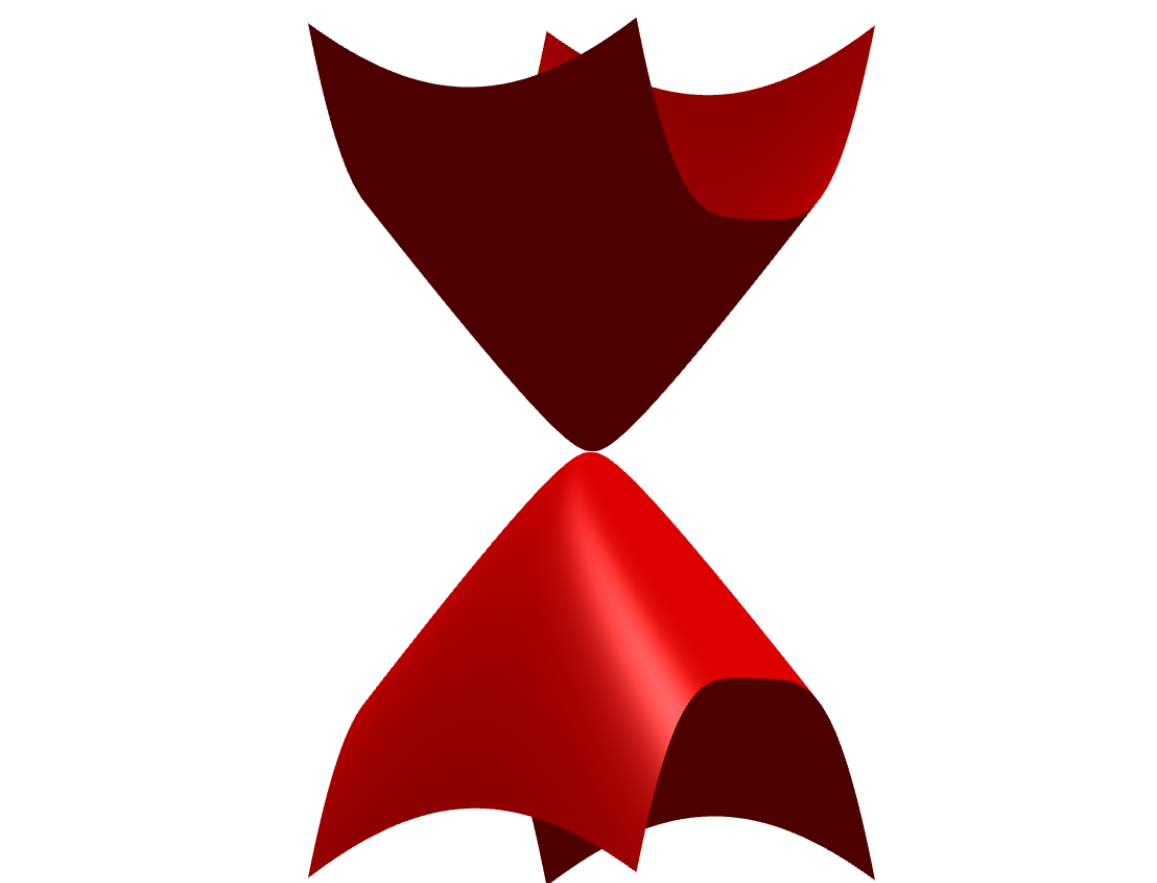}\label{fig:semi_diracbnd}}
     \subfloat[Titled Dirac cones]{\includegraphics[scale =0.25,trim={1cm 8cm 1cm 7cm},clip]{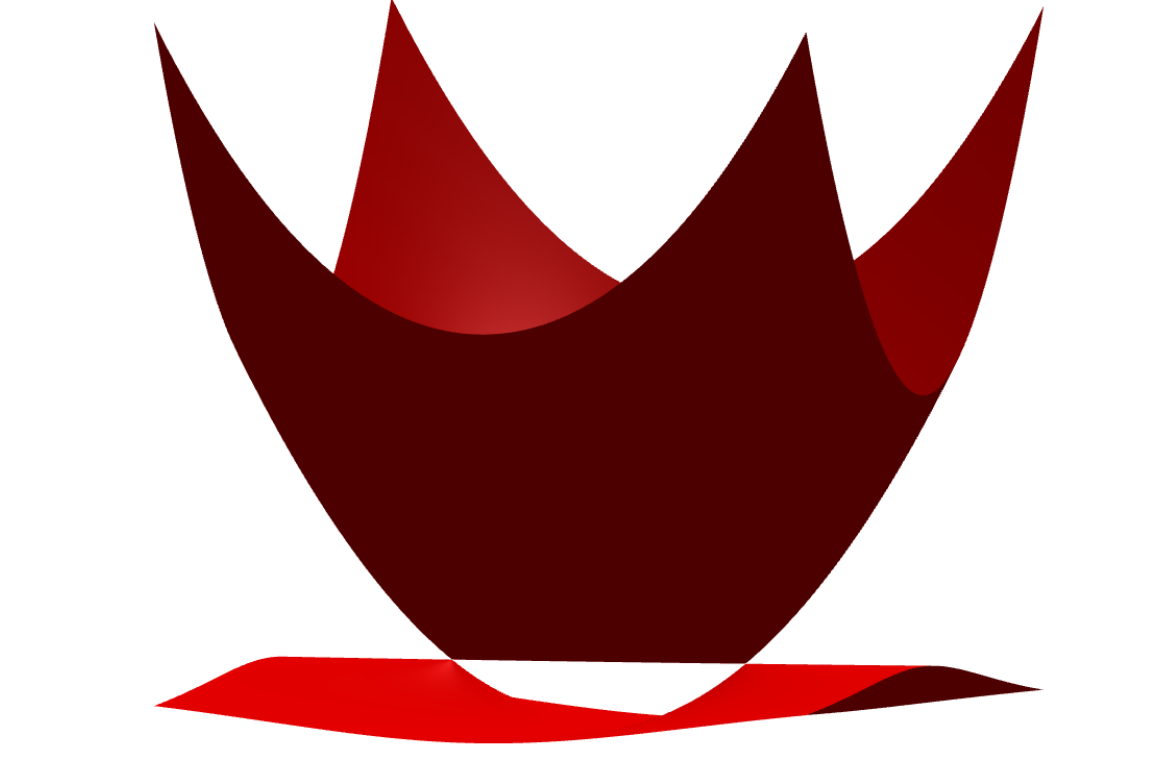}\label{fig:tilted_diracbnd}}
    \caption{(a) Schematic of a phase diagram in parametric space showing phases that can appear in different lattices generated using line graph $L(\mathcal{X})$ and split $S(\mathcal{X})$ graph operations. The red line denotes the Dirac semimetallic state, $\nu$ is the $\Z_2$ index, and $\alpha_{c_1}, \alpha_{c_2}$ are the critical points (strain or system parameters) where the bandgap vanishes.  Note that not all lattices show the phases highlighted here, and the order of appearance of the phases can also be different from the depiction above. Illustration of (b) ordinary semimetal, (c) Dirac semimetal, (d) semi-Dirac bands and (e) tilted Dirac cones.}
    \label{fig:topo_classication}
\end{figure}

In many lattices (including almost all the ones considered here), on introducing intrinsic SOC, the gap between the bulk energy bands can open up, and  topologically protected states manifest at the edges of quasi-1D nanoribbons. To visualize such edge states, we considered our tight binding model on nanoribbons and calculated the 1D spectrum, which can clearly reveal the crossing of bands in the bulk gap. The topological state of the system can also be characterized by calculating the topological index $\Z_2$, denoted as $\nu$, which can be obtained by tracking the evolution of Wannier charge centers (we used the Z2Pack code \cite{soluyanov2011computing,gresch2017z2pack} for this). When $\nu = 1$, the lattice supports the topological edge states in the bulk gap, whereas, afort $\nu = 0$, the lattice will be a trivial insulator.

Apart from this, the 2D lattices can exhibit diverse electronic phases, such as 2D Dirac semimetal (DSM), semi and tilted Dirac bands,  and ordinary semimetal (OSM). The DSM phase is associated with massless relativistic fermions, where the conduction and valence bands touch locally in a conical manner with energy dispersion $E_{\pm}(\bfk)\propto \pm |\bfk|$, with $\bfk$ denoting the quasi-momentum (Fig.~\ref{fig:DSM}). These bands carry a Berry (winding) phase of $\pm \pi$ that keeps the Dirac nodes locally stable \cite{PhysRevLett.89.077002,bernevig2013topological,burkov2016topological}. The semi-Dirac bands features hybrid dispersion, being linear in one direction and quadratic in other, as represented in Fig.~\ref{fig:semi_diracbnd}. On the other hand, the tilted Dirac bands shown in Fig.~\ref{fig:tilted_diracbnd} consist of a tilted cone characterized by massless fermions, accompanied by directional flat bands where electrons have very high effective mass. Finally, in the OSM phase, the valence and the conduction bands cross the Fermi level without overlapping each other, maintaining distinct energies across all quasi-momentum space as illustrated in Fig.~\ref{fig:OSM}. In our calculations, the above specific features were used to characterize and identify the specific electronic phases obtained, as the model parameters were varied.

The filling fraction denotes how many bands are fully occupied in the system. By tuning the magnitude and direction of the applied strain along with the system parameters (such as hopping amplitudes, on-site energies, and intrinsic SOC), the material can be made to undergo TPTs at different filling fractions. Specifically, we observed 2D (Dirac) semimetallic \cite{ahn2017unconventional}, topological nontrivial ($\nu = 1$)  or trivial ($\nu = 0$) phases (not all lattices show all phases or in the same order) in the lattices as illustrated by a phase diagram shown in Fig.~\ref{fig:topo_classication}. In the next section, we systematically analyze the effects of strain and the variations in a system's parameters, on the band structure and  and topological phases in  each of the lattices shown in Fig.~\ref{fig:rep_sq_hx}.

\section{Results}
\label{sec:results}
In this section, we discuss the effect of strain and system parameters on electronic and topological properties of the lattices shown in Fig.~\ref{fig:rep_sq_hx}.  We start with the root graphs --- square and honeycomb lattices --- where the effect of deformation on quantum properties is negligible owing to the simplicity of the lattice. We then discuss first generation lattices, i.e., split graphs (Lieb and honeycomb split graph lattices) and line graphs (checkerboard and kagome lattices). Finally, we focus on second generation lattices, namely line graphs of split graphs (square-octagon, decorated honeycomb lattices) and split graphs of line graphs (split graph of checkerboard and triangular-kagome). Altogether, $10$ lattices over three generations are studied.
\subsection{Parent Graphs}
\begin{figure}[!htbp]
    \centering
    \subfloat[]{\includegraphics[scale =0.24,trim={6cm 1cm 8cm 2cm},clip]{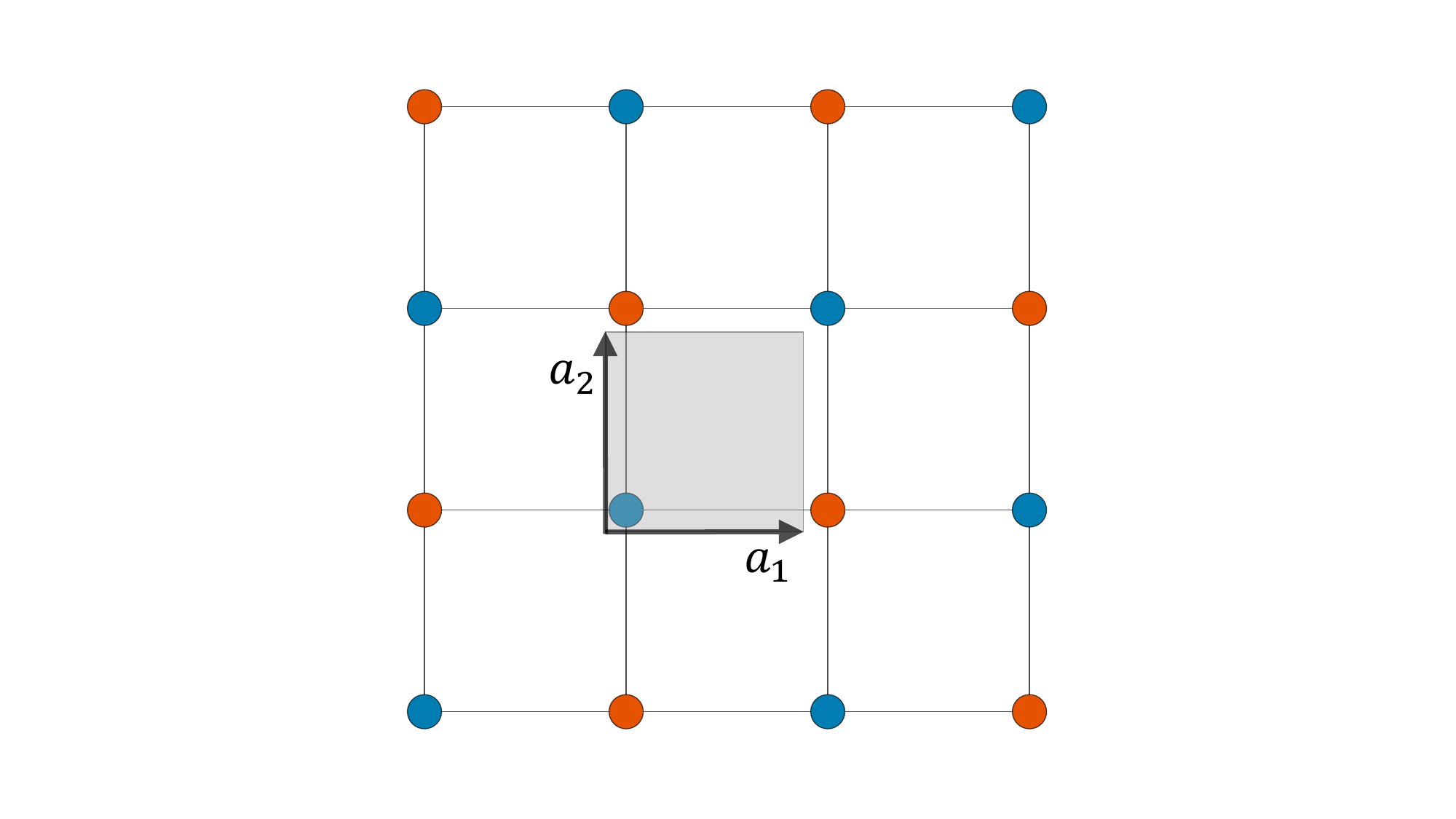}\label{fig:square_lattice}}
    \subfloat[]{\includegraphics[scale =0.28,trim={0cm 6.5cm 0cm 7cm},clip]{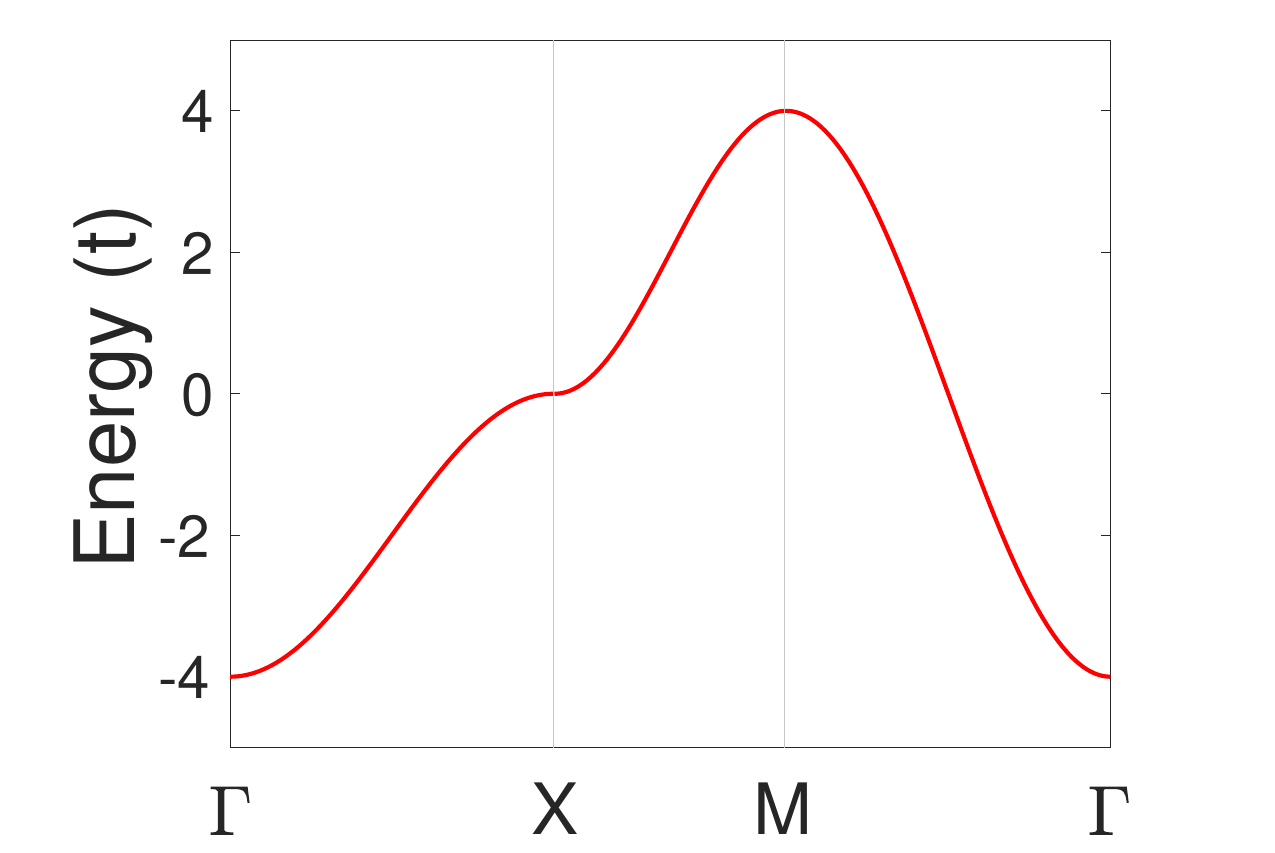} \label{fig:square_2dband}} 
    \caption{(a) The schematic of bipartite square lattice, $\mathcal{X}_4$. The black arrows indicate the lattice vectors $\bfa_1$ and $\bfa_2$, and the gray region is the unit cell. (b) Tight-binding band diagram without SOC.  }
    \label{fig:sq_latt_w_bnds}
\end{figure}

As seen in the section \ref{sec:overview}, the Euclidean parent (root) graphs form the foundation of many complex topological and flat-band systems. The simplest example is a square lattice ($\mathcal{X}_4$), formed by all integer linear combinations of two linearly independent vectors in $\R^2$, shown in Fig.~\ref{fig:square_lattice}. Conventionally, in the square lattice only one atom is considered in the unit cell. Due to the interesting features that can be obtained in a simple setting, the square lattice has become a playground for exploring exotic phenomena such as magnetism and unconventional superconductivity \cite{nevidomskyy2008magnetism,asadzadeh2014superconductivity,kariyado2019flat}. The explicit form of the TB Hamiltonian of $\mathcal{X}_4$:
\beq
H_{\mathcal{X}_4}(\bfk) = -2t(\cos \bfk\cdot \bfa_1+ \cos\bfk\cdot \bfa_2).
\label{eq:sq_hamil}
\eeq
 The band structure plotted in Fig.~\ref{fig:square_2dband} exhibits cosine-like dependence on the momentum space. At half-filling, the lattice has a gapless metallic phase with saddle point at $X-$point in BZ with associated van-Hove singularity in the density of states. Upon application of strain, there is not a particularly pronounced effect on the single band characterizing the electronic structure.

\begin{figure}[!htbp]
    \centering
    \subfloat[]{\includegraphics[scale=0.26,trim={8cm 1cm 8cm 1cm},clip]{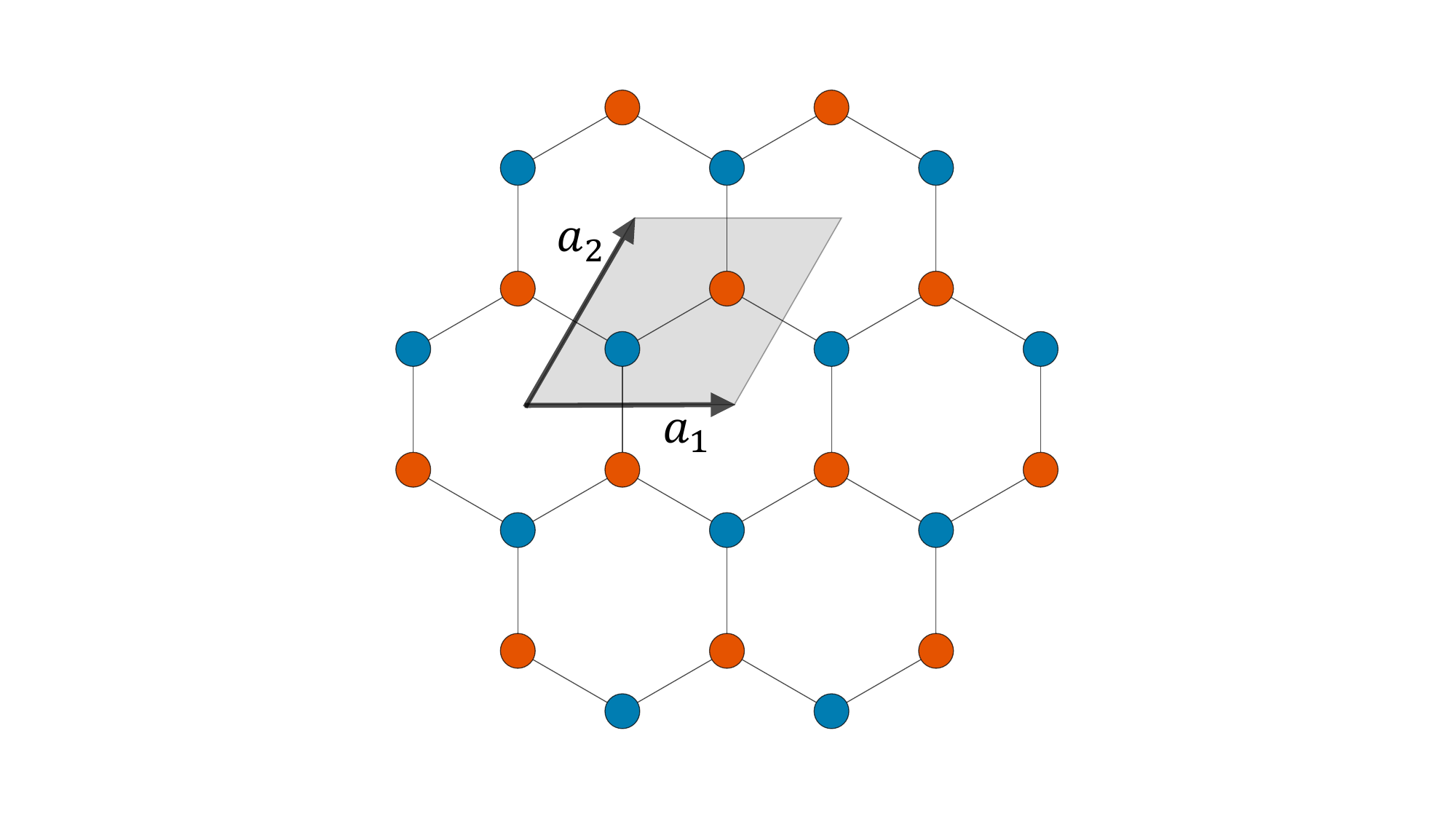}\label{fig:graphene_lattice}}
    \subfloat[]{\includegraphics[scale =0.28,trim={0cm 6cm 2cm 7cm},clip]{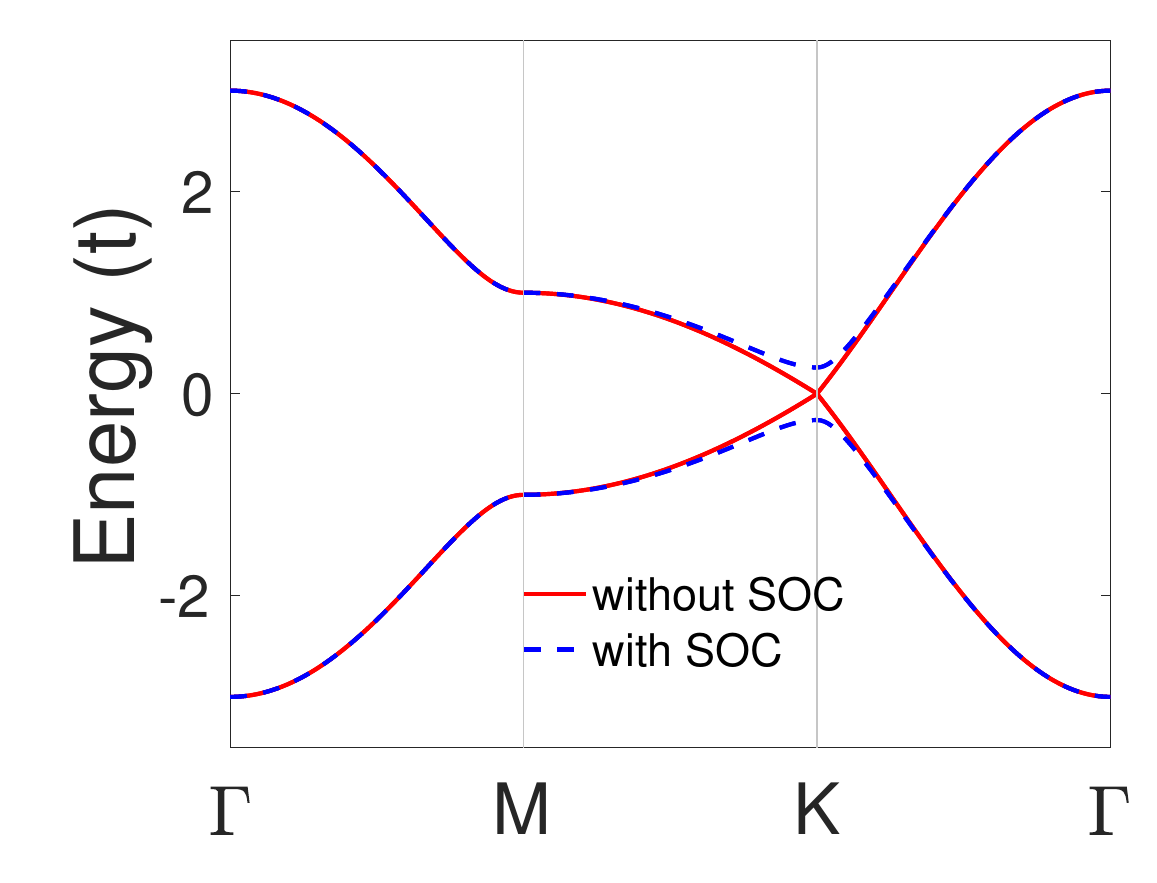} \label{fig:graphene_2dband}} \quad
    \subfloat[]{\includegraphics[scale =0.28,trim={3.1cm 6.5cm 0cm 7cm},clip]{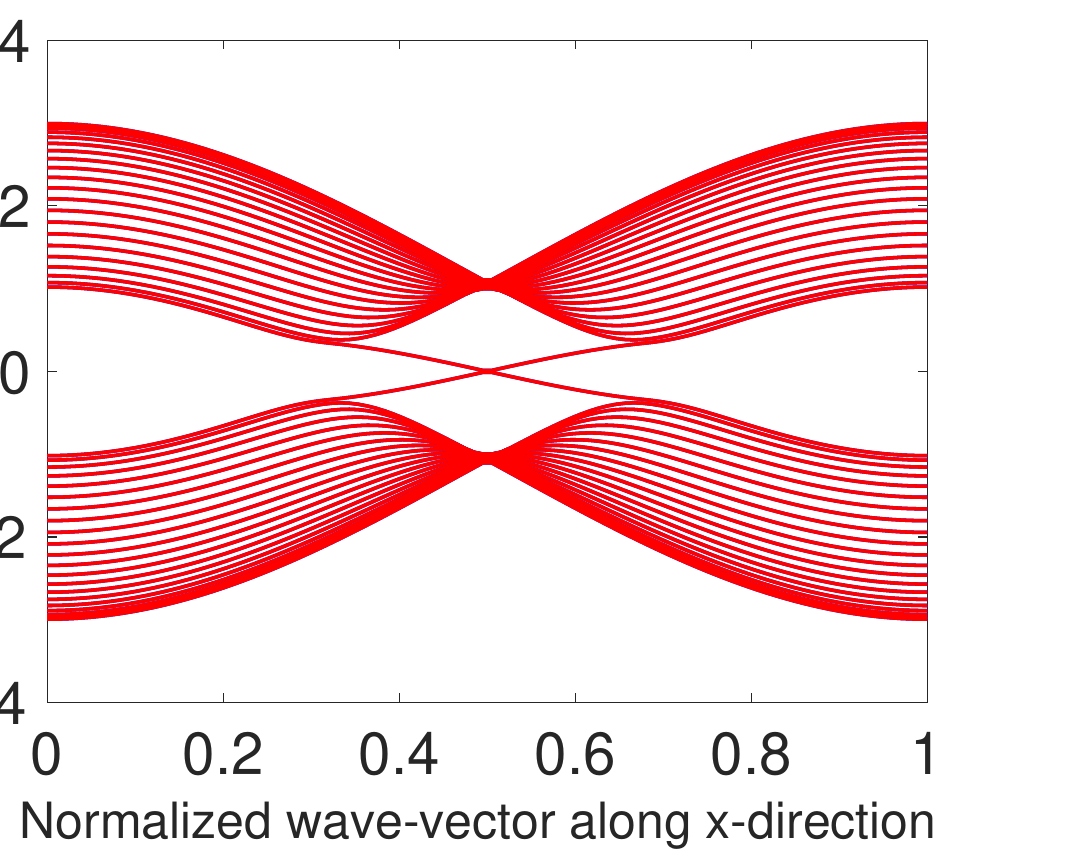} \label{fig:graphene_2dribbon}}
        \caption{ (a)The schematic of bipartite honeycomb lattice, $\mathcal{X}_6$. The black arrows indicate the lattice vectors $\bfa_1$ and $\bfa_2$, and the gray region is the unit cell. The tight binding band diagrams of (b) 2D lattice without (red solid line) and with (blue broken line) SOC ($\lambda_I = 0.1t$), and (c) 1D zigzag hexagonal honeycomb nanoribbon with $\lambda_I = 0.1t$.   }
    \label{fig:graphene_latt_w_bnds}
\end{figure}

The parent graph for hexagonal lattices is the honeycomb lattice, which is the periodic translations of two points (sublattices) in two dimensions (Fig.~\ref{fig:graphene_lattice}). This atomic arrangement renders honeycomb lattice with unique electronic properties characterized by massless Dirac fermions at the $K-$points where the conduction and valence bands meet linearly (see the red band diagram in Fig.~\ref{fig:graphene_2dband}, see Fig.~\ref{fig:rep_sq_hx} for identification of symmetry points in the Brillouin Zone). The spinful Hamiltonian  with SOC in Fourier space is given as:
\beqs
H_{\mathcal{X}_6} = \Bigg(1 + \sum_{i=1}^3t_i\cos\bfk\cdot\bfa_i\Bigg)\tau_x\otimes\sigma_0 &+&\left(\sum_{i=1}^3t_i\sin\bfk\cdot\bfa_i\right)\tau_y\otimes\sigma_0\nonumber\\
&+&\lambda_I\left(\sum_{i=1}^3(-1)^{i+1}\sin\bfk\cdot\bfa_i\right)\tau_z\otimes\sigma_z
\eeqs

In the absence of SOC, the TB Hamiltonian of the honeycomb lattice exhibits the well-known  electronic band structure of pristine graphene \cite{castro2009electronic}. Here, $C_3$ symmetry in addition to inversion and time reversal symmetries induces global stability to the Dirac nodes by making them fixed at $K$-points and gapless. Breaking the $C_3$ symmetry by applying strain moves the Dirac cones away from the $K$-points but it remains gapless for small strain. Under high strain, the hopping parameters for the bonds on each atom can become all distinct (high anisotropic limit), and the band structure becomes gapped  without inducing topological character \cite{pereira2009tight}. A nonzero SOC parameter can trigger gapless edge states in the quasi-1D nanoribbons with gapped bulk bands (broken blue lines in Fig.~\ref{fig:graphene_2dband}) and nontrivial $\Z_2$ index as shown in Fig.~\ref{fig:graphene_2dribbon}.   

For the purposes of this paper, we have considered strains till the reasonable limit of $10\%$ (for graphene, the failure strain can be around $15 - 25\%$, depending on direction \cite{cao2020elastic}). Under this limit the two parent lattices described above do not show any significant change in the topological characteristics (graphene is known to have TPTs under very high anisotropic limits \cite{ahn2017unconventional}).  Subsequent generations of these parent graphs show fascinating quantum phase transitions, which we systematically study below.
\subsection{First generation graphs}
The first generation lattices comprise a total of four lattices formed by split graph, $\mathcal{S(\mathcal{X})}$, and line graph, $L(\mathcal{X})$ operations on each parent lattice.

\subsubsection{Split graphs \texorpdfstring{$\mathcal{S}(\mathcal{X})$}{Lg}: Lieb and Honeycomb splitgraph lattices }
Split graphs are $d,2-$biregular Euclidean lattices where additional sites are placed at the edge center of parent graphs. The energy spectrum of these structures, given by eq.~\ref{eq:spectrum},  are particle-hole symmetric and consist of a flat band at zero energy which touches dispersive bands at the Dirac point, thus forming a threefold degenerate point  \cite{ma2020spin}. Near the point of degeneracy, the low energy Hamiltonian of this set of three  bands can be described by the $3\times 3$ matrices that form spin-1 representation of SU(2). 

\begin{figure}[!htbp]
    \centering
  \subfloat[]{\includegraphics[scale = 0.24, trim={8cm 1cm 6cm 1cm}, clip]{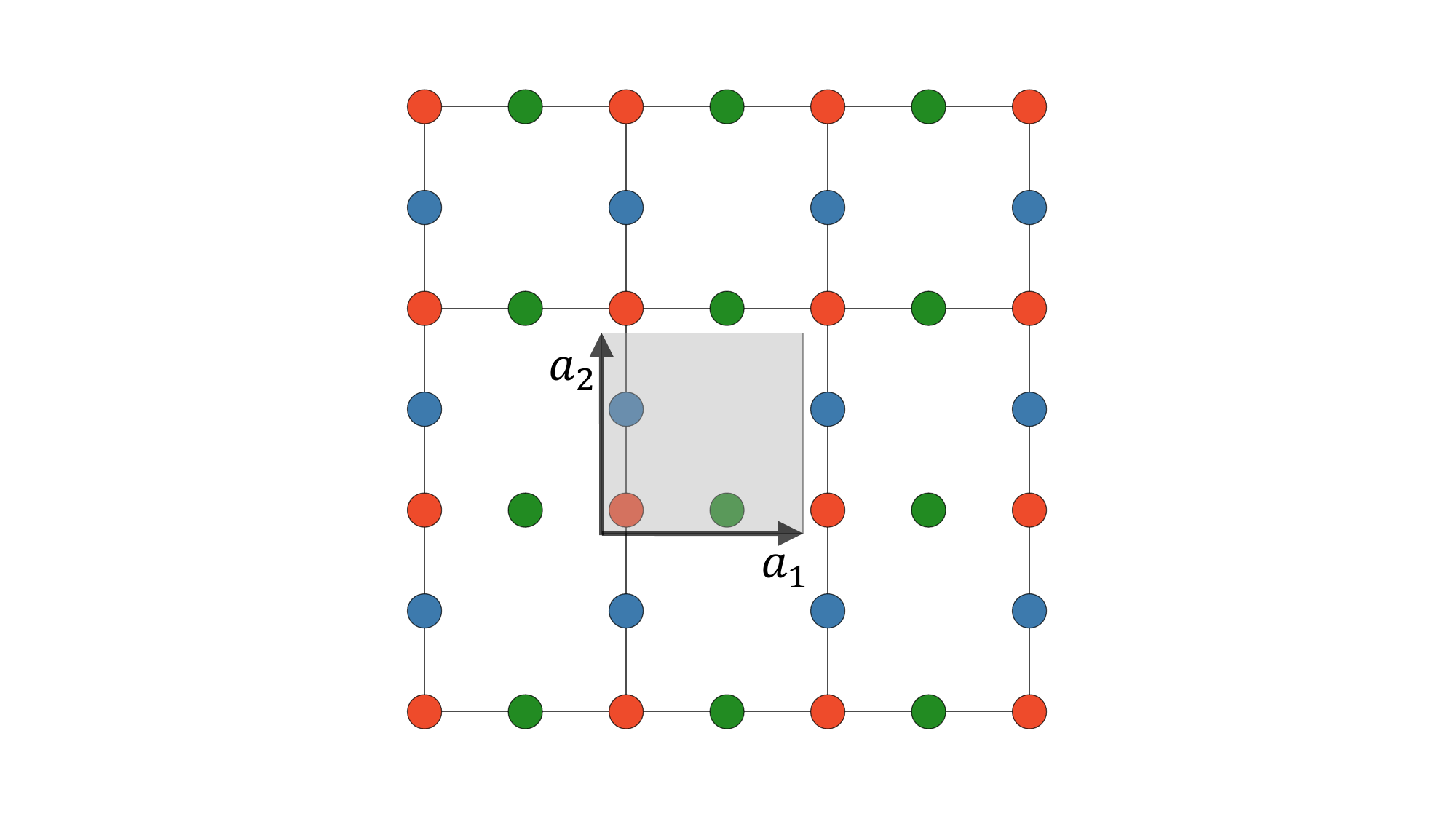}
    \label{fig:Lieb_lattice}}
    \subfloat[]{\includegraphics[scale =0.28,trim={0cm 6.7cm 0cm 7cm},clip]{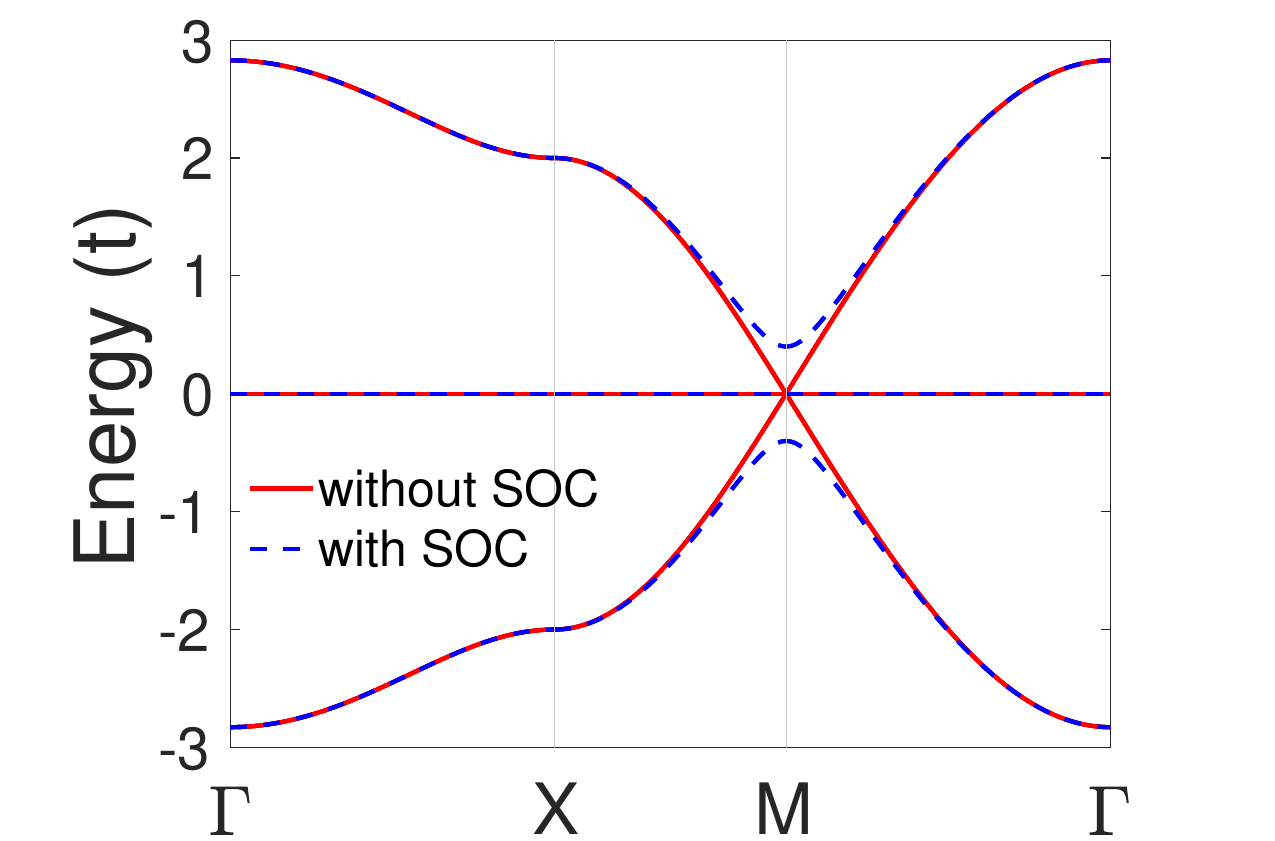} \label{fig:lieblattice_2dband_SOC}}
    \subfloat[]{\includegraphics[scale =0.28,trim={3.1cm 6.5cm 0cm 7cm},clip]{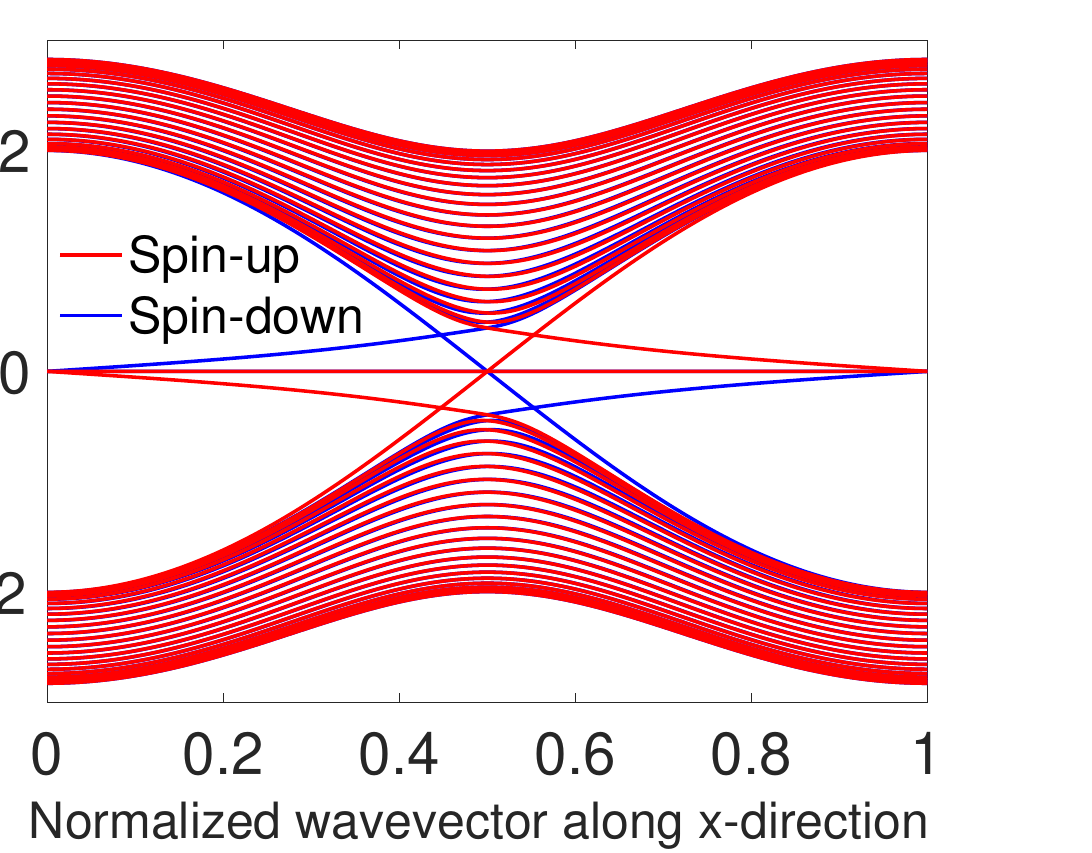} \label{fig:lieb_nanoribbon}}
        \caption{ (a) The schematic of split graph of bipartite square graph (Lieb lattice, $S(\mathcal{X}_4)$). The black arrows indicate the lattice vectors $\bfa_1$ and $\bfa_2$, and the gray region is the unit cell. The tight binding band diagram of (b) 2D lattice without (red solid line) and with (blue broken line) SOC ($\lambda_I = 0.1t$) and (c) 1D zigzag Lieb lattice nanoribbon at $\lambda_I = 0.1t$. The red and blue lines show the counter-propagating spin-up and spin-down states, respectively.}
    \label{fig:Lieb_latt_w_bnds}
\end{figure}

\underline{Lieb lattice:} A particularly well-known split graph with square geometry is the Lieb lattice \cite{weeks2010topological} (Fig.~\ref{fig:Lieb_lattice}). The explicit form of the spinful TB Hamiltonian of the Lieb lattice, $S(\mathcal{X}_4)$ is:
\beqs
H_{S(\mathcal{X}_4)}(\bfk) = & &\sum_{i=1}^{3}\varepsilon_iS_{ii}\otimes \sigma_o + \left(\sum_{i=1}^2 t_i(1+\cos\bfk\cdot\bfa_i)\Lambda_i -\sum_{i=4}^5t_{i-3}\sin\bfk\cdot\bfa_{i-3}\Lambda_i\right)\otimes\sigma_o \nonumber \\
&-&\lambda_I(1-e^{i\bfk\cdot \bfa_1}-e^{i\bfk\cdot \bfa_2}+e^{i\bfk\cdot(\bfa_2-\bfa_1)})\Lambda_3\otimes \sigma_z,
\eeqs
where, $S_i$ is defined as $\bfe_i\otimes\bfe_i$ and matrices $\Lambda_i$ are Gell-Mann matrices matrices defining the orbital space (see appendix \ref{appendix_1}). In a pristine system, with equal on-site energies, without SOC and strain, the band structure is represented by the solid red lines in Fig.\ref{fig:lieblattice_2dband_SOC}, where the flat band lies in between two linearly dispersive bands.  This non-dispersive band is degenerate with the Dirac bands at the $M$-point in the square BZ (see Fig.~\ref{fig:rep_sq_hx} for identification of symmetry points in the Brillouin Zone). With inclusion of SOC, the bands become isolated with a gap of $\Delta = 4|\lambda_I|$ at the Dirac point (shown by broken blue lines in Fig.~\ref{fig:lieblattice_2dband_SOC}). Additionally, the flat band transforms into a topological non-trivial state  with corresponding counter-propagating spin currents represented in the quasi-1D ribbon band diagram (Fig.~\ref{fig:lieb_nanoribbon}).

    The triple degeneracy at the $M-$point is protected by the rotational symmetries of the pristine lattice. However, several translational symmetry preserving perturbations (e.g. strains) and different on-site energies can break this rotational symmetry. This can influence the stability of the flat band and can introduce additional features to the band structure. Within the NN TB model however, application of strain alone does not induce any distortions to the flat band \cite{lim2020dirac,jiang2019topological}.  To observe interesting transitions induced by strain, next-nearest-neighbor interactions \citep{jiang2019topological} or unequal on-site energies have to be introduced, the latter of which we pursue.
    
\begin{figure}[!htbp]
    \centering
    \subfloat[]{\includegraphics[scale =0.2,trim={8cm 2cm 8cm 2cm},clip]{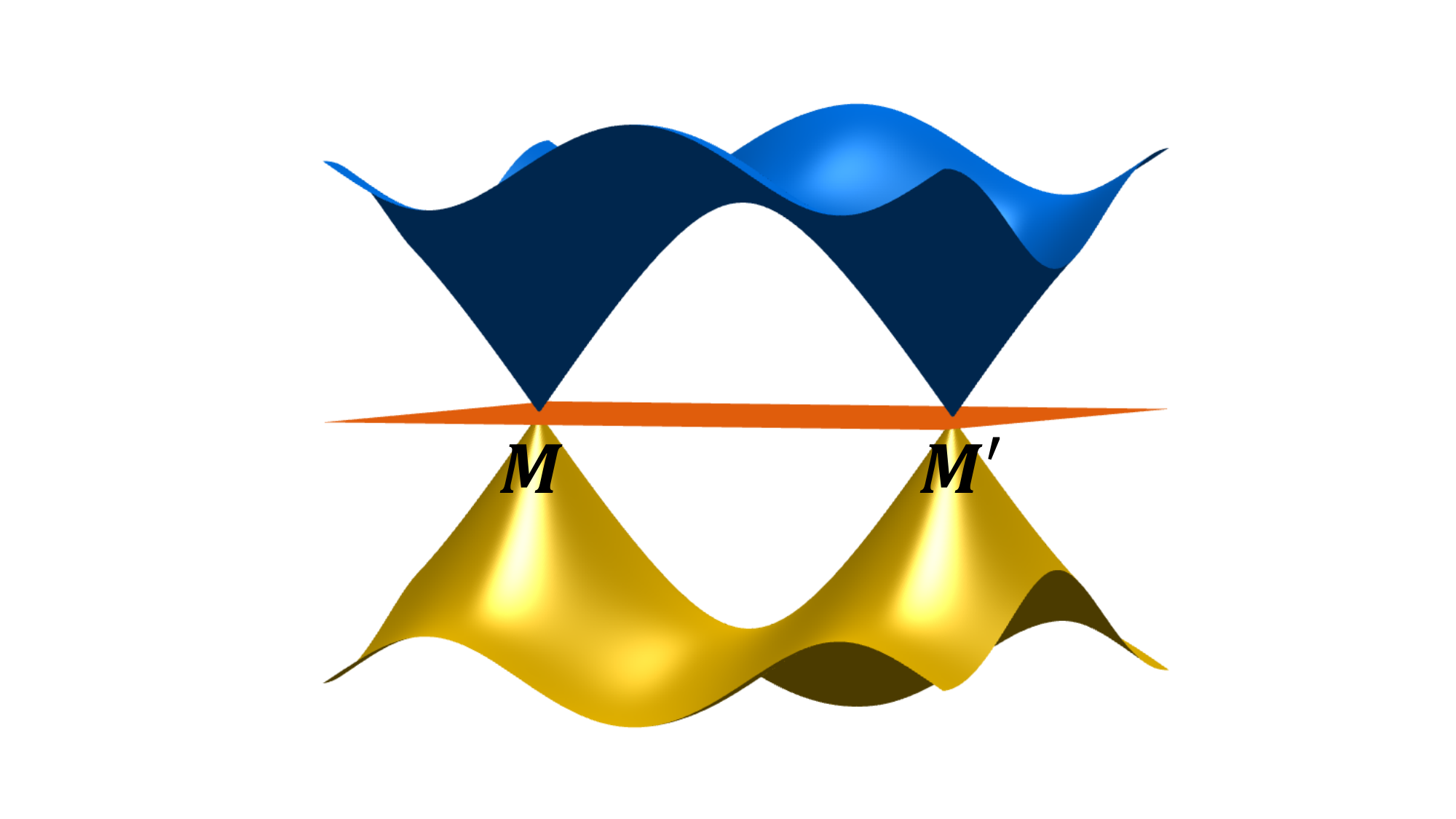} \label{fig:lieb_trans_eb_0_0}} \quad
    \subfloat[]{\includegraphics[scale =0.2,trim={5cm 2cm 5cm 2cm},clip]{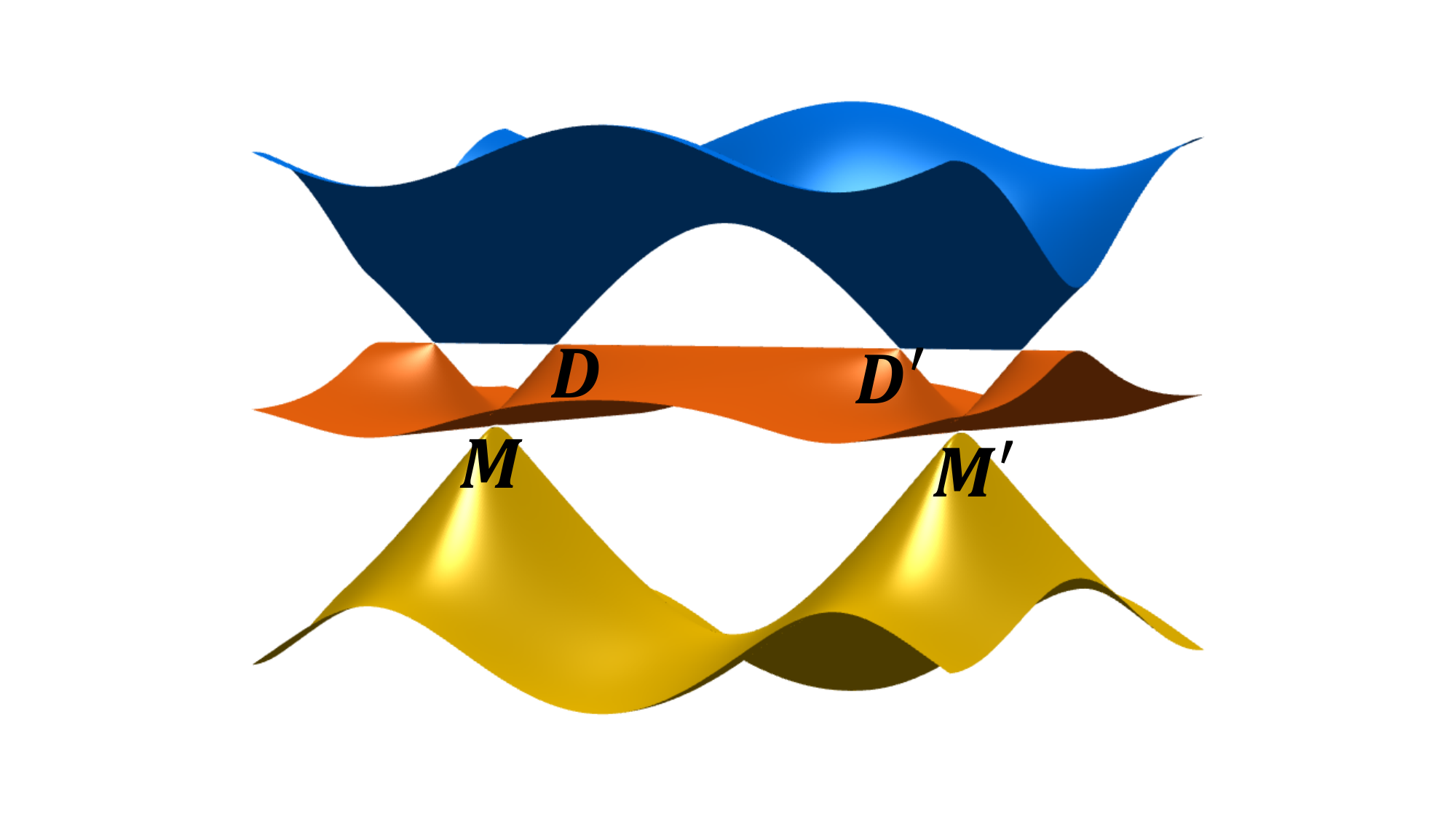} \label{fig:lieb_trans_eb_0_8}}
    \subfloat[]{\includegraphics[scale =0.2,trim={5cm 2cm 5cm 2cm},clip]{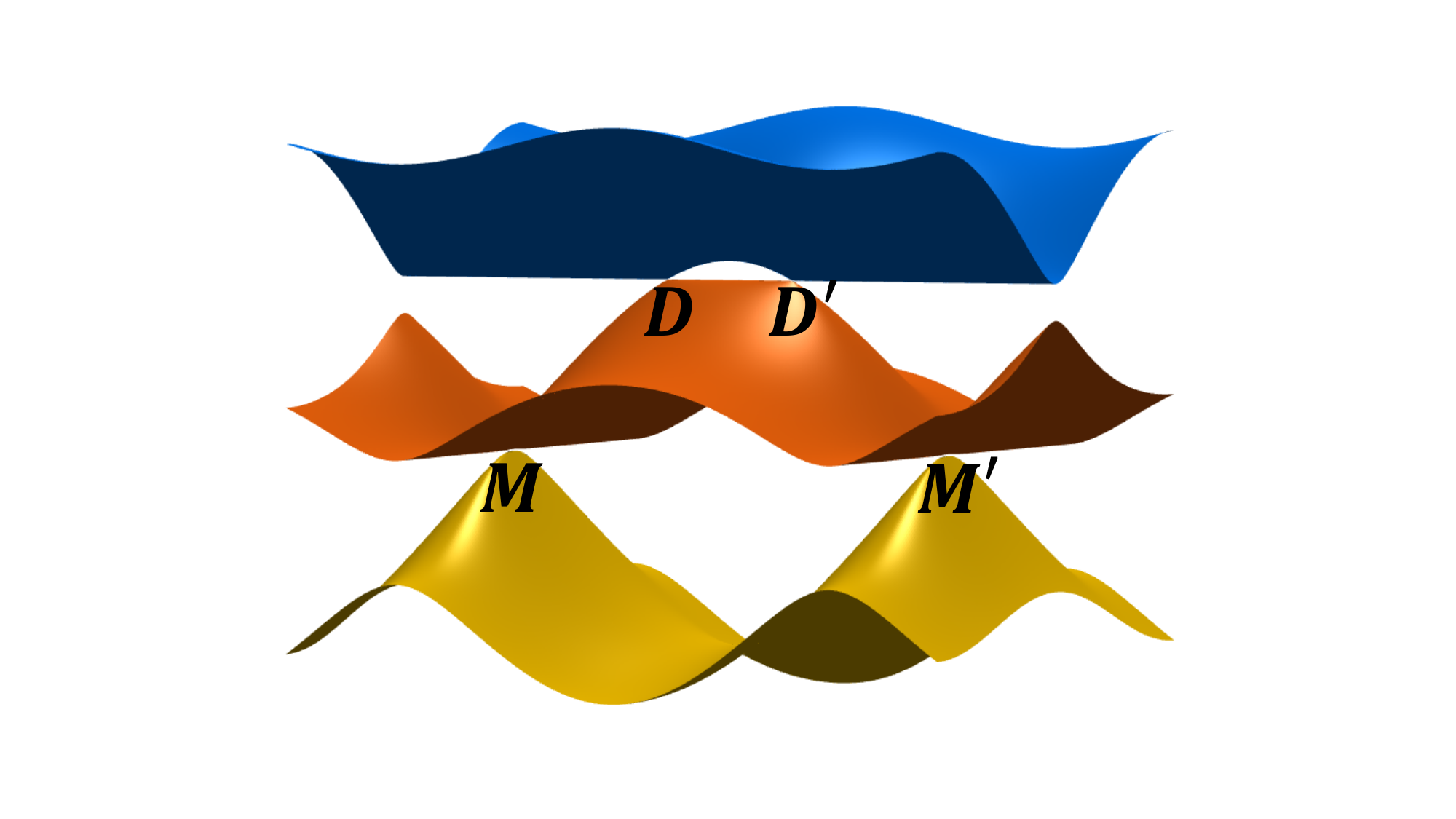} \label{fig:lieb_trans_eb_1_8}}\\
    \subfloat[]{\includegraphics[scale =0.2,trim={5cm 2cm 5cm 2cm},clip]{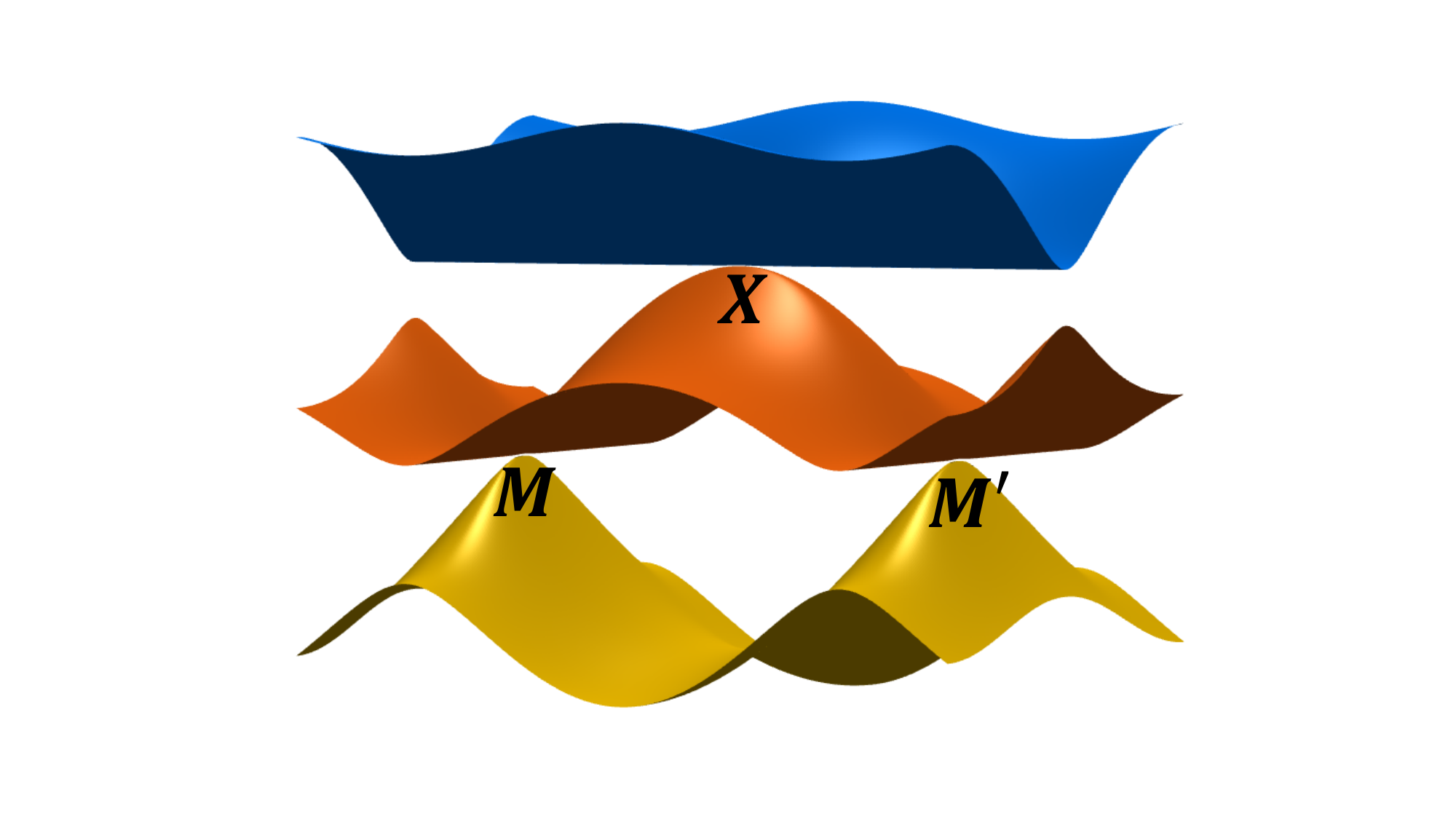} \label{fig:lieb_trans_eb_2_0}}
    \subfloat[]{\includegraphics[scale =0.2,trim={5cm 2cm 5cm 2cm},clip]{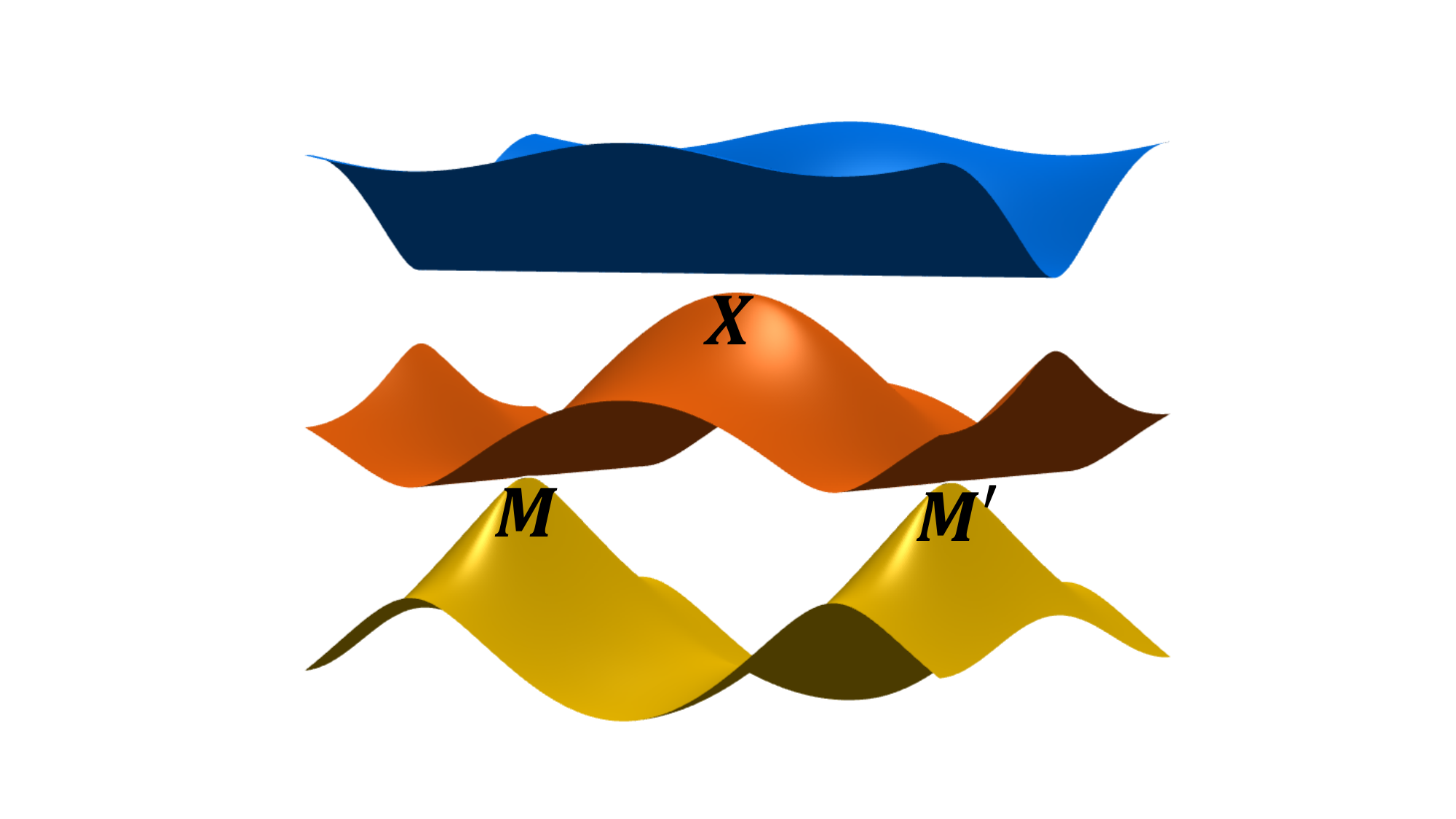} \label{fig:lieb_trans_eb_2_2}}
    \caption{Emergence and motion of two pairs of tilted Dirac cones ($D,D^\prime$) at $(M,M^\prime)$ points in Brillouin zone in the top band from triply degenerate point when on-site potential $\varepsilon_b$ is varied (a) $\varepsilon_b = 0$ , (b) $\varepsilon_b = 0.8$, (c) $\varepsilon_b = 1.8$, (d) $\varepsilon_b = 2$ and (e) $\varepsilon_b = 2.2$. At $\varepsilon_b = 2$, the tilted Dirac cones with opposite winding numbers ($+1, -1$) merges to form semi-Dirac band at point $X$ with zero winding number, shown in (d) and when $\varepsilon_b>2$  the top two bands become gapped as shown in (e).  } 
    \label{fig:Lieb_trans}
\end{figure}

     A non-zero on-site energy ($\varepsilon_b$) at the edge atoms (e.g., the green atoms in Fig.~\ref{fig:Lieb_lattice}) transforms the upper Dirac and flat bands into two tilted Dirac cones ($D,D^\prime$) (see Fig.~\ref{fig:lieb_trans_eb_0_8}) with opposite winding numbers ($+1,-1$). The lower Dirac cone emerges into a semi-Dirac band, linear in one direction and quadratic in the other, which touches the middle band at the $M-$point ( Fig.~\ref{fig:lieb_trans_eb_0_8}). On further increasing the on-site potential, the $D$ and $D^\prime$ Dirac points approach toward each other starting from  $M$ and $M^\prime$ points, respectively, as shown in Fig.~\ref{fig:lieb_trans_eb_1_8}. When $\varepsilon_b=2$,  Dirac points merge into a semi-Dirac band at $X$ with total winding number zero and the top band become directionally flat (Fig.~\ref{fig:lieb_trans_eb_2_0}). At $\varepsilon_b>2$ the gap opens up between the middle and top band but the bottom band never gets isolated. The evolution of the two top bands can be captured simply by the generalized Hamiltonian in the vicinity of the $X-$point \cite{montambaux2018winding}: 
\beq
\mathcal{H}(\bfk) = \left(\delta + \frac{k_x^2}{2m}\right)\tau_a  + ck_y\tau_b,
\label{eq:merge_semidirac}
\eeq
where, $\tau_a$ can be $\tau_x$ (or $\tau_z$), $\tau_b$ is $\tau_y$, and $\delta$, $m$ and $c$ are parameters. Here, $\delta$ plays the role of $\varepsilon_b$. The tilted Dirac cone phase corresponds to $\delta <0$,  at $\delta = 0$ it transforms to semi-Dirac bands and the gapped state corresponds to $\delta > 0$.

\begin{figure}[!htbp]
    \centering
    \subfloat{\includegraphics[scale=0.3,trim={1cm 6.8cm 1cm 7.3cm},clip]{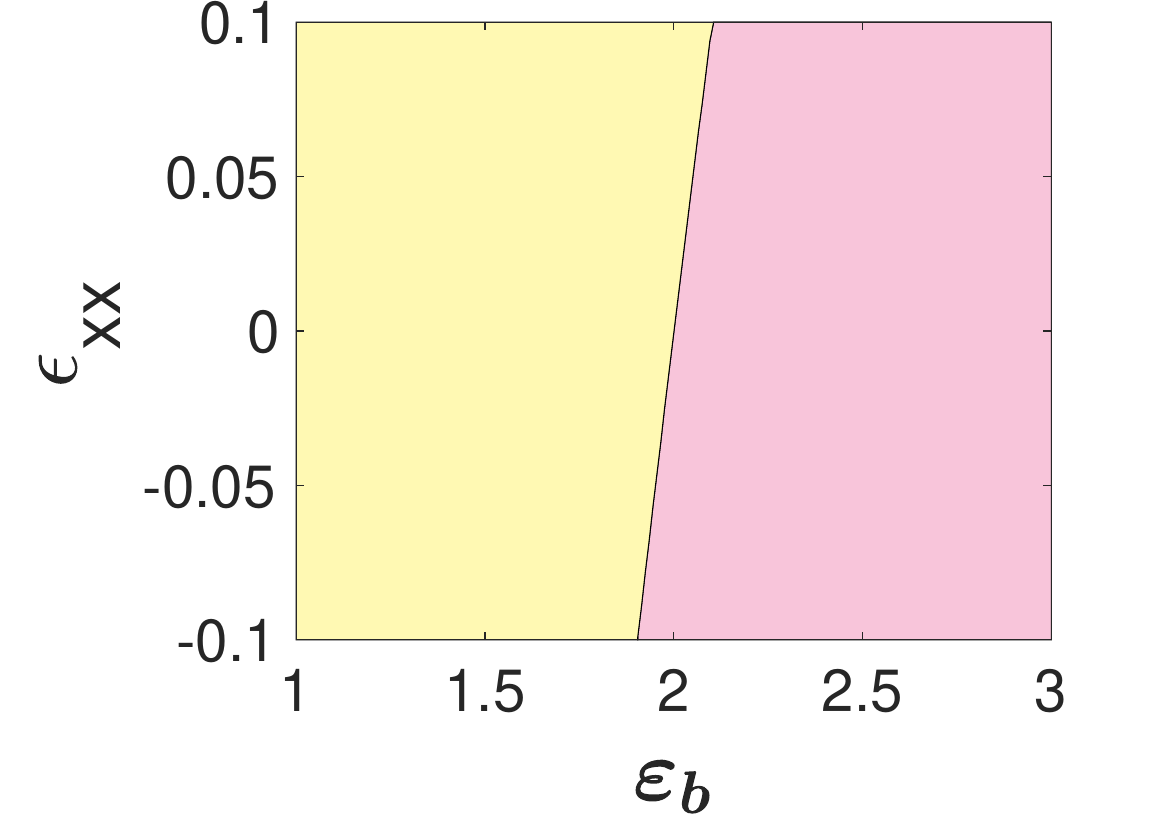}\label{fig:Lieb_phase_lam_0p2_b3}}
    \caption{Topological phase diagram of Lieb lattice at $2/3$ filling as the function of strain in x-direction ($\epsilon_{xx}$) and on-site energy ($\varepsilon_b$) at $\lambda_I = 0.2t$. The phases are distinguished by the colors as follows. Yellow  (\tiyellow{}): topological band insulator and pink (\bipink{}): band insulator. }
    \label{fig:Lieb_lattice_phase}
\end{figure}

\begin{figure}[!htbp]
    \centering
    \subfloat[]{\includegraphics[scale=0.3,trim={1cm 6.8cm 1cm 7.3cm},clip]{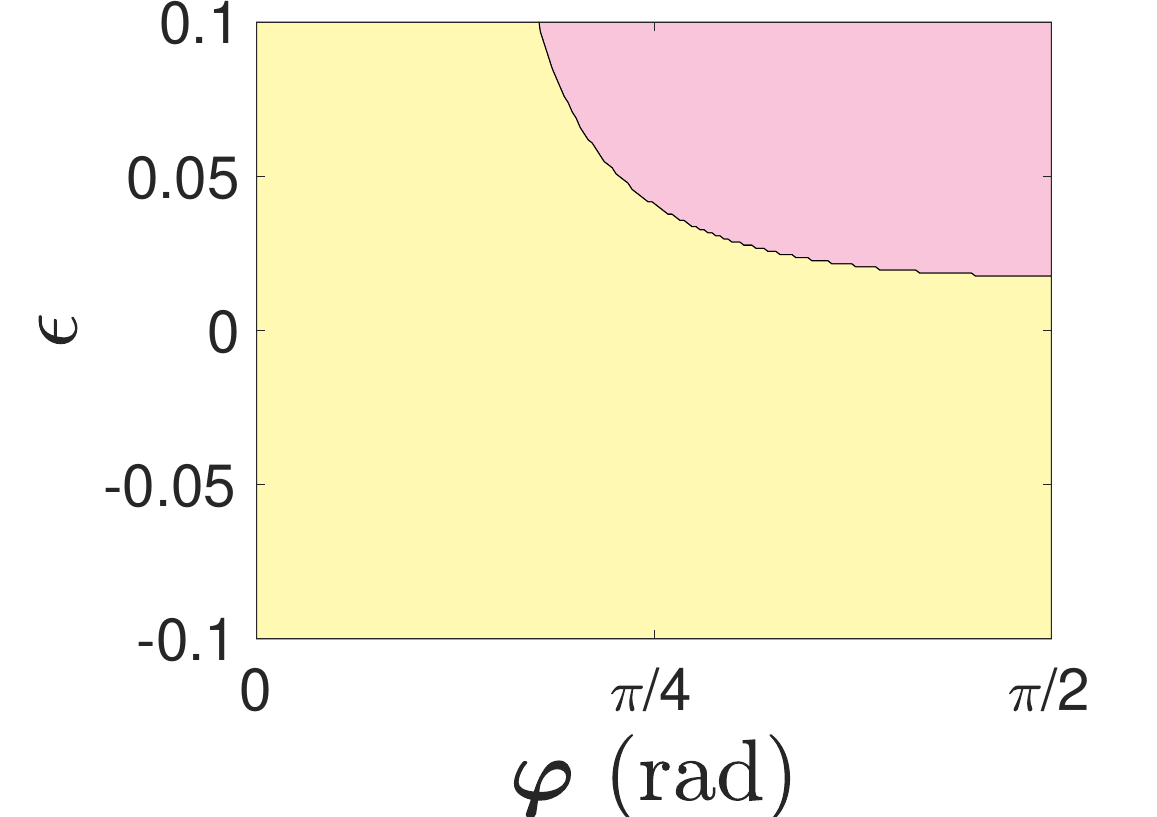}\label{fig:Lieb_phase_lam_0p2_ec_1p9_b3}}
    \subfloat[]{\includegraphics[scale=0.3,trim={2.8cm 6.8cm 1cm 7.3cm},clip]{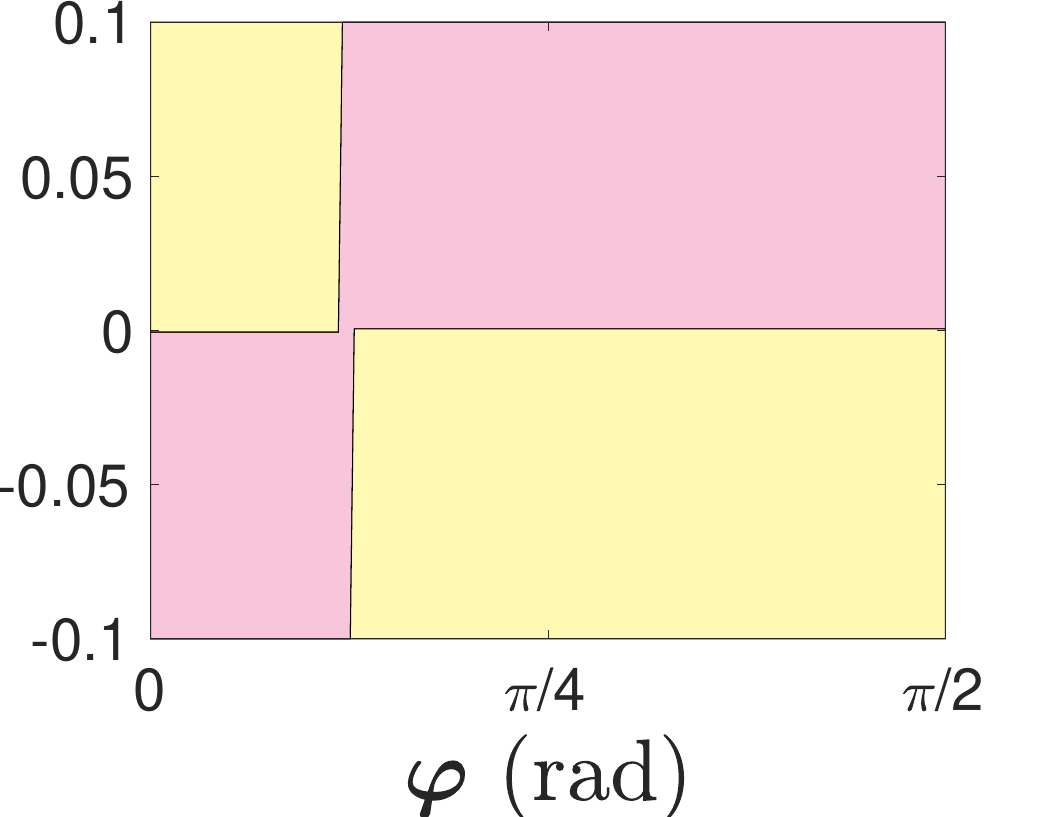}\label{fig:Lieb_phase_lam_0p2_ec_2p0_b3}}
    \subfloat[]{\includegraphics[scale=0.3,trim={2.8cm 6.8cm 1cm 7.3cm},clip]{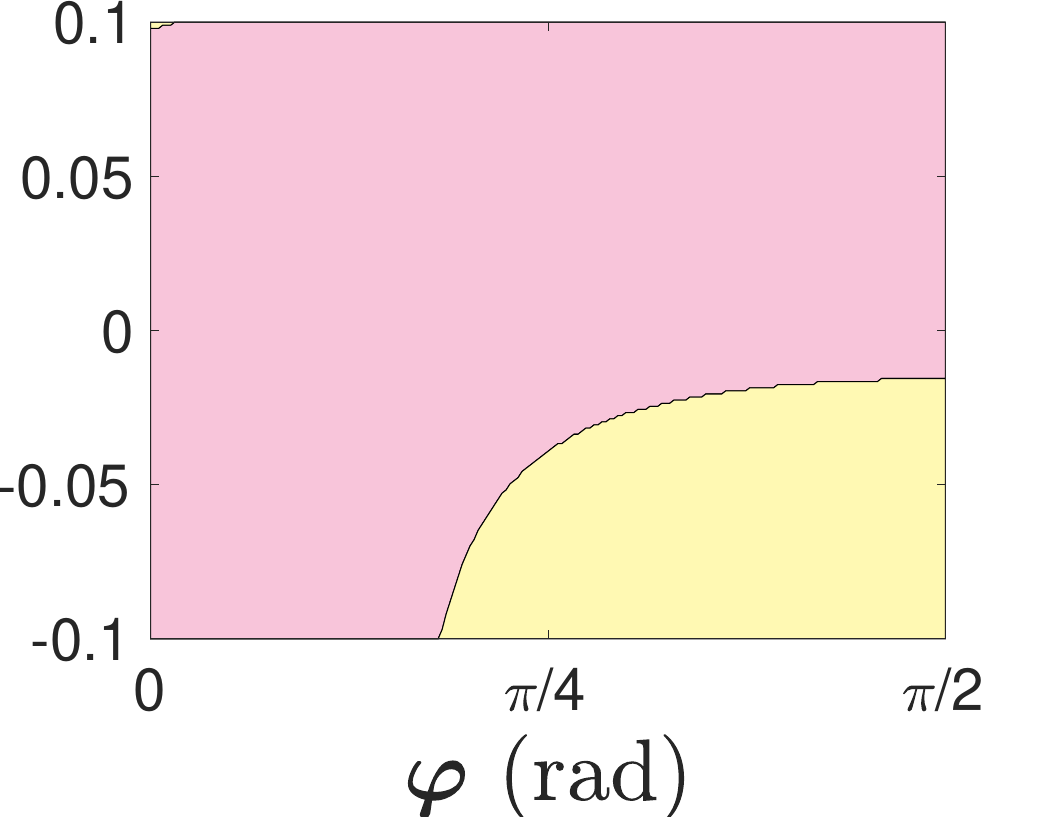}\label{fig:Lieb_phase_lam_0p2_ec_2p1_b3}}
    \caption{Topological phase diagram of Lieb lattice at $2/3$ filling as the function of magnitude $\epsilon$ and the direction $\varphi$ of applied strain for on-site energy (a) $\varepsilon_b = 1.9$, (b) $\varepsilon_b = 2$ and (c) $\varepsilon_b = 2.1$ . The phases are distinguished by the colors as follows. Yellow  (\tiyellow{}): topological band insulator and pink (\bipink{}): band insulator. }
    \label{fig:Lieb_lattice_phase_phi}
\end{figure}

A uniform strain combined with site asymmetry generates intriguing topological phase diagrams. At filling fraction 2/3, with fixed intrinsic SOC parameter, we observed that varying the strain along x-direction ($\epsilon_{xx}$) and the on-site energy ($\varepsilon_b$), results in a linear phase boundary between the trivial and topological phases. A prototypical example (with $\lambda_I = 0.2t$) is shown in  Fig.~\ref{fig:Lieb_lattice_phase} where, increasing $\epsilon_{xx}$ from $-10\%$ to $10\%$ the phase boundary vary linearly between $\varepsilon_b=1.9$ and  $\varepsilon_b=2.1$. In other words, at $\varepsilon_b = 2 $, the uniaxial distortion drives the flat band in the Lieb lattice from the trivial phase to the topological phase. 
Additionally, varying the orientation along with the magnitude of the applied strain also influences the system's phases. A notable effect occurs near the transition point $\varepsilon_b=2$, where the middle band undergoes transition from the tilted Dirac cones to a gapped phase passing through the semi-Dirac band phase. The phase diagrams in the $\epsilon - \varphi$ space in Fig.~\ref{fig:Lieb_lattice_phase_phi} show that at $\varepsilon_b = 1.9$ the majority region is the topological phase, except the top right (pink region), where the strain is large and positive, making the system a trivial insulator (Fig.~\ref{fig:Lieb_phase_lam_0p2_ec_1p9_b3}). In contrast, at $\varepsilon_b = 2.1$, the system is largely in a trivial phase except the area under compressive strain as represented in Fig.~\ref{fig:Lieb_phase_lam_0p2_ec_2p1_b3}. At the transition point, $\varepsilon_b = 2$, there are equal diagonal and antidiagonal patches of topological and trivial phases, shown in Fig.~\ref{fig:Lieb_phase_lam_0p2_ec_2p0_b3}. The nature of the distribution of phases is directed by the evolution of the shape of the middle band as discussed above.

\begin{figure}[!htbp]
    \centering
    \subfloat[]{\includegraphics[scale=0.24,trim={9cm 1cm 6cm 1cm},clip]{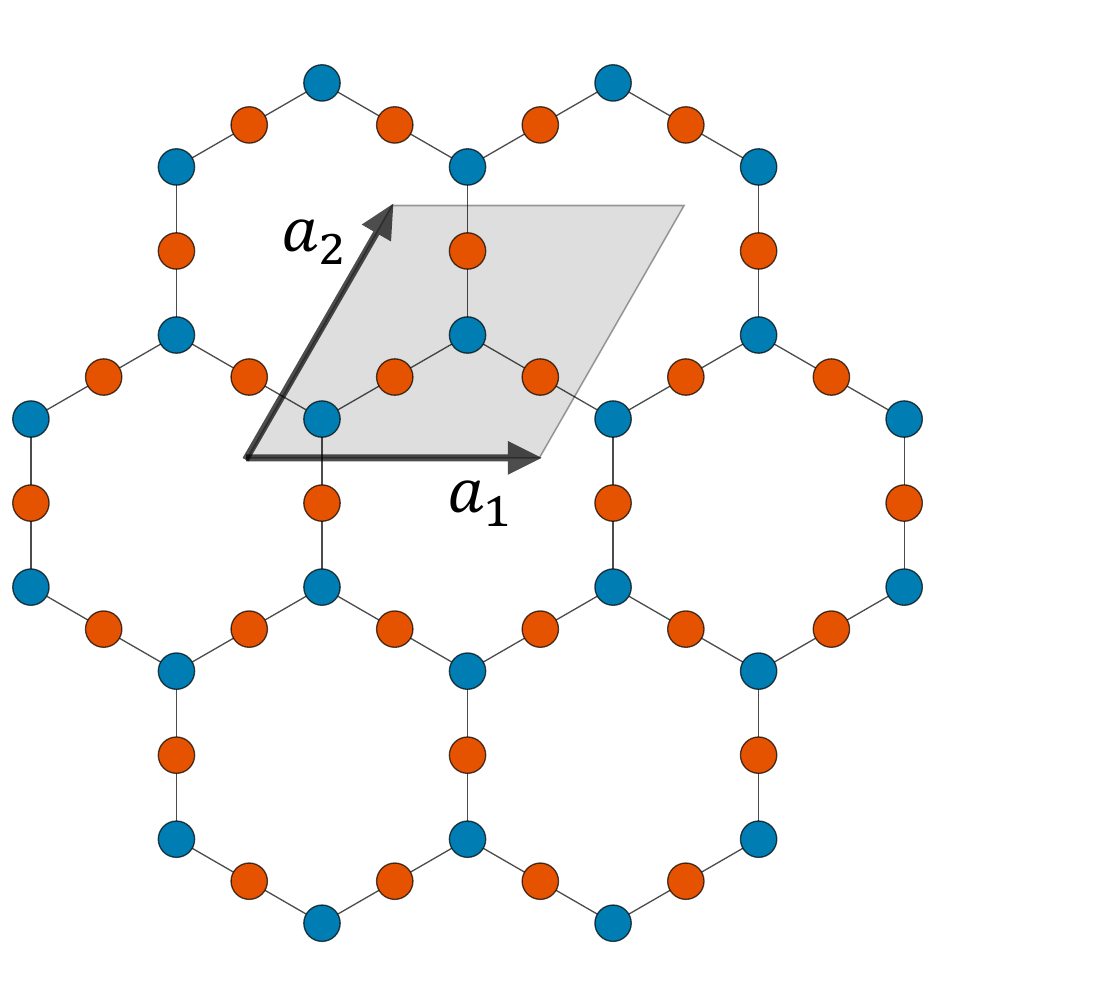}\label{fig:honeycomb_split_graph}} 
    \subfloat[]{\includegraphics[scale =0.28,trim={0cm 6.7cm 0cm 7cm},clip]{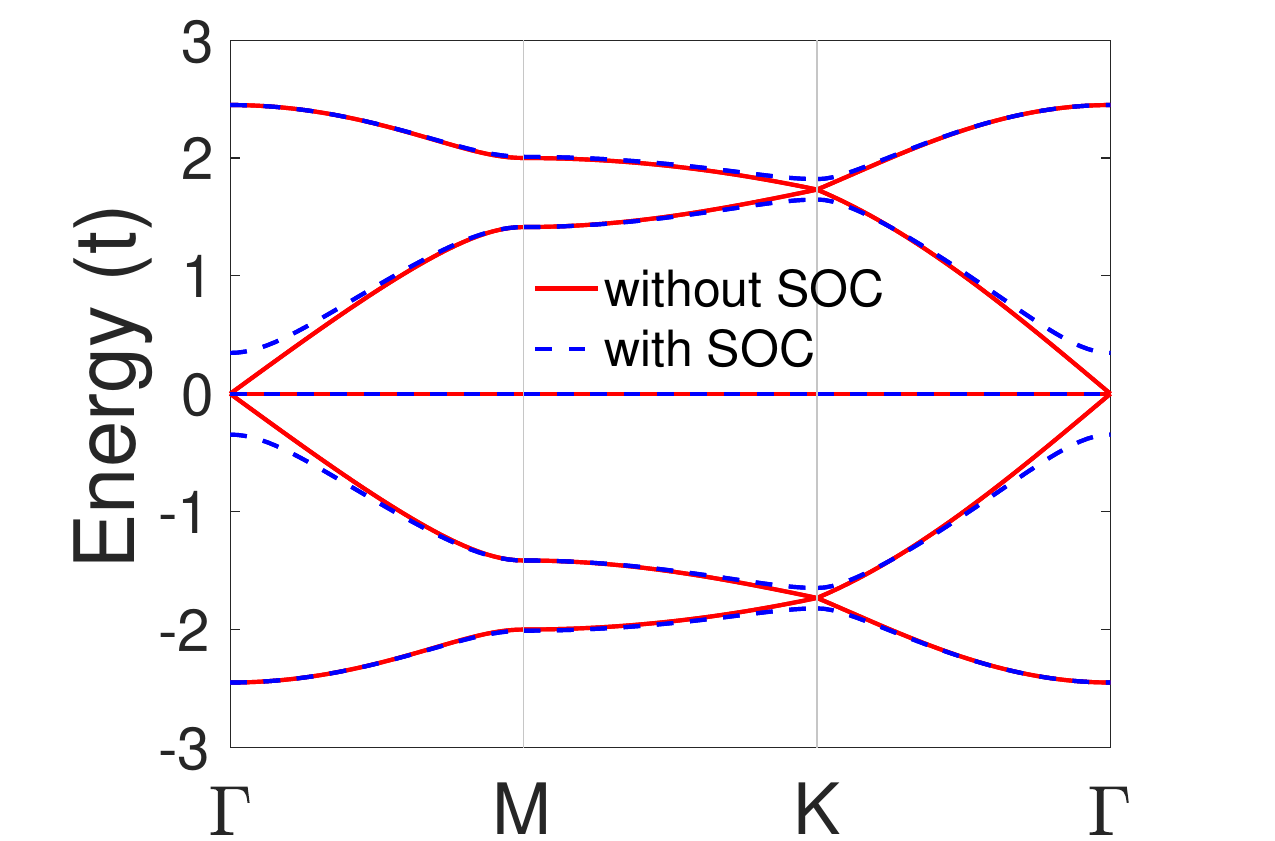} \label{fig:honeycomb_linegraph_2dband}}
     \subfloat[]{\includegraphics[scale =0.28,trim={3.1cm 6.5cm 0cm 7cm},clip]{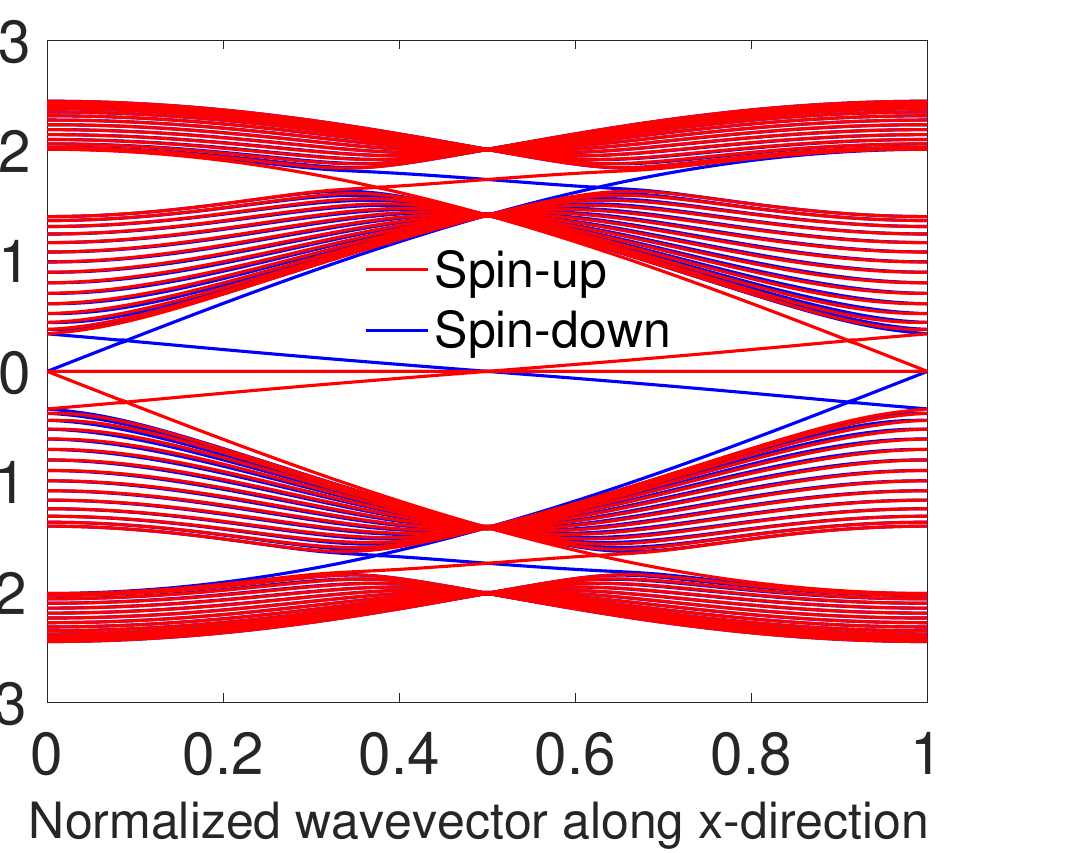} \label{fig:honeyvcom_line_ribbon}}
    \caption{(a) The splitgraph of honeycomb lattice $S(\mathcal{X}_6)$. The black arrows indicate the lattice vectors $\bfa_1$ and $\bfa_2$, and the gray region is the unit cell. The tight binding band diagram of (b) 2D lattice without (red solid line) and with (blue broken line) SOC ($\lambda_I = 0.1t$) and (c) 1D zigzag lattice nanoribbon with $\lambda_I = 0.1t$. The red and blue lines show the counter-propagating spin-up and spin-down states, respectively. }
    \label{fig:hnycmb_spltgrph_latt_w_bnds}
\end{figure}

\underline{Honeycomb Kagome Lattice:} For the hexagonal graph $\mathcal{X}_6$, the split graph $\mathcal{S}(\mathcal{X}_6)$ is referred to as the Honeycomb-Kagome (HK) lattice since it combines structural and electronic features of both of Honeycomb and Kagome lattices \cite{mizoguchi2020square,mizoguchi2021square}. Specifically, the unit cell consists of five atoms: the red sites, shown in Fig.~\ref{fig:honeycomb_split_graph} form a Kagome lattice, while the blue atoms occupy the corners and form a hexagonal lattice. The HK lattice has found to be stable for various species of atoms in 2D material morphologies \cite{pan2018half,liu2018hexagonal, hashmi2020ising,mellaerts2021two} as well as in quasi-1D form \cite{ShivamP2C3}.  The spinful TB Hamiltonian for this structure can be expressed as:

\beq
H_{S(\mathcal{X}_6)}(\bfk) = H^O_{S(\mathcal{X}_6)}(\bfk)\otimes\sigma_0 + H^{SO}_{S(\mathcal{X}_6)}(\bfk)\otimes\sigma_z, 
\eeq
where,
\beq
H^{O}_{S(\mathcal{X}_6)}(\bfk) = \begin{pmatrix}
    \mathcal{O}_{2\times 2} && \Phi^\dagger(\bfk) \\
    \Phi(\bfk) && \calO_{3\times 3} 
\end{pmatrix}, \quad \text{with} \;\;
\Phi(\bfk) = \begin{pmatrix}
    t && t \\
    t && te^{-i\bfk\cdot \bfa_1} \\
    t && te^{-i\bfk\cdot\bfa_2}
\end{pmatrix},
\eeq
and 
\beq
H^{SO}_{S(\mathcal{X}_6)}(\bfk) = \lambda_I\Bigg(\tau_0\oplus\left(\sum_{j=1}^3(-1)^j\sin\bfk\cdot\bfa_{j}\Lambda_j)-\sum_{j=4}^6(1+\cos\bfk\cdot\bfa_{j-3})\Lambda_j\right)\Bigg)
\eeq
The corresponding TB band diagram shown in Fig.~\ref{fig:honeycomb_linegraph_2dband}, is symmetric about zero energy and contains a flat band which is triply degenerate with the linearly dispersive band at the $\Gamma$ point. The non-zero $\lambda_I$ isolates all the bands represented by broken blue lines in Fig.~\ref{fig:honeycomb_linegraph_2dband} and the system becomes non-trivial with edge states, which can be visualized in Fig.~\ref{fig:honeyvcom_line_ribbon}. The magnitude of the gap between the flat band and the Dirac band is given by $\Delta = 2\sqrt{3}|\lambda_I|$. 

\begin{figure}[!htbp]
    \centering
    \subfloat[]{\includegraphics[scale=0.27,trim={3cm 9cm 2cm 8cm},clip]{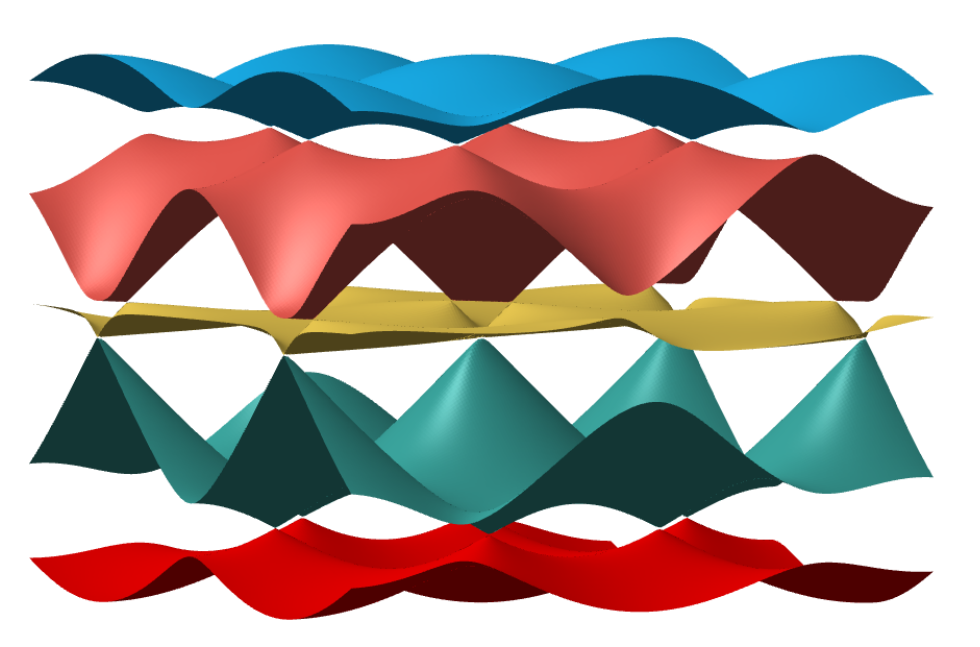}\label{fig:honeycomb_splitgrph_ec0p5}} \quad
    \subfloat[]{\includegraphics[scale=0.27,trim={3cm 9cm 2cm 8cm},clip]{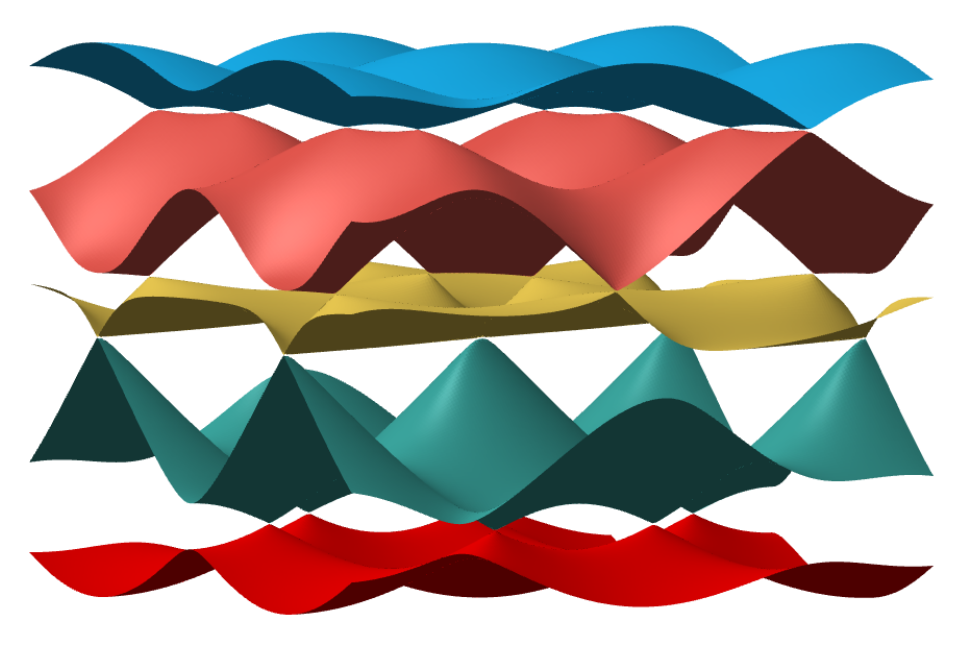}\label{fig:honeycomb_splitgrph_ec0p9}}\\
    \subfloat[]{\includegraphics[scale=0.27,trim={3cm 9cm 2cm 8cm},clip]{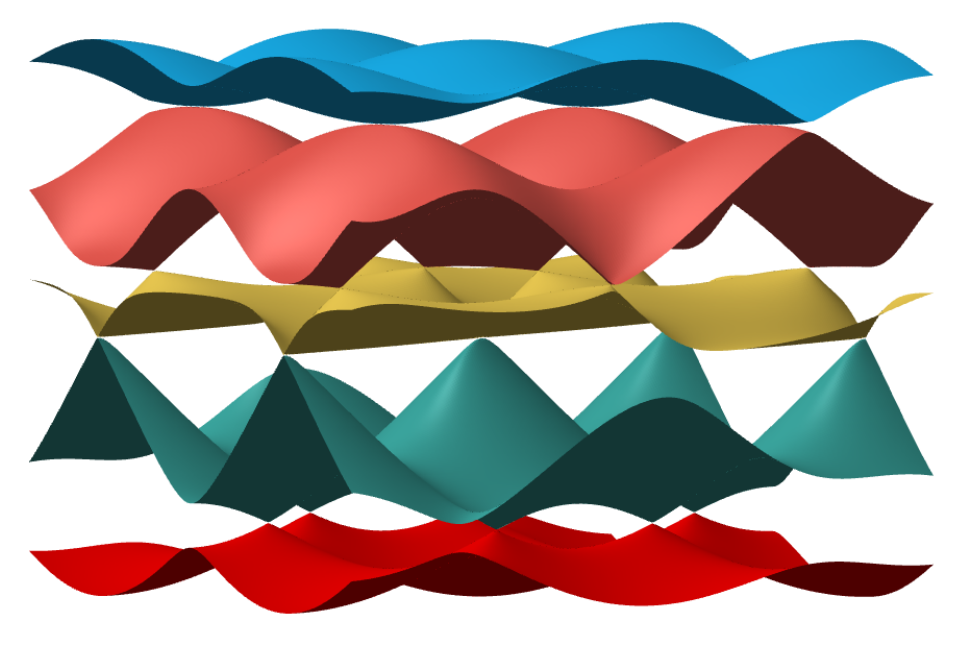}\label{fig:honeycomb_splitgrph_ec1p0}} \quad
    \subfloat[]{\includegraphics[scale=0.27,trim={3cm 9cm 2cm 8cm},clip]{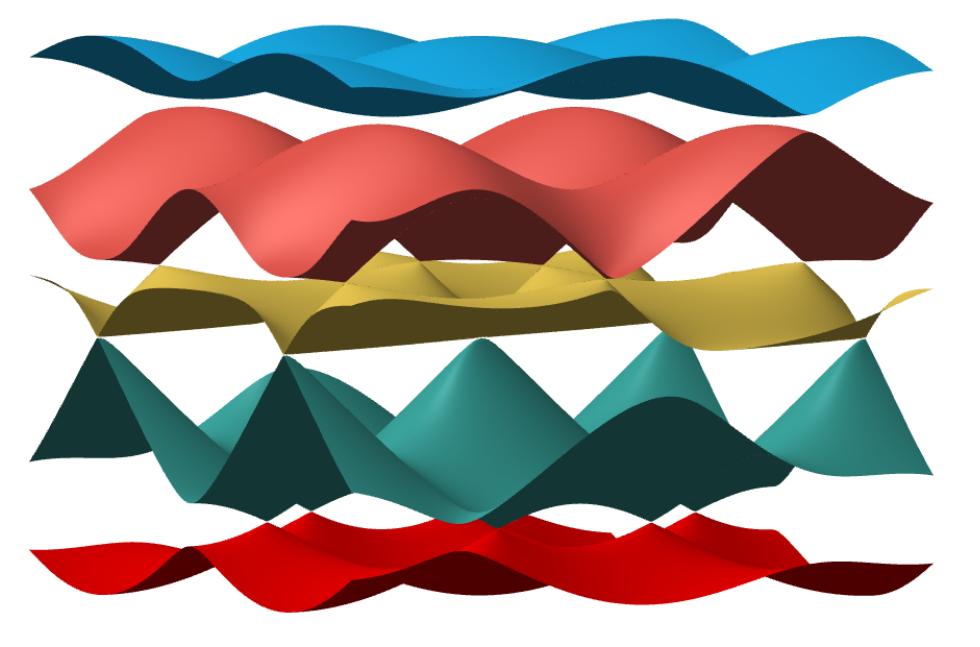}\label{fig:honeycomb_splitgrph_ec1p1}}
    \caption{Evolution of Dirac cones of opposite winding numbers at 4/9 filling fraction starting from $K$ and $K^\prime$ points when on-site potential $\varepsilon_c$ is varied. (a) $\varepsilon_c = 0.5t$ (b) $\varepsilon_c = 0.9t$ (c) $\varepsilon_c = 1.05t$ (d) $\varepsilon_c = 1.1t$. The Dirac cones merges at $\varepsilon_c = 1.05t$ and form semi-Dirac cones and finally separates at $\varepsilon_c = 1.1t$ }
    \label{fig:honeycomb_splitgrph_ec}
\end{figure}

\begin{figure}[!htbp]
    \centering
    \subfloat[]{\includegraphics[scale=0.3,trim={1cm 6.5cm 1cm 7cm},clip]{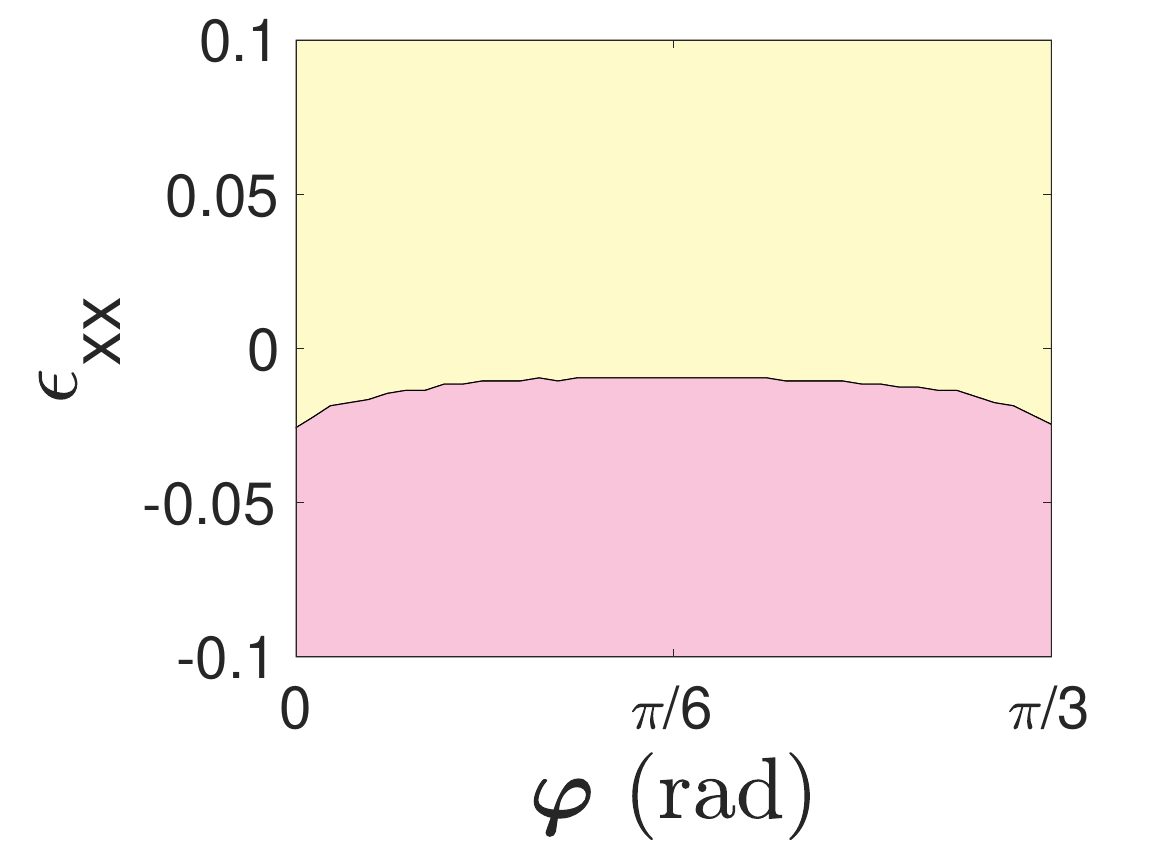}\label{fig:honeycomb_splitgrph_phase_lam_0p2_ec_1p0_b4}}
    \subfloat[]{\includegraphics[scale=0.3,trim={3.1cm 6.5cm 1cm 7cm},clip]{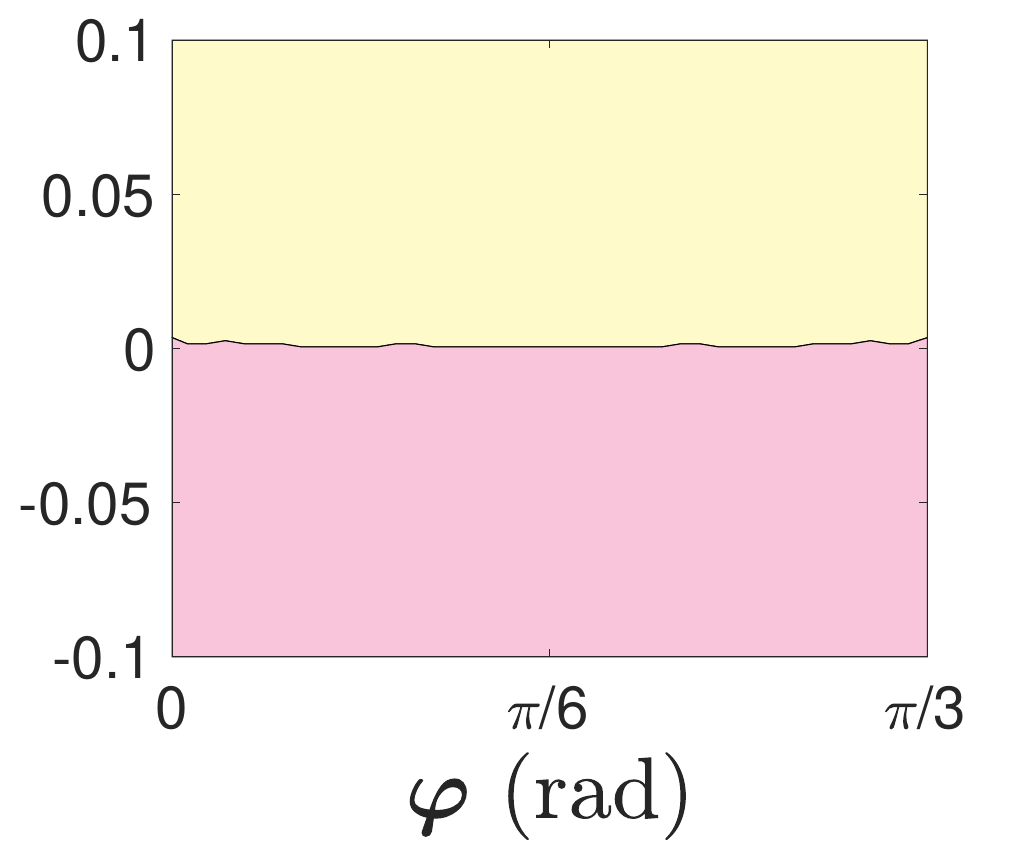}\label{fig:honeycomb_splitgrph_phase_lam_0p2_ec_1p05_b4}}
    \subfloat[]{\includegraphics[scale=0.3,trim={3.1cm 6.5cm 1cm 7cm},clip]{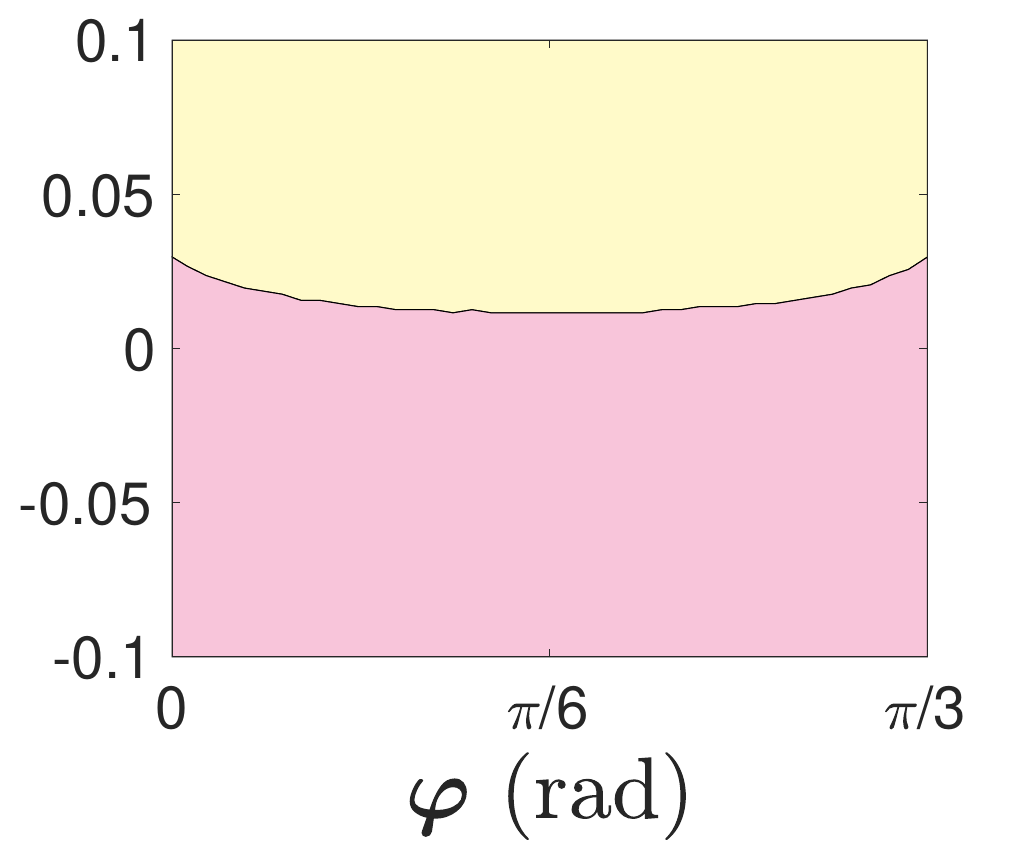}\label{fig:honeycomb_splitgrph_phase_lam_0p2_ec_1p10_b4}}
    \caption{Topological phase diagram of honeycomb splitgraph lattice at $4/5$ filling as the function of strain in x-direction ($\epsilon_{xx}$) and on-site potential ($\varepsilon_c$) at $\lambda_I = 0.2t$ and (a) $\varepsilon = 1$ (b) $\varepsilon = 1.05$  and (c)  $\varepsilon = 1.10$. The different phases are distinguished by the colors as follows. Yellow  (\tiyellow{}): topological band insulator and pink (\bipink{}): trivial band insulator. }
    \label{fig:honeycomb_splitgraph_phase_angle}
\end{figure}

Similar to the Lieb lattice, the flat band in the HK lattice also remains dispersionless under deformation within the NN TB model. Analogously, site asymmetry produces fascinating effects in the electronic properties of the HK lattice, even in the absence of SOC. First, we focus on the electronic band structure (Fig.~\ref{fig:honeycomb_splitgrph_ec}). Introducing a non-zero on-site energy ($\varepsilon_c$) at one of the red atoms, splits the upper linearly dispersive and flat band into tilted Dirac spectrum near the $\Gamma$ point whereas the lower Dirac band (at $2/5$ filling) converts into a semi-Dirac state, as shown in Fig.~\ref{fig:honeycomb_splitgrph_ec0p5}. Unlike the Lieb lattice, increasing $\varepsilon_c$ causes the separation between the titled Dirac cones to widen but they never merge with Dirac nodes entering from the neighboring BZ. Intriguingly, at $\varepsilon_c = 1.05$ and 4/5 filling fraction, the Dirac points at $K$ and $K^\prime$ with opposite winding numbers $\{1,-1\}$ annihilate into semi-Dirac bands, as represented in Fig.\ref{fig:honeycomb_splitgrph_ec1p0}. When $\varepsilon_c > 1.05$, the the fifth band becomes isolated (see Fig.~\ref{fig:honeycomb_splitgrph_ec1p1}).  
\begin{figure}[!htbp]
    \centering
    \subfloat[]{\includegraphics[scale=0.3,trim={1cm 6.6cm 1cm 7cm},clip]{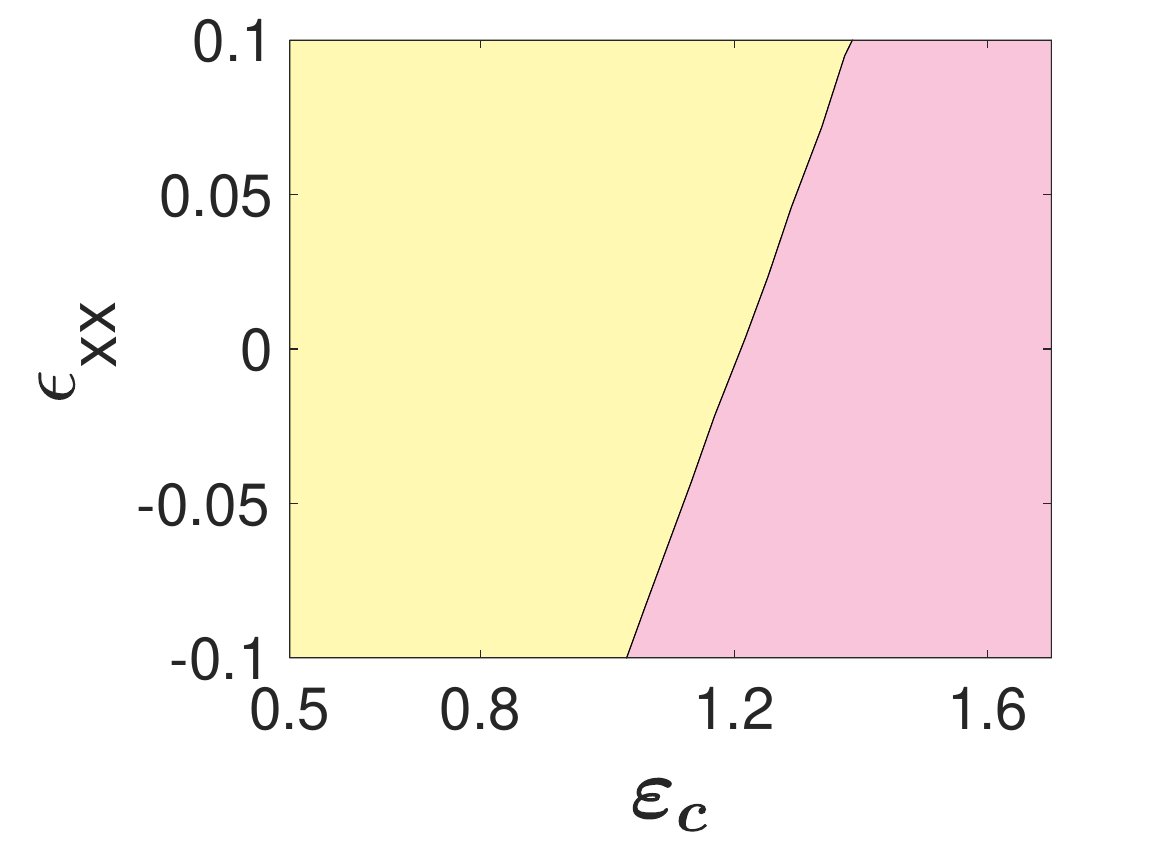}\label{fig:honeycomb_splitgrph_phase_lam_0p4_b4}}
    \subfloat[]{\includegraphics[scale=0.3,trim={2.97cm 6.6cm 1cm 7cm},clip]{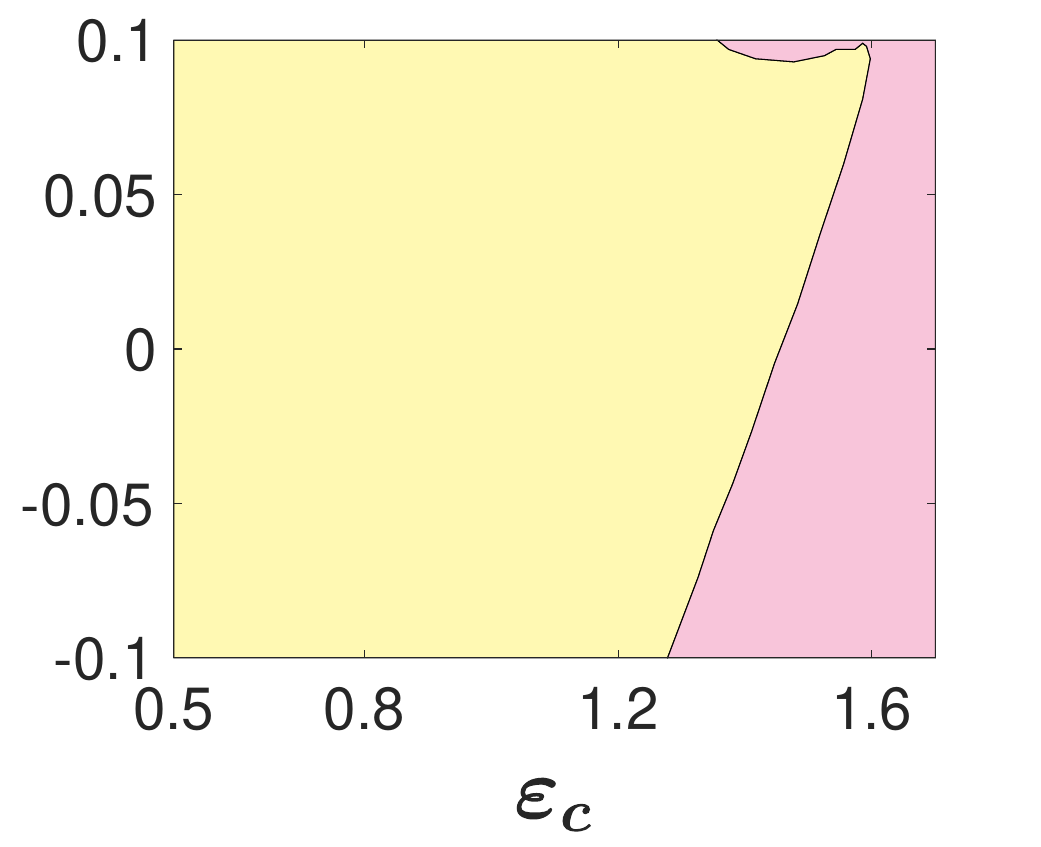}\label{fig:honeycomb_splitgrph_phase_lam_0p6_b4}}
    \subfloat[]{\includegraphics[scale=0.3,trim={2.99cm 6.6cm 1cm 7cm},clip]{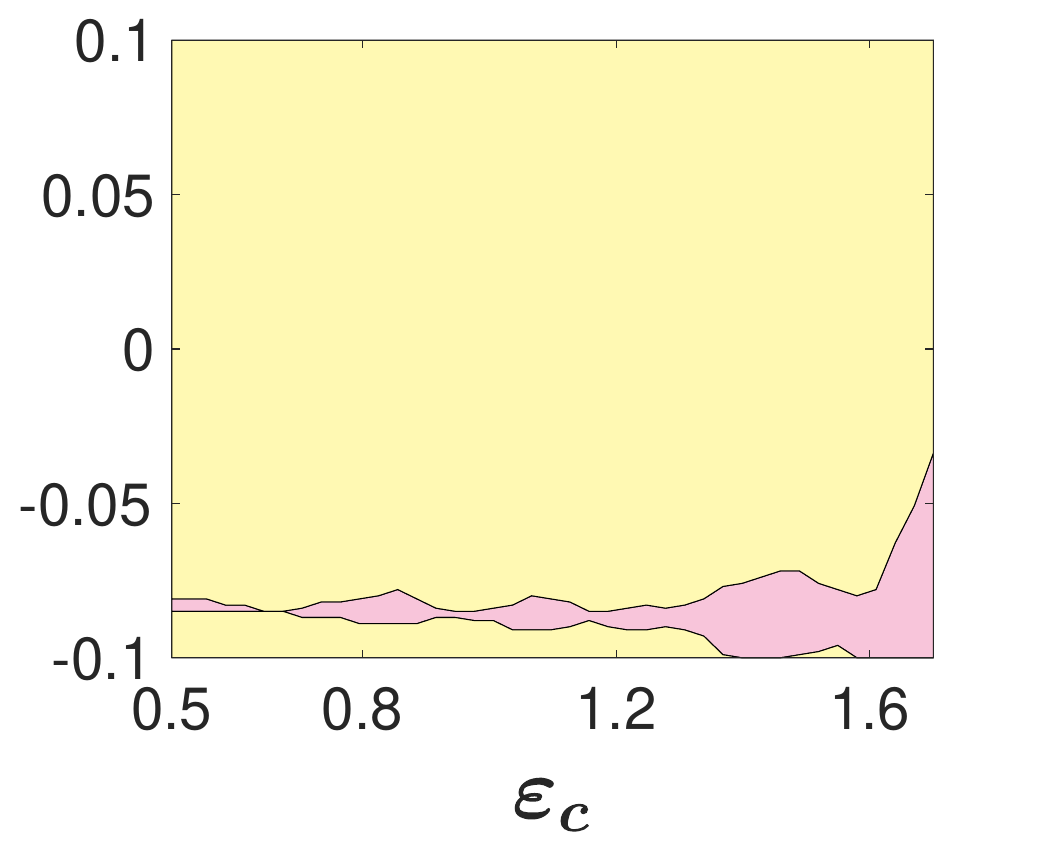}\label{fig:honeycomb_splitgrph_phase_lam_0p8_b4}}
    \caption{Topological phase diagram of honeycomb splitgraph lattice at $4/5$ filling as the function of strain in x-direction ($\epsilon_{xx}$) and on-site potential ($\varepsilon_c$) at (a) $\lambda_I = 0.4t$ (b) $\lambda_I= 0.6t$  and (c) $\lambda_I = 0.8t$. The different phases are distinguished by the colors as follows.  Yellow  (\tiyellow{}): topological band insulator and pink (\bipink{}): band insulator. }
    \label{fig:honeycomb_splitgraph_phase}
\end{figure}
Fig.~\ref{fig:honeycomb_splitgraph_phase} and Fig.~\ref{fig:honeycomb_splitgraph_phase_angle} presents a few noteworthy topological phase diagrams at a filling fraction of $4/5$, illustrating how the phases vary as a function of the magnitude of uniaxial deformation ($\epsilon_{xx}$), it's orientation ($\varphi$) and the on-site energy ($\varepsilon_c$). When $\lambda_I=0.4t$, the phase boundary shifts linearly as both $\epsilon_{xx}$ and $\varepsilon_c$ are increased (Fig.\ref{fig:honeycomb_splitgrph_phase_lam_0p4_b4}). The $\lambda_I = 0.4t$ case can also be particularity interesting because the middle three bands transform into isolated non-dispersive bands at equal onsite energies. Further, increasing the SOC to $\lambda_I=0.6t$ in Fig.~\ref{fig:honeycomb_splitgrph_phase_lam_0p6_b4}, the boundary shifts to higher value of $\varepsilon_c$ and also becomes non-linear at higher strain.  The phase boundary become completely non-trivial  at $\lambda = 0.8t$, where the trivial phase only appears at the compressive strain, as shown in \ref{fig:honeycomb_splitgrph_phase_lam_0p8_b4}.

In the HK lattice, altering the angle of the applied strain also influences the gap closing points. Interestingly, at the transition point $\varepsilon_c = 1.05$, the phase boundary is nearly parallel to x-axis, showing little to no dependence on the magnitude of the applied strain (Fig.~\ref{fig:honeycomb_splitgrph_phase_lam_0p2_ec_1p05_b4}). However, in the vicinity of $\varepsilon_c = 2$, the separation between the two phases changes it curvature from concave downwards (Fig.~\ref{fig:honeycomb_splitgrph_phase_lam_0p2_ec_1p0_b4}, at $\varepsilon_c = 1$) to convex (Fig.~\ref{fig:honeycomb_splitgrph_phase_lam_0p2_ec_1p10_b4}, at $\varepsilon_c = 1.1$).

\subsubsection{Line graphs \texorpdfstring{$\mathcal{L}(\mathcal{X})$}{Lg}: Checkerboard and Kagome lattices}
\begin{figure}
    \centering
   \subfloat[]{\includegraphics[scale =0.24,trim={9cm 1cm 6cm 1cm},clip]{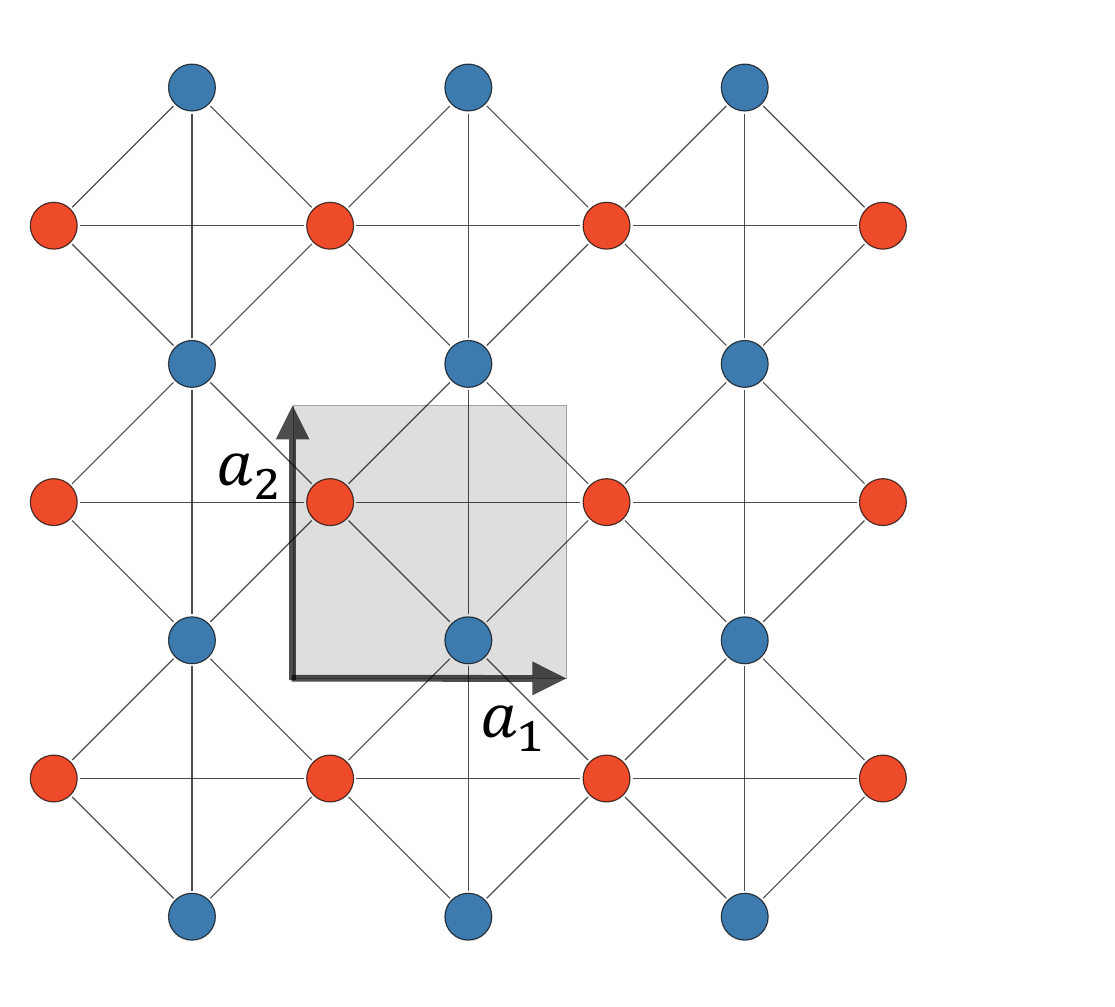}\label{fig:checkerboard_lattice}}
   \subfloat[]{\includegraphics[scale =0.28,trim={0cm 6.7cm 0cm 7cm},clip]{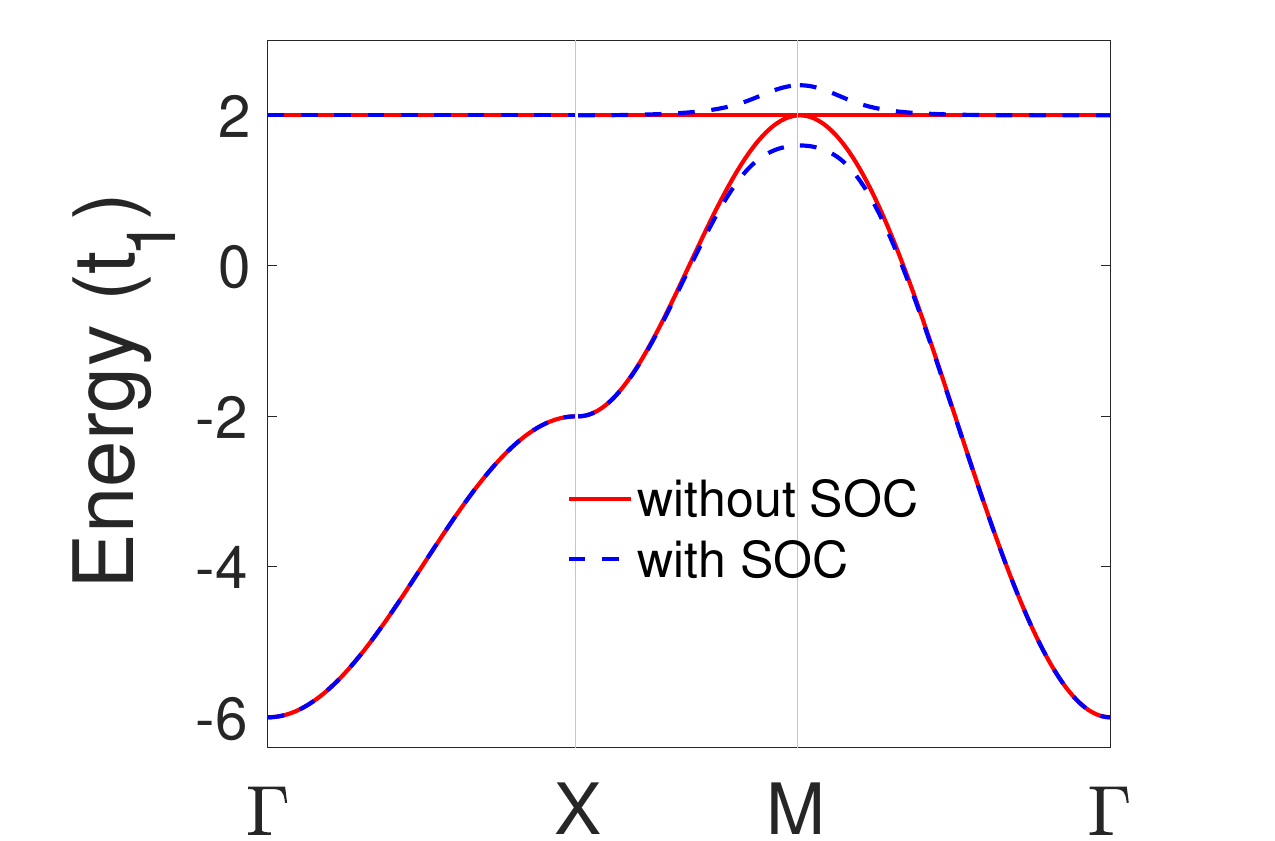} \label{fig:checkerboard_2dband}}
      \subfloat[]{\includegraphics[scale =0.28,trim={3.1cm 6.5cm 0cm 7cm},clip]{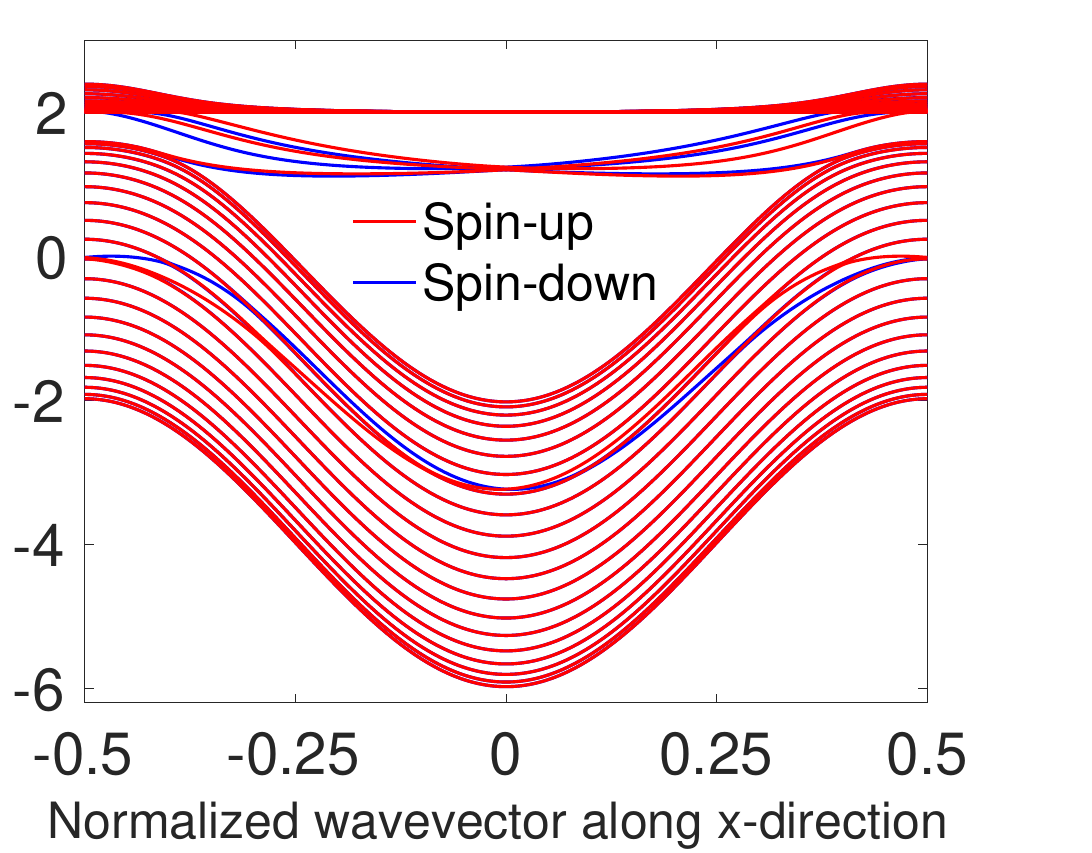} \label{fig:checkerboard_ribbon_soc}}
        \caption{ (a) The line graph of the square lattice, i.,e. the checkerboard lattice ($L(\mathcal{X}_4)$). The black arrows indicate the lattice vectors $\bfa_1$ and $\bfa_2$, and the gray region is the unit cell. The tight binding band diagram of (b) 2D lattice without (red solid line) and with (blue broken line) SOC ($\lambda_I = 0.1t_1$) and (c) 1D zigzag lattice nanoribbon with $\lambda_I = 0.1t_1$. The red and blue lines show the counter-propagating spin-up and spin-down states, respectively. }
    \label{fig:checkerboard_latt_w_bnds}
\end{figure}

\begin{figure}[!htbp]
    \centering
    \subfloat[]{\includegraphics[scale=0.24,trim={8cm 1cm 6cm 1cm},clip]{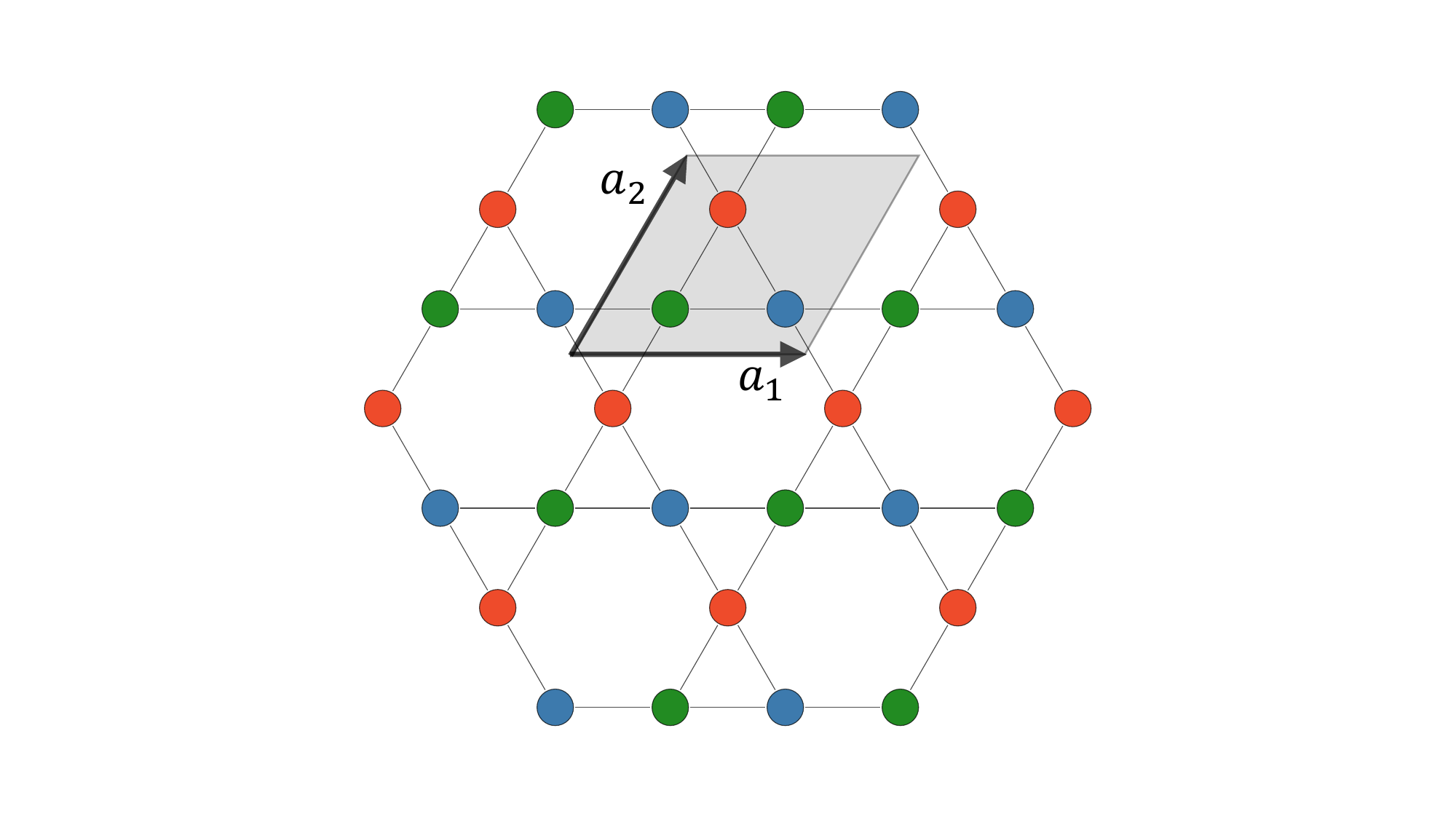}\label{fig:kagome_lattice}} 
    \subfloat[]{\includegraphics[scale =0.28,trim={0cm 6.7cm 0cm 7cm},clip]{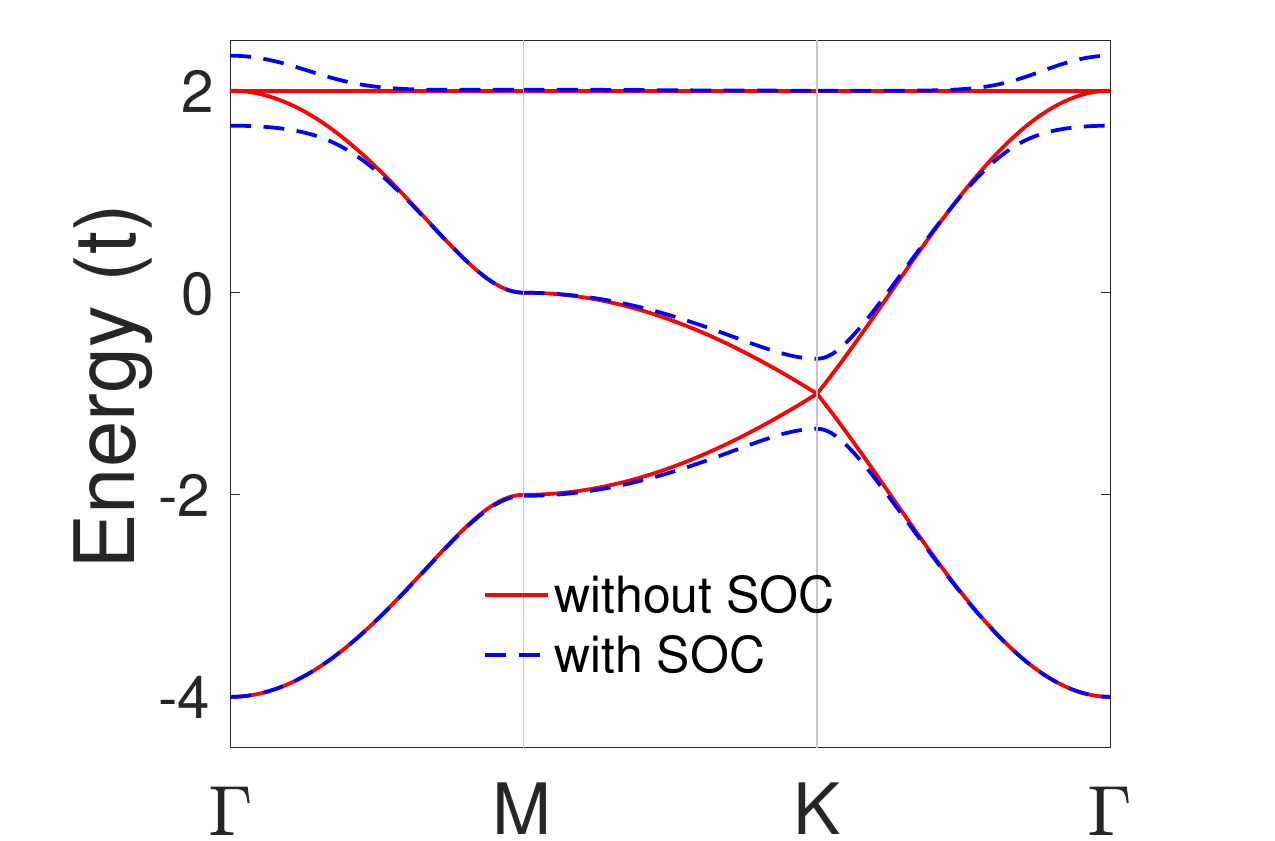} \label{fig:kagome_2dband}}
    \subfloat[]{\includegraphics[scale =0.28,trim={3.1cm 6.5cm 0cm 7cm},clip]{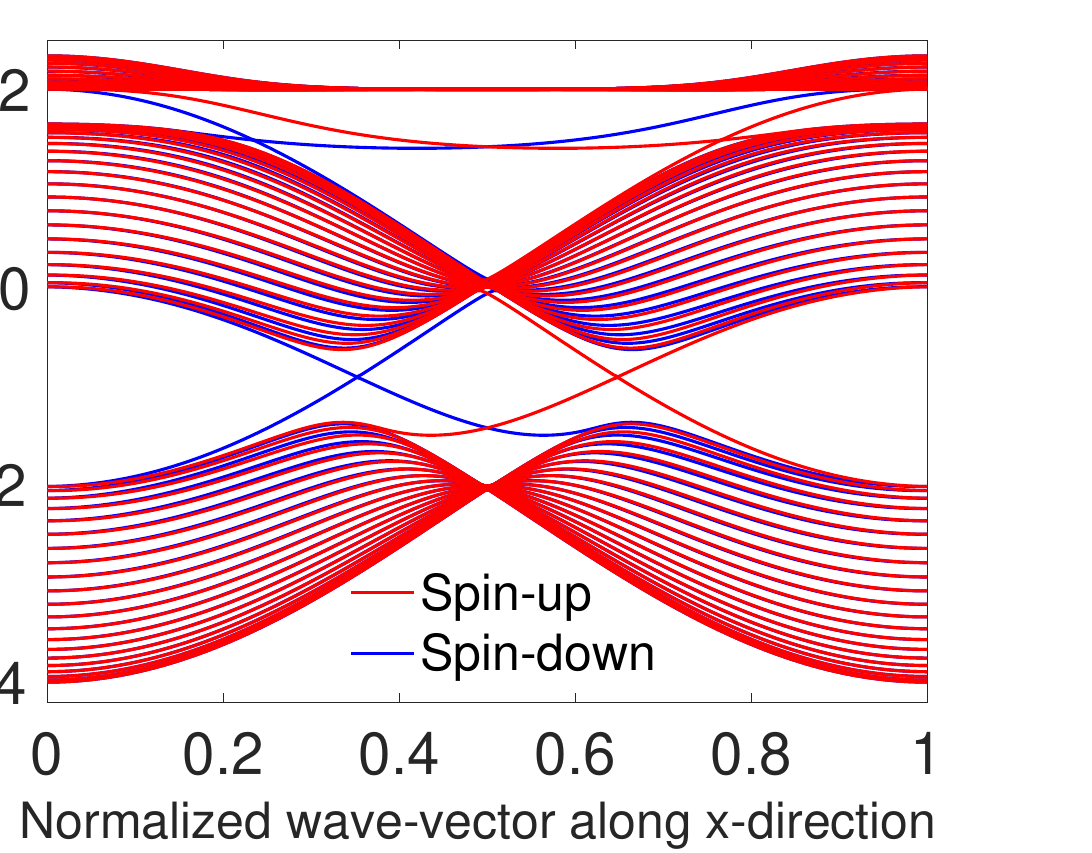} \label{fig:kagome_ribbon_soc0p1}}
    \caption{(a) The line graph of honeycomb lattice $L(\mathcal{X}_6)$ also known as Kagome lattice. The black arrows indicate the lattice vectors $\bfa_1$ and $\bfa_2$, and the gray region is the unit cell. The tight binding band diagram of (b) 2D lattice without (red solid line) and with (blue broken line) SOC ($\lambda_I = 0.1t$) and (c) 1D zigzag lattice nanoribbon with $\lambda_I = 0.1t$. The red and blue lines show the counter-propagating spin-up and spin-down states, respectively. }
    \label{fig:Kagome_latt_w_bnds}
\end{figure}

 Checkerboard ($\mathcal{L}(\mathcal{X}_4)$) and Kagome lattices ($\mathcal{L}(\mathcal{X}_6$)), which can be derived as line graphs of the parent lattices (${X}_4$ and  ${X}_6$) discussed above, have traditionally been a playground for understanding strongly correlated and topological properties of flat band materials. The coordination number of these graphs is $2d-2$ and we depict them here in Fig.~\ref{fig:checkerboard_lattice} and Fig.~\ref{fig:kagome_lattice}, respectively.
For completeness, we will now briefly review the electronic features of each lattice, although many of these results appear in the literature \cite{mojarro2023strain,sun2009topological,sun2011nearly,tang2011high,neupert2011fractional,jiang2019topological}. The energy spectrum of these lattices, given by the first equation in \ref{eq:spectrum}, has a flat band at $E = 2$ which touches a dispersive band parabolically. The touching point is stabilized by the $C_4$ and $C_6$ rotational symmetry of the checkerboard and Kagome unit cell, respectively. The low-energy Hamiltonian at this point can be written as \cite{sun2009topological,mojarro2023strain,rhim2019classification}:
\beq
\mathcal{H}(\bfk) = (k_x^2 + k_y^2)\mathbb{I} + 2k_xk_y\tau_x + (k_x^2 - k_y^2)\tau_z.
\label{eq:qbtp}
\eeq
Any symmetry breaking perturbations such as lattice distortion either creates a gap between these two bands or converts the quadratic touching point into tilted Dirac bands. For instance, the isolated bands in the above Hamiltonian can be generated by adding $m\tau_x$ to the Hamiltonian, whereas tilted Dirac cones appear due to the addition of $m\tau_z$ term ($m$ is the mass).

\underline{Checkerboard lattice:} The checkerboard lattice is bipartite with two atoms in the unit cell, as illustrated in Fig.~\ref{fig:checkerboard_lattice}. Due to its structural simplicity, it is usually employed as a toy model to study the fractional quantum hall effect \citep{li2014fractional,sun2011nearly, sheng2011fractional}. The four-band spinful Hamiltonian can be written as: 

\beqs
H_{L(\mathcal{X}_4)} = & &\left(\frac{t_1}{2}(1+\tau_z)\cos\bfk\cdot \bfa_1 + \frac{t_1}{2}(1-\tau_z)\cos \bfk\cdot \bfa_2 \right)\otimes\sigma_o\nonumber \\
&+& \left(\left(1+\sum_{i=1}^3\cos \bfk\cdot \bfa_i\right)\tau_x+\left(\sum_{i=1}^{3}(-1)^i\sin\bfk\cdot \bfa_i \right)\tau_y \right)\otimes(t_2\sigma_o+i\lambda_I\sigma_z).
\eeqs

Here, $t_1$ and $t_2$ are the hopping amplitude between the same and different atomic species, respectively. When $t_1$ and $t_2$ are equal, the spectrum contains a flat band across the entire BZ with quadratic band touching at the $M-$point. The dispersive band here is reminiscent of the square lattice spectrum (see Figs.~\ref{fig:checkerboard_latt_w_bnds} and \ref{fig:square_2dband}).  The band structure, in this case, is shown by red lines in Fig.~\ref{fig:checkerboard_2dband}. In contrast, when $t_1 \neq t_2$ a directional flat band appears from $X$ to the $M$ point. With non-zero $\lambda_I$, the bands become isolated and the flat band transform into a  quasi-flatband as represented by broken blue lines. The dispersion graph of checkerboard ribbon is drawn in Fig.~\ref{fig:checkerboard_ribbon_soc} at $\lambda_I = 0.2t_1$, where non-trivial edge states emerge in quasi-1D structure. We have not found any significant strain-induced effects on the topological properties, likely due to the simplicity of the checkerboard lattice. 

\underline{Kagome lattice:} The Kagome lattice consists of three sublattices arranged in the pattern of corner sharing triangles, as shown in Fig.~\ref{fig:kagome_lattice}. The explicit form of the TB Hamiltonain of the Kagome lattice, with spin degrees of freedom reads:
\beqs
H_{S(\mathcal{X}_4)}(\bfk)=& &\sum_{i=1}^{3}\varepsilon_iS_i\otimes\sigma_o + \Bigg(\sum_{i=1}^3 t_i(1+\cos\bfk\cdot\bfa_i)\Lambda_i -\sum_{i=4}^6t_{i-3}\sin\bfk\cdot\bfa_{i-3}\Lambda_i\Bigg)\otimes\sigma_o \nonumber \\
&+& \lambda_I\Big(\sum_{i=4}^6(\cos\bfk\cdot\bfa_{i+2}+\cos\bfk\cdot\bfa_{i+1})\Lambda_{i} + \sum_{i=1}^3((-1)^i\sin\bfk\cdot\bfa_{i+2} \nonumber\\ 
&+&\sin\bfk\cdot\bfa_{i+1})\Lambda_{i}\Big)\otimes\sigma_z,\quad \quad
\eeqs

In the spectrum of the Kagome lattice dispersionless and Dirac bands --- inherited from  honeycomb lattice --- coexist, as depicted by the red lines in Fig.~\ref{fig:kagome_2dband}. The flat band touches the Dirac band quadratically at the $\Gamma-$point at $E = 2$. With the inclusion of SOC, a gap opens up between all three bands with a magnitude of $\Delta = 4\sqrt{3}|\lambda_I|$ and the system turns topological. The flat band also becomes slightly dispersive, as represented by the broken blue line in Fig.~\ref{fig:kagome_2dband}. The corresponding quasi-1D Kagome nanostrip band diagram at $\lambda_I = 0.2t$ is shown in Fig.~\ref{fig:kagome_ribbon_soc0p1}, which clearly reveals helical edge states crossing the gap at $1/3$ and $2/3$ fillings. 

The coexistence of these two types of bands (dispersionless and linear dispersion) makes the Kagome lattice intriguing, and much research has been dedicated to exploring a range of emergent fundamental physical phenomena such as superconductivity, fractional quantum hall effect, quantum spin liquid and topological phases, in this lattice. The effect of strain on the electronic and topological properties of the Kagome lattice has also been rigorously studied. We guide the reader to the literature  \cite{sun2009topological,mojarro2023strain,bolens2019topological,rhim2019classification}, instead of presenting these known results here again. Notably, the structural and electronic properties of line graph of hexagonal graph (Kagome lattice) and split graph of square lattice (Lieb lattice) are inter-convertible by applying strain in the diagonal direction \cite{jiang2019topological}.

\subsection{Second generation lattices}
Second generation lattices are generated by applying a combination of split and line graph operations to the parent graph. Here, we have analyzed line of split graphs $L(S(\mathcal{X}))$ and split of line graph $S(L(\mathcal{X}))$.For each configuration, we construct two distinctive structures based on $4$-regular square and 
$3$-regular hexagonal cases. These lattices inherit many of the electronic and topological features from their precursors. However, due to bigger unit cell and more system parameters to tune, the phase diagrams of second generation lattices are richer than structures considered earlier. For the sake of brevity, we have not pursued investigations into line-of-line-graph $L(L(\mathcal{X}))$ or split-of-split-graph $S(S(\mathcal{X}))$ lattices, although we present their band diagrams in  \cref{appendix_2}. These configurations do not appear as frequently in the literature (although we are aware of some earlier contributions \citep{liu2021localization}), making them worthy of future investigations. 
\subsubsection{Line graph of split graphs  \texorpdfstring{$L(\mathcal{S}(\mathcal{X}))$}{Lg} : Square-Octagon and decorated honeycomb lattices}

\begin{figure}[!htbp]
{
    \centering
     \subfloat[]{\includegraphics[scale=0.26,trim={9cm 1cm 6cm 1cm},clip]{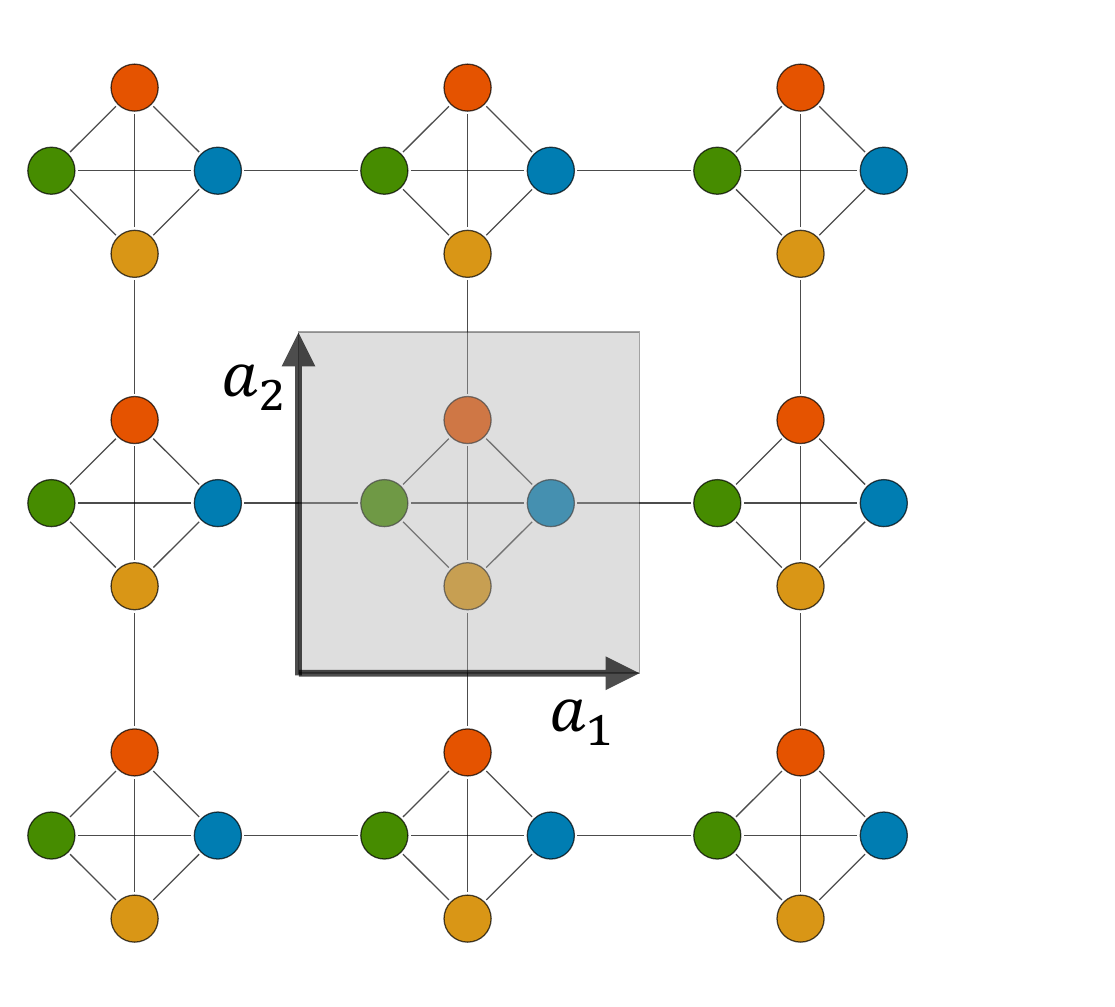}\label{fig:sq_oct}}
     \subfloat[]{\includegraphics[scale =0.28,trim={0cm 6.7cm 0cm 7cm},clip]{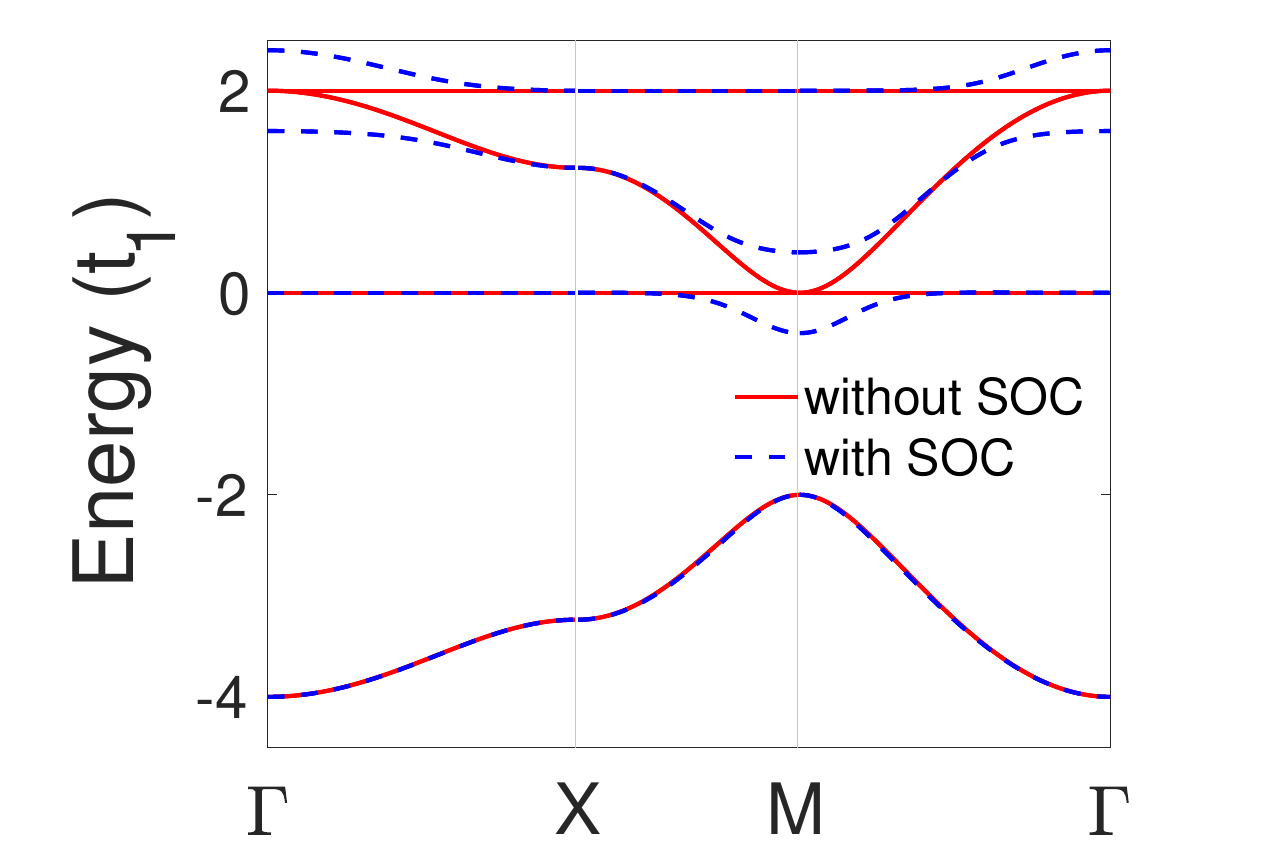}
\label{fig:square_octagon_2dband}}
    \subfloat[]{\includegraphics[scale =0.28,trim={3.1cm 6.5cm 0cm 7cm},clip]{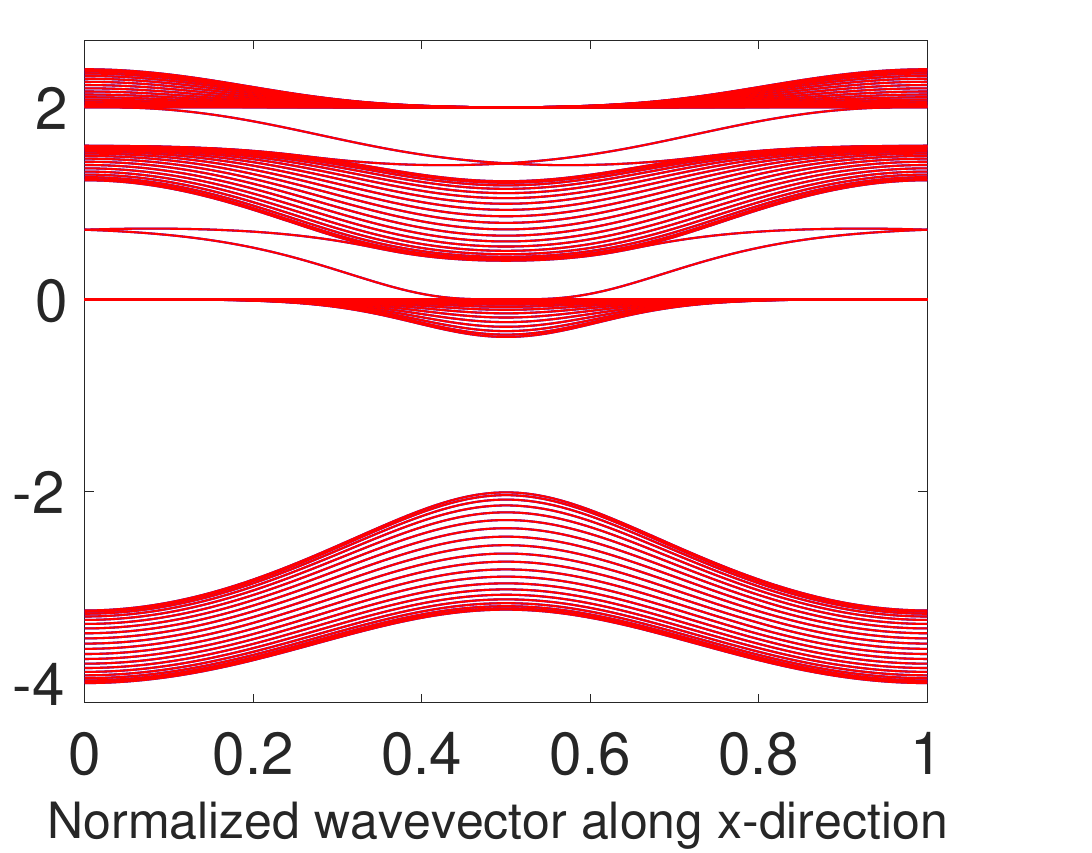}
\label{fig:square_octagon_ribbon}}
}
\caption{(a) The line graph of the Lieb lattice is the square octagon lattice, $L(S(\mathcal{X}_4))$. The black arrows indicate the lattice vectors $\bfa_1$ and $\bfa_2$, and the gray region is the unit cell. The tight binding band diagram of (b) 2D lattice without (red solid line) and with (blue broken line) SOC ($\lambda_I = 0.1t$) with $t = t_1 = t_2 = t_3$ and (c) 1D zigzag lattice nanoribbon with $\lambda_I = 0.1t$. The red and blue lines show the counter-propagating spin-up and spin-down states, respectively. }
    \label{fig:sq-oct_latt_w_bnds}
\end{figure}

\begin{figure}[!htbp]
    \centering
    \subfloat[]{\includegraphics[scale=0.26,trim={6cm 1cm 6cm 1cm},clip]{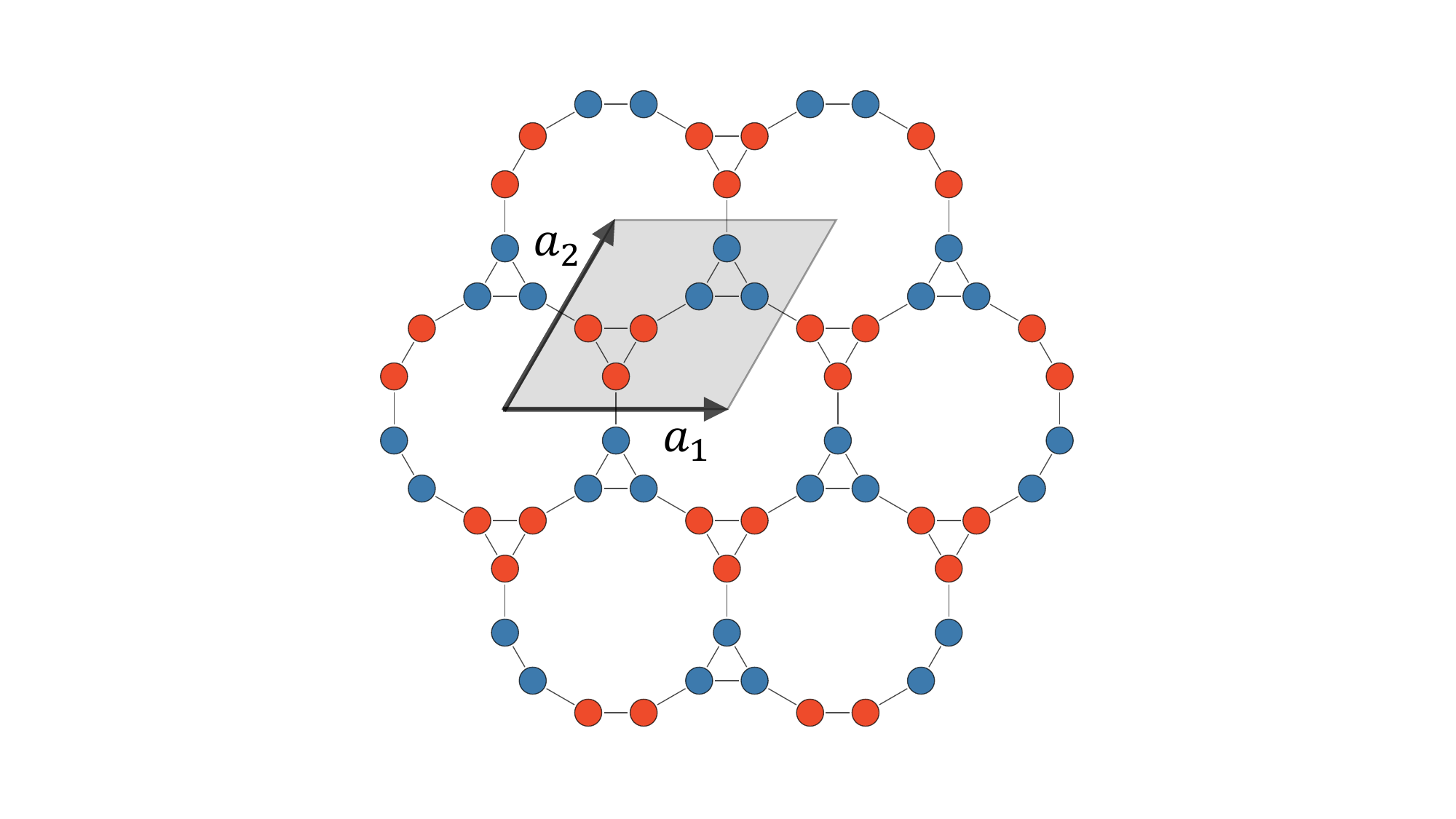}\label{fig:decorated_honeycomb_lattice}} 
    \subfloat[]{\includegraphics[scale =0.28,trim={1cm 6.7cm 1cm 7cm},clip]{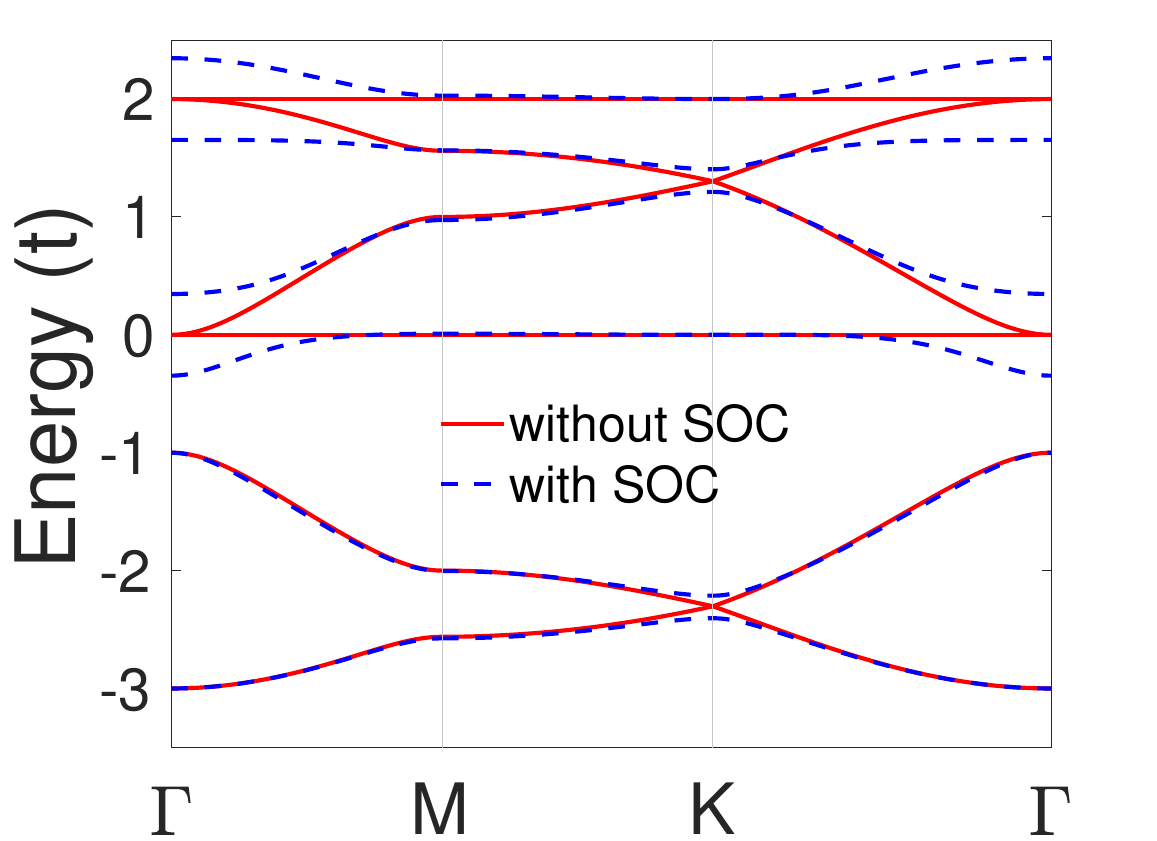}
\label{fig:decorated_honeycomb_2dband}} 
    \subfloat[]{\includegraphics[scale =0.28,trim={3.1cm 6.5cm 1cm 7cm},clip]{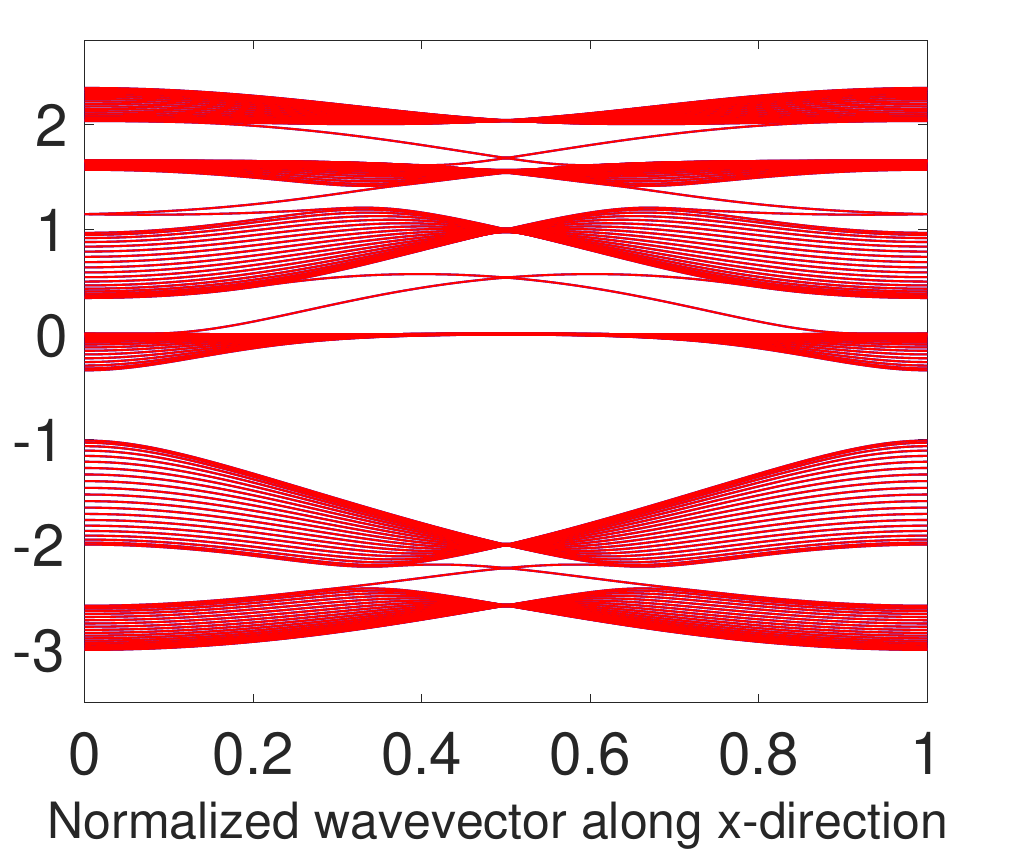}
\label{fig:decorated_honeycomb_ribbon}} 
        \caption{(a) Line graph of honeycomb split graph $L(S(\mathcal{X}_6))$, also called the decorated honeycomb lattice. The black arrows indicate the lattice vectors $\bfa_1$ and $\bfa_2$, and the gray region is the unit cell. The tight binding band diagram of (b) 2D lattice without (red solid line) and with (blue broken line) SOC ($\lambda_I = 0.1t$) and (c) 1D zigzag lattice nanoribbon with $\lambda_I = 0.1t$.  Here $t = u = v$.}
    \label{fig:decorated_hnycmb_latt_w_bnds}
\end{figure}

Such graphs are realized by starting from the 
$d$-regular parent graphs $\mathcal{X}$, then taking the $d,2$-biregular graph $\mathcal{S}(\mathcal{X})$ and finally applying line graph operations to get $L(\mathcal{S}(\mathcal{X}))$, which is also  $d$-regular. The schematics of $L(\mathcal{S} (\mathcal{X}))$ lattices, namely square-octagon and decorated honeycomb lattices, are illustrated in Fig.~\ref{fig:sq_oct} and Fig.~\ref{fig:decorated_honeycomb_lattice}, respectively. 

The spectrum of these graphs is given by last equation in \ref{eq:spectrum}. They possess gapless flat bands at $E = 0$, while $E = 2$ touches the dispersive band parabolically. Additional bands are also inherited from the respective predecessors. Here too, the quadratic band touching point is protected by the lattice's rotational symmetry and the description surrounding equation \ref{eq:qbtp} also applies. 

\underline{Square Octagon lattice:} First, we analyze the square-octagon lattice, which is the line graph of the checkerboard lattice, $L(S(\mathcal{X}_4))$. The basic building block of the lattice is a square plaquette consisting of four atoms, as illustrated in Fig.~\ref{fig:sq_oct}. Recently, the lattice has attracted significant attention due to a plethora of novel phases predicted in this system, including quantum magnetic phases\cite{bao2014quantum}, topological insulating phases \cite{kargarian2010topological,wang2021topological,yang2018topological, he2023dirac} and high-temperature superconductivity states \cite{kang2019single}. The spinful single-orbital per site TB Hamiltonian of the square-octagon lattice takes the form:
\begin{align}
\nonumber
H_{L(S(\mathcal{X}_4))}(\bfk) &= \Big(t_1(\Gamma_{01} + \Gamma_{11})+\frac{1}{2}\big((t_3+t_2\cos\bfk\cdot\bfa_1)(\Gamma_{10}+\Gamma_{13}) \\
& +(t_3+t_2\cos\bfk\cdot\bfa_2)(\Gamma_{10}-\Gamma_{13})) \nonumber \\ 
&-\sin\bfk\cdot\bfa_1(\Gamma_{20}+\Gamma_{23})+\sin\bfk\cdot\bfa_2(\Gamma_{20}-\Gamma_{23})\Big)\otimes\sigma_0 \nonumber \\
&+ i\lambda_I\big((e^{i\bfk\cdot\bfa_1}+\cos\bfk\cdot\bfa_2)(\Gamma_{12}-\Gamma_{02})-i\sin\bfk\cdot\bfa_2\Gamma_{12}\big)\otimes\sigma_z\,.
\end{align}
Here, the matrices $\Gamma_{ij} = \tau_i\otimes\gamma_j$, and the hopping parameters are as follows: $t_1$ is along the sides of the square, $t_2$ is between two adjacent squares and $t_3$ is the hopping amplitude connecting the diagonals of the squares. As discussed earlier, the band structure of $L(S(\mathcal{X}_4))$ in the ideal isotropic case ($t_1 = t_2 = t_3$) contains a couple of flat bands at $E = 0$ and $E = 2$, as well as dispersive bands. One of the dispersive bands is sandwiched between the two flat bands, touching quadratically the lower and upper flat band at $M$ and $\Gamma$ points, respectively (Fig.~\ref{fig:square_octagon_2dband}). The other dispersive band at $1/4$ filling, resembles that of the parent square lattice ($\mathcal{X}_4$), and is separated from the $E = 0$ flat band by the gap of $\Delta = 2t_1$. 

\begin{figure}[!htbp]
    \centering
     \subfloat[]{\includegraphics[scale=0.3,trim={0.2cm 6.5cm 1.2cm 7cm},clip]{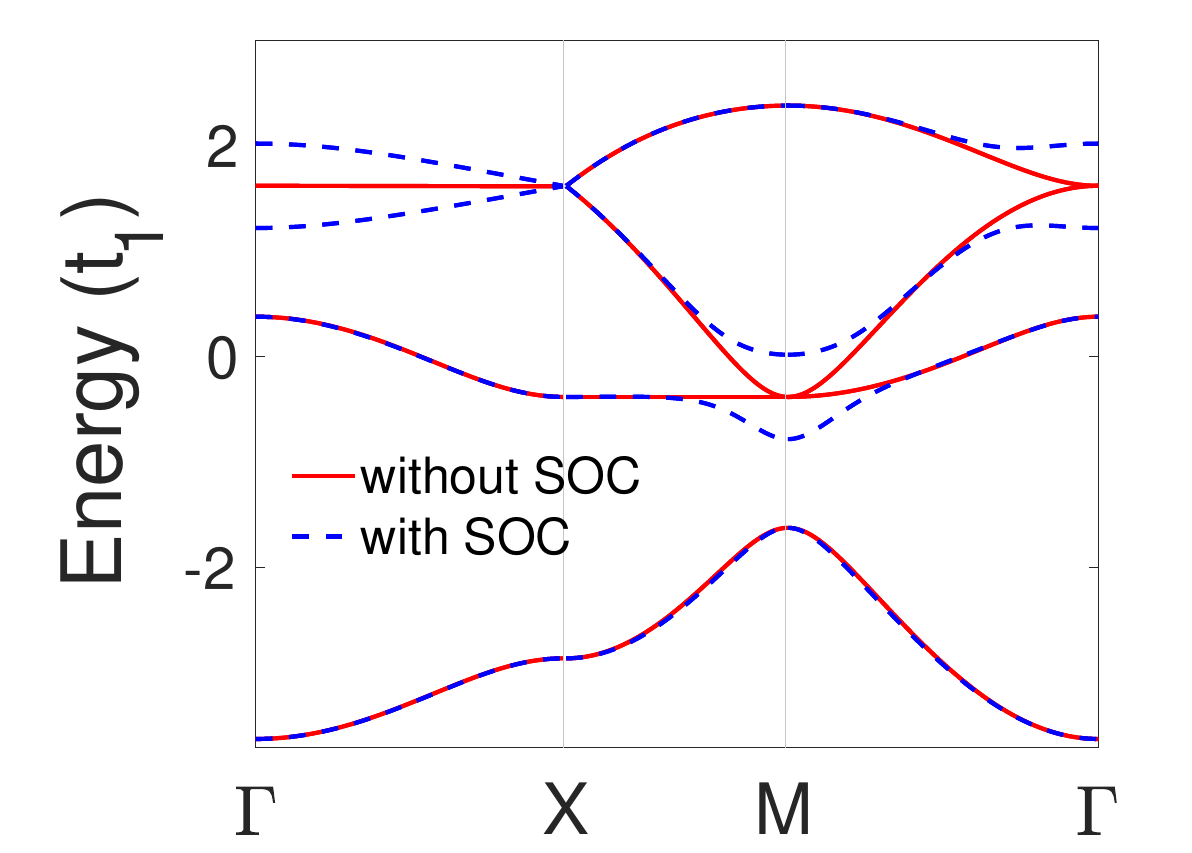}\label{fig:sq_oct_band_0p62}}
     \subfloat[]{\includegraphics[scale =0.3,trim={3cm 6.5cm 1.2cm 7cm},clip]{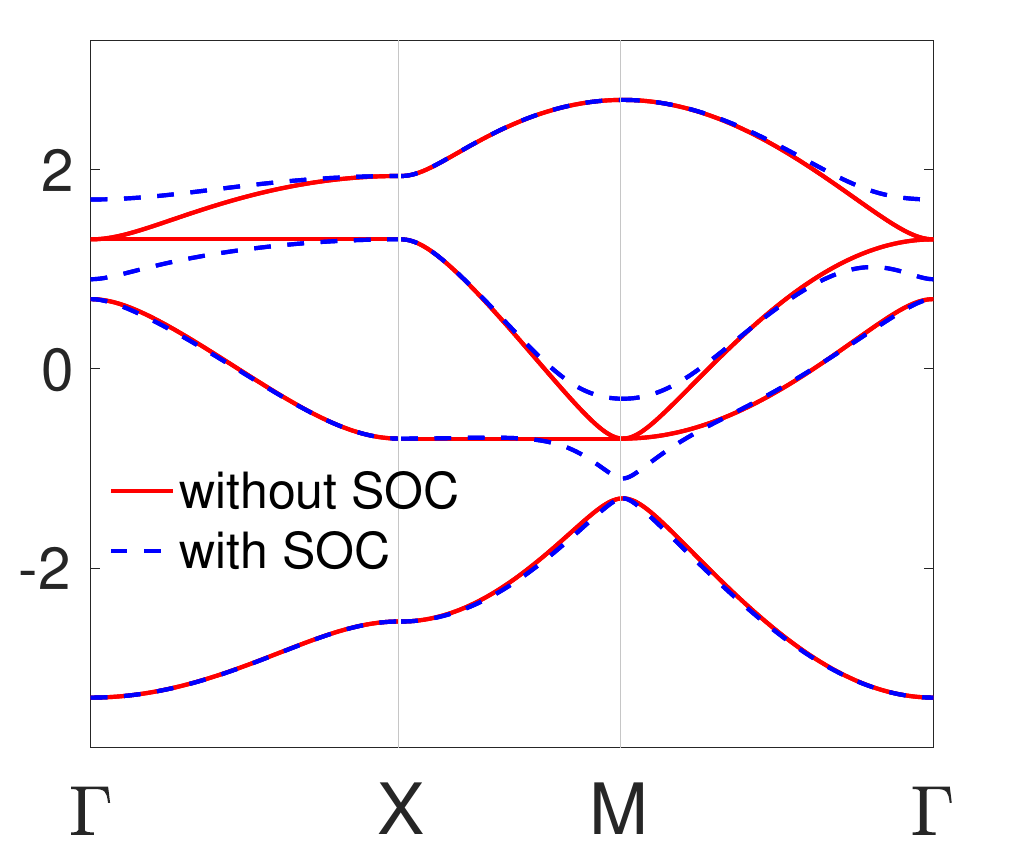}
\label{fig:sq_oct_band_0p3}}
    \subfloat[]{\includegraphics[scale =0.3,trim={3cm 6.7cm 1.2cm 7cm},clip]{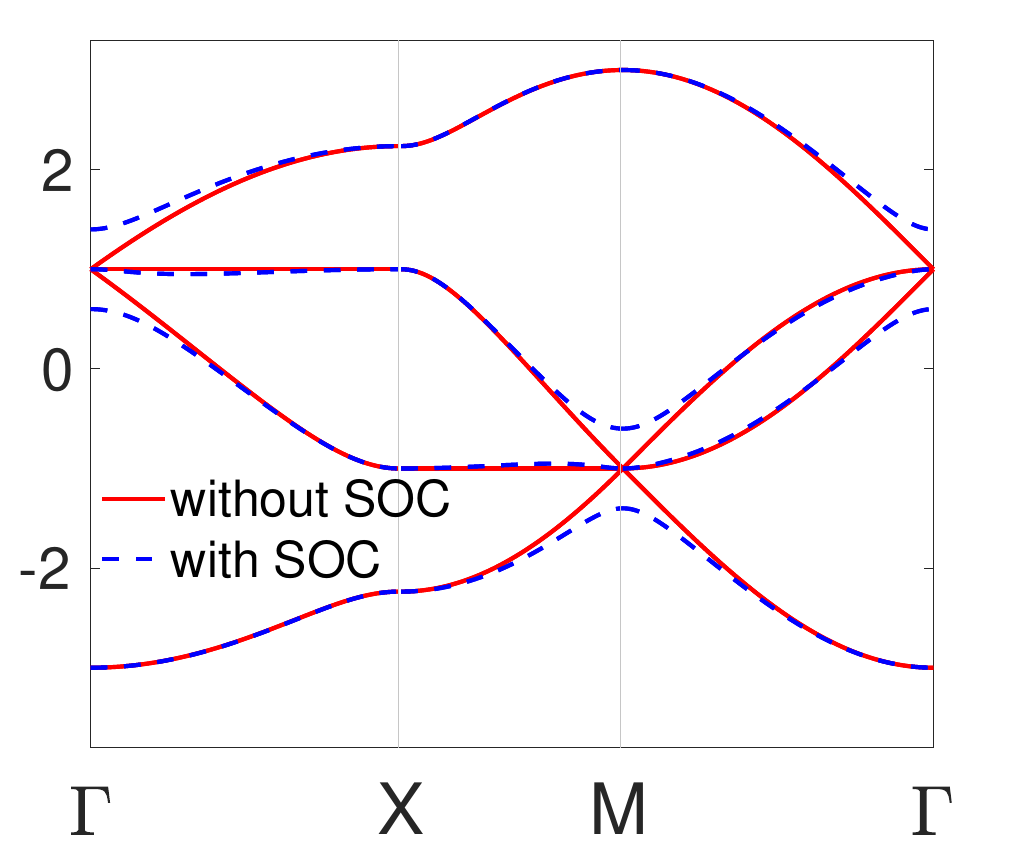}
\label{fig:sq_oct_band_0p0}}\\
     \subfloat[]{\includegraphics[scale=0.3,trim={0.2cm 6.5cm 1.2cm 7cm},clip]{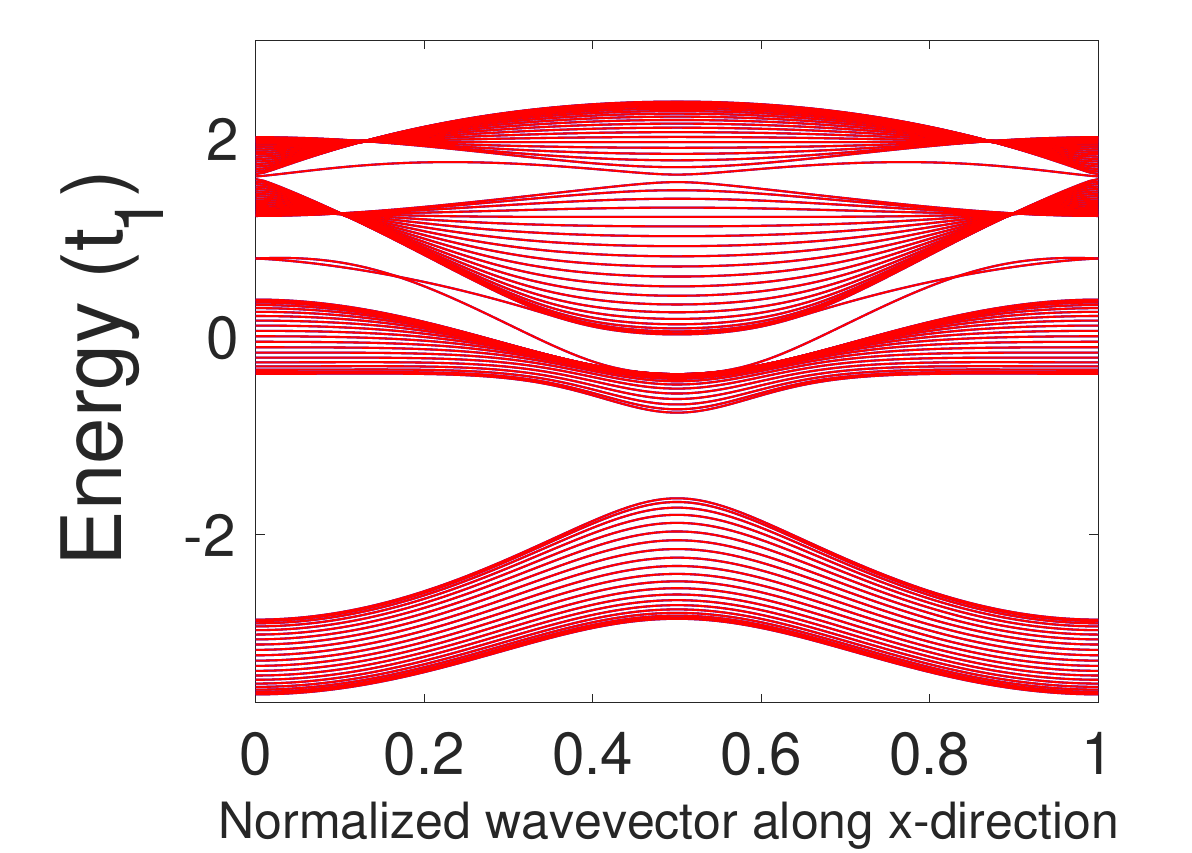}\label{fig:sq_oct_band_ribbon_0p7}}
     \subfloat[]{\includegraphics[scale =0.3,trim={3cm 6.5cm 1.2cm 7cm},clip]{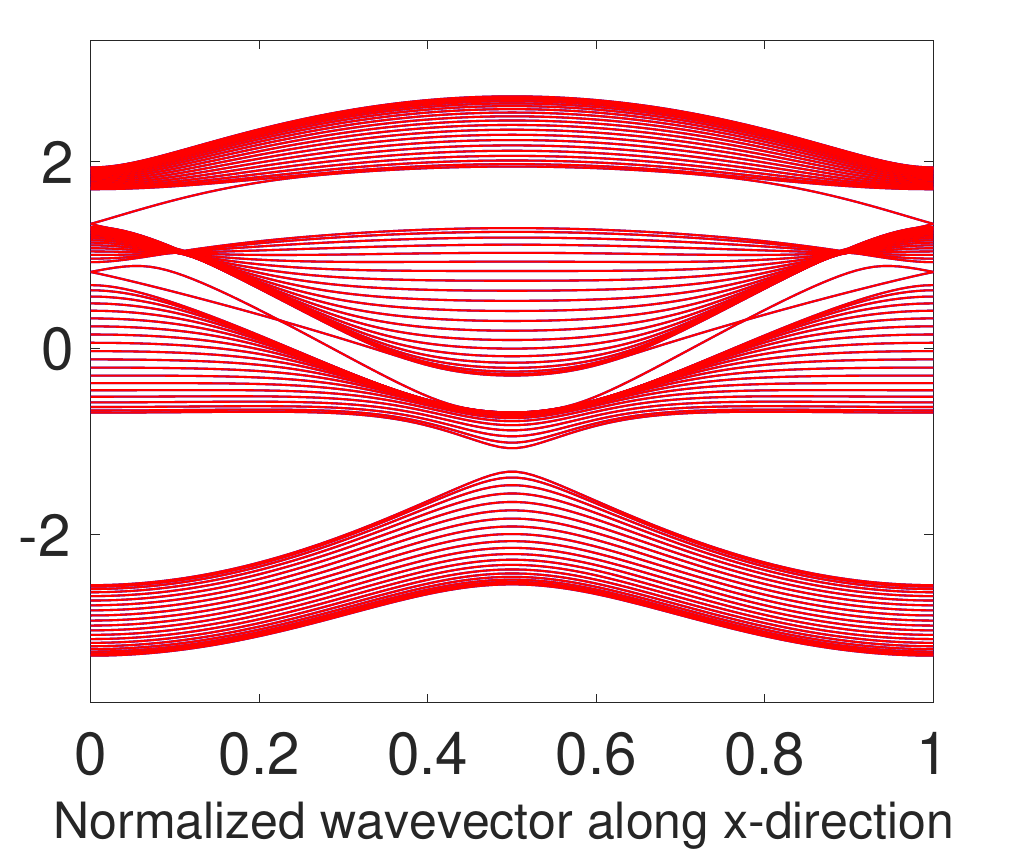}
\label{fig:sq_oct_band_ribbon_0p3}}
    \subfloat[]{\includegraphics[scale =0.3,trim={03cm 6.5cm 1.2cm 7cm},clip]{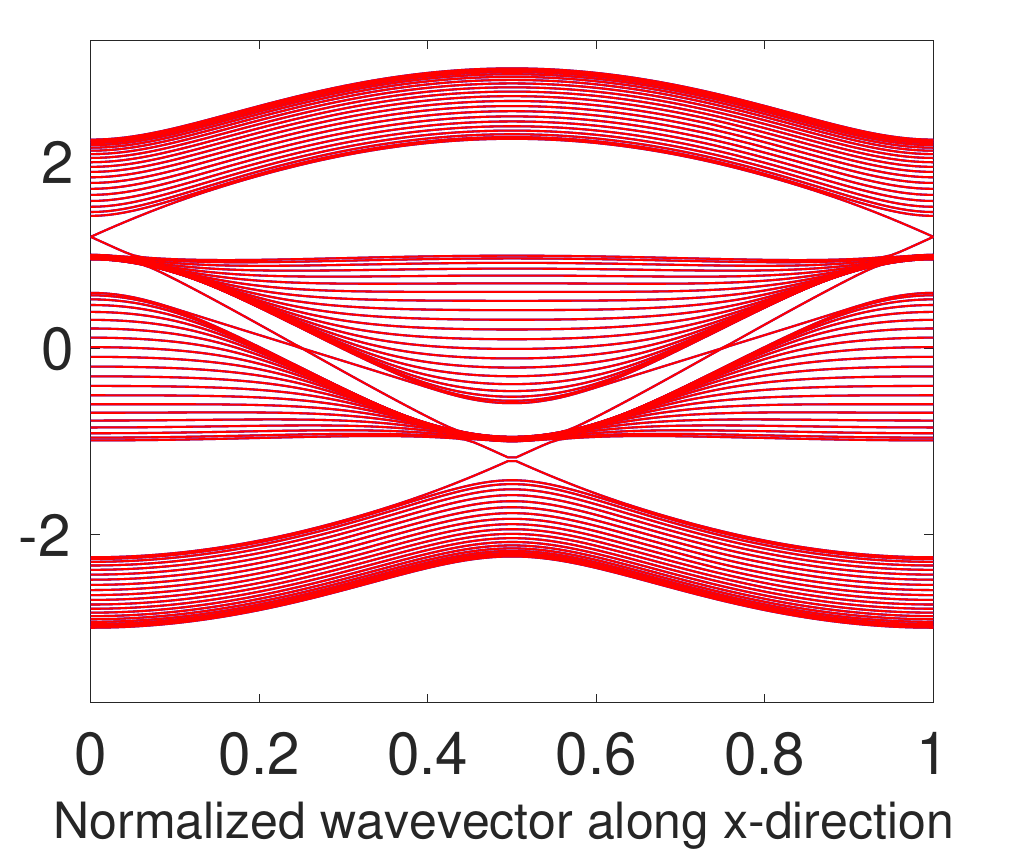}
\label{fig:sq_oct_band_ribbon_0p0}}
    \caption{Evolution of diagonal connection square-octagon to no interaction between the diagonals. The 2D band diagrams with (rid lines) and without SOC (broken blue lines) at (a) $t_3 = 0.62 t_1$, (b) $t_3 = 0.3 t_1$ and (c) $t_3 = 0$. (d) to (f) Corresponding quasi-1D band ribbon band diagrams. The spin orbit coupling parameter is $\lambda_I = 0.1t_1$.}
    \label{fig:sq-oct_latt_w_bnds_evo}
\end{figure}

For the square octagon lattice, the inclusion of diagonal hopping along with isotropic hopping energies is critical for generating a flat band across the whole BZ. Interestingly, when $t_3\neq t_1$, the band structure features various emergent  quantum phases, including Lieb lattice like pseudospin-1 Dirac cones with directional flat bands. The cases when $t_3<t_1$ are particularly intriguing (Fig.~\ref{fig:sq-oct_latt_w_bnds_evo}). For example, at $t_3 = 0.62 t_1$,  the third and fourth bands become flat and degenerate along the entire $\Gamma-X$ path, while the second band is dispersionless along $X-M$, with a parabolic touching with the  third band at $M$ (Fig.~\ref{fig:sq_oct_band_0p62}). As $t_3$ decreases further, the third band become directionally dispersionless, as shown in Fig.~\ref{fig:sq_oct_band_0p3}. Finally, when $t_3=0$, the band structure becomes  chiral symmetric as the triply degenerate Dirac band crossing at the $\Gamma$ point also appears at the $M$ point (Fig.~\ref{fig:sq_oct_band_0p0}). In this limiting case, the Hamiltonian satisfies the symmetry relation $ H_{L(S(\mathcal{X}_4))}[k_x+\pi, k_y + \pi] = - H_{L(S(\mathcal{X}_4))}[k_x,k_y]$. A mathematical analysis of the low energy expressions of band structures in the two limits are given in \cite{yamashita20133,oriekhov2021orbital,sil2019emergence}.

With the inclusion of SOC, the gap between the degenerate bands open up, e.g. the case of $\lambda_I = 0.1t_1$ is shown in \cref{fig:square_octagon_2dband,fig:sq_oct_band_0p62,fig:sq_oct_band_0p3,fig:sq_oct_band_0p0}, and is represented by dashed blue lines. In the isotropic limit, the lattice at $1/2$ and $3/4$ filling fractions show a QSH state while the first band remains a trivial insulator.  Fig.~\ref{fig:square_octagon_ribbon} shows the  spectrum of square-octagon strip geometry where the helical edge states can be observed at time-reversal invariant momentum $0$ and $\pi$, at $1/2$ and $3/4$ filling fractions, respectively. On decreasing $t_3$, we observed that the second band transforms into a semimetallic state and at $t_3 = 0$ the edge states appear only in the lowest band and third band, as depicted in \cref{fig:sq_oct_band_ribbon_0p7,fig:sq_oct_band_ribbon_0p3,fig:sq_oct_band_ribbon_0p0}. 

\begin{figure}[!htbp]
    \centering
    \subfloat[]{\includegraphics[scale=0.3,trim={1cm 7cm 1cm 7cm},clip]{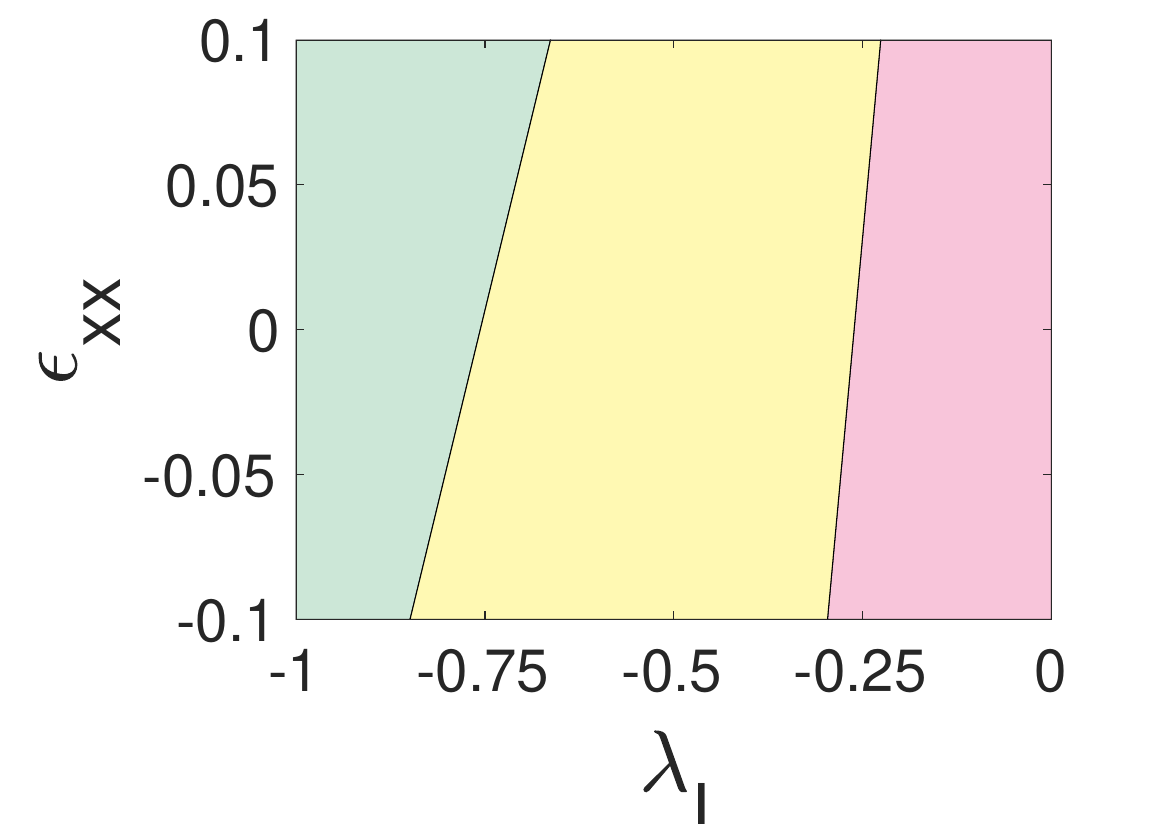}\label{fig:sq_oct_phase_t2_0p5}}
    \subfloat[]{\includegraphics[scale=0.3,trim={3.1cm 7cm 1cm 7cm},clip]{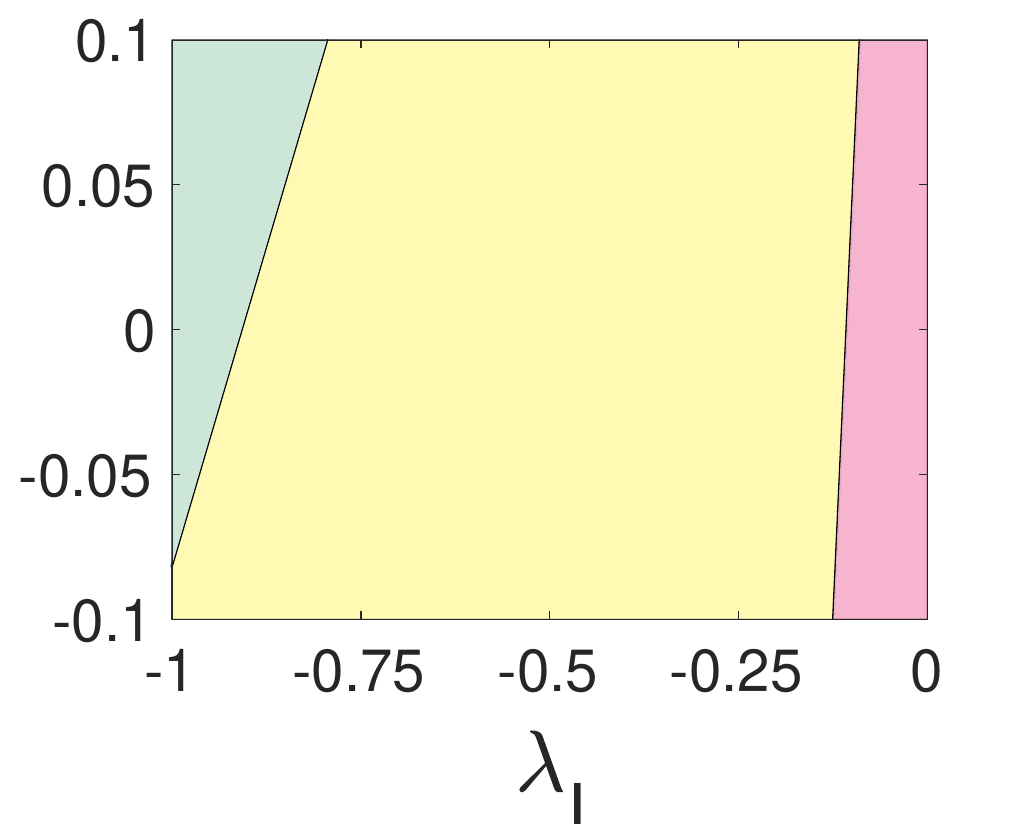}\label{fig:sq_oct_phase_t2_0p8}}
   \subfloat[]{\includegraphics[scale=0.3,trim={3.1cm 7cm 1cm 7cm},clip]{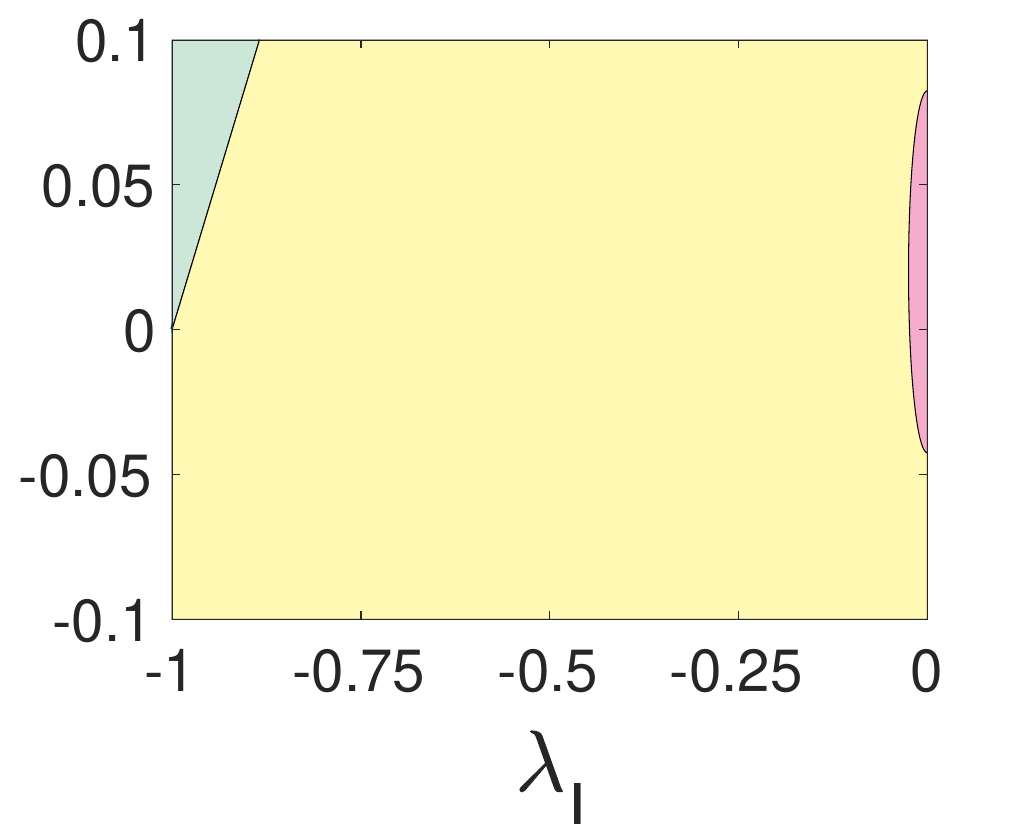}\label{fig:sq_oct_phase_t2_1p0}}\\
     \subfloat[]{\includegraphics[scale=0.3,trim={1cm 6cm 1cm 7cm},clip]{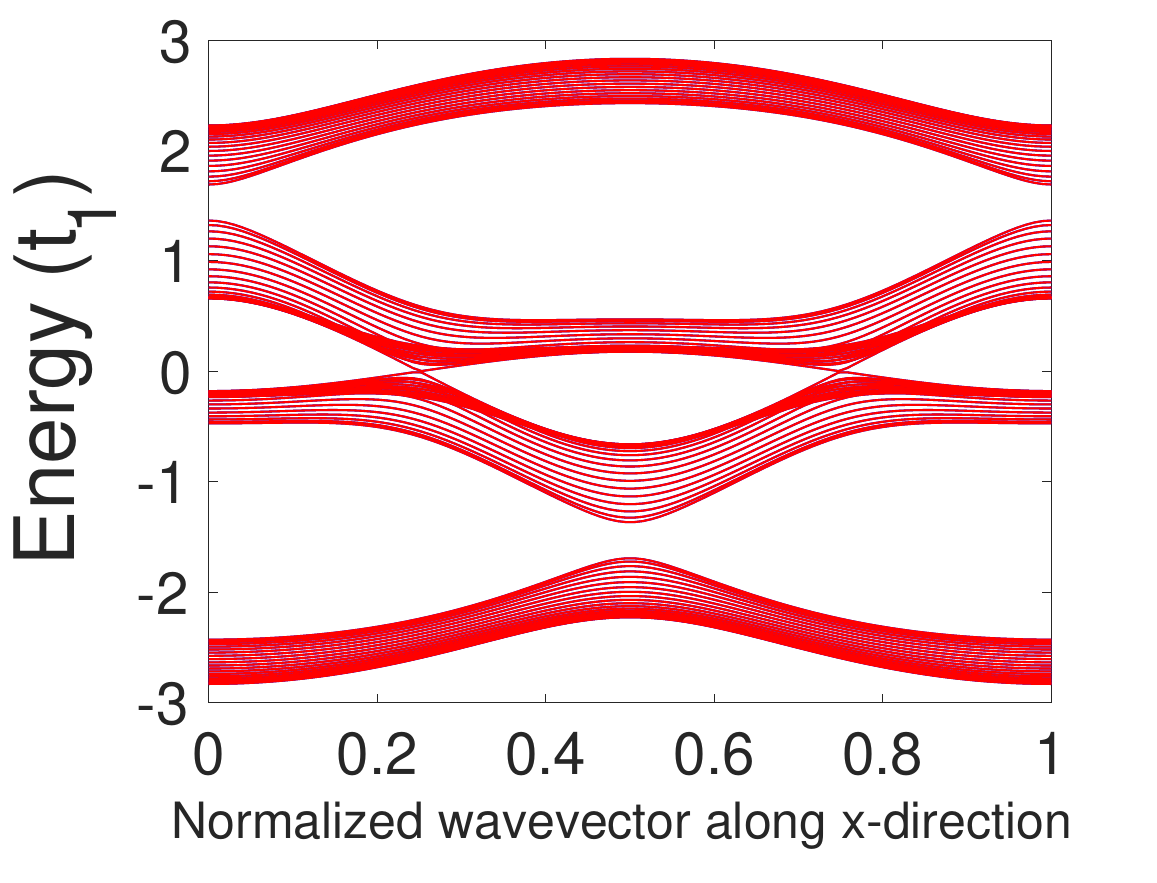}\label{fig:sq_oct_band_t2_0p5}}
    \subfloat[]{\includegraphics[scale=0.3,trim={3.1cm 6cm 1cm 7cm},clip]{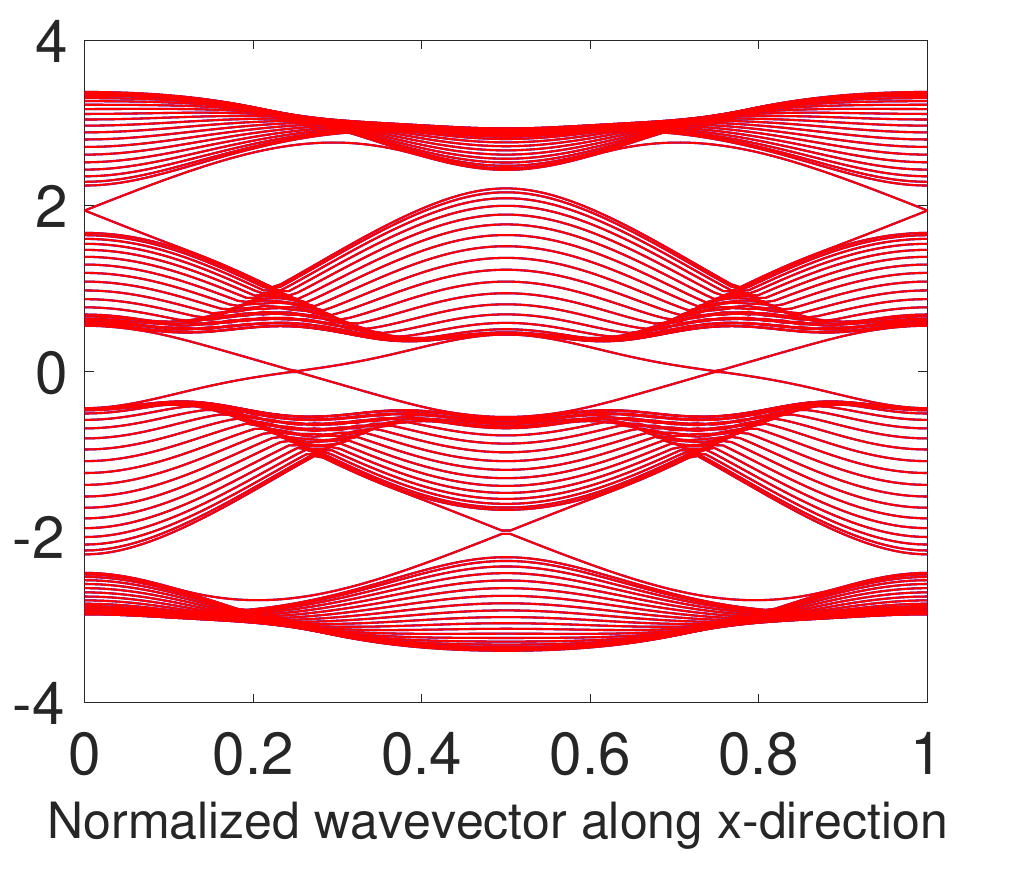}\label{fig:sq_oct_band_t2_0p5_lam_0p7_exx_neg0p1}}
    \subfloat[]{\includegraphics[scale=0.3,trim={3.1cm 6cm 1cm 7cm},clip]{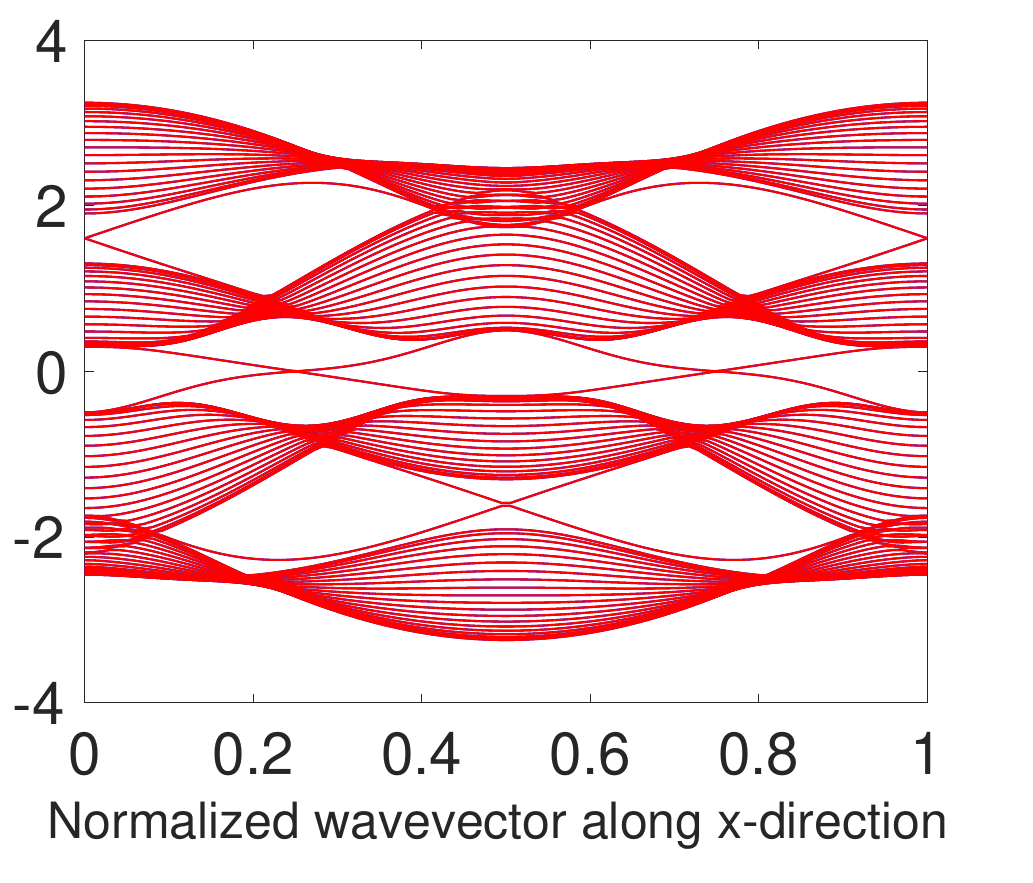}\label{fig:sq_oct_band_t2_0p5_lam_07_exx_0p1}}
    \caption{Topological phase diagram of square octagon lattice at $1/4$ filling in uniaxial strain ($\epsilon_{xx}$) and intrinsic SOC ($\lambda_I$) space with (a) $t_2 = 0.5t_1$ (b) $t_2 = 0.8t_1$  and (c) $t_2 = t_1$. The different phases are distinguished by the colors as follows. Pale mint (\semicyan{}): semimetal, yellow  (\tiyellow{}): topological band insulator and pink (\bipink{}): band insulator. From (d) to (f) the square octagon nonoribbon band diagrams at  $t_2 = 0.5$  and (d) $\epsilon_{xx} = -0.1$, $\lambda_{I} = -0.2$, (e) $\epsilon_{xx} = -0.1$, $\lambda_I = -0.7$, and (e) $\epsilon_{xx} = 0.1$, $\lambda_I = -0.7$.  }
    \label{fig:enter-label}
\end{figure}

Interestingly, as shown in \cref{fig:sq_oct_phase_t2_0p5,fig:sq_oct_phase_t2_1p0,fig:sq_oct_phase_t2_0p8}, the square octagon lattice exhibits three distinct quantum phases at $1/4$ filling in the strain-SOC space for different values of $t_2$, while $t_3$ is set to $0$. The phase diagram at $3/4$ filling is the same due to chiral symmetry. The phase boundaries vary linearly with the strain. In the pristine case ($\epsilon_{xx} = 0$), with $t_1 = 1$, SOC opens up an energy gap of value $\Delta = |4\lambda_I + 2t_2 -2|$, indicating that for $t_2 = 1$ the lower band is topological for any non-zero value of $\lambda_I$. However, under tensile strain, a semimetallic phase emerges at higher values of $\lambda_I$. This is replaced by the appearance of a trivial insulating phase at smaller values of $\lambda_I$, as represented in  Fig.~\ref{fig:sq_oct_phase_t2_1p0}.

To observe the topological phases, we calculated the ribbon band diagram. \cref{fig:sq_oct_band_t2_0p5,fig:sq_oct_band_t2_0p5_lam_0p7_exx_neg0p1,fig:sq_oct_band_t2_0p5_lam_07_exx_0p1} show the behavior of edge modes at $1/4$ filling and $t_2 = 0.5t_1$, through the transition between phases. An example of the trivial phase, i.e., one without edge modes, is shown in Fig. \ref{fig:sq_oct_band_t2_0p5}, with uniaxial compression $\epsilon_{xx} = -0.1$ and $\lambda_I = -0.2$. However, as expected in the $\nu = 1$ phase, edge states appear in the bulk gap at the same strain and (see Fig.~\ref{fig:sq_oct_band_t2_0p5_lam_0p7_exx_neg0p1} for an example with $\epsilon_{xx} = -0.1$ and $\lambda_I = -0.7$). On increasing the strain, the edge protected modes have bulk conduction and valance bands merged into each other at the same value of $\lambda_I$, as depicted in Fig.~\ref{fig:sq_oct_band_t2_0p5_lam_07_exx_0p1}. Even in this case, we can observe that there are edge modes at the time reversal invariant point $\pi$  and these can be shown to be localized along the boundaries of the nanoribbon. However, they are not robust against the disorder and can mix with the bulk states, and hence not ``topologically protected''.

\underline{Decorated Honeycomb lattice:} The decorated Honeycomb lattice  $(L(S(\mathcal{X}_6)))$ can be obtained by applying line-graph operation on the Honeycomb-Kagome lattice. Fig.~\ref{fig:decorated_honeycomb_lattice} shows the unit cell of the decorated honeycomb lattice containing six atoms arranged to form two equilateral triangle plaquettes symmetrically positioned with respect to the inversion center located at the center of the unit cell. The spinful Hamiltonian is:
\begin{align}
H_{L(S(\mathcal{X}_6))}(\bfk) =   &\left(\sum_{j=1}^3u_{j}\tau_0\otimes\Lambda_i + \sum_{j=1}^3v_{j}(\cos\bfk\cdot\bfa_{i-1}\tau_x - \sin \bfk\cdot\bfa_{j-1}\tau_y)\otimes S_{jj}\right)\otimes\sigma_0 \nonumber \\
& +\lambda_I \Bigg(\Big(1+\sum_{j=4}^5(-1)^{j-1}(\cos\bfk\cdot\bfa_{j-3}+i(-1)^j\sin\bfk\cdot\bfa_{i-3})\Big)\tau_x \otimes \Lambda_j \nonumber \\
& + \sum_{j=1}^2(-\cos\bfk\cdot\bfa_{j}+\sin\bfk\cdot\bfa_j)\tau_x\otimes\Lambda_6\Bigg)\sigma_z,    
\end{align}

where, $u_j$ and $v_j$ denote the intra- and inter-triangular hopping amplitudes, respectively. The band structure, inherits in features from honeycomb and the kagome lattices, and consists of two flat bands at $E = 0$ and $E = 2$ along with two sets of Dirac bands (Fig.~\ref{fig:decorated_honeycomb_2dband}). In the isotropic hopping parameter regime, one set of Dirac bands is sandwiched between the flat bands and exhibits quadratic band touching at the $\Gamma$-point, while the other set of Dirac bands remains isolated.  As before, the states at the $\Gamma$-point are not accidental and protected by the $C_6$ point group symmetry of the unit cell.  At $v_j = 1.5 u_j$ for all $j$, the spectrum exhibits pseudospin-1 Dirac crossing at $\Gamma$-point like its immediate parent $\mathcal{S}(\mathcal{X}_6)$ graph. Due to its unique properties and the possibility of being realized as stable materials in the form of elemental allotropes, various studies have been performed on the lattice in ``realistic"  nanomaterial morphologies \cite{chen2014carbon,chen2018ferromagnetism,sarikavak2020structural, yu2024carbon, Chenhaoyue_SiKL}. 

As usual, inclusion of SOC isolates the degenerate bands (blue dashed lines shown in Fig.~\ref{fig:decorated_honeycomb_2dband}) and drives the system into a QSH state at all filling fractions. This is also evident from the spectrum of a decorated honeycomb nonoribbon plotted in Fig.~\ref{fig:decorated_honeycomb_ribbon}, where the band crosses the bulk gap at time reversal invariant momenta $\pi$ and $0$.

\begin{figure}[!htbp]
    \centering
    \subfloat[]{\includegraphics[scale=0.3,trim={1cm 7cm 1cm 7cm},clip]{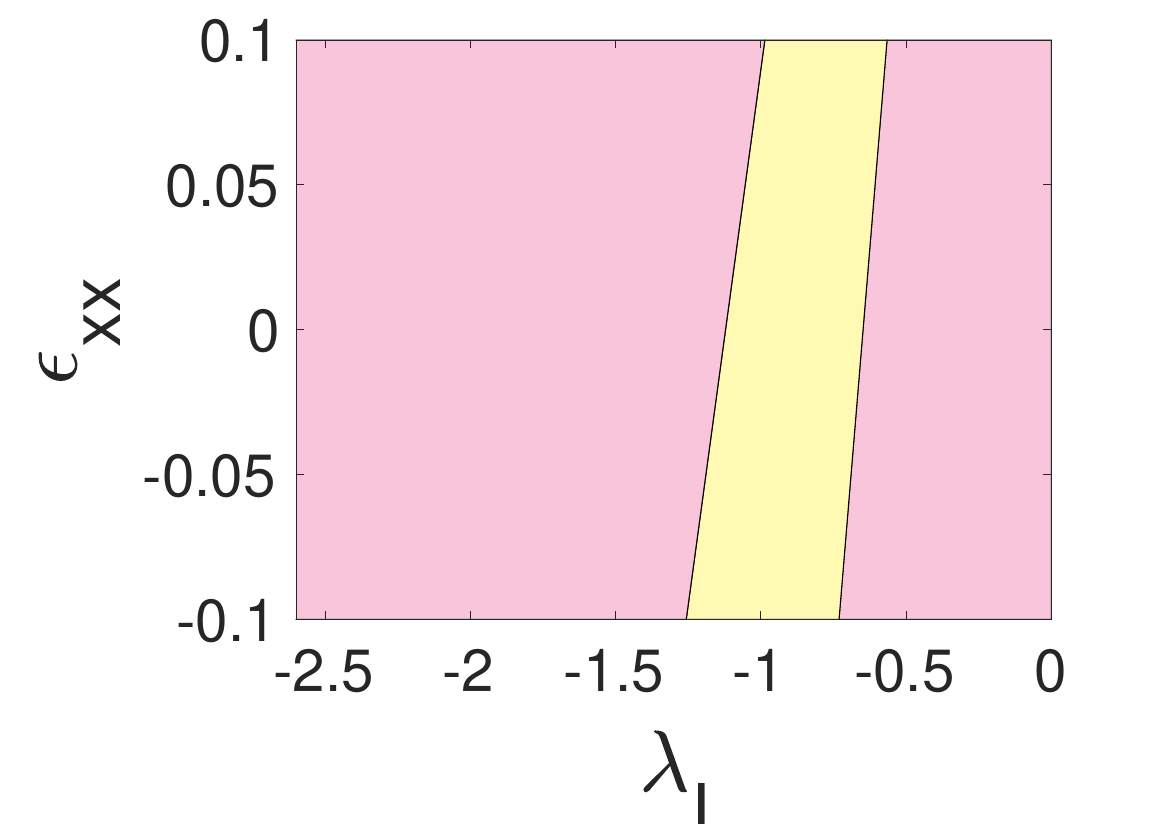}\label{fig:decorated_hny_phase_t2_0p4}}
     \subfloat[]{\includegraphics[scale=0.3,trim={3.1cm 7cm 1cm 7cm},clip]{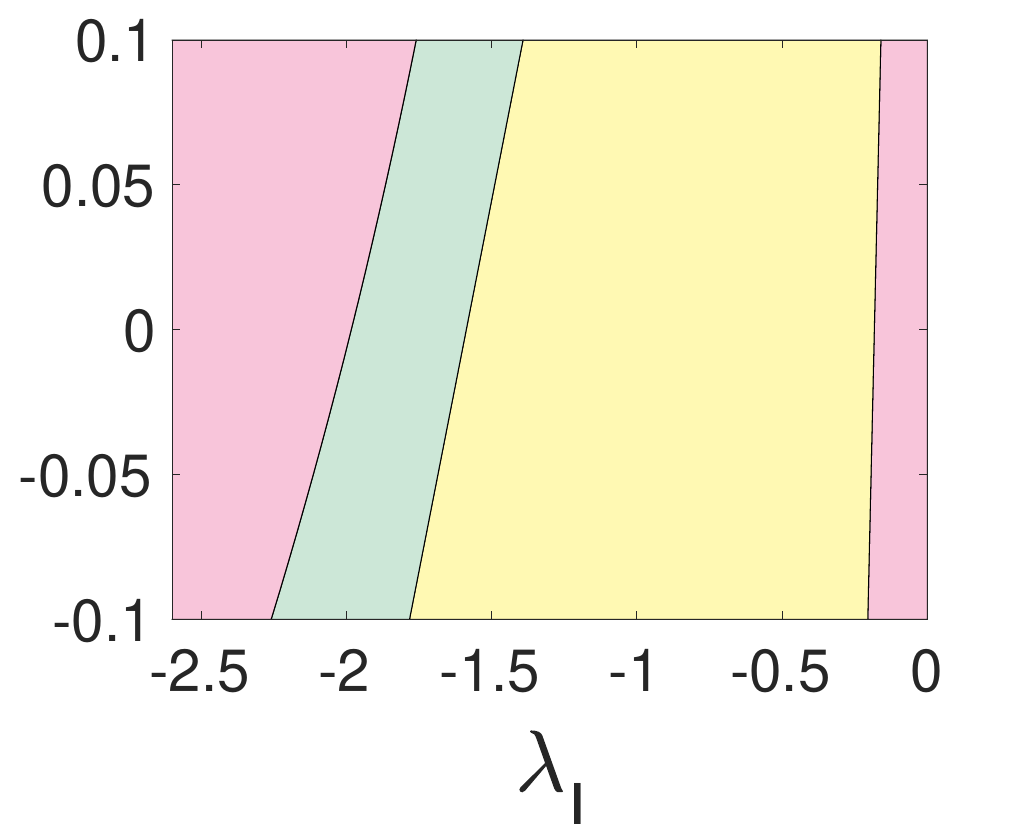}\label{fig:decorated_hny_phase_t2_1p2}}
   \subfloat[]{\includegraphics[scale=0.3,trim={3.1cm 7cm 1cm 7cm},clip]{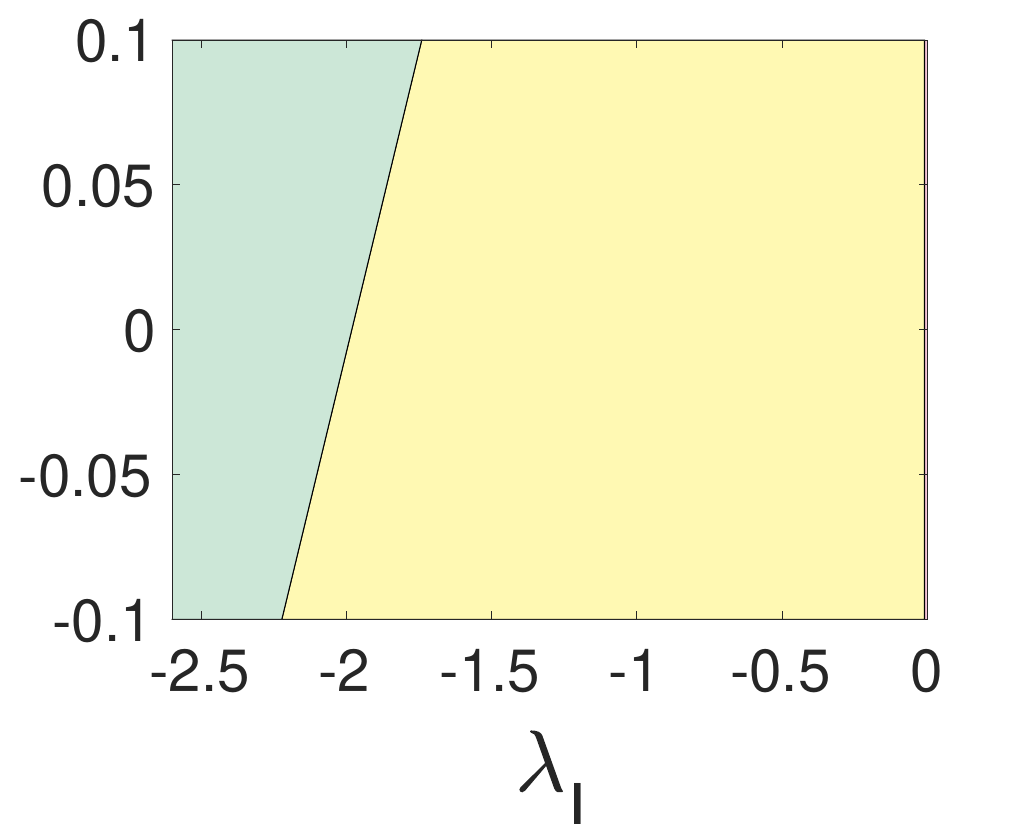}\label{fig:decorated_hny_phase_t2_2p4}}    
    \caption{Topological phase diagram of decorated honeycomb lattice, $L(S(\mathcal{X}_6))$, at $1/3$ filling as a function of strain ($\epsilon_{xx}$) and intrinsic spin orbit coupling ($\lambda_I$) at: (a) $v = 0.4u$, (b) $v = 1.2u$  and (c) $v = 2.4u$. The different phases are distinguished by the colors as follows. Pale mint (\semicyan{}): semimetal, yellow  (\tiyellow{}): topological band insulator and pink (\bipink{}): band insulator. }
    \label{fig:decorated_honeycomb_phasediag}
\end{figure}

\begin{figure}[!htbp]
    \centering
    \subfloat[]{\includegraphics[scale=0.3,trim={1cm 7cm 1cm 7cm},clip]{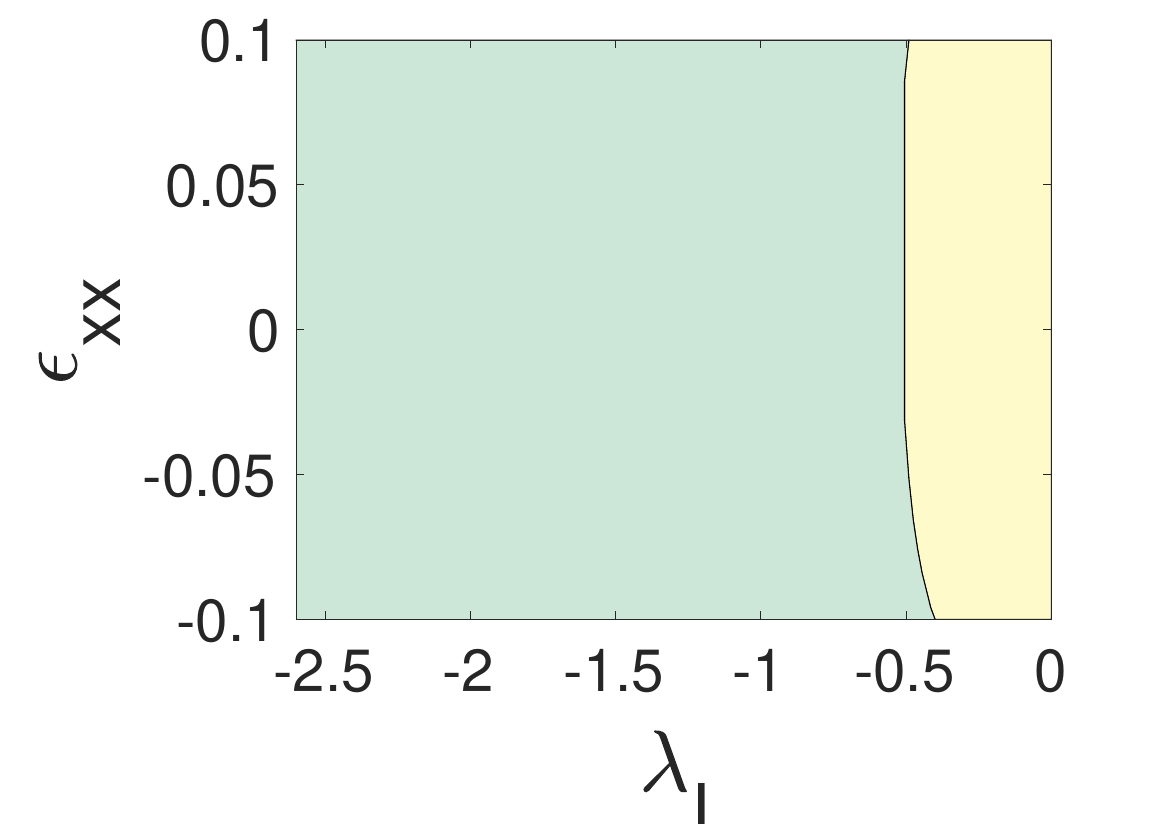}\label{fig:decorated_hny_phase_t2_0p4_band3}}
     \subfloat[]{\includegraphics[scale=0.3,trim={3.1cm 7cm 1cm 7cm},clip]{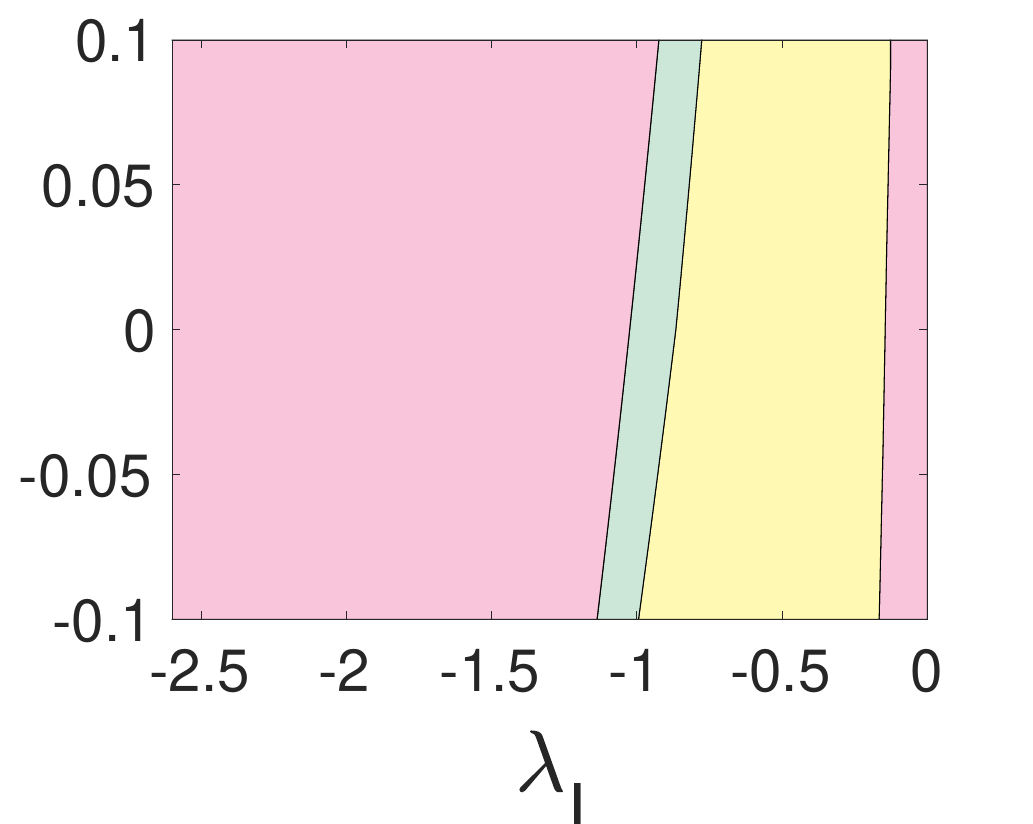}\label{fig:decorated_hny_phase_t2_1p8_band_3}}
   \subfloat[]{\includegraphics[scale=0.3,trim={3.1cm 7cm 1cm 7cm},clip]{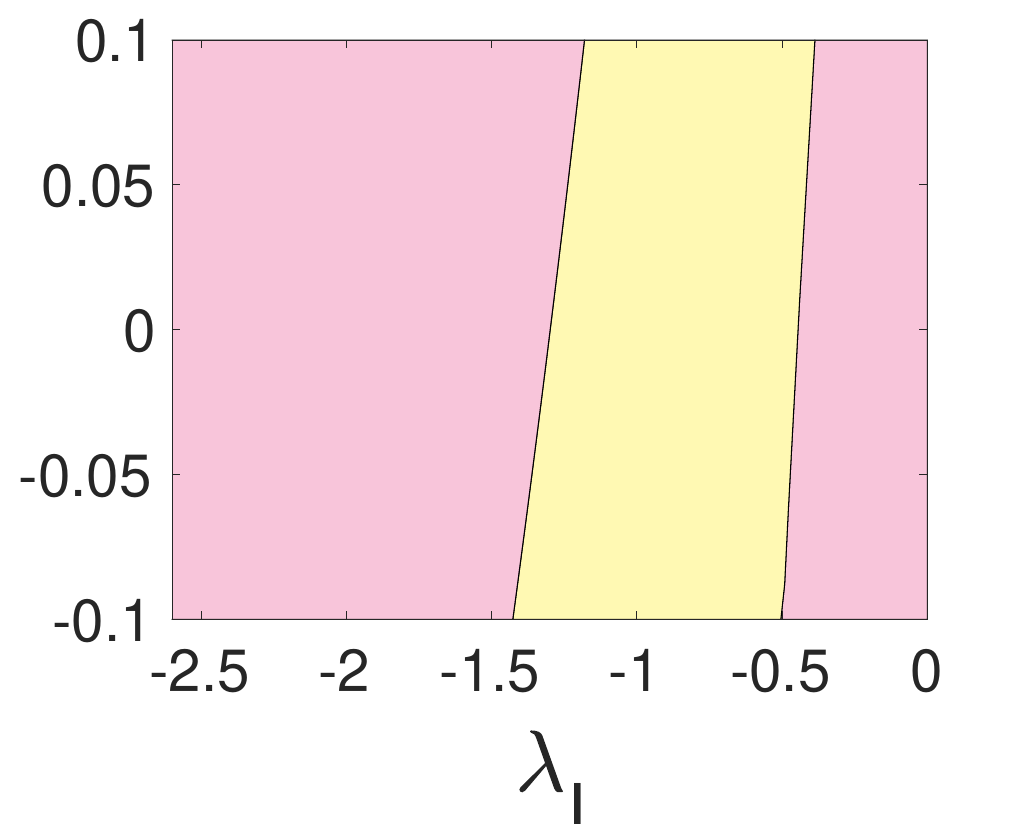}\label{fig:decorated_hny_phase_t2_2p4_band3}}    
    \caption{Topological phase diagram of the decorated honeycomb lattice, $L(S(\mathcal{X}_6))$, at $1/2$ filling as a function of strain ($\epsilon_{xx}$) and intrinsic spin orbit coupling ($\lambda_I$) at: (a) $v = 0.4u$ (b)$v = 1.8u$  and (c) $v = 2.4u$. The different phases are distinguished by the colors as follows. Pale mint (\semicyan{}): semimetal, yellow  (\tiyellow{}): topological band insulator and pink (\bipink{}): band insulator. }
    \label{fig:decorated_honeycomb_phasediag_band3}
\end{figure}

The lattice exhibits a rich variety of phases at different energy levels. Here, we highlight some of the most interesting phase diagrams. \Cref{fig:decorated_honeycomb_phasediag,fig:decorated_honeycomb_phasediag_band3} show topological and quantum phases in strain-SOC space of the second Dirac band from the bottom and of the flat band at $E=0$, respectively. Three different values of $v$ are considered. At $v<u$ the trivial insulating phase dominates, with only a narrow window of the non-trivial state (Fig.~\ref{fig:decorated_hny_phase_t2_0p4}). On the other hand, when $v>u$, an OSM phase emerges (Fig.~\ref{fig:decorated_hny_phase_t2_1p2}). At larger values of $v$ in this scenario, (e.g. at $v = 2.4u$, Fig.~\ref{fig:decorated_hny_phase_t2_2p4}) only the semimetal and topological insulator phases persist. In contrast, the flat band at $E=0$ behaves differently, as shown in Fig.~\ref{fig:decorated_honeycomb_phasediag_band3}. We observed that at $v = 0.4u$ flat band transitions directly into the OSM phase from the non-trivial phase with a non-linear phase boundary. However, as $v$ increases, the OSM phase diminishes and the phase diagram becomes dominated by the trivial insulating phase, as illustrated in \cref{fig:decorated_hny_phase_t2_1p8_band_3,fig:decorated_hny_phase_t2_2p4_band3}.   

\subsubsection{Split graph of line graph lattices \texorpdfstring{$\mathcal{S}(\mathcal{L}(\mathcal{X}))$}{Lg}: Checkerboard split graph and triangular Kagome lattices}
\label{subsec:SLX}
\begin{figure}[!htbp]
    \centering
      \subfloat[]{\includegraphics[scale=0.26,trim={8cm 1cm 8cm 1cm},clip]{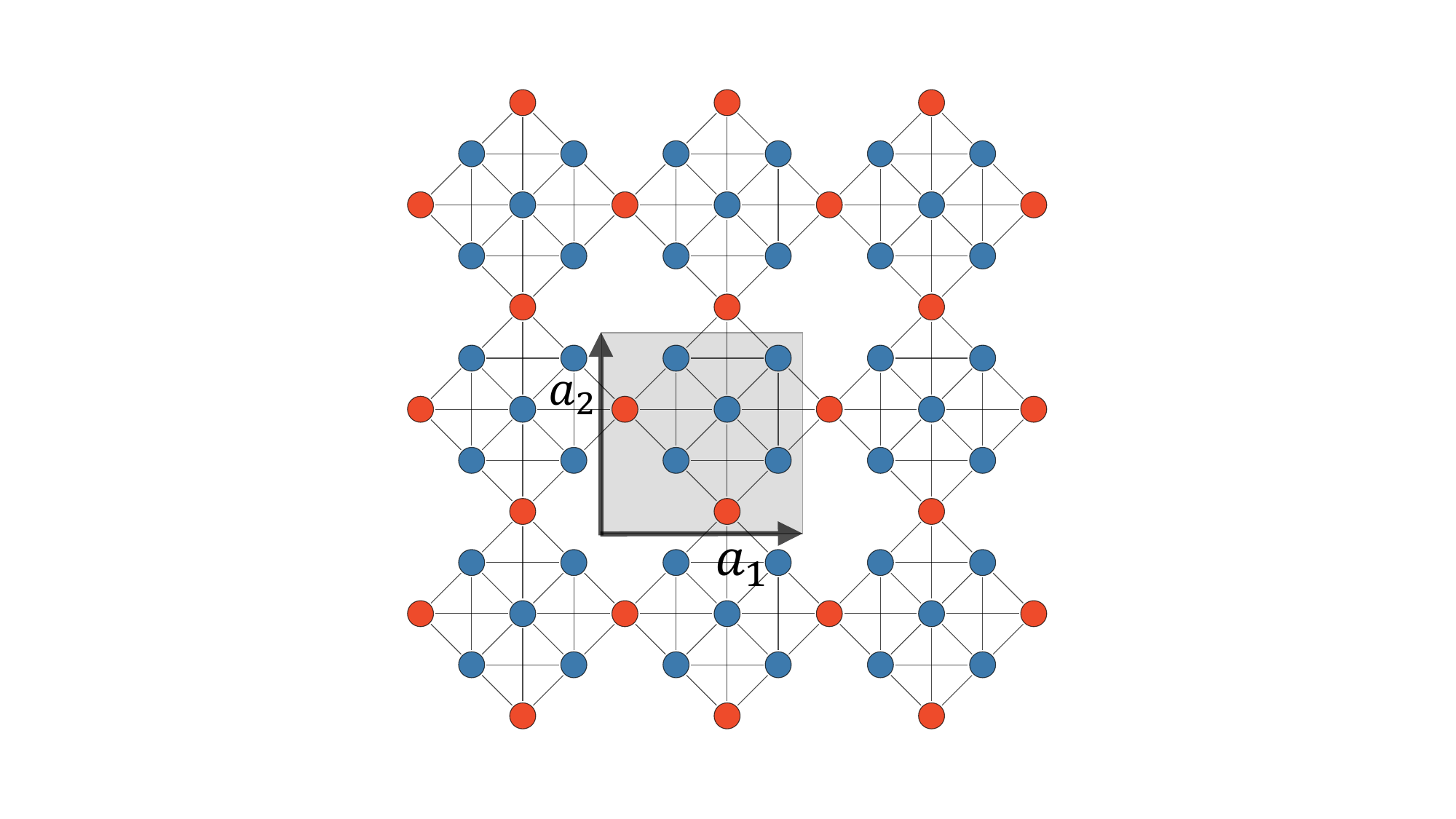}\label{fig:split_checkbrd}}
      \subfloat[]{\includegraphics[scale =0.28,trim={0cm 6.7cm 1cm 7cm},clip]{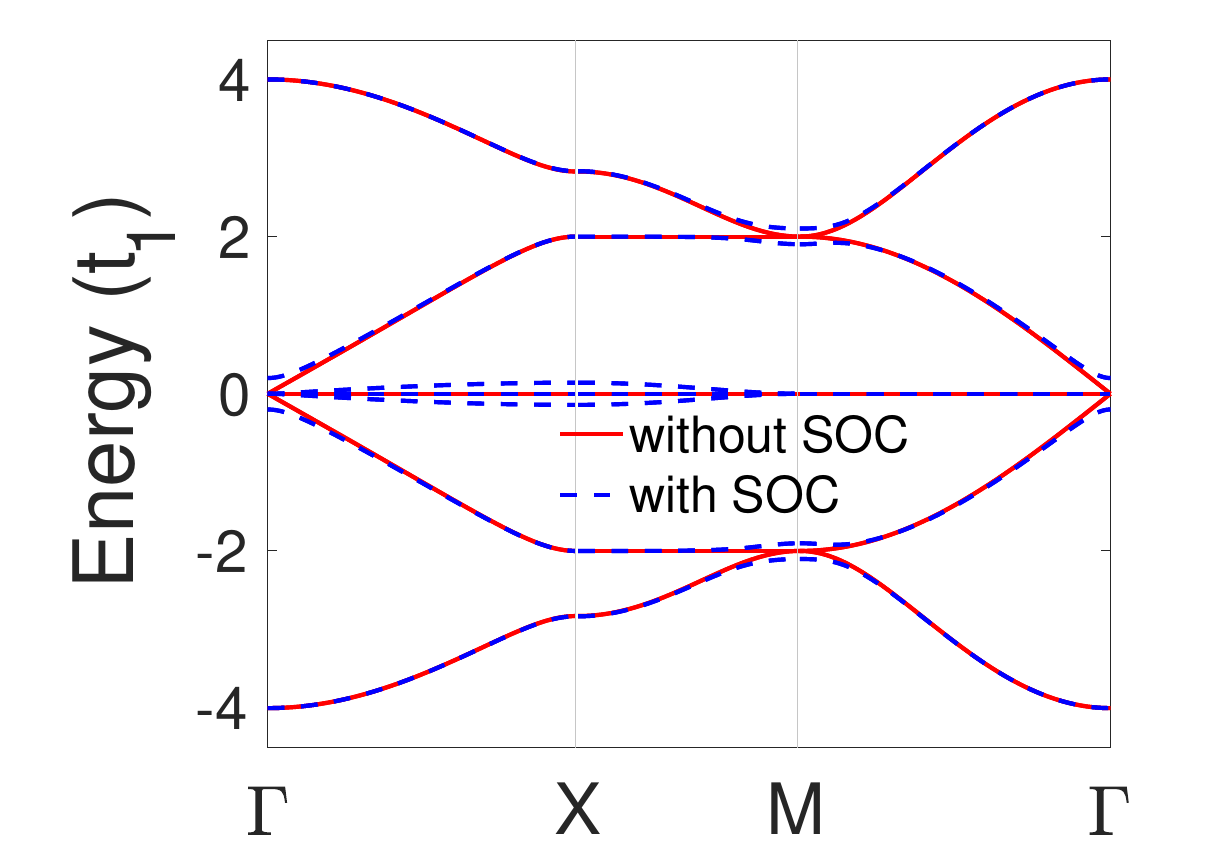}
\label{fig:split_checkerboard_2dband}}
      \subfloat[]{\includegraphics[scale =0.28,trim={3.1cm 6.7cm 0cm 7cm},clip]{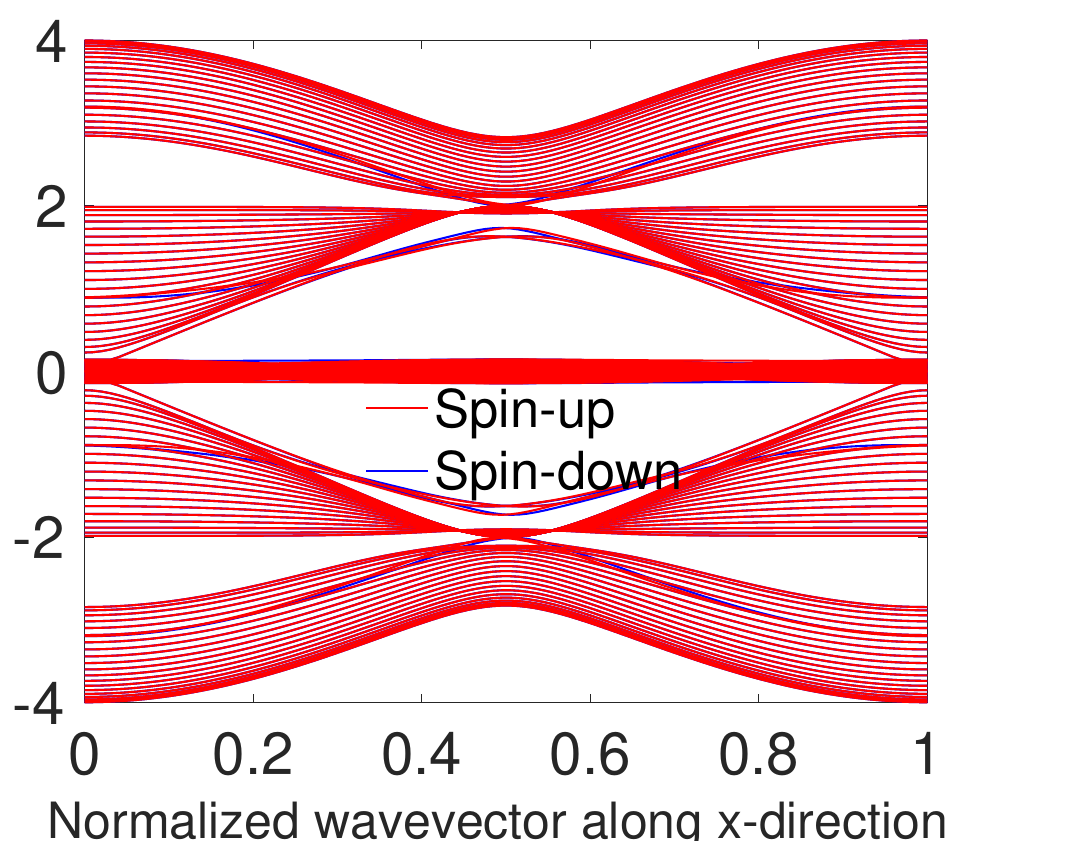}
\label{fig:split_checkerboard_ribbon}}
\caption{(a) The split graph of the checkerboard lattice gives $S(L(\mathcal{X}_4))$. The black arrows indicate the lattice vectors $\bfa_1$ and $\bfa_2$, and the gray region is the unit cell. The tight binding band diagram ($t_3 = 0$) of (b) 2D lattice without (red solid line) and with (blue broken line) SOC ($\lambda_I = 0.1t_1$) and (c) 1D zigzag lattice nanoribbon with $\lambda_I = 0.1t_1$. The red and blue lines show the counter-propagating spin-up and spin-down states, respectively. }
    \label{fig:split_checkbrd_latt_w_bnds}
\end{figure}

\begin{figure}[!htbp]
    \centering
      \subfloat[]{\includegraphics[scale=0.26,trim={7cm 1cm 7cm 1cm},clip]{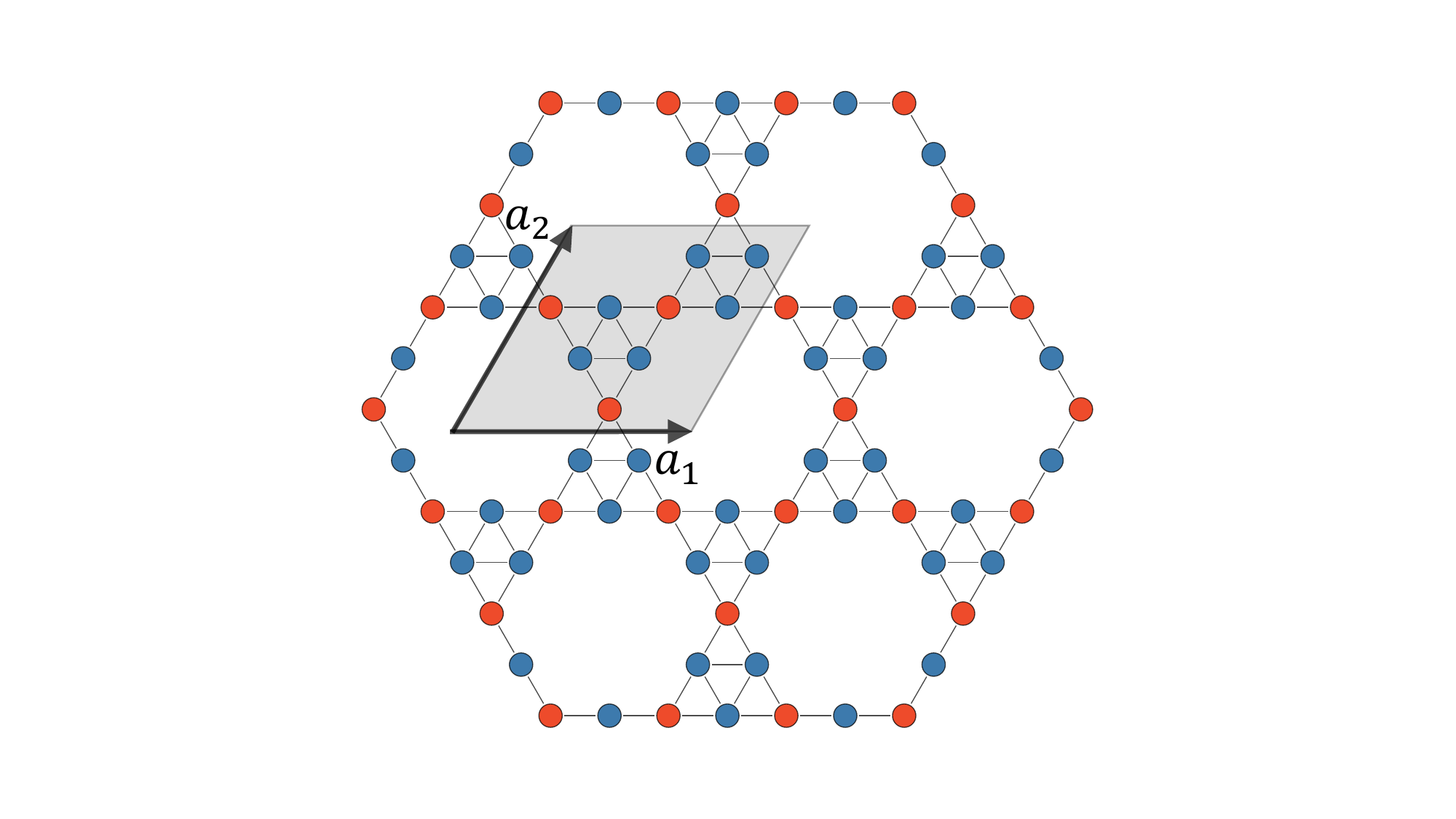}\label{fig:triangle_triangle_lattice}}
      \subfloat[]{\includegraphics[scale =0.28,trim={0cm 6.7cm 1cm 7cm},clip]{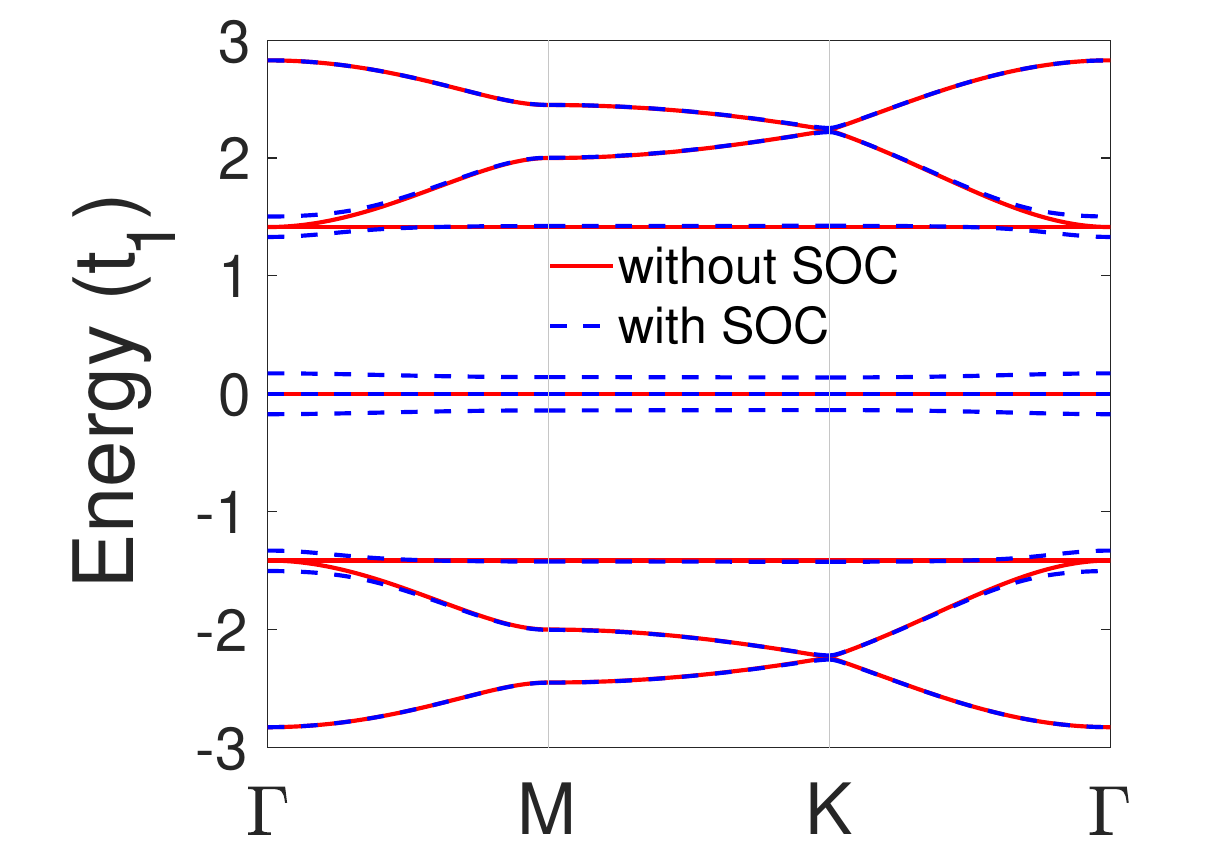}
\label{fig:triangular_kagome_2dband}}
      \subfloat[]{\includegraphics[scale =0.28,trim={3.1cm 6.7cm 0cm 7cm},clip]{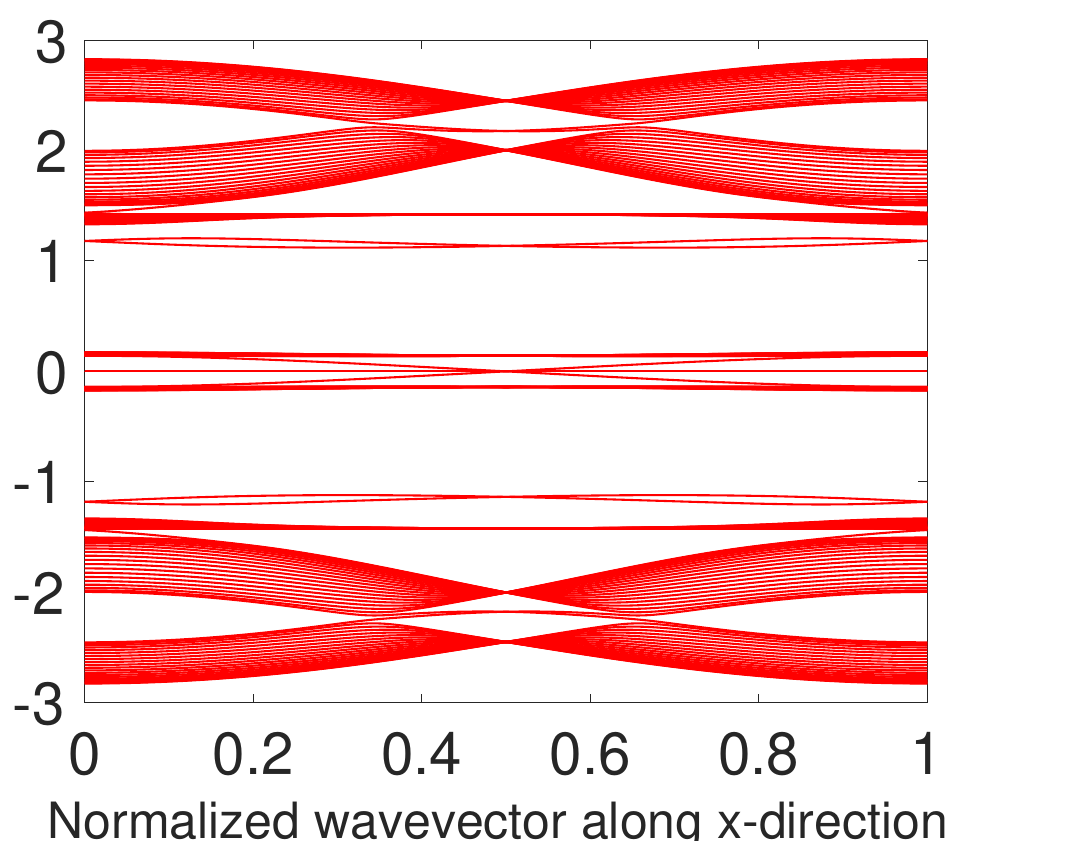}
\label{fig:triangular_kagome_ribbon}}
\caption{(a) Triangular kagome lattice (TKL) obtained by applying the split graph operation on the kagome lattice, $S(L(\mathcal{X}_6))$. The black arrows indicate the lattice vectors $\bfa_1$ and $\bfa_2$, and the gray region is the unit cell. The tight binding band diagram ($t_{bb} = 0$) of (b) 2D lattice without (red solid line) and with (blue broken line) SOC ($\lambda_I = 0.1 t$) and (c) 1D zigzag lattice nanoribbon with $\lambda_I = 0.1 t$. The red and blue lines show the counter-propagating spin-up and spin-down states, respectively. }
\label{fig:trianglr_kag_latt_w_bnds}
\end{figure}

The subdivision of the $2d-2$-regular line graphs $L(\mathcal{X}_i)$, where $i\in \{4,6\}$, yields bipartite $(2d-2,2)$-biregular graphs $\mathcal{S}(L(\mathcal{X}_i))$. These structures introduce additional  sites positioned at the midpoint of the edges of the line graphs, as shown via blue circles in \cref{fig:split_checkbrd,fig:triangle_triangle_lattice}. By definition, the blue atoms are not directly connected to each other in either lattice. However, in this study, we have also taken into account the interactions between these blue sites to explore their influence on the electronic and topological properties, as has also been done previously in the literature \citep{wang2018coexistence}. 

\underline{Checkerboard split-graph lattice:} The schematic of the split graph of the checkerboard lattice $S(L(\mathcal{X}_4))$ is shown in Fig.~\ref{fig:split_checkbrd}. This structure can be interpreted as checkerboard lattice decorated with $90^\circ$-rotated squares, each hosting an atom at the center. Interestingly, while the electronic and topological properties of some of the lattices described above have been studied in the literature, the $S(L(\mathcal{X}_4))$ graph remains  unexplored heretofore.

The spinful Hamiltonian for the split graph of the checkerboard lattice, $S(L(\mathcal{X}_4))$ can be written as:
\beq
H_{S(L(\mathcal{X}_4))}(\bfk) =  H^O_{S(L(\mathcal{X}_4))}(\bfk)\otimes\sigma_0 + H^{SO}_{S(L(\mathcal{X}_4))}(\bfk)\otimes\sigma_z,
\eeq
where,
\beq
H^O_{S(L(\mathcal{X}_4))}(\bfk) = \begin{pmatrix}
    \mathcal{O}_{2\times 2} && \Phi^\dagger(\bfk)\\
    \Phi(\bfk) && \mathcal{M}_{5\times5}(\bfk)
\end{pmatrix}, \quad H^{SO}_{S(L(\mathcal{X}_4))}(\bfk) =\begin{pmatrix}
    \mathcal{O}_{2\times 2} && \mathcal{O}_{2\times 5}\\
    \mathcal{O}_{5\times 2} && \mathcal{S}_{5\times5}(\bfk)
\end{pmatrix}
\eeq
where,
\beqs
\Phi(\bfk) &=& \begin{pmatrix}
    t_1 && t_1 \\
    t_1e^{i\bfk\cdot\bfa_2} && t_1 \\
    t_1e^{i\bfk\cdot(\bfa_1+\bfa_2)} && t_1 \\
    t_1e^{i\bfk\cdot\bfa_1} && t_1\\
    t_2(1+e^{i\bfk\cdot\bfa_1}) && t_2(1+e^{-i\bfk\cdot\bfa_2})
\end{pmatrix},  \nonumber \\
\mathcal{M}_{5\times 5}(\bfk) &=& t_3\begin{pmatrix}
    0 && e^{i\bfk\cdot\bfa_2} && 0 && 1 && 1\\
    e^{-i\bfk\cdot\bfa_2} && 0 && 1 && 0 && e^{-i\bfk\cdot\bfa_2}\\
    0 &&1 && 0 && e^{-i\bfk\cdot\bfa_2} && e^{-i\bfk\cdot\bfa_2} \\
   1 && 0 && e^{i\bfk\cdot\bfa_2} && 0 && 1\\
    1 && e^{i\bfk\cdot\bfa_2} && e^{i\bfk\cdot\bfa_2} && 1&& 0
\end{pmatrix} \quad \text{and} \nonumber \\
\mathcal{S}_{5\times5}(\bfk) &=& i\lambda_I\begin{pmatrix}
    0 && -1 && 0 && e^{i\bfk\cdot\bfa_1} && 0 \\
    1 && 0 && -e^{i\bfk\cdot\bfa_1} && 0 && 0 \\
    0 && e^{-i\bfk\cdot\bfa_1} && 0 && -1 && 0 \\
    -e^{-i\bfk\cdot\bfa_1} && 0 && 1 && 0 && 0\\
    0 && 0 && 0 && 0 && 0  
\end{pmatrix}. 
\eeqs

\begin{figure}[!htbp]
    \centering
     \subfloat[]{\includegraphics[scale=0.28,trim={0cm 6.7cm 1cm 7cm},clip]{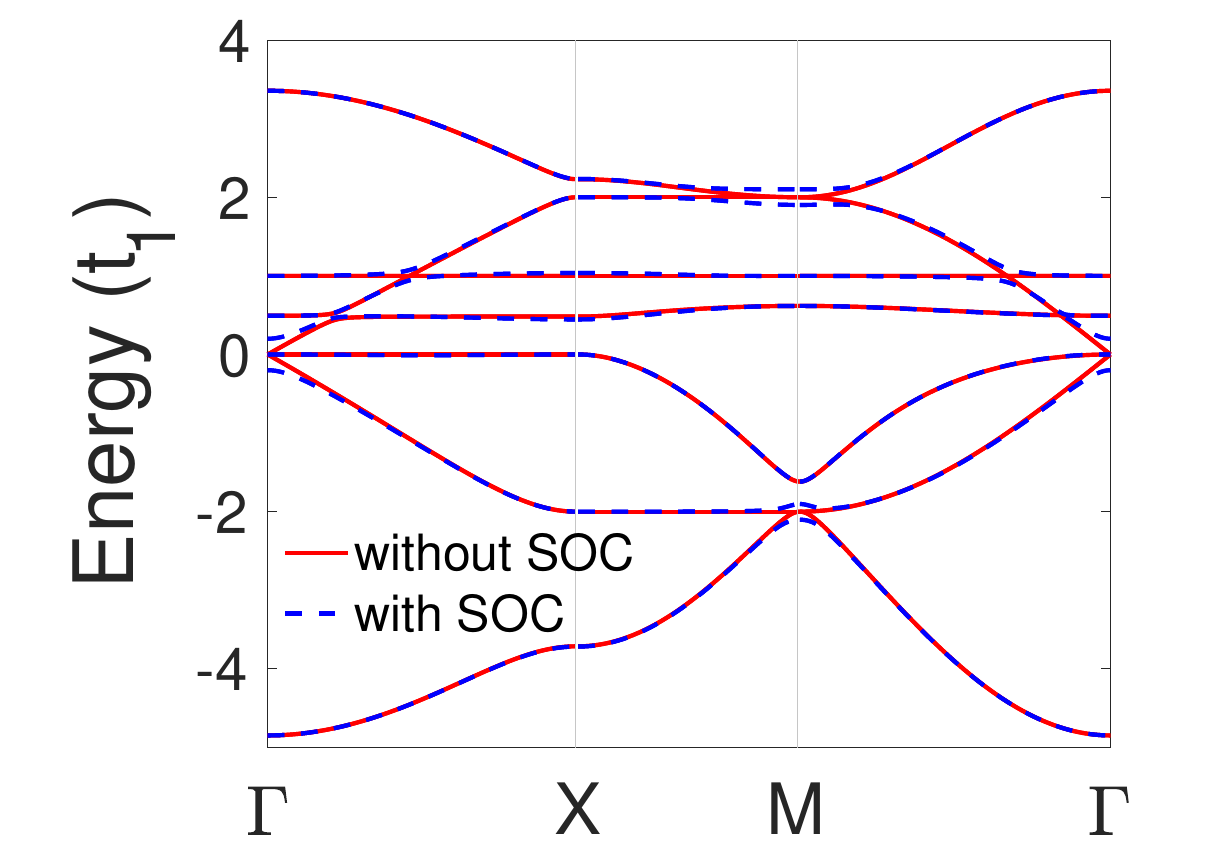}\label{fig:split_checkbrd_t3_0p5}}
      \subfloat[]{\includegraphics[scale=0.28,trim={3cm 6.7cm 1cm 7cm},clip]{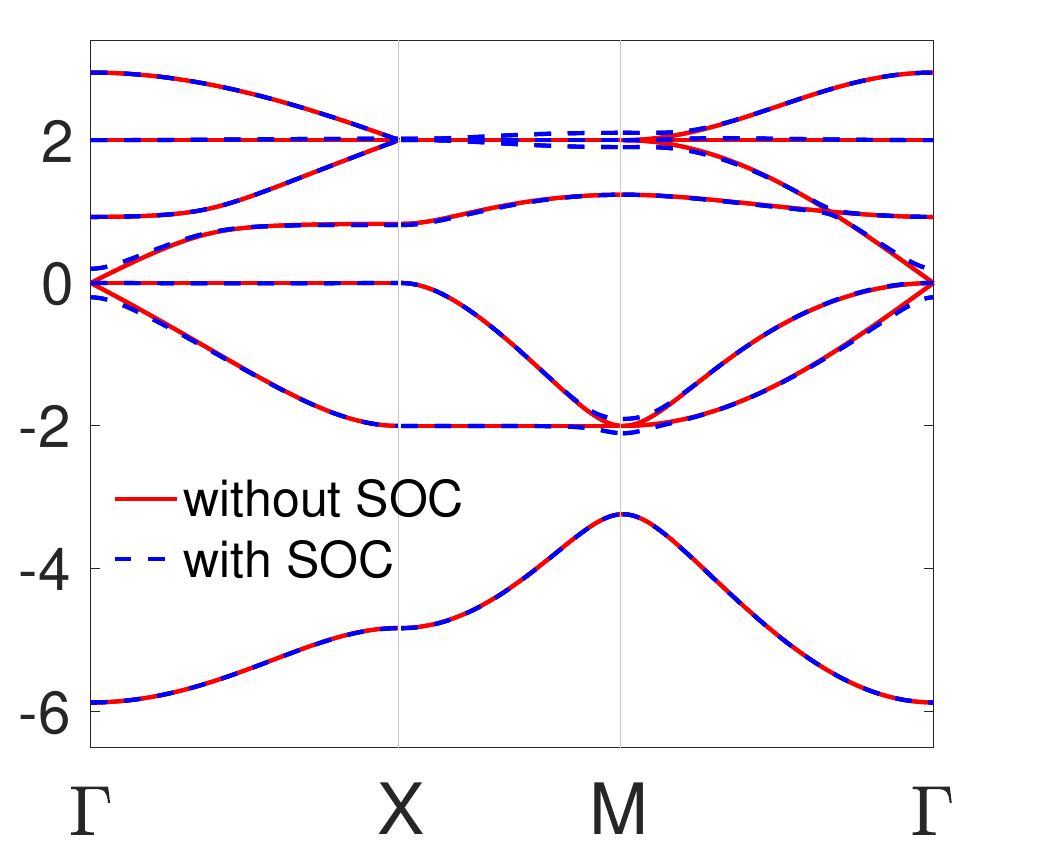}\label{fig:split_checkbrd_t3_1}}
\caption{The tight binding band diagram of checkerboard split graph  $S(L(\mathcal{X}_4))$ at different values of $t_3$. (a) $t_3 = 0.5 t_1$ (b) $t_3 = t_1$. The solid red color lines is when $\lambda_I = 0$ and dashed blue lines when $\lambda_I = 0.1t_1$. }
    \label{fig:split_checkbrd_bands_t3}
\end{figure}

Here, the hopping parameters $t_1$ and $t_2$ describe the interactions between a red atom and its neighboring blue atoms, located at the edges and at the intersections of the diagonals, respectively. The third hopping parameter $t_3$ connects the blue atoms, and can be set to $0$ for a simplified model. While the checkerboard lattice has two sites in the unit cell, the $S(L(\mathcal{X}_4))$ lattice has seven sites, resulting in seven electronic bands. The typical band structure with $t_3 = 0$ is shown in Fig.~\ref{fig:split_checkerboard_2dband}. We find that the band diagram is mirror symmetric about the zero energy level. Remarkably, there are three overlapping flat bands which are degenerate with Dirac cones at the $\Gamma$-point, similar to the $\mathcal{S}(\mathcal{X}_4)$ graph (Lieb lattice). The Dirac bands are also degenerate with the parabolic band at $M$-point. Upon tuning $t_3$, the three overlapping dispersionless bands splits apart into partially dispersive bands with the onset of tilted Dirac cones as shown in Fig.~\ref{fig:split_checkbrd_t3_0p5}. Finally, when all the hopping parameters are equal, the lowest energy band is isolated (see Fig.~\ref{fig:split_checkbrd_t3_1}), resembling the band diagram of the parent graph, $\mathcal{X}_4$. 

Usually, inclusion of SOC isolates the bands and transforms the flat band into quasi-flat. Intriguingly, we observed that at $t_3 = 0$, two out of the three dispersionless bands at zero energy turn into non-degenerate quasi-flat bands along the $\Gamma-X-M$ path (see dashed blue lines in Fig.~\ref{fig:split_checkerboard_2dband}). However, the central flat band shows no dispersion in the entire BZ even when SOC is included. 

\begin{figure}[!htbp]
    \centering
    \subfloat[]{\includegraphics[scale=0.3,trim={1cm 7cm 1cm 7cm},clip]{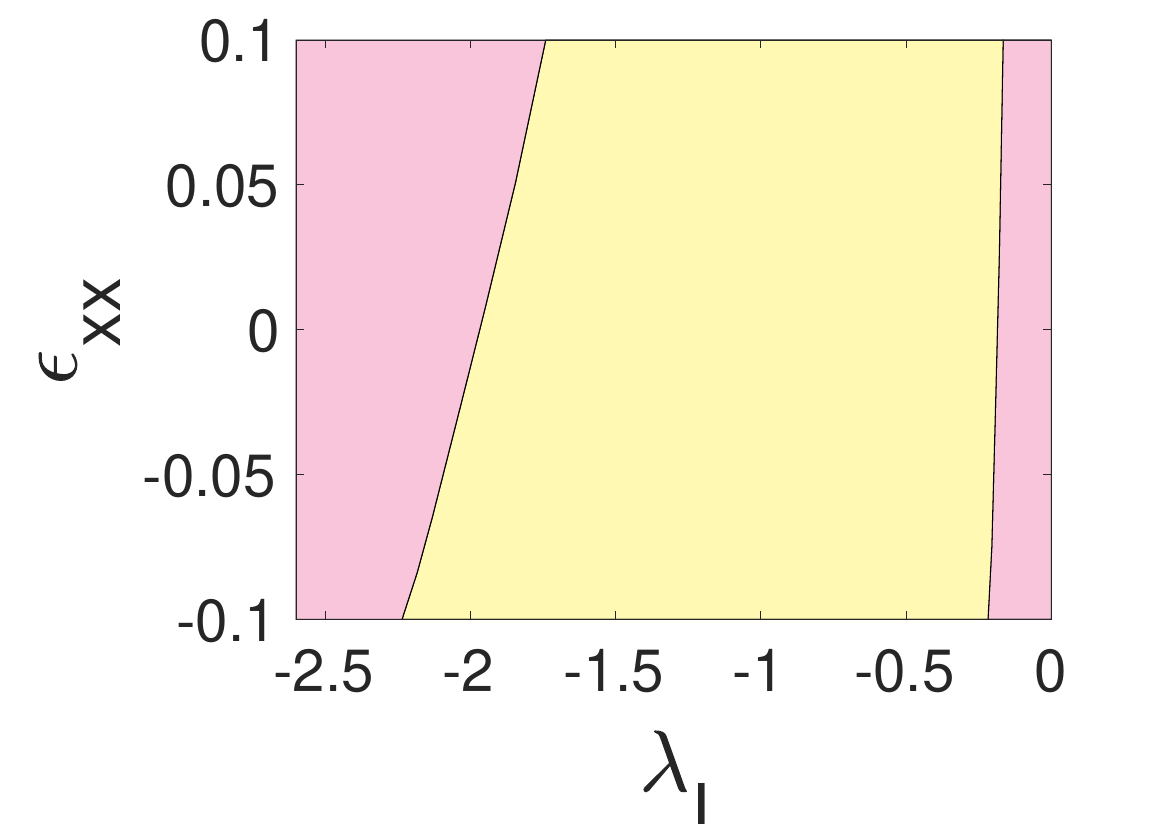}\label{fig:split_checker_kag_phase_t2_0p2_band1}}
     \subfloat[]{\includegraphics[scale=0.3,trim={3.1cm 7cm 1cm 7cm},clip]{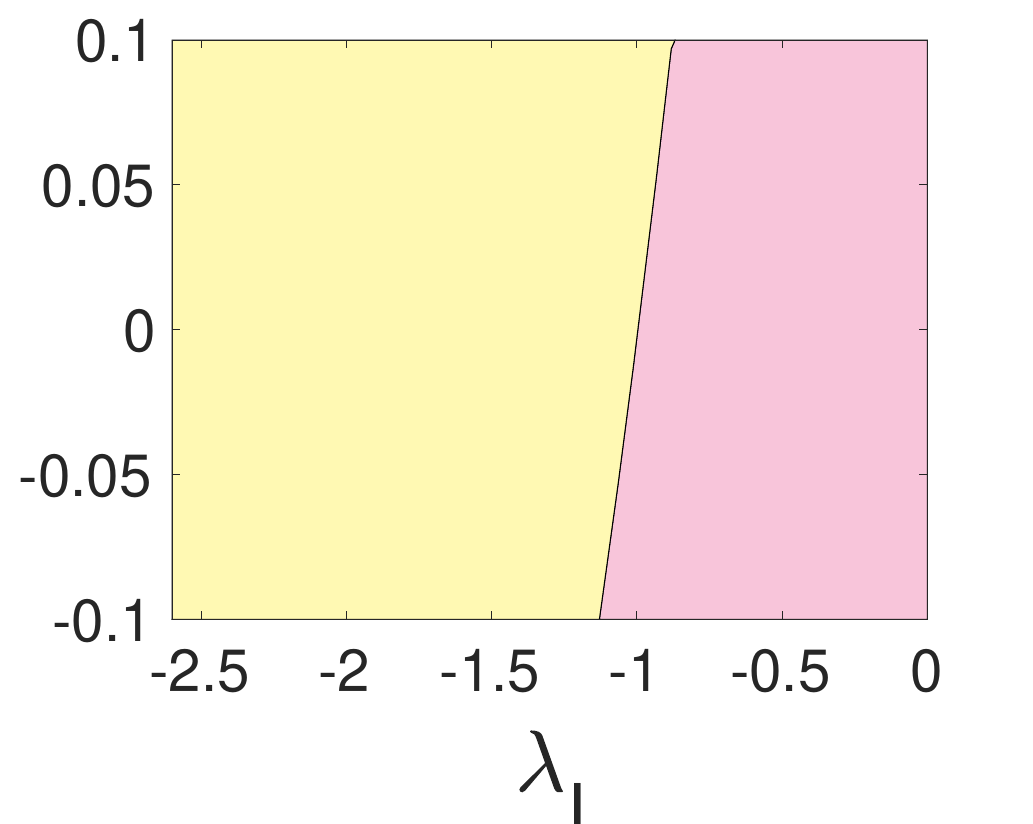}\label{fig:split_checker_kag_phase_t2_1p0_band1}}
      \subfloat[]{\includegraphics[scale=0.3,trim={3.1cm 7cm 1cm 7cm},clip]{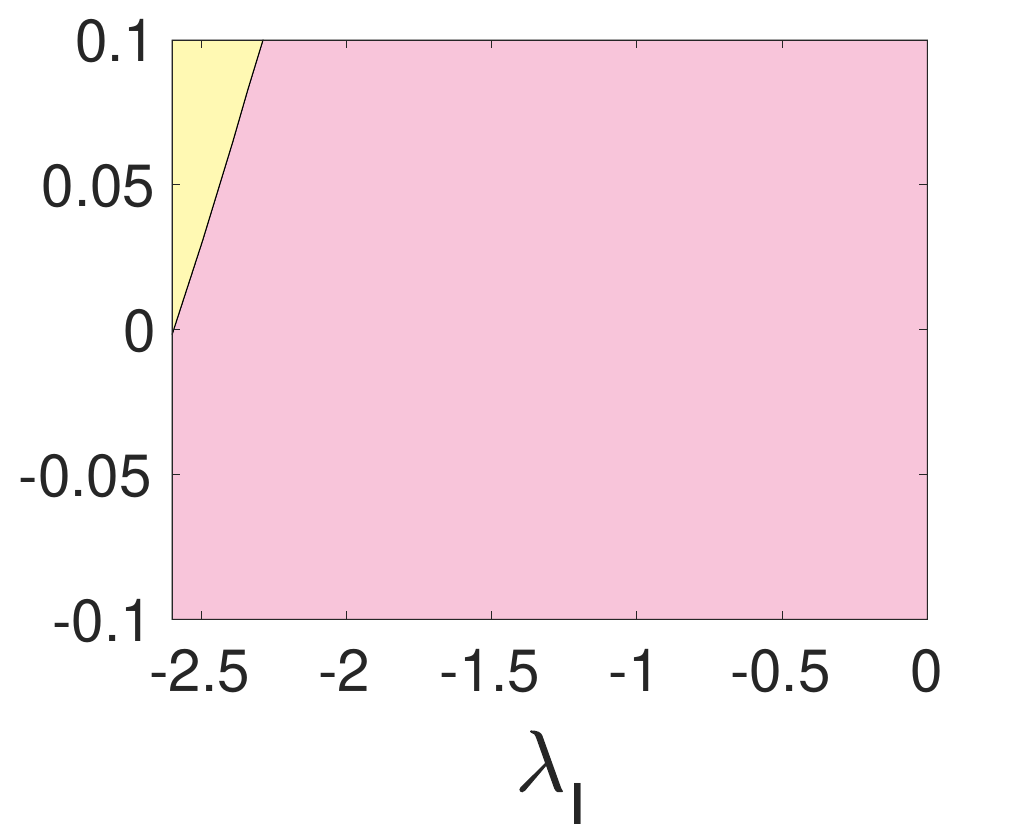}\label{fig:split_checker_kag_phase_t2_2p6_band1}}
    \caption{Splitgraph of checkerboard lattice, $S(L(\mathcal{X}_4))$, at $1/7$ filling as the function of strain ($\epsilon_{xx}$) and intrinsic spin orbit coupling ($\lambda_I$) at (a) $t_2 = 0.2 t_1$ (b)$t_2 = t_1$, and (c) $t_2 = 2.6 t_1$. The different phases are distinguished by the colors as follows. Yellow  (\tiyellow{}): topological band insulator and pink (\bipink{}): band insulator.}
  \label{fig:split_checker_phase_band1}
\end{figure}

Fig.~\ref{fig:split_checker_phase_band1} shows the topological phases at $1/7$ filling fraction in the strain-SOC space (i.e., $\epsilon_{xx}$ vs. $\lambda_I$)  for different values of $t_3$. The phase boundaries are seen to vary linearly with $\epsilon_{xx}$. At $t_3=0.2 t_1$, the topologically non-trivial region sandwiched between the trivial insulator dominates the phase space as shown in Fig.~\ref{fig:split_checker_kag_phase_t2_0p2_band1} . With increasing $t_3$, the non-trivial area shifts towards the right and almost covers the entire phase diagram, as represented in Fig.~\ref{fig:split_checker_kag_phase_t2_2p6_band1}. At other filling fractions, we do not observe any significant changes in the topological properties.

\underline{Triangular kagome lattice:} The split graph of the Kagome lattice $\mathcal{S}(L(\mathcal{X}_6))$ is composed of two sublattices, as depicted in Fig.~\ref{fig:triangle_triangle_lattice}. It contains nine atoms in the unit cell and can be viewed as triangles within triangles where red atoms sits at the Kagome sites and blue atoms form smaller equilateral triangles. By definition of split graphs, the blue atoms are not connected to each other, although as before, we consider a non-zero hopping between them in our analysis below, to explore a richer physics. The magnetic properties of this lattice has been widely studied \cite{isoda2012consistent,loh2008thermodynamics,zhang2023topological,zhou2018xxz, yao2008xxz} but its electronic and topological properties have not been well explored \cite{wang2018coexistence,chen2023fragile}. Notably, certain lattices with fractal geometries \citep{pal2018flat}, appear to have connections with this lattice.

The TB Hamiltonian of the triangular Kagome lattice (TKL), i.e.,  $S(L(\mathcal{X}_6))$ is:
\beq
H_{S(L(\mathcal{X}_6))}(\bfk) = 
\begin{pmatrix}
    \mathcal{O}_{3\times3} &&  \mathcal{A}(\bfk) &&  \mathcal{B}(\bfk) \\
    \mathcal{A}^\dagger(\bfk) &&  \mathcal{C}(\bfk) && \mathcal{O}_{3\times 3}\\
     \mathcal{B}^\dagger(\bfk) && \mathcal{O}_{3\times 3} &&  \mathcal{D}(\bfk)
\end{pmatrix}\otimes\sigma_0 + \begin{pmatrix}
        \mathcal{O}_{3\times3} && \mathcal{O}_{3\times3} && \mathcal{O}_{3\times3} \\
    \mathcal{O}_{3\times3} &&  \mathpzc{E}(\bfk) && \mathcal{O}_{3\times 3}\\
    \mathcal{O}_{3\times3}&& \mathcal{O}_{3\times 3} &&  \mathcal{F}(\bfk)
\end{pmatrix}\otimes\sigma_z,
\eeq
where,
\beqs
 \mathcal{A}(\bfk) &=& t_{rb}(\Lambda_1+\Lambda_3+S_{11}+S_{33}),\nonumber \\
 \mathcal{B}(\bfk) &=& t_{rb}\left(e^{i\bfk\cdot\bfa_1}(S_{11}+S_{33}),+e^{\bfk\cdot\bfa_2}(S_{12}+S_{23})+S_{21}+S_{32}\right) \nonumber \\
 \mathcal{C}(\bfk) &=& t_{bb}\sum_{i=1}^3\Lambda_{i},  \nonumber \\
 \mathcal{D}(\bfk) &=& t_{bb}\left(\sum_{i=1}^3(\cos\bfk\cdot\bfa_{4-i})\Lambda_i - \sin\bfk\cdot\bfa_{3}\Lambda_4+\sin\bfk\cdot\bfa_{2}\Lambda_5+\sin\bfk\cdot\bfa_{1}\Lambda_6\right), \nonumber\\
 \mathpzc{E}(\bfk) &=&\lambda_I\sum_{i=4}^6 (-1)^{i-1}\Lambda_i, \quad \text{and} \nonumber \\
 \mathcal{F}(\bfk) &=& \lambda_I\left(-\cos\bfk\cdot\bfa_3\Lambda_4+\cos\bfk\cdot\bfa_2\Lambda_5-\cos\bfk\cdot\bfa_1\Lambda_6+\sum_{i=1}^3(-1)^{i}\sin\bfk\cdot\bfa_{4-i}\Lambda_i\right).
\eeqs
Here, $t_{rb}$ and $t_{bb}$ are the hopping energies between red-blue and blue-blue atoms, respectively. Compared to the Kagome lattice, the TKL structure exhibits  enhanced frustration which makes it advantageous in hosting various dispersionless bands and topological states.  When $t_{bb} = 0$ and $\lambda_I = 0$, the spectrum of $\mathcal{S}(L(\mathcal{X}_6))$ contains three degenerate isolated flat bands at $E=0$ and two groups of Kagome-type bands that are mirror symmetric around $E=0$, as depicted in Fig.~\ref{fig:triangular_kagome_2dband}. These gapped perfect flat bands that do not interact with any dispersive bands are known to  enhance the Coulomb interaction, thus leading to strongly correlated behavior \cite{tarnopolsky2019origin,li2021isolated} (e.g., superconductivity in moir\'e superlattices in magic angle twisted bilayer graphene \cite{cao2018correlated,cao2018unconventional}).  In contrast to some other materials platforms, which require some engineering or precise control of system parameters (e.g.  twist angle, magnetic field strength, symmetry breaking potentials and SOC to achieve such behaviors \citep{green2010isolated,huda2020designer,chen2024isolated}), the TKL hosts such desirable electronic states natively. Therefore, realization of this lattice as a natural atomic allotrope is an especially fascinating prospect. Moreover, we found that the $E=0$ flat bands are resilient against deformation and are likely to preserve strongly correlated states. 

\begin{figure}[!htbp]
    \centering
      \subfloat[]{\includegraphics[scale=0.28,trim={0cm 6.7cm 1cm 7cm},clip]{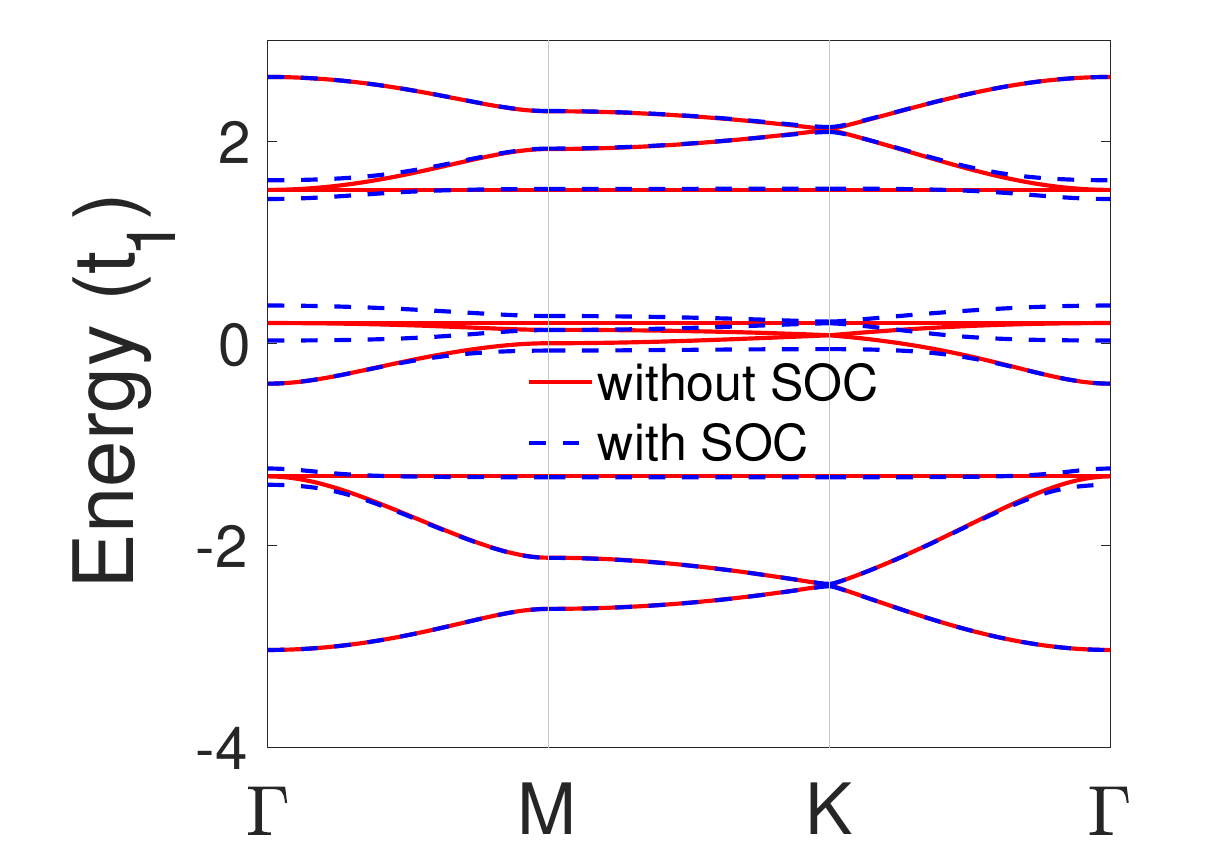}\label{fig:triangle_triangle_lattice_0p2}}
      \subfloat[]{\includegraphics[scale =0.28,trim={3cm 6.7cm 1cm 7cm},clip]{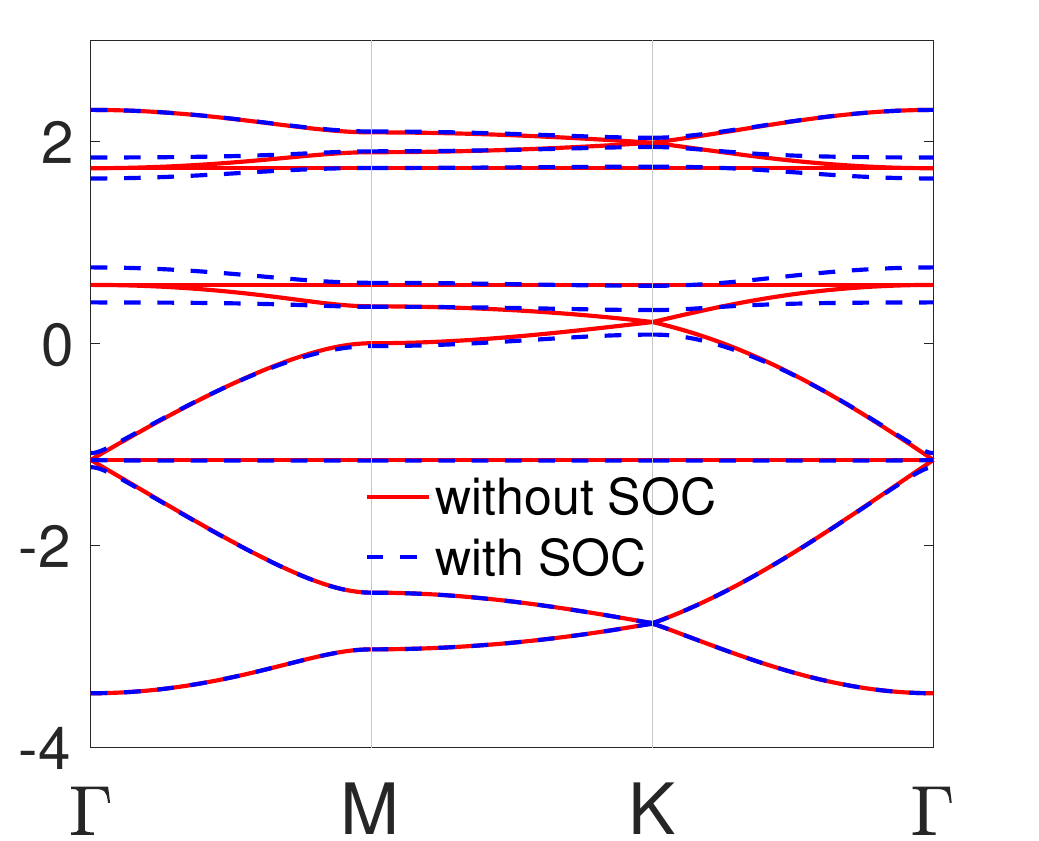}\label{fig:triangular_kagome_2dband_sqrt3}}
      \subfloat[]{\includegraphics[scale =0.28,trim={3cm 6.7cm 1cm 7cm},clip]{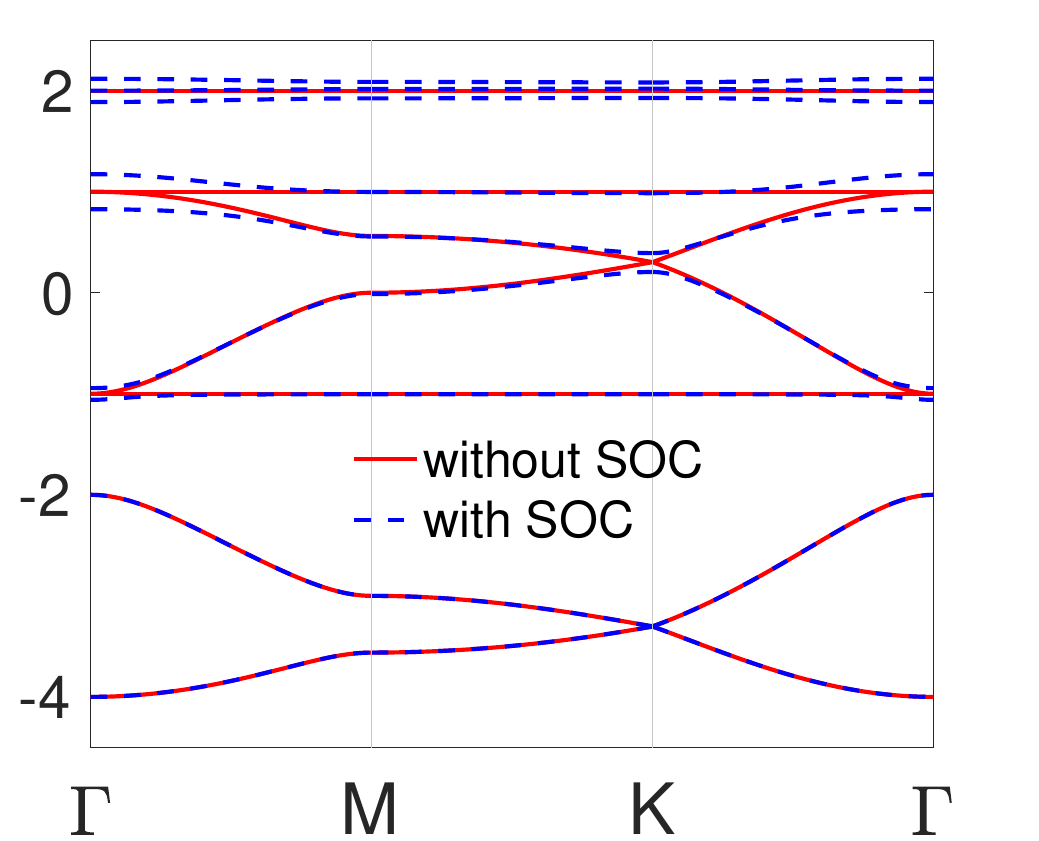}\label{fig:triangular_kagome_2dband_1}}
\caption{The tight binding band structure of triangular Kagome lattice $L(L(\mathcal{S}(\mathcal{X}_6)))$ at different values of $t_{bb}$. (a) $t_{bb} = 0.2 t_{rb}$, (b) $t_{bb} = \frac{1}{\sqrt{3}} t_{rb}$ and (c) $t_{bb} = t_{rb}$. The solid red color lines are for $\lambda_I = 0$, while the dashed blue lines are when $\lambda_I = 0.1 t_1$.}
\label{fig:trianglr_kag_latt_w_bnds_tbb}
\end{figure}

 Upon changing $t_{bb}$ to no-zero values, a richer set of electronic properties can be observed. In particular, tuning $t_{bb}=0.2 t_{rb}$, introduces three sets of Kagome-like bands (see Fig.~\ref{fig:triangle_triangle_lattice_0p2}) which are known to exhibit tilted Dirac cones under lattice deformation. During the tuning process, when $t_{bb} = \frac{1}{\sqrt{3}}t_{rb}$, the three bands touch at the $\Gamma$-point and form HK lattice ($\mathcal{S}(\mathcal{X}_6)$)-like  pseudospin-1 states, as shown in Fig.\ref{fig:triangular_kagome_2dband}.  At the $t_{bb} = 1$ limit, the three isolated overlapping flat bands appear again but at the top of the band structure (Fig.~\ref{fig:triangular_kagome_ribbon}). 

\begin{figure}[!htbp]
    \centering
     \subfloat[]{\includegraphics[scale=0.3,trim={1cm 7cm 1cm 7cm},clip]{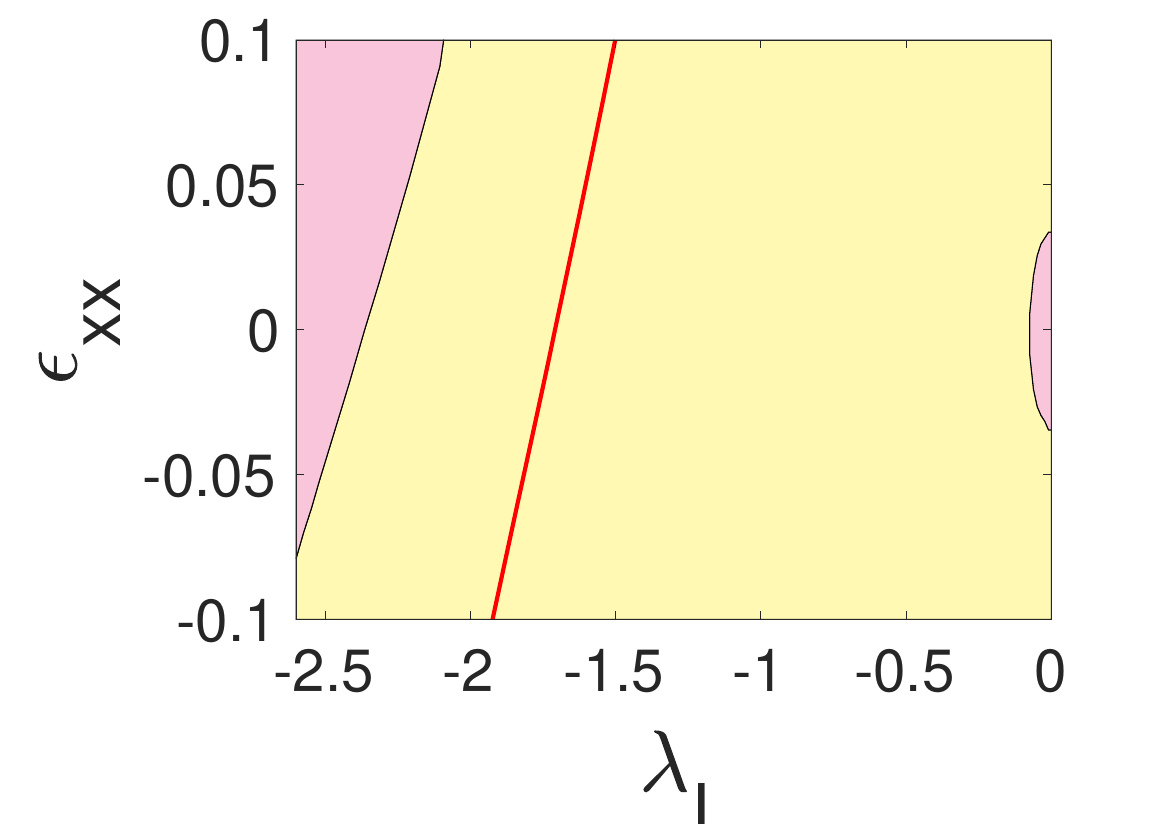}\label{fig:triangular_kag_phase_t2_0p6_band2}}
   \subfloat[]{\includegraphics[scale=0.3,trim={3.1cm 7cm 1cm 7cm},clip]{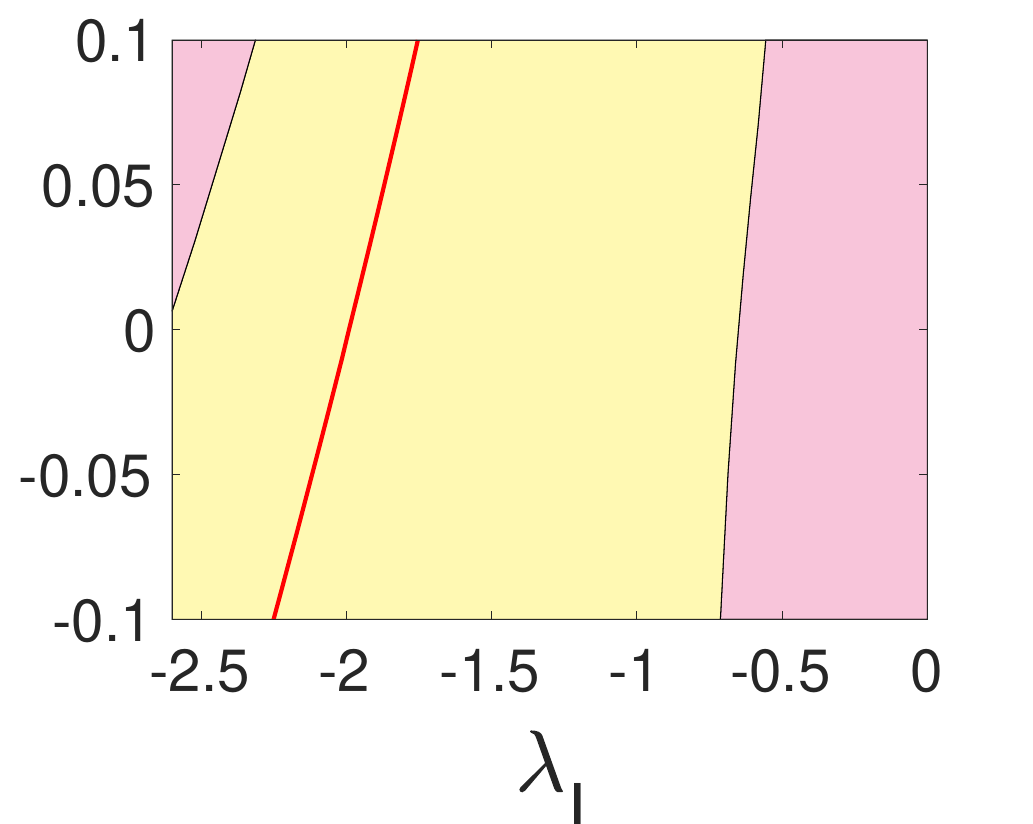}\label{fig:triangular_kag_phase_t2_0p8_band2}} 
    \subfloat[]{\includegraphics[scale=0.3,trim={3.1cm 7cm 1cm 7cm},clip]{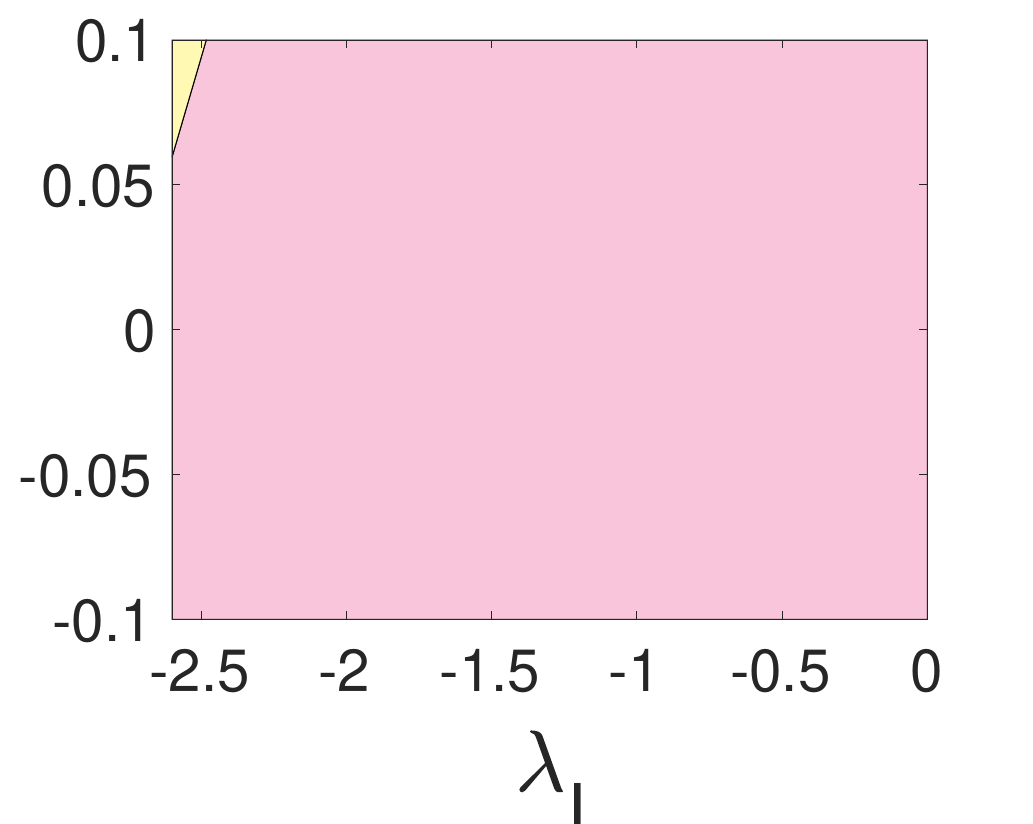}\label{fig:triangular_kag_phase_t2_1p8_band2}}  
    \caption{Topological phase diagram of triangular kagome lattice, $S(L(\mathcal{X}_6))$, at $2/9$ filling as the function of strain ($\epsilon_{xx}$) and intrinsic spin orbit coupling ($\lambda_I$) at  (a)$t_{bb}  = 0.6 t_{rb} $, (b) $t_{bb}  = 0.8 t_{rb}$,  and (c)  $t_{bb}  = 1.8 t_{rb}$. The different phases are distinguished by the colors as follows. Pale mint (\semicyan{}): semimetal, yellow  (\tiyellow{}): topological band insulator, pink (\bipink{}): band insulator and the red line denotes the DSM phase. }
    \label{fig:triangular_kag_phasediag_band2}
\end{figure}

\begin{figure}[!htbp]
    \centering
    \subfloat[]{\includegraphics[scale=0.3,trim={1cm 7cm 1cm 7cm},clip]{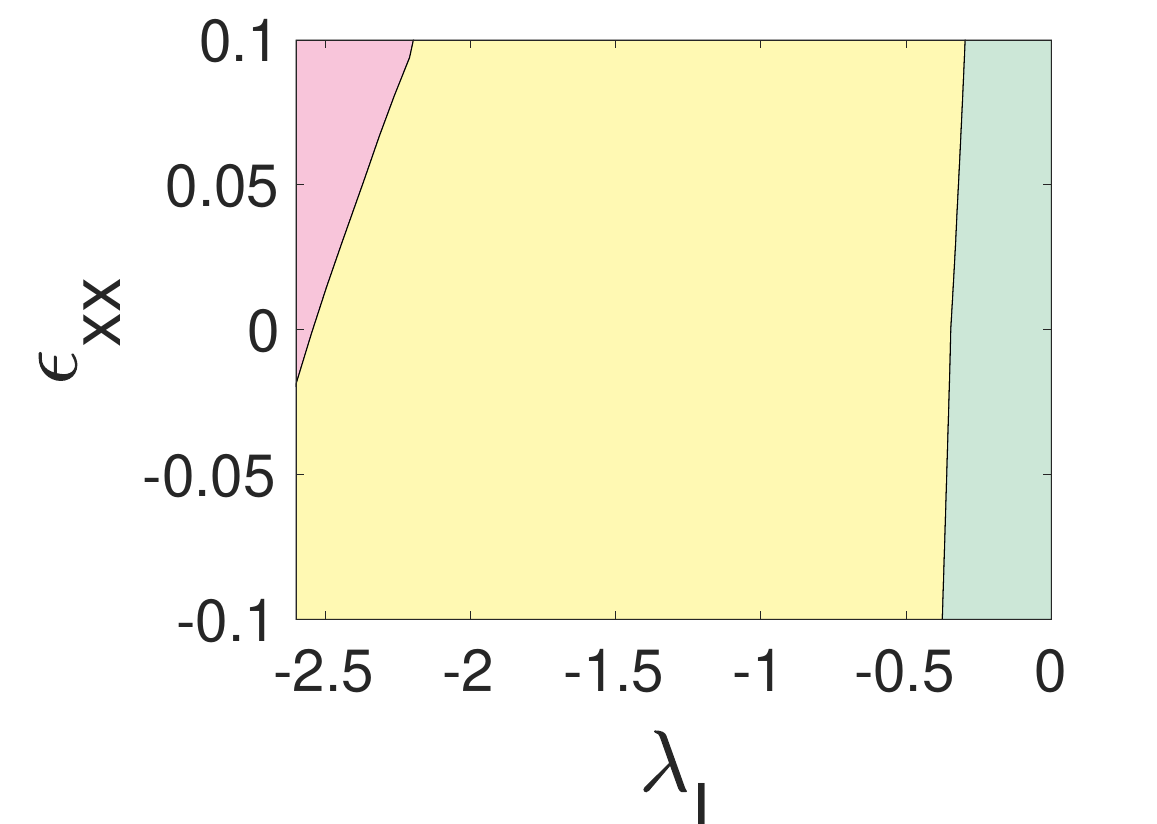}\label{fig:triangular_kag_phase_t2_0p2_band4}}
    \subfloat[]{\includegraphics[scale=0.3,trim={3.1cm 7cm 1cm 7cm},clip]{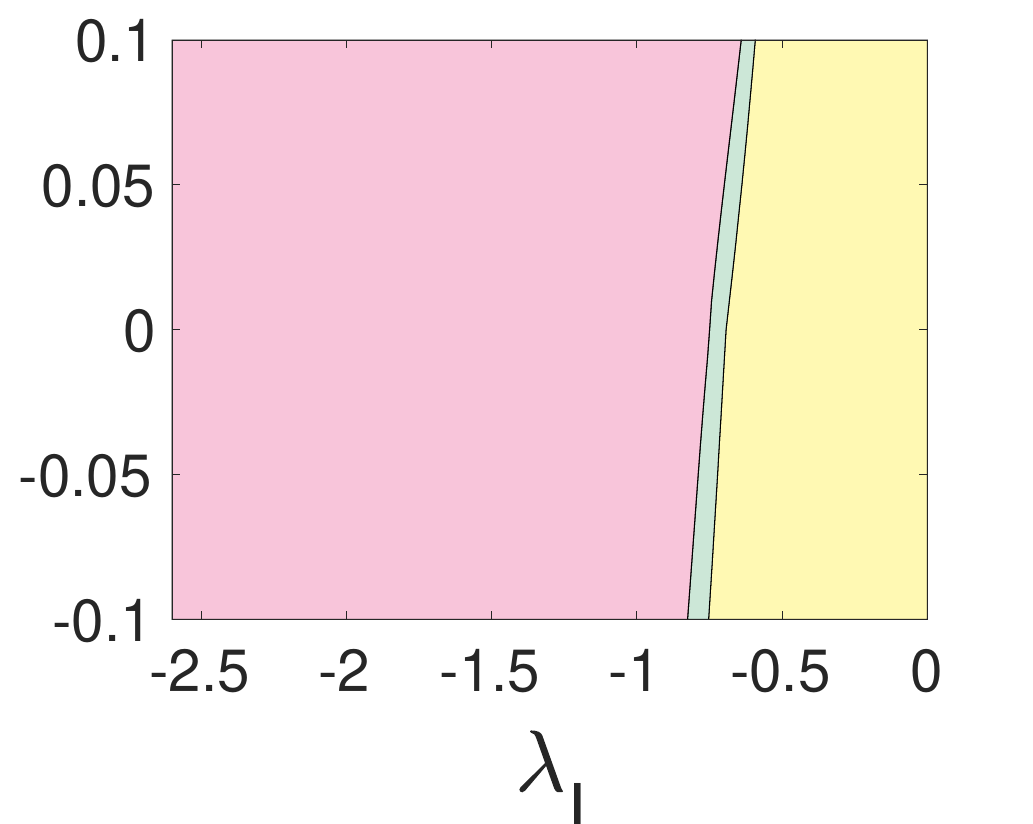}\label{fig:triangular_kag_phase_t2_0p4_band4}}
    \subfloat[]{\includegraphics[scale=0.3,trim={3.1cm 7cm 1cm 7cm},clip]{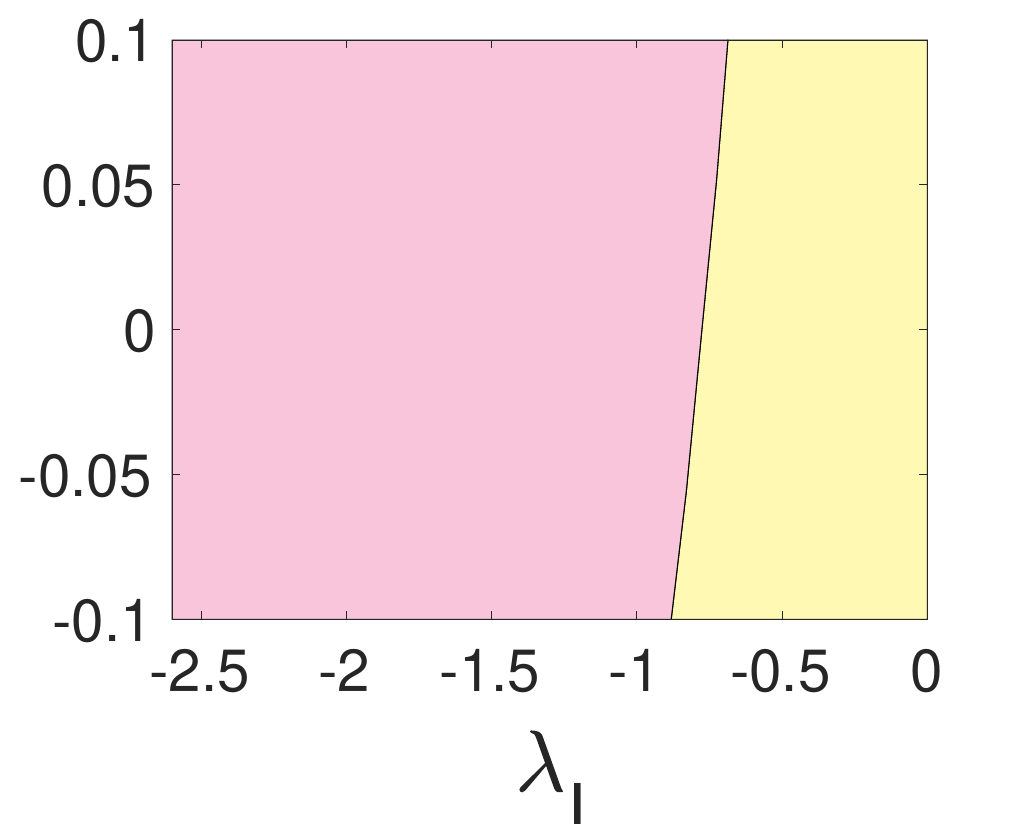}\label{fig:triangular_kag_phase_t2_0p6_band4}}
    \caption{Topological phase diagram of triangular kagome lattice, $S(L(\mathcal{X}_6))$, at $4/9$ filling as the function of strain ($\epsilon_{xx}$) and intrinsic spin orbit coupling ($\lambda_I$) at (a) $t_{bb} = 0.2 t_{rb}$ (b)$t_{bb} = 0.4 t_{rb}$, and (c) $t_{bb} = 0.6 t_{rb}$. The different phases are distinguished by the colors as follows. Yellow  (\tiyellow{}): topological band insulator and pink (\bipink{}): band insulator. }
  \label{fig:decorated_honeycomb_phasediag_band4}
\end{figure}

\begin{figure}[!htbp]
    \centering
    \subfloat[]{\includegraphics[scale=0.3,trim={1cm 7cm 1cm 7cm},clip]{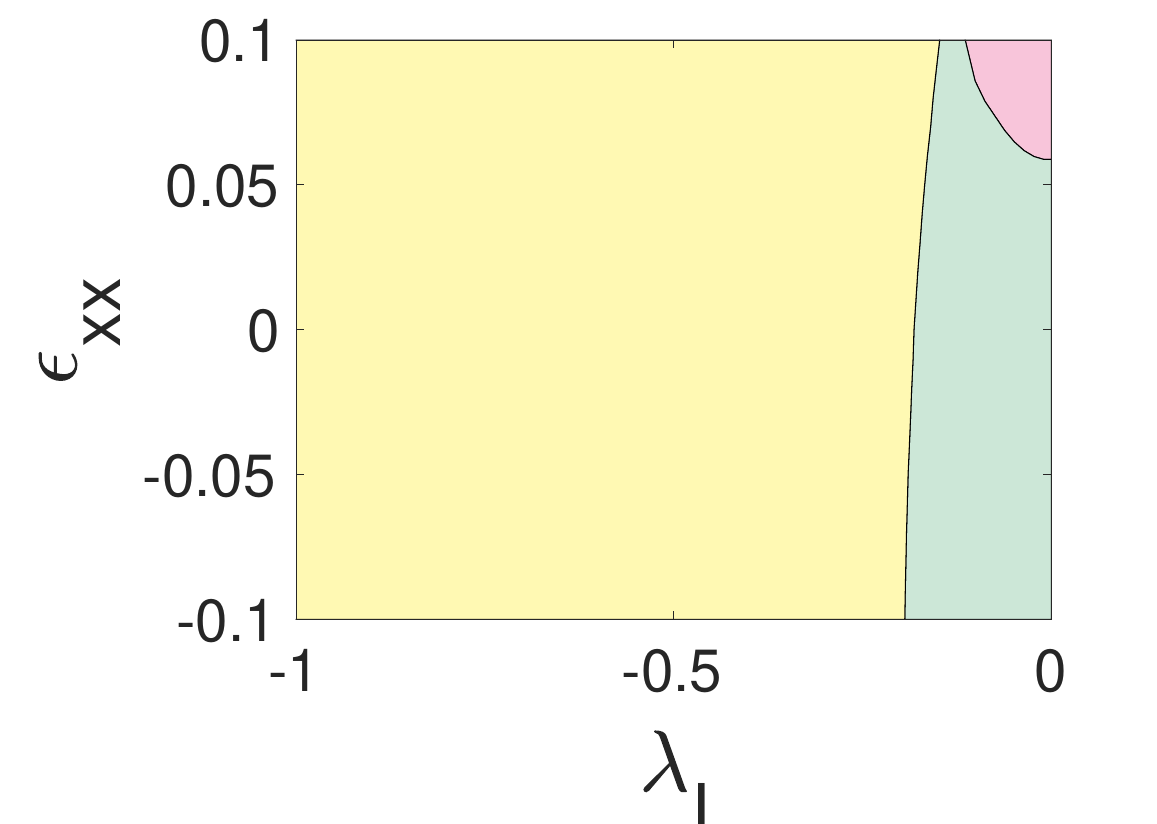}\label{fig:triangular_kag_phase_t2_0p4_band5}}
    \subfloat[]{\includegraphics[scale=0.3,trim={3.1cm 7cm 1cm 7cm},clip]{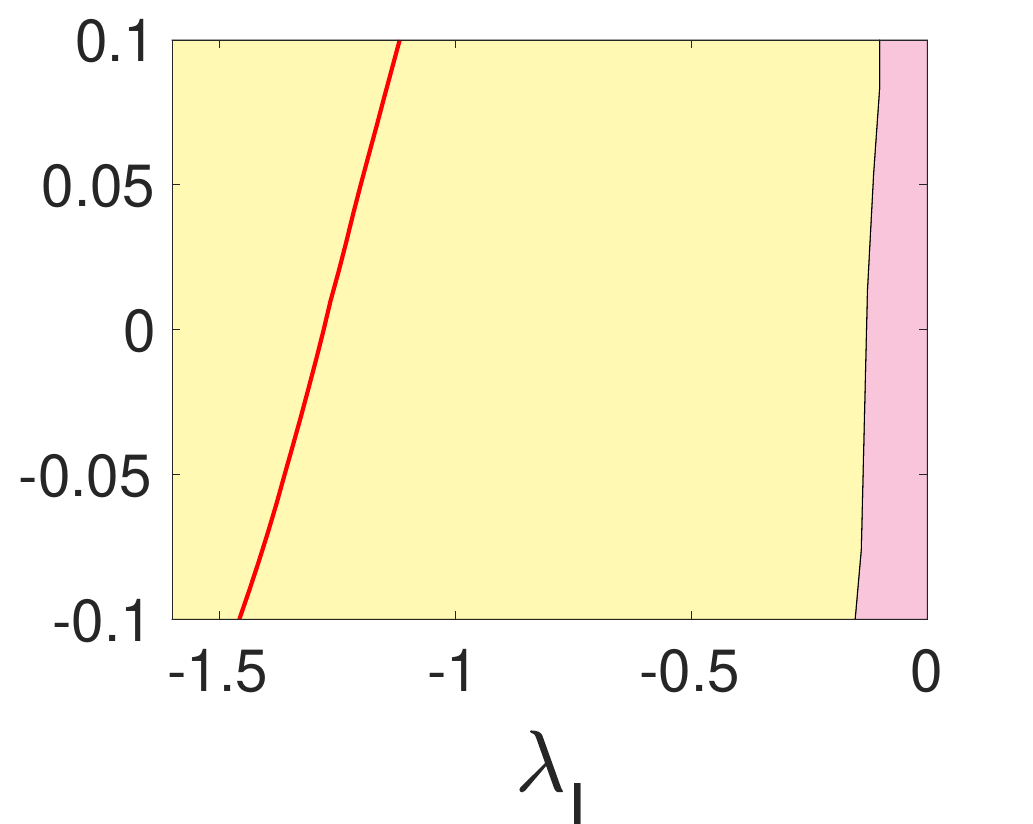}\label{fig:triangular_kag_phase_t2_1p8_band5}}
    \caption{Topological phase diagram of triangular kagome lattice, $S(L(\mathcal{X}_6))$, at $5/9$ filling as the function of strain ($\epsilon_{xx}$) and intrinsic spin orbit coupling ($\lambda_I$) at (a) $t_{bb} = 0.4 t_{rb}$ and (b) $t_{bb} = 1.8 t_{rb}$. The different phases are distinguished by the colors as follows. Pale mint (\semicyan{}): semimetal, yellow  (\tiyellow{}): topological band insulator, pink (\bipink{}): band insulator and the red line denotes the DSM phase.}
 \label{fig:decorated_honeycomb_phasediag_band5}
\end{figure}

\begin{figure}[!htbp]
    \centering
    \subfloat[]{\includegraphics[scale=0.3,trim={1cm 7cm 1cm 7cm},clip]{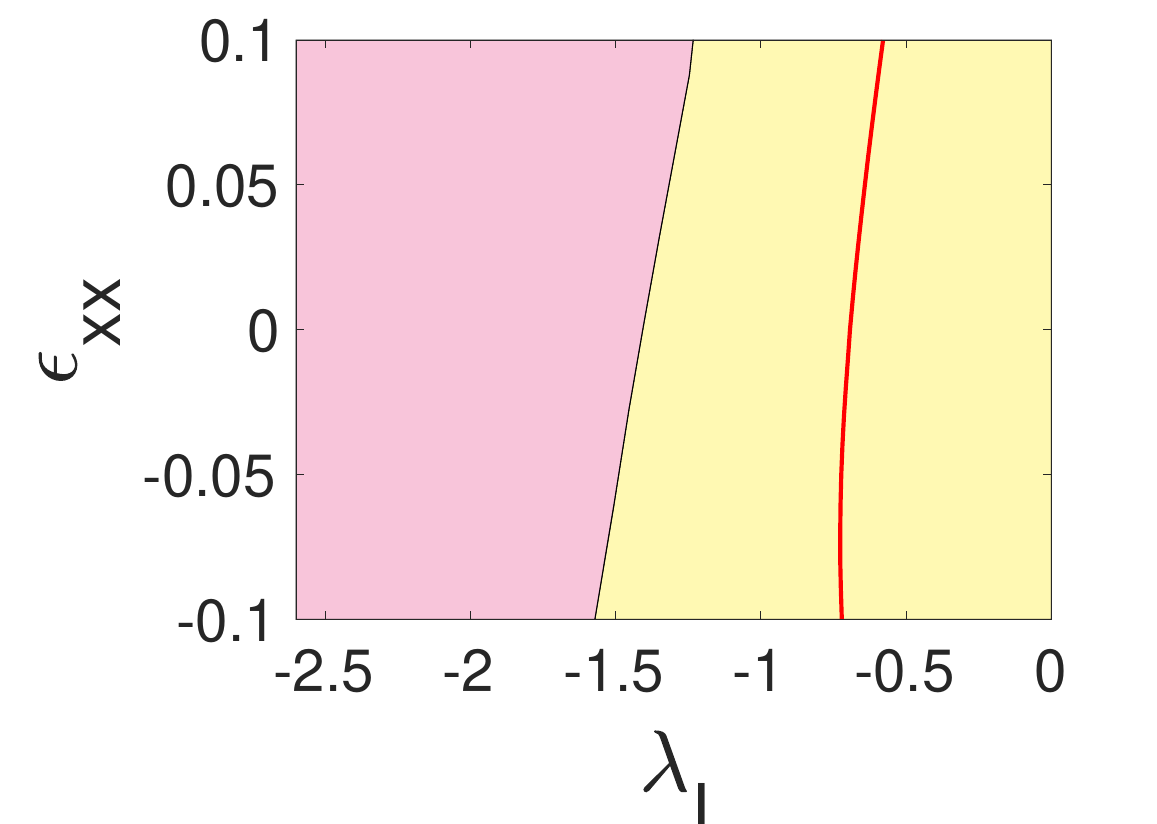}\label{fig:triangular_kag_phase_t2_0p2_band7}}
     \subfloat[]{\includegraphics[scale=0.3,trim={3.1cm 7cm 1cm 7cm},clip]{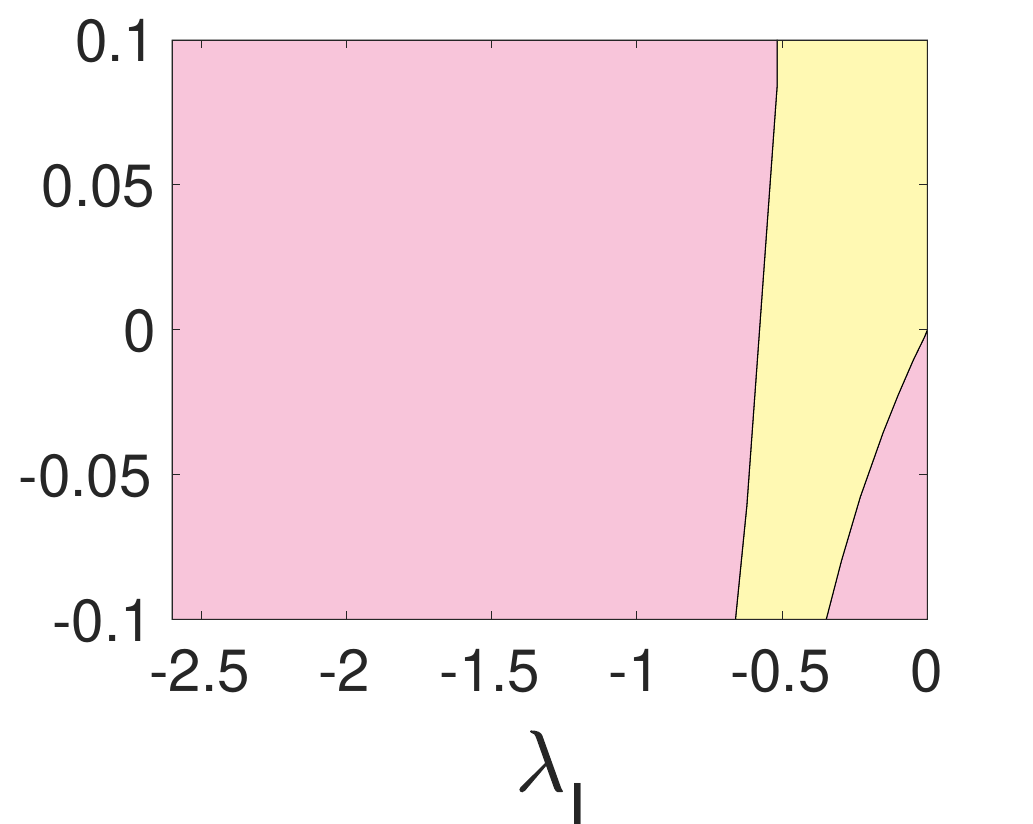}\label{fig:triangular_kag_phase_t2_1p0_band7}}
    \subfloat[]{\includegraphics[scale=0.3,trim={3.1cm 7.cm 1cm 7cm},clip]{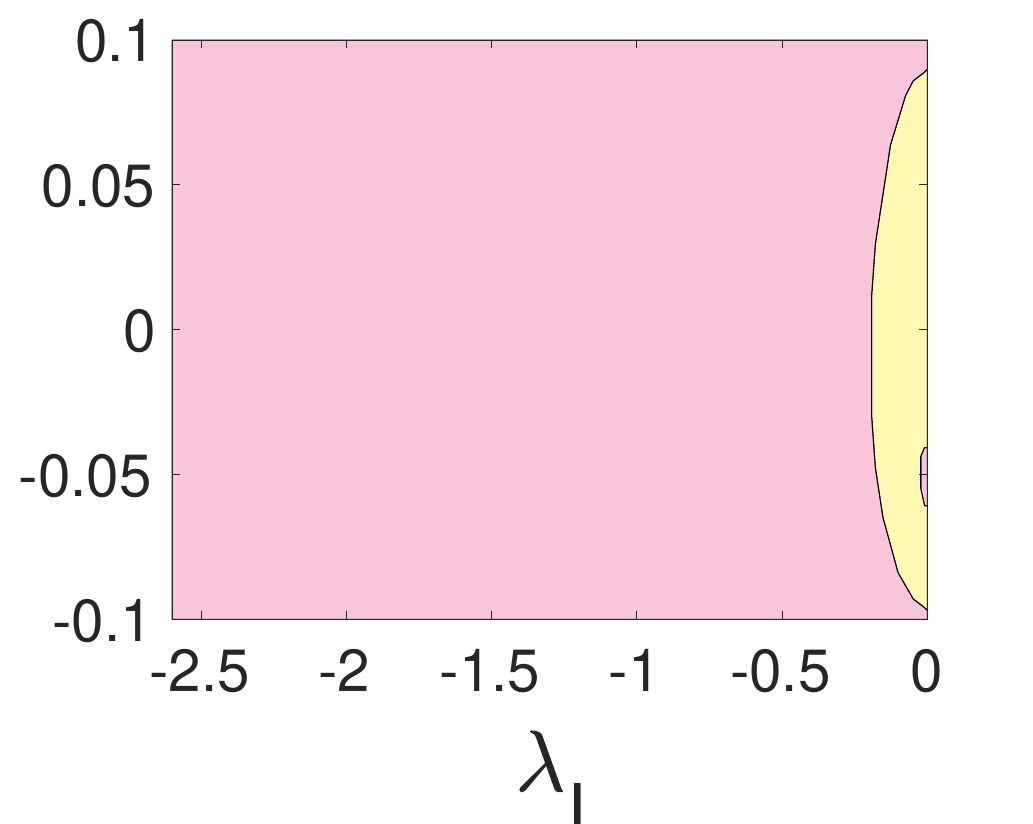}\label{fig:triangular_kag_phase_t2_2p6_band7}}
    \caption{Topological phase diagram of triangular kagome lattice, $S(L(\mathcal{X}_6))$, at $7/9$ filling as the function of strain ($\epsilon_{xx}$) and intrinsic spin orbit coupling ($\lambda_I$) at (a) $t_{bb} = 0.2t_{rb}$, (b) $t_{bb} = 1t_{rb}$ and (c) $t_{bb} = 2.6t_{rb}$ . The different phases are distinguished by the colors as follows. Yellow  (\tiyellow{}): topological band insulator, pink (\bipink{}): band insulator and the red line denotes the DSM phase.}
 \label{fig:decorated_honeycomb_phasediag_band7}
\end{figure}

The TKL exhibits a larger a larger number of bands than the kagome lattice, and its electronic phase diagram under strain subsumes all of the phases also shown by its parent lattice (discussed above). \Cref{fig:triangular_kag_phasediag_band2,fig:decorated_honeycomb_phasediag_band4,fig:decorated_honeycomb_phasediag_band5,fig:decorated_honeycomb_phasediag_band7} show the strain-SOC phase diagrams at different filling fractions for different values of $t_{bb}$. In general, the phase boundaries show a linear relationship with the phase variables. By definition, two topologically distinct phases are separated by the bandgap closing point. Remarkably, in some cases, we noticed that for $2/9$ (Fig.~\ref{fig:triangular_kag_phasediag_band2}), $5/9$ (Fig.~\ref{fig:decorated_honeycomb_phasediag_band5}) and $7/9$ (Fig.~\ref{fig:decorated_honeycomb_phasediag_band7}) filling fractions, there is line showing the appearance of the DSM phase (shown in red) lying in the topological regions. At this line, the band gap vanishes but the system does not undergo topological phase transition. There is an onset of an ordinary semimetal phase at the fourth and fifth bands at different values of $t_{bb}$, as depicted in \cref{fig:decorated_honeycomb_phasediag_band4,fig:decorated_honeycomb_phasediag_band5}, respectively.


\section{Conclusion}
\label{sec:conclusion}
In summary, we have demonstrated the rich landscape of topological phase transitions in 2D flat-band lattices, as induced by strain. We have leveraged graph-theoretic transformations on root bipartite lattices, and systematically explored how split and line graph operations can yield unique (and sometimes unexplored) 2D lattices with tunable electronic and topological properties. We have also explored the role of on-site energies and the spin-orbit coupling (SOC) parameter on these properties. Overall, our work  aims to provide comprehensive and foundational understanding that could inform experimental pursuits aimed at synthesizing 2D materials with tunable topological characteristics. 

Our calculations reveal that strategically applied strain induces various phase transitions in these lattices, including those between trivial and topological insulators, as well as semimetallic and Dirac phases. In particular, our results highlight how strain-induced distortion of lattice symmetries can result in the introduction of new Dirac points, the tilting and merging of Dirac cones, and the formation of semi-Dirac phases. These behaviors underscore the critical role of structural deformations in modulating electronic band structures in these systems. Indeed, it is easily conceivable that in experimental settings, such strains can be applied through substrates or device contacts, thus enabling  or suppressing such transitions in a controllable manner. Our study further illustrates the utility of graph-theoretic frameworks in constructing and understanding higher-generation lattices from existing ones, thus providing deterministic recipes of generating structures with desirable electronic properties. The topological robustness observed in these systems under certain conditions (specifically, with the inclusion of SOC), opens potential pathways for applications in quantum spintronics and topologically protected quantum transport. At the same time, the ability to switch between different electronic states (e.g. type-I and type-II Dirac fermions) through strain suggests new avenues for realizing exotic quasiparticles with tailored dispersion properties. The computational discovery and characterization of realistic materials featuring such properties, especially with the help of high-throughput first principles techniques \citep{jia2022metallic, miyamachione,wang2025structure} is an attractive near-future research direction. Incorporation of interaction into the models (e.g. the Hubbard $U$ term) and appropriate solutions of the resulting equations forms yet another research direction.

\begin{center}
---
\end{center}
\appendix
\section{Gell-Mann matrices}
\label{appendix_1}
The Gell-Mann matrices are:
\beqs
\Lambda_1 = \begin{pmatrix}
0 & 1 & 0\\
1 & 0 & 0 \\
0 & 0 & 0
\end{pmatrix}, \quad \Lambda_2 = \begin{pmatrix}
    0 & 0 & 1 \\
    0 & 0 & 0\\
    1 & 0 & 0
\end{pmatrix}, \quad  \Lambda_3 = \begin{pmatrix}
    0 & 0 & 0\\
    0 & 0 & 1\\
    0 & 1 & 0
\end{pmatrix}, \quad \Lambda_4 = \begin{pmatrix}
    0 & -i & 0\\
    i & 0 & 0\\
    0 & 0 & 0
\end{pmatrix}, \nonumber \\
\Lambda_5 = \begin{pmatrix}
    0 & 0 & -i \\
    0 & 0 & 0\\
    i & 0 & 0
\end{pmatrix}, \quad \Lambda_6 = \begin{pmatrix}
    0 & 0 & 0\\
    0 & 0 & -i\\
    0 & i & 0
\end{pmatrix}, \quad \Lambda_7 = \begin{pmatrix}
    1 & 0 & 0\\
    0 & -1 & 0\\
    0 & 0 & 0
\end{pmatrix}, \quad \Lambda_8 = \frac{1}{\sqrt{3}}\begin{pmatrix}
    1 & 0 & 0\\
    0 & 1 & 0\\
    0 & 0 & -2
\end{pmatrix}. 
\eeqs





\section{A few more second generation lattices}
\label{appendix_2}
Here, for the sake of completeness, we present a few more second generation lattices and their corresponding band diagrams. Specifically, we consider repeated split and line-graph operations on the parent lattices, but no combinations thereof. Thus, the lattices $L(L(\mathcal{X}_6))$, $S(S(\mathcal{X}_6))$, $L(L(\mathcal{X}_4))$ and $S(S(\mathcal{X}_4))$ are considered. Some of these (e.g. the extended Lieb lattice $S(S(\mathcal{X}_4))$), have been investigated earlier \citep{bhattacharya2019flat}, while others (e.g. $L(L(\mathcal{X}_4))$ and $L(L(\mathcal{X}_6))$) remain completely unexplored.  As shown in the figures below, many of these lattices also feature flat bands and Dirac cones. An extensive study of these systems, particularly strain-induced topological phase transitions in them, in a manner similar to the ones laid out above, is the scope of future work.

\begin{figure}[!htbp]
    \centering
    \subfloat[]{\includegraphics[scale=0.26,trim={9cm 1cm 6cm 1cm},clip]{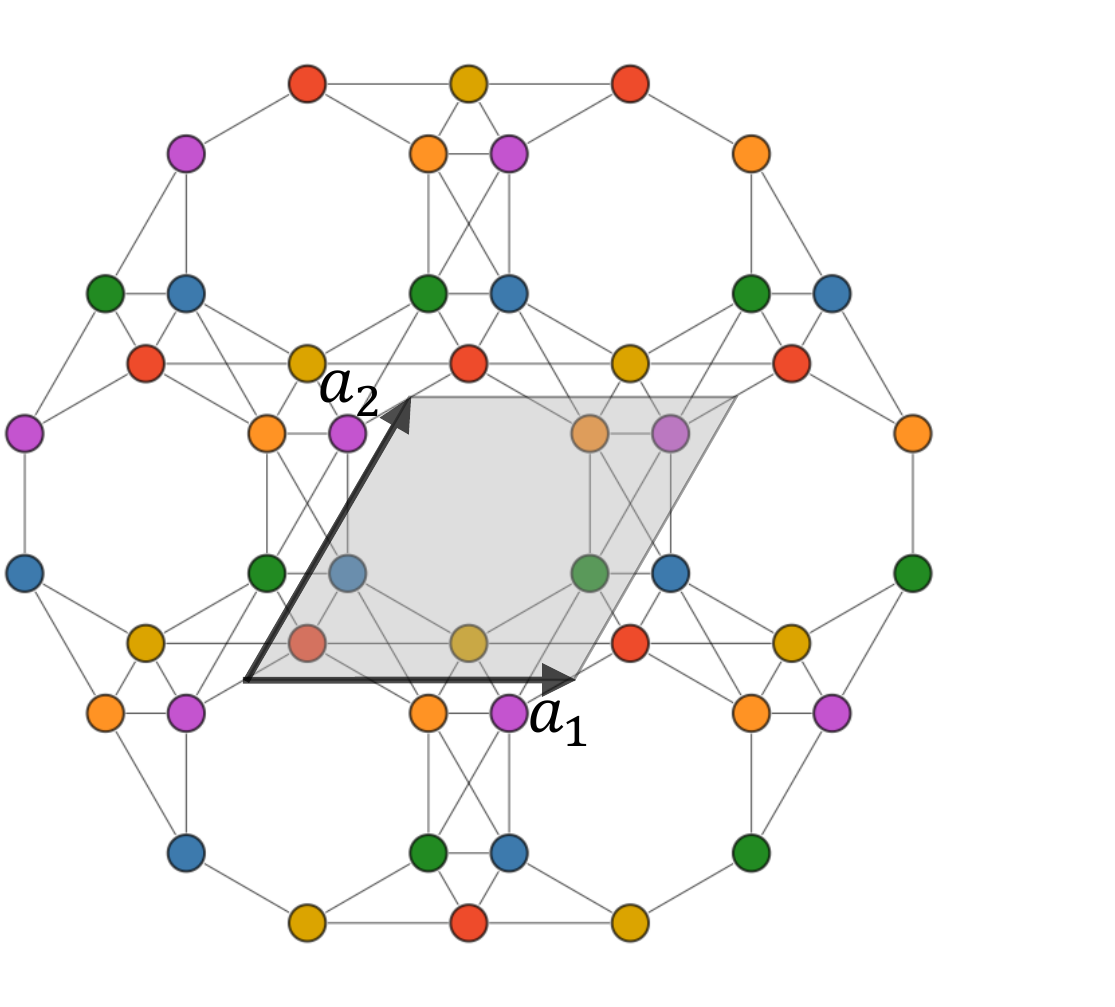}\label{fig:line_kagome}} \quad
    \subfloat[]{\includegraphics[scale=0.28,trim={1cm 6cm 2cm 7cm},clip]{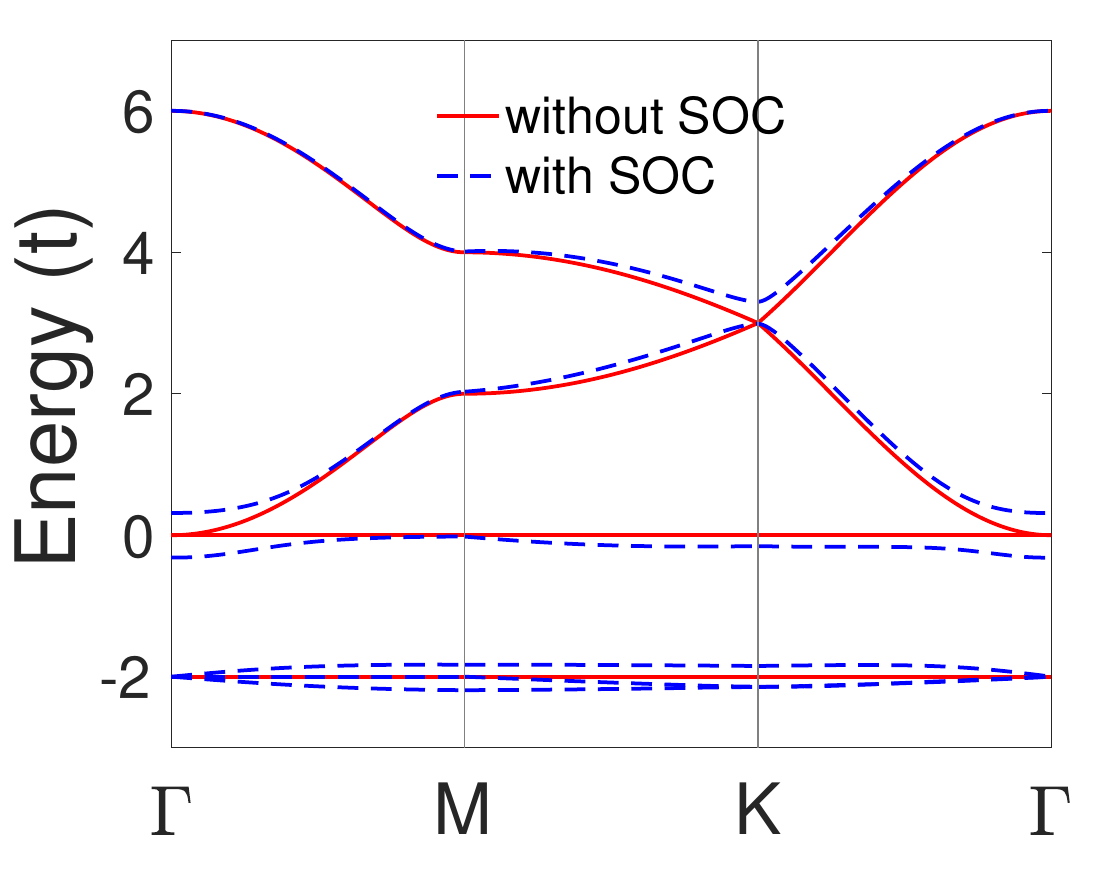}\label{fig:line_kagome_banddiagram}}   
    \caption{(a) Line graph of Kagome lattice, $L(L(\mathcal{X}_6))$. The black arrows indicate the lattice vectors $\bfa_1$ and $\bfa_2$, and the gray region is the unit cell. (b) The tight binding band diagram  of  2D lattice without (red solid line) and with (blue broken line) SOC ($\lambda_I = 0.1t$) }
\label{fig:kagome_line}
\end{figure}

\begin{figure}[!htbp]
    \centering
    \subfloat[]{\includegraphics[scale=0.26,trim={9cm 1cm 6cm 1cm},clip]{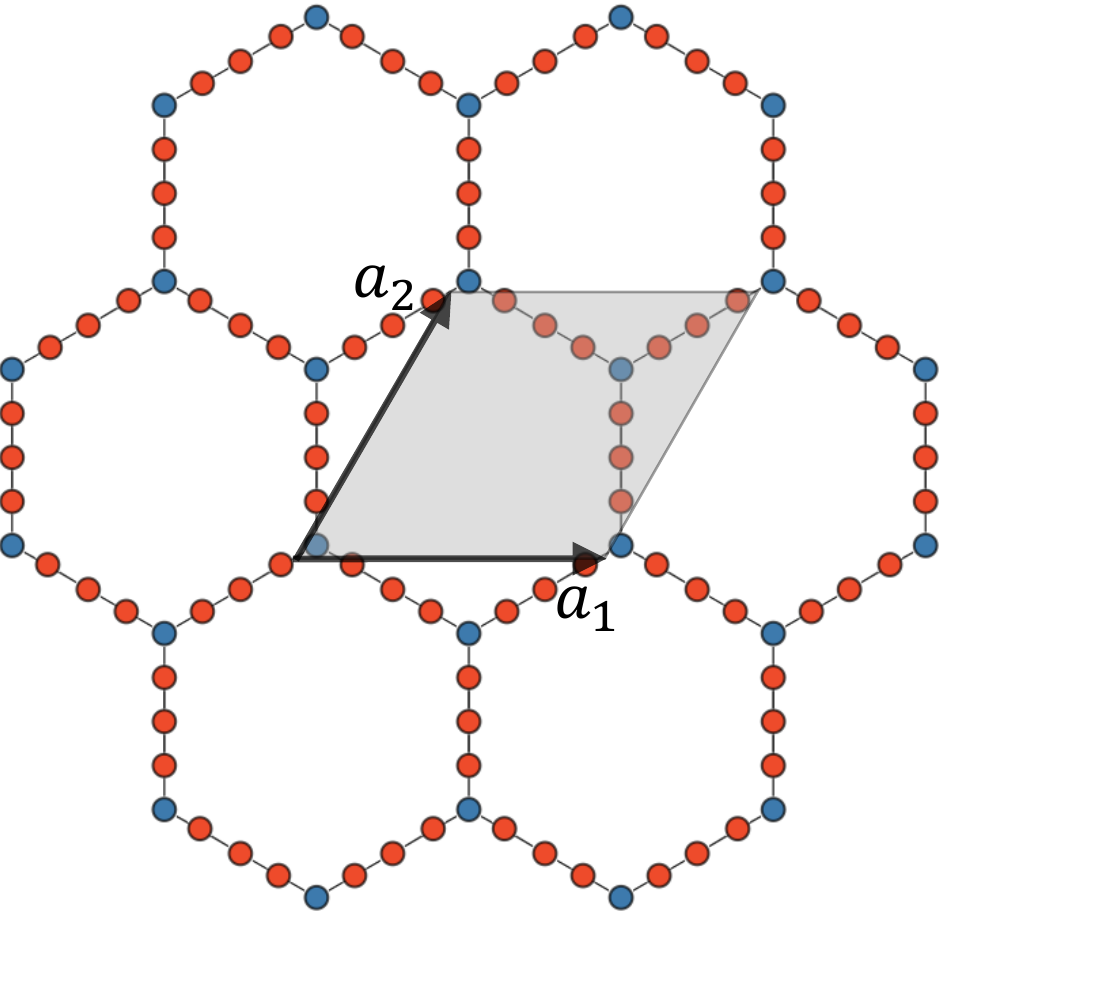}\label{fig:split_split_honeycomb}} \quad
    \subfloat[]{\includegraphics[scale=0.28,trim={1cm 6cm 2cm 7cm},clip]{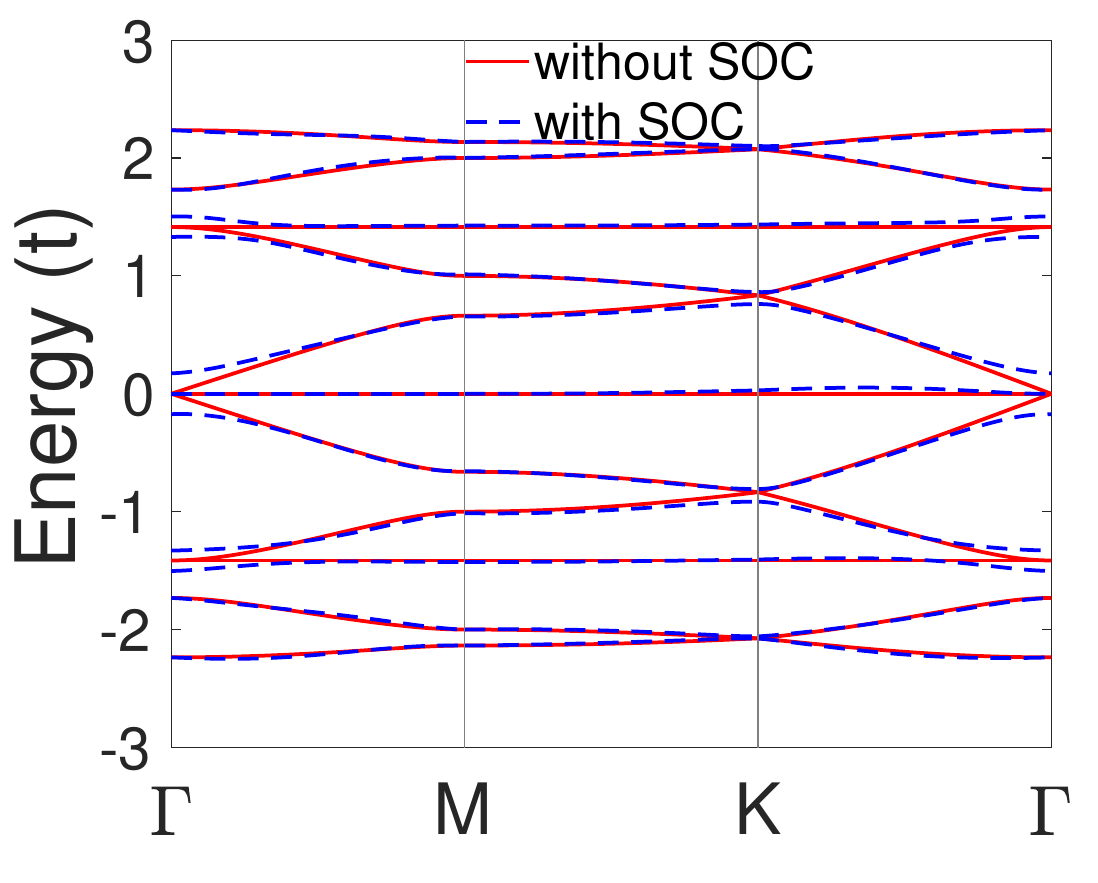}\label{fig:split_split_honeycomb_band}}
     \caption{(a) Split graph of Honeycomb-Kagome lattice, $S(S(\mathcal{X}_6))$. The black arrows indicate the lattice vectors $\bfa_1$ and $\bfa_2$, and the gray region is the unit cell. (b) The tight binding band diagram  of  2D lattice without (red solid line) and with (blue broken line) SOC ($\lambda_I = 0.1t$) }
\label{fig:split_HK}
\end{figure}
\begin{figure}[!htbp]
    \centering
    \subfloat[]{\includegraphics[scale=0.26,trim={9cm 1cm 6cm 1cm},clip]{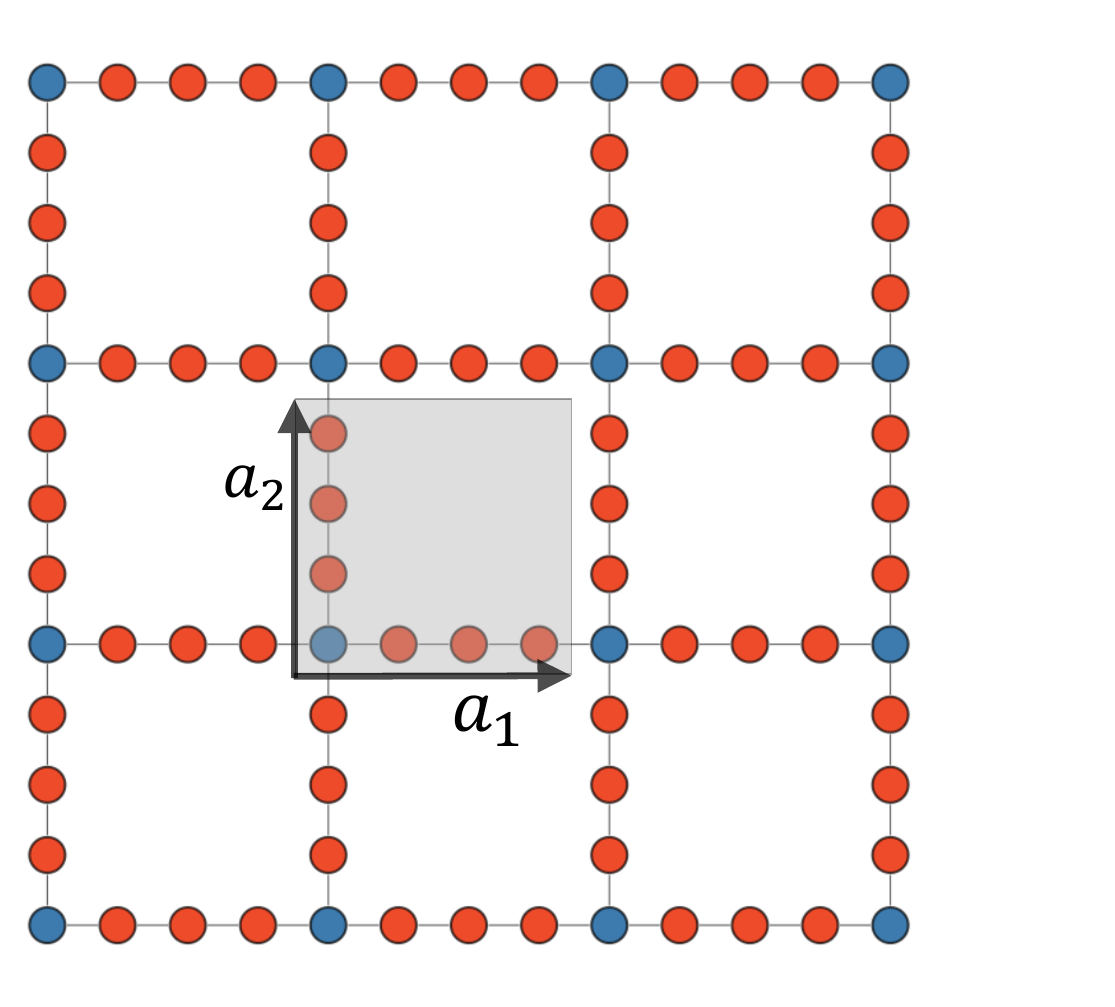}\label{fig:split_lieb}} \quad
    \subfloat[]{\includegraphics[scale=0.28,trim={1cm 6cm 2cm 7cm},clip]{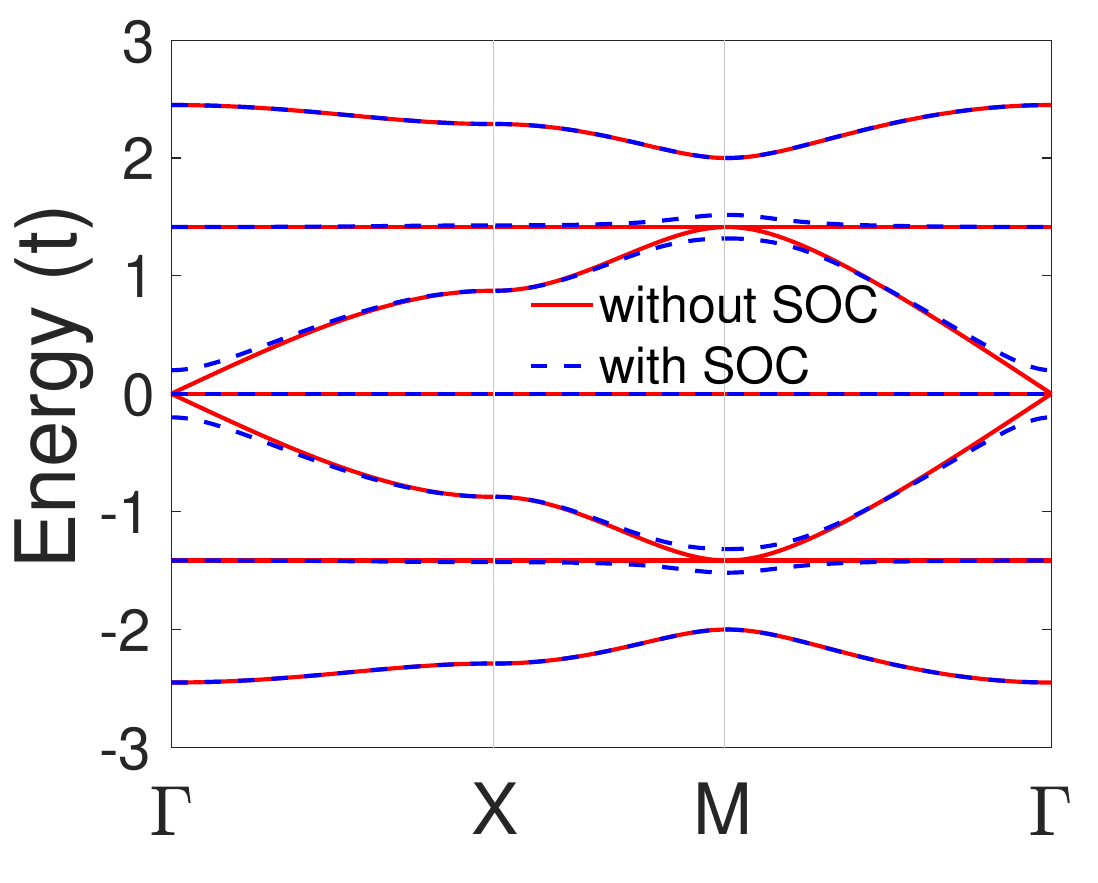}\label{fig:split_lieb_band}}
         \caption{(a) Split graph of Lieb lattice, $S(S(\mathcal{X}_4))$. The black arrows indicate the lattice vectors $\bfa_1$ and $\bfa_2$, and the gray region is the unit cell. (b) The tight binding band diagram  of  2D lattice without (red solid line) and with (blue broken line) SOC ($\lambda_I = 0.1t$) }
\label{fig:split_Lieb}
\end{figure}
\begin{figure}[!htbp]
    \centering
    \subfloat[]{\includegraphics[scale=0.26,trim={9cm 1cm 6cm 1cm},clip]{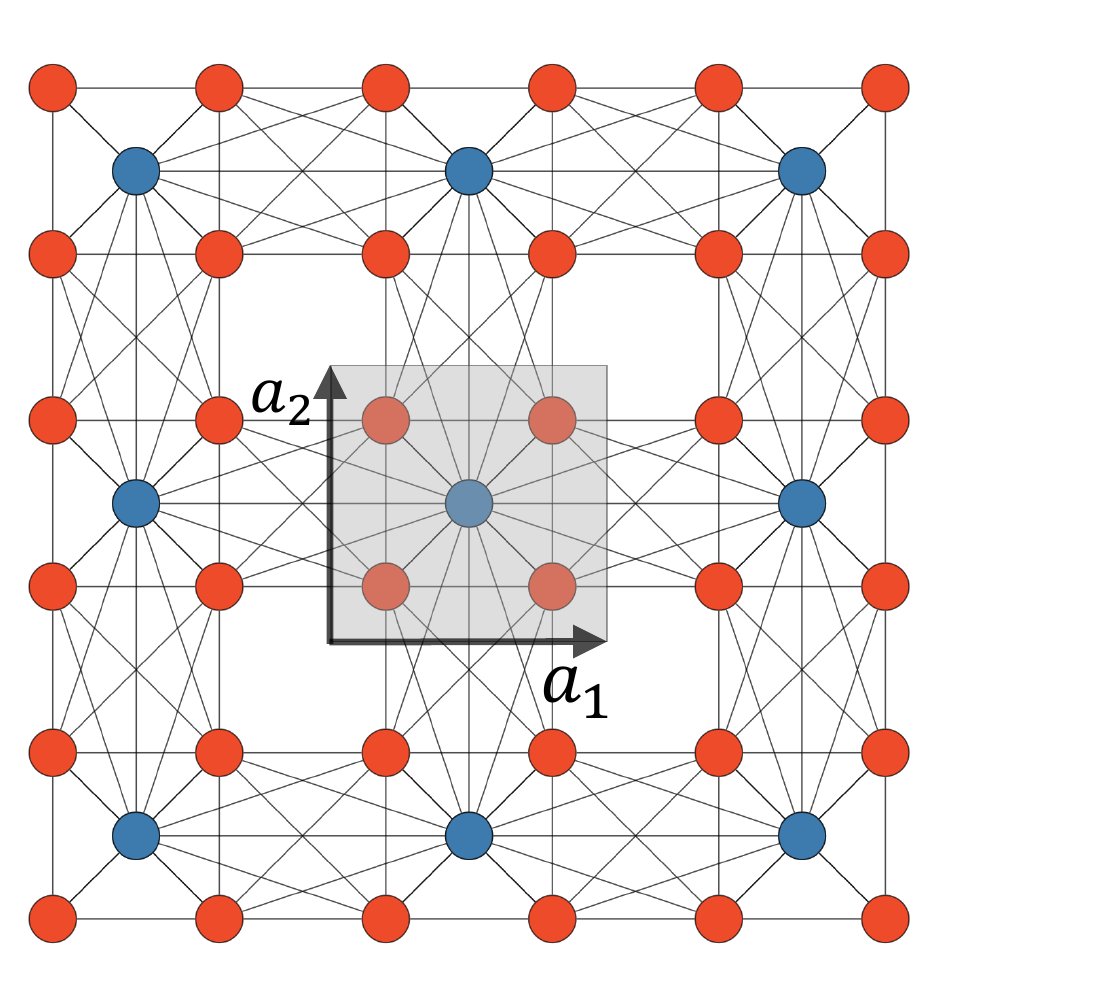}\label{fig:chkrbrd_line_lattice}} \quad
    \subfloat[]{\includegraphics[scale=0.28,trim={1cm 6cm 2cm 7cm},clip]{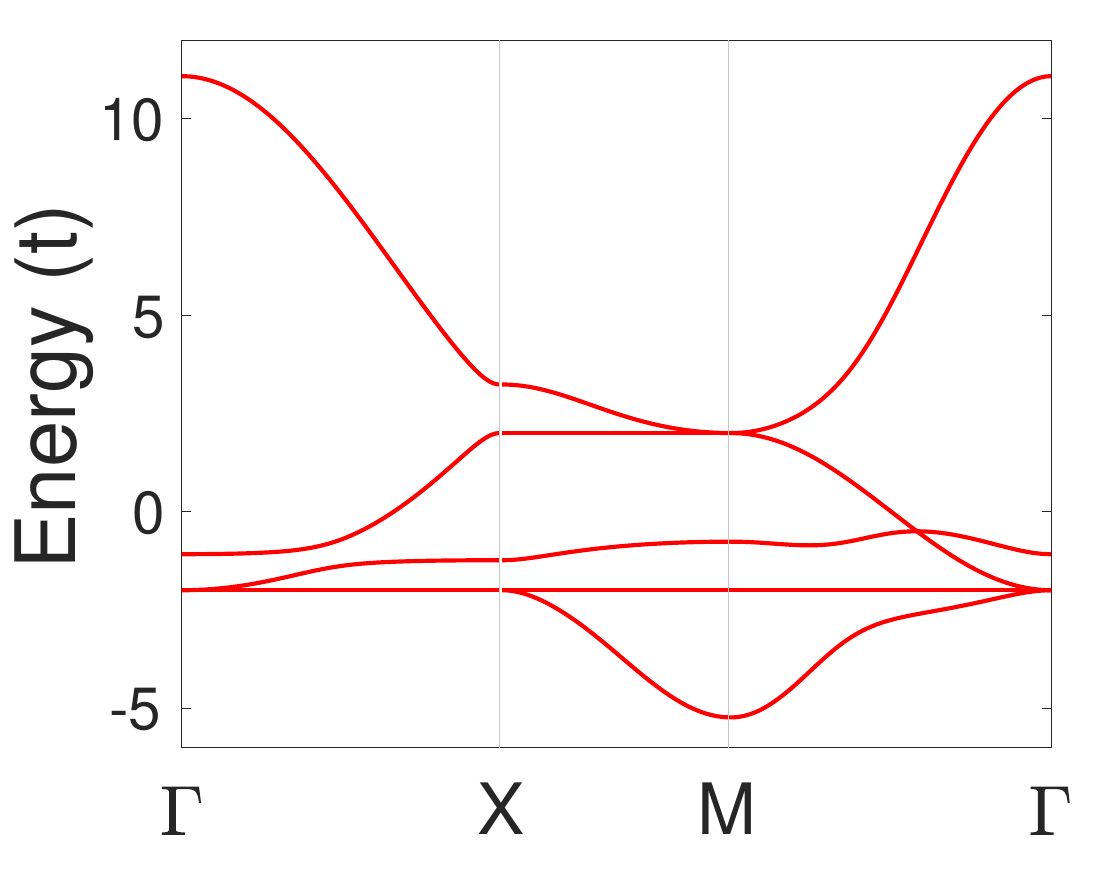}\label{fig:chkrbrd_line_band}}
         \caption{(a) Line graph of checkerboard lattice, $L(L(\mathcal{X}_4))$. The black arrows indicate the lattice vectors $\bfa_1$ and $\bfa_2$, and the gray region is the unit cell. (b) The tight binding band diagram  of  2D lattice. }
\label{fig:chkrbrd_line}
\end{figure}
\section*{Acknowledgements}
This work was primarily supported by grant DE-SC0023432 funded by the U.S. Department of Energy, Office of Science. This research used resources of the National Energy Research Scientific Computing Center, a DOE Office of Science User Facility supported by the Office of Science of the U.S. Department of Energy under Contract No.~DE-AC02-05CH11231, using NERSC awards BES-ERCAP0033206, BES-ERCAP0033206, BES-ERCAP0025205, BES-ERCAP0025168, and BES-ERCAP0028072. ASB  acknowledges startup support from the Samueli School Of Engineering at UCLA, as well as support through a Faculty Career Development Award from UCLA. SS would like to thank Richard D. James (University of Minnesota) for helpful discussions and for providing support through the Vannevar Bush Faculty Fellowship (Grant No. N00014-19-1-2623) and the Air Force Defense Research Sciences Program Grant No. FA9550-24-1-0344.
The authors acknowledge the use of the GPT-4o (OpenAI) model to polish the language and edit grammatical errors in some sections of this manuscript. The authors subsequently inspected, validated and edited the text generated by the AI model, before incorporation.
\bibliography{main}
\end{document}